\documentclass[BCOR=10mm,DIV=9,fontsize=11pt,twoside,captions=tableheading]{scrreprt}

\makeatletter
\DeclareOldFontCommand{\rm}{\normalfont\rmfamily}{\mathrm}
\DeclareOldFontCommand{\sf}{\normalfont\sffamily}{\mathsf}
\DeclareOldFontCommand{\tt}{\normalfont\ttfamily}{\mathtt}
\DeclareOldFontCommand{\bf}{\normalfont\bfseries}{\mathbf}
\DeclareOldFontCommand{\it}{\normalfont\itshape}{\mathit}
\DeclareOldFontCommand{\sl}{\normalfont\slshape}{\@nomath\sl}
\DeclareOldFontCommand{\sc}{\normalfont\scshape}{\@nomath\sc}
\makeatother

\KOMAoptions{titlepage=true,abstract=true}
\KOMAoption{toc}{listof,index,bibliography}
\KOMAoption{headsepline}{true}
\KOMAoption{numbers}{noendperiod}
\usepackage[mode=buildnew]{standalone}
\usepackage[T1]{fontenc} 
\usepackage[utf8]{inputenc}

\usepackage[ngerman,english]{babel}

\usepackage[autostyle]{csquotes}

\usepackage[letterspace=200]{microtype}

\usepackage[usenames,dvipsnames,svgnames,x11names,table]{xcolor}

\usepackage{circuitikz}

\usepackage{tikz}
\usetikzlibrary{shapes.geometric}
\usetikzlibrary{decorations.pathreplacing}
\usetikzlibrary{patterns}

\usepackage{scrhack}

\usepackage{subfig}

\usepackage{multirow}
\usepackage{varioref}

\usepackage{booktabs}

\usepackage{tabularx}

\usepackage{longtable}

\usepackage{tabu}

\usepackage[tableposition=top]{caption}

\usepackage{url}

\usepackage{listings}
\definecolor{darkgray}{gray}{0.4}
\usepackage{xspace}

\usepackage{graphicx}

\usepackage{multicol}

\usepackage{amsmath}
\usepackage{amsfonts}
\usepackage{amscd}
\usepackage{amssymb}
\usepackage{amsthm}
\usepackage{verbatim} 
\usepackage{amsbsy}
\newtheorem{theorem}{Theorem}

\usepackage{stmaryrd}

\usepackage{textcomp}

\usepackage{dirtree}

\usepackage{siunitx}
\DeclareSIUnit{\molar}{\textsc{M}}

\usepackage{colortbl}

\usepackage{esdiff}

\usepackage{bm}
\usepackage{soul}

\usepackage{mdframed}

\usepackage{geometry}

\usepackage[singlespacing]{setspace}
\usepackage[disable,colorinlistoftodos]{todonotes}

\usepackage[
  natbib=true,
  style=numeric,
  backref=true,
  backrefstyle=three,
  backend=biber,
  firstinits=true,
  maxnames=10,
  minnames=10,
  url=true,
  doi=true,
  eprint=true
]{biblatex}
\usepackage{thesis}

\usepackage{thesiscolors}

\usepackage{imakeidx}
\makeindex[intoc]

\newcommand{\pdflinkcolor}{niceblue}

\usepackage[pdftex,
  colorlinks=true,		%
  bookmarks=true,
  bookmarksnumbered=true,	%
  bookmarksopen=true,		%
  pdfstartview=FitH,	%
  linkcolor=\pdflinkcolor, %
  anchorcolor=\pdflinkcolor,
  citecolor=\pdflinkcolor, %
  urlcolor=\pdflinkcolor, %
  filecolor=\pdflinkcolor,
  hyperindex=true,		%
  plainpages=false,	%
  breaklinks=true,		%
  pdfpagelabels, %
  hypertexnames=true,
]{hyperref}
\usepackage{bookmark}

\usepackage{cleveref}
\usepackage[acronym,toc,nomain]{glossaries}
\makeglossaries
\newacronym{LSA}{LSA}{line source approximation}
\newacronym{PNP}{PNP}{Poisson-Nernst-Planck}
\newacronym{ED}{ED}{electrodiffusion}
\newacronym{LFP}{LFP}{local field potential}
\newacronym{DOFs}{DOFs}{degrees of freedom}
\newacronym{AP}{AP}{action potential}
\newacronym{HH}{HH}{Hodgkin-Huxley}
\newacronym{ODE}{ODE}{ordinary differential equation}
\newacronym{PDE}{PDE}{partial differential equation}
\newacronym{FEM}{FEM}{finite element method}
\newacronym{cG}{cG}{continuous Galerkin}
\newacronym{dG}{dG}{discontinuous Galerkin}
\newacronym{GHK}{GHK}{Goldman-Hodgkin-Katz}
\newacronym{PCM}{PCM}{parallel conductance model}
\newacronym{ILU}{ILU}{inexact LU}
\newacronym{BiCGStab}{BiCGStab}{stabilized biconjugate gradient}
\newacronym{AMG}{AMG}{algebraic multigrid}
\newacronym{GMRes}{GMRes}{generalized minimal residual}
\newacronym{TMP}{TMP}{template meta program}
\newacronym[\glslongpluralkey={degrees of freedom}]{DOF}{DOF}{degree of freedom}
\newacronym{ISTL}{ISTL}{iterative solver template library}
\newacronym{EEG}{EEG}{electroencephalography}
\newacronym{ES}{ES}{extracellular space}
\newacronym{VC}{VC}{volume conductor}
\newacronym{HPC}{HPC}{high performance computing}
\newacronym{LS}{LS}{linear solver}
\newacronym{EDL}{EDL}{electric double layer}
\newacronym{CRTP}{CRTP}{curiously recurring template pattern}
\newacronym{IO}{IO}{input/output}
\newacronym{EAP}{EAP}{extracellular action potential}

\titlehead{}
\newcommand{\mytitle}{Electrodiffusion Models of Axon and Extracellular Space Using the 
Poisson-Nernst-Planck Equations}
\title{\mytitle}
\date{}
\subtitle{\vspace{5cm}\includegraphics[width=0.9\textwidth]{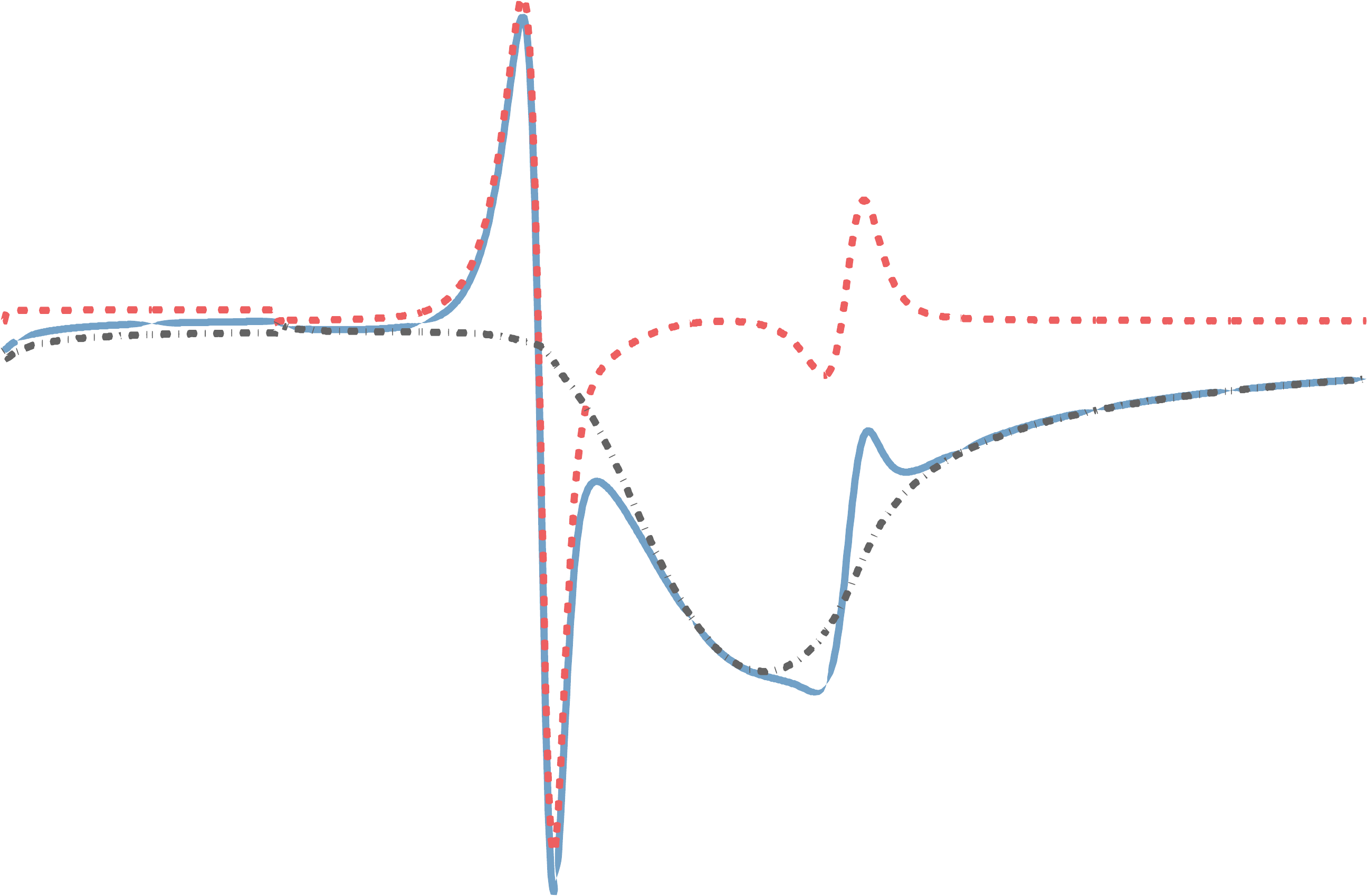}\vspace{3cm}}
\author{}
\publishers{Gutachter: Prof.~Dr.~Peter Bastian}

\newcommand{\keywords}{Axon, Extracellular Signal, Local Field Potential,
Membrane, Electrolyte, Numerical Simulation, PNP, NPP, Electrodiffusion, Debye layer,
Myelin, Node of Ranvier, Ephaptic Coupling}
\uppertitleback{\vspace{25cm}
\small Cover art by \href{http://www.jantijssen.com}{Janneman} based on simulation data from this work}
\makeatletter
\hypersetup{
  pdftitle={\mytitle},
  pdfauthor={Jurgis Pods},
  pdfsubject={\@subject},
  pdfkeywords={\keywords},
  pdfproducer={\@author~using \LaTeX}
}
\makeatother

\makeatletter
\renewcommand{\fps@figure}{htbp}
\renewcommand{\fps@table}{htbp}
\newcommand*{\rom}[1]{\expandafter\@slowromancap\romannumeral #1@}
\makeatother

\graphicspath{{./img/ax1_bj/}{./img/external/}{./img/matlab_temp/}}

\newcommand{\mycaption}[2][]{\caption[#1]{\textbf{#1}. #2}}
\begin{document}
\interfootnotelinepenalty=10000

\clubpenalty = 1000 %
\widowpenalty = 10000 %
\pagestyle{empty}

\extratitle{
\vspace{2cm}
\begin{center}
\LARGE\textsc{\textls{Inaugural--Dissertation}}\\
\Large zur\\
Erlangung der Doktorw\"urde\\
der\\
Naturwissenschaftlich-Mathematischen Gesamtfakult\"at\\
der\\
Ruprecht -- Karls -- Universit\"at\\
Heidelberg\\
\vspace{13.5cm}
\Large vorgelegt von\\
Diplom-Informatiker Jurgis Jonas Pods\\
aus Preetz\\
\vspace{0.3cm}
Tag der m\"undlichen Pr\"ufung: 4.~Juli~2014\\
\vspace{0.8cm}
\mdfsetup{%
middlelinecolor=red,
middlelinewidth=2pt,
backgroundcolor=red!10,
roundcorner=10pt}
\begin{minipage}{0.8\textwidth}
\begin{mdframed}
\centering\textsf{NOTE: This is a slightly modified version with some minor corrections with respect to the dissertation handed in.
It was last edited on \today. The original version can be found here: \url{http://www.ub.uni-heidelberg.de/archiv/17128}}
\end{mdframed}
\end{minipage}
\end{center}}
\dedication{\normalsize\raggedleft F\"ur Paul\\
\vspace{0.2cm}\textit{\guillemotleft F\"ur deine Verh\"altnisse nicht schlecht.\guillemotright}}

\newgeometry{textheight=28cm,left=2cm,right=2cm}
\maketitle
\restoregeometry
\KOMAoptions{DIV=last}

\begin{abstract}
In studies of the brain and the nervous system, extracellular signals -- as measured by local field potentials (LFPs)
or electroencephalography (EEG) -- are of capital importance, as they allow to simultaneously obtain data from multiple neurons.
The exact biophysical basis of these signals is, however, still not fully understood.
Most models for the extracellular potential today are based on volume conductor theory, which assumes that the extracellular fluid is electroneutral and that the only contributions to the electric field are given by membrane currents, which can be imposed as boundary conditions in the mathematical model.
This neglects a second, possibly important contributor to the extracellular field: the time- and position-dependent concentrations of ions in the intra- and extracellular fluids.

In this thesis, a 3D model of a single axon in extracellular fluid is presented
based on the Poisson-Nernst-Planck (PNP) equations of electrodiffusion.
This fundamental model includes not only the potential, but also the concentrations of all participating ion concentrations in a self-consistent way.
This enables us to study the propagation of an action potential (AP) along the axonal membrane based on first principles by means of numerical simulations.

By exploiting the cylinder symmetry of this geometry, the problem can be reduced to two dimensions.
The numerical solution is implemented in a flexible and efficient way, using the \Dune framework.
A suitable mesh generation strategy and a parallelization of the algorithm allow to solve the problem in reasonable time, with a high spatial and temporal resolution.
The methods and programming techniques used to deal with the numerical challenges of this multi-scale problem are presented in detail.

Special attention is paid to the Debye layer, the region with strong concentration gradients close to the membrane, which is explicitly resolved by the computational mesh.
The focus lies on the evolution of the extracellular electric potential at different membrane distances. 
Roughly, the extracellular space can be divided into three distinct regions: first, the distant farfield, which exhibits a characteristic
triphasic waveform in response to an action potential traveling along the membrane. This is consistent with previous modeling efforts and experiments.
Secondly, the Debye layer close to the membrane, which shows a completely different extracellular response in the form of an ``AP echo'', which is also observed in juxtacellular recordings.
Finally, there is the intermediate or diffusion layer located in between, which shows a gradual transition from the Debye layer potential towards the farfield potential.
Both of these potential regions show marked deviations from volume conductor models, which can be attributed to the redistribution of concentrations and associated ion fluxes.
These differences are explained by analyzing the capacitive and ionic components of the potential.

In an extension, we also include myelination into the model, which has a significant impact on the extracellular field.
Again, the numerical results are compared to volume conductor models.

Finally, a model study is carried out to assess the magnitude of ephaptic effects, i.e.~the influence of the electric field of one cell on a neighboring cell, in a somewhat artificial geometry.
While the results probably can not be interpreted quantitatively in the majority of physiological situations, the qualitative behavior shows interesting effects.
An axon can elicit an action potential in a surrounding bundle of axons, given that the distance is small enough and the resistivity of the extracellular medium is significantly increased.
Further results of this study are extremely large extracellular potentials with amplitudes up to \SI{100}{\milli\volt} and an unusual neuronal firing mode in which the cell is not depolarized by an increase in the intracellular potential, but by a decrease in the extracellular potential.
Some literature references are given that show that these observations are consistent with previous studies.
\end{abstract}

\selectlanguage{ngerman}
\begin{abstract}
In Studien des Gehirns und des Nervensystems sind extrazelluläre Signale -- gemessen in der Form von Local Field Potentials (LFPs) oder als Elektroenzephalografie (EEG) -- von großer Bedeutung, da sie die simultane Erhebung von Daten mehrerer Neuronen erlauben.
Die genaue biophysikalische Grundlage dieser Signale ist jedoch noch immer nicht vollständig verstanden.
Heutzutage basieren die meisten Modelle für das extrazelluläre Potential auf dem Konzept von ``Volume Conductors'' (Volumenleitern), bei denen angenommen wird, dass die extrazelluläre Flüssigkeit elektroneutral ist und dass die einzigen Beiträge zum extrazellulären Feld von Membranströmen stammen, die als Randbedingungen ins mathematische Modell Einzug finden.
Dies vernachlässigt einen zweiten, möglicherweise bedeutenden Beitrag zum extrazellulären Feld: Die zeit- und ortsabhängigen Konzentrationen der in den intra- und extrazellulären Fluiden befindlichen Ionen.

In dieser Arbeit wird ein 3D-Modell eines Axons in extrallulärer Flüssigkeit auf Basis der Poisson-Nernst-Planck Gleichungen der Elektrodiffusion präsentiert.
Dieses fundamentale Modell beinhaltet das Potential und zudem die Konzentrationen aller beteiligten Ionen in einer selbstkonsistenten Art und Weise.
Dies ermöglicht es, die Ausbreitung eines Aktionspotentials (AP) entlang der axonalen Membran auf Basis grundlegender physikalischer Gesetze mit den Mitteln der numerischen Simulation zu studieren.

Durch das Ausnutzen der Zylindersymmetrie der vorliegenden Geometrie kann das Problem auf zwei Dimensionen reduziert werden.
Die numerische Lösung ist in einer flexiblen und effizienten Weise unter Verwendung des \Dune-Frameworks implementiert.
Eine geeignete Gittergenerierungsstrategie und eine Parallelisierung des numerischen Algorithmus erlauben es, das Problem in angemessener Zeit und mit einer hohen räumlichen und zeitlichen Auflösung zu lösen.
Die verwendeten Methoden und Programmiertechniken zur Überwindung der numerischen Herausforderungen dieses Multiskalen-Problems sind detailliert dargestellt.

Besondere Beachtung gilt dabei dem Debye-Layer, der Region nahe der Membran, innerhalb der die Konzentrationen starke Gradienten aufweisen. Diese Region ist im Rechengitter explizit aufgelöst.
Das Hauptaugenmerk liegt auf der Entwicklung des extrazellulären elektrischen Potentials bei unterschiedlichen Abständen von der Membran.
Der Extrazellulärraum kann grob in drei verschiedene Bereiche aufgeteilt werden: Erstens das Fernfeld, in welchem die Wellenform einen charakteristischen dreiphasigen Verlauf aufweist. Dies ist konsistent mit vorhergehenden Modellierungsversuchen.
Zweitens der Debye-Layer nahe der Membran, welcher eine gänzlich andere extrazelluläre Antwort in Form eines ``AP-Echos'' aufweist, welches auch
in juxtazellulären Messungen beobachtet wird.
Zwischen diesen beiden Regionen befindet sich die intermediäre oder Diffusions-Schicht, welche einen graduellen Übergang vom Potential des Debye-Layers  zu dem des Fernfeldes zeigt.
Diese beiden letzten Regionen zeigen deutliche Abweichungen von Volume Conductor-Modellen, die auf Konzentrationsumverteilungen und die damit einhergehenden Ionenflüsse zurückgeführt werden können.
Diese Unterschiede werden durch Analyse der einzelnen, kapazitiven und ionischen Beiträge zum Potential erklärt.

In einer Erweiterung wird auch eine Myelinisierung und deren signifikanter Einfluss auf das extrazelluläre Feld im Modell berücksichtigt.
Die Resultate werden wiederum mit Volume-Conductor-Modellen verglichen.

Schließlich wird eine weitere Modellstudie durchgeführt, die der Beurteilung von ephaptischen Effekten dient, also der Beeinflussbarkeit einer Zelle durch das elektrische Feld einer benachbarten Zelle, was unter Benutzung einer recht artifiziellen Geometrie simuliert wird.
Obwohl die Ergebnisse für die Mehrzahl der physiologischen Situationen wahrscheinlich nicht quantitativ interpretiert werden können, zeigen sie dennoch interessante qualitative Effekte.
Ein Axon kann ein Aktionspotential in einem umgebenden Axonbündel auslösen, solange die Distanz zwischen beiden klein genug und die Resistivität des Extrazellulärraums stark erhöht ist.
Weitere Ergebnisse dieser Studie sind extrem große extrazelluläre Potentiale mit Amplituden bis zu \SI{100}{\milli\volt} und ein ungewöhnlicher Feuermodus, bei dem eine Zelle nicht durch die Zunahme des intrazellulären Potentials, sondern durch die Abnahme des extrazellulären Potentials depolarisiert wird.
Dazu werden einige Referenzen aus der Literatur angegeben, die zeigen, dass diese Beobachtungen konsistent mit vorausgehenden Studien sind.
\end{abstract}
\selectlanguage{english}

\pagestyle{plain}
\setcounter{page}{1}
\pagenumbering{Roman}

\pdfbookmark{\contentsname}{toc}
\tableofcontents
\listoffigures
\listoftables

\setchapterpreamble[u]{%
\dictum[Albert Einstein]{If we knew what it was we were doing, it would not be called research, would it?}\bigskip}
\chapter{Introduction}
\setcounter{page}{1}
\pagenumbering{arabic}
\pagestyle{headings}

\section{Motivation}
When opening \url{google.com} and typing in the word ``computational'', the first two occurrences the auto-complete feature gives you are ``computational fluid dynamics'' and ``computational biology'' (as of December 19, 2013).
While this might in no way be representative, it hints at the popularity the application of computational methods in biology has gained.
Recently, the field of \emph{computational neuroscience} has been established, which strives to use methods from computational sciences for the understanding of the brain and the nervous system. 

Why is this?
Biology is a particular example of a field that has always been dominated by experimental methods.
This is not astonishing, since it deals with the study of living organisms, which are complex and are hard to describe in a theoretical framework.
Observations have therefore been the primary source of knowledge until today.
However, for certain biological disciplines like systems biology or molecular biology, mathematical models have been quite successful in explaining and predicting the behavior of the underlying system, thereby complementing experiments as a second pillar.

When it comes to complicated systems, the corresponding mathematical models can not be evaluated by hand, either because the model -- mostly in the form of differential equations -- is too complex, or simply because there is no known analytical solution.
For these cases, the computer has proven to be a powerful tool for solving even the most difficult problems.

Computers can be used to carry out complicated calculations within a fraction of a second, which in former times would have taken great mathematicians years.
Consequently, the computer has become the single most important tool in the majority of mathematical divisions. 
Computer simulations based on the rigorous mathematical foundation of numerical analysis have established as a third pillar in addition to theory and experiment.

In biology, the importance of numerical simulations is even reflected in the terminology, where the phrases  \emph{in vivo} (in the living organism) and \emph{in vitro} (in the test tube) for the description of an experiment have been extended by \emph{in silico} (in the computer), showing that simulations are regarded as another form of experiment.

Neuroscience, as a special case, imposes additional difficulties on experiments, as the nervous system and particularly the human brain are especially complex systems.
Neurons, the main building blocks, show complicated morphologies, extending and intertwining and eventually forming connections between each other -- the synapses.
It is estimated that the average human brain consists of about 85 billion neurons, about 20 billion of which are located in the neocortex, the outer layers of the brain. 
On average, each of these neurons forms 7000 connections to other neurons, resulting in the impressive number of 1.5 trillion synapses for the neocortex alone, in a volume of about \SI{1.5}{\litre} \cite{drachman2005we}.
This makes the brain a very inaccessible organ rendering experimental investigations extremely difficult. 

It is also notable that the availability of intact human brain tissue naturally is very restricted, so most of the experimental studies have been carried out in other mammals such as mice and rats, whose brains show similarities to the human one.

Considering all this, it is clear why \emph{in silico} investigations in the form of computer simulations have become increasingly popular.
But there is more to it: the brain computes, as does the computer -- although in a very different way.
Both share the computation on an abstract level, i.e.~transforming a certain set of inputs into a set of outputs.
While the circuit between input and output may look completely different for a set of neurons and a desktop computer, both have the fundamental property of implementing a transfer function from input to output data, with an underlying \emph{encoding} of the data.
While this might seem trivial on first sight, it has shown to be the hardest question of all: how is information encoded in the brain?

For a computer, we know the numeral system is binary, we know how to map sequences of bits to data types, how to (approximately) represent real numbers as a bit string, and how to tell the processor to direct data through transistor gates to implement a certain function.
This is everything but trivial, but the process is completely transparent, as human beings invented the concepts and built them up from scratch, documenting structure and function down to the lowest level.
The brain, in contrast, was built by nature, so its structure and function have to be reverse-engineered.

In fact, experiments can tell us something about the transfer function between input and output, but, to stay in the picture of a digital computer, telling that an input bit sequence of \lstinline!1010001010! is being transformed to a sequence of \lstinline!100101! does not provide us with any knowledge, as long as we do not know what those sequences \emph{mean}.
The application of information theory and data analysis on experimental data, although still in its infancy, has already proven extremely useful, and it can be expected that these methods will strongly shape the future of neuroscience.

At present, the field of neuroscience is booming, reflected in billion-dollar initiatives like the European Human Brain Project \cite{markram2012human} or its American counterpart BRAIN \cite{insel2013nih}.
Still, the code of the brain has not been deciphered until now.
Joint efforts involving experimental and theoretical work will be required to systematically uncover structure and function of the brain, starting at the elementary unit, the single neuron.

In this thesis we will focus on the \emph{axon}, the part of the neuron responsible for generating the fundamental data unit of the brain, the \gls{AP}.
We can think of it as a bit for now, but -- as we will see later -- it is not exactly binary, as it can be shaped to a large degree by several physiological mechanisms.
In this work we are, however, not primarily interested in the action potential itself, but rather in the electric field it generates extracellularly.
The extracellular potential is of special interest, as many experiments today rely on extracellular recordings which allow to capture signals called \glspl{LFP} of a moderately large amount of nearby cells without the need to place the electrode into the soma of a neuron, as it is done in intracellular recordings.
Extracellular recordings are very useful for the investigation of the interplay of a network of neurons, particularly ``brain waves'' emerging from synchronized firing of a large number of neurons.
These network oscillations can be observed \emph{in vivo} as well as \emph{in vitro}, but not only by (invasive) electrode measurements, but also by noninvasive \gls{EEG}.
They are most commonly classified according to their frequency, e.g.~theta (\SIrange{4}{7}{\hertz}), alpha (\SIrange{8}{13}{\hertz}), beta (\SIrange{12}{30}{\hertz}), and gamma (\SIrange{25}{100}{\hertz}).

Since these rhythms are used very frequently to explain network behavior of the brain -- and, as such, are also involved in deciphering the code of the brain -- it is crucial to understand the origin of these signals.
This thesis strives to elucidate how extracellular signals are generated based on first principles by setting up a detailed model involving the movement of ions, the elementary charge carriers responsible for any electric activity in living organisms.

\section{Challenges for Scientific Computing in Computational Neuroscience}
Above, we introduced the aspiring field of computational neuroscience.
The name implies the application of computational methods to the field of neuroscience.
But it does not pinpoint the actual character of the methods involved.
In fact, many projects in neuroscience are being relabeled and equipped with the trendy tag ``computational'', thus acknowledging that the used software for carrying out statistical evaluations indeed involves computational methods.
However, the use of truly new computational methods in the sense of scientific computing involves conceptual mathematic modeling and computer simulations, which -- depending on the complexity of these models -- requires special numerical methods.
It is clear that the numerical simulation of complex biological models is not covered by the domain of biology anymore, since it requires deeper knowledge in e.g.~mathematics, physics, and computer science.

The number of publications that use computer simulations as a complement to experiments is steeply increasing.
Still, many of these models are simplistic and based on the availability of simulation tools such as NEURON \cite{hines1997neuron}.
Sometimes it seems the models were merely fitted to reproduce the experimental data, and not for their ability to explore situations which are not feasible in experimental setups, or to make predictions which can be validated or falsified in experiments.
Likely, this is caused by the lack of expertise in scientific computing, therefore limiting the complexity of models and the accuracy of simulations.

There are surprisingly few projects that actually employ \gls{HPC}, the discipline that deals with parallelized simulations running on supercomputers.
The underlying mathematical models often have systems with millions or even billions of \glspl{DOF}, such that they are not solvable sequentially in reasonable time.
This is the boundary that many neuroscientific investigations do not dare to cross, as the parallelization of numerical codes is cumbersome and requires special knowledge in both numerics and programming.
One of the few \gls{HPC} projects in neuroscience is the Blue Brain Project \cite{markram2006blue}, which is now part of the EU initiative Human Brain Project.
IBM sponsored one of their Blue Gene supercomputers for the purpose of building accurate models of the human brain based on a parallelized version of the NEURON simulator.
This project can therefore be seen as a lighthouse project which dares to approach the challenge of tackling simulations with over a hundred million connected neurons. 
The list of comparable projects using current methods from scientific computing is small. 

The low visibility of scientific computing in neuroscience -- as opposed to many disciplines in physics, chemistry or engineering, like molecular dynamics, groundwater flow or aerodynamics -- should be seen as an encouragement.
It is not that methods of scientific computing are not needed in biology and neuroscience, the opposite is the case: the nervous system is elaborate and delicate, with many levels of abstraction from the macroscopic level of brain regions down to the subcellular molecular level.
On each of these levels, models of different complexity may be employed.
Time-dependent nonlinear \glspl{PDE} in three dimensions usually require parallelization in order to achieve reasonable accuracy and resolution of the given domain, as the number of \glspl{DOF} quickly reaches the order of one million or more, where a sequential solution would take too long.
This involves techniques such as domain decomposition, suitable parallel linear solvers and preconditioners, and parallel \gls{IO} for writing out the 
results.
A lot of experience and knowledge has been acquired in scientific computing in other application domains, and it seems beneficial to transfer the applied methods to the field of neuroscience.

In the present work, we will deal with a nonlinearly coupled system of \glspl{PDE} in a three-dimensional setting.
Even after the introduction of a cylinder geometry and the resulting complexity reduction to a 2D problem, large numbers of unknowns ranging roughly between $10^5$ and $10^7$ are required. This necessitates an efficient parallelization scheme to achieve acceptable simulation times.
We will see that not only the large number of \glspl{DOF}, but also the inherent physical properties of the \gls{PDE} system provide substantial difficulties for the numerical solution of the model.
One central point will be the electrical property of the membrane acting as a capacitance, which necessitates a fine spatial resolution in perpendicular direction, while the direction parallel to the membrane may be resolved to a much coarser degree.
This results in a trade-off between the effort to reduce the number of unknowns and the emerging anisotropy in the spatial resolution.
It is known that an anisotropic grid imposes severe difficulties on the numerical solution, as the linear system to be solved involves dealing with a 
badly conditioned matrix.
This holds true even more in a parallel computation, since interior domain borders (marking processor boundaries) need to be treated with special care.

For all these reasons, the development of a proper mathematical model together with an accurate and efficient numerical solution is a topic on its own and very much worth investigating in the context of scientific computing.
Hopefully, the insights gained throughout this thesis will not only contribute to the understanding of neuronal signal propagation in the neuroscientific context, but also to the field of scientific and high performance computing, as the methods used might also be applied to related models with similar physical and numerical properties.
Jointly, and most importantly from a personal point of view, we hope that this thesis might serve as a small step and future motivation to bring the fields of neuroscience and scientific computing closer together, as in the future many open problems in neuroscience might be solved by the application of scientific computing and its powerful computational methods.

\section{Related Work}\label{sec:intro.related}
Simulations of neuronal signal propagation have a long history and are well established in neuroscience as a useful tool for the study of brain function.
Most models today are based on the seminal work of Hodgkin and Huxley \cite{hodgkin1952quantitative} for the dynamics of membrane currents and the
application of \emph{cable theory} \cite{rall1989cable} to account for the tree-like neuron morphology.
In essence, the neuronal geometry is reduced to a one-dimensional structure of line segments, and the evolution of the membrane potential in time is given by a one-dimensional \gls{PDE} -- the \emph{cable equation} -- which incorporates the Hodgkin-Huxley model of membrane currents in each compartment.
Two well-known simulators for these kinds of models are NEURON \cite{hines1997neuron} and GENESIS \cite{bower1995book}.
There is a vast number of studies using cable equation models, and it is impossible to even list the most important ones\footnote{As of March 14 2014, Google Scholar showed over 1400 citations of the original NEURON paper by \citeauthor{hines1997neuron}.}.

Several approximations and assumptions are made by the cable equation, which are explained in detail in \cite{holt1997critical}.
These assumptions are briefly summarized below.
\begin{enumerate}
\item The neuron cross section is small compared to its length, so variations in radial direction can be neglected; the neuron can be accurately represented as a 1D structure.
\item The extracellular space is isopotential (grounded) and does not have a reverse effect on the neuron.
\item Magnetic fields are negligible.
\item All quantities, particularly membrane currents, can be described by continuous variables, i.e.~stochastic effects can be neglected, since the number of participating ions is large enough and the ion diameter is sufficiently small compared to the relevant physical length scales.
\item Intra- and extracellular spaces can be represented as a homogeneous medium with effective physical parameters, e.g.~diffusion coefficients and resistivity.
\item Ion concentrations do no vary significantly, so they can be regarded as being constant.
\item The activity of a neuron is completely determined by its synaptic inputs and intrinsic states, i.e.~the electric activity of other neurons in the vicinity does not have a direct effect.
\item The neuronal membrane can be modeled as a capacitor in parallel with a conductance.
\end{enumerate} 
This is quite a number of assumptions. While some are certainly justified, others can be questioned.
In this context, particularly the assumptions of an isopotential extracellular space and its independence of the intracellular space as well as neglecting any concentration changes and their effects on the electric potential may be seen critically.
There have been some interesting modeling efforts that address one or more of these approximations in order to come up with a more accurate model.

In \cite{qian1989electro}, concentration dynamics were included by coupling the Nernst-Planck equation to the cable equation.
Intracellular concentrations were allowed to change, while extracellular concentrations were kept constant.
The authors reported deviations from the cable equation in small structures like dendritic spines, since in restricted volumes concentrations can vary significantly.
Furthermore, an adjustment of the cable equation was proposed, where the constant ionic batteries in the Hodgkin-Huxley model were replaced by concentration-dependent Nernst potentials, which can be calculated from a separate set of equations for the  ionic concentrations.
In a second step, also the single constant intracellular resistances in each compartment were replaced by parallel resistances for each ion species, which led to a better agreement with the Nernst-Planck cable model.

One central problem of the cable equation is the exclusion of the extracellular space, which is the region of interest in this work.
Most models for the extracellular potential are based on \gls{VC} theory \cite{plonsey1995volume}, where the extracellular space is assumed to be electroneutral (and in most cases also homogeneous), i.e. any concentration effects by ionic charges are neglected.
The relevant parameter for the extracellular medium in these models is the conductivity $\sigma$ (or equivalently, its inverse, the resistivity $\rho$).
The problem is reduced to the solution of the electrostatic part of Maxwell's equations, where the membrane is the only current source.
The local changes in ion concentrations caused by drift and diffusion and their contributions to the electric field are not considered.

Instead, current sources are imposed as boundary conditions.
This way, the electric potential can be calculated by Laplace's equation.
A special case is the \gls{LSA} \cite{holt1997critical}, where the membrane surface is collapsed to a line source, for which an analytical solution can be expressed in cylinder coordinates.
This avoids the need for a numerical solution and is computationally tractable, since one only has to compute it at the points of interest.
It has also shown to give quite accurate results at distances larger than \SI{1}{\micro\metre} from the membrane in an experimental comparison \cite{gold2006origin}.

Refined models consider an inhomogeneous extracellular space \cite{bedard2004modeling} that account for effects like frequency filtering \cite{PhysRevE.73.051911,bedard2009macroscopic}.

An interesting technique in this context is the application of inverse methods to these kinds of models, enabling the estimation of current source densities from \gls{LFP} measurements \cite{pettersen2008estimation}.

Another aspect of volume conductor models is that they are one-way models which take their inputs from a cable equation simulation, but the result is not fed back into the cable equation, such that intra- and extracellular space are still completely decoupled.
In a recent work \cite{agudelo2013numerical}, this is accounted for by solving the governing equations not only on the extracellular domain, but on both intra- and extracellular domains, coupled by Hodgkin-Huxley-type interface conditions. 

More detailed models explicitly consider the dynamics of ionic concentrations and their effects on the electric field, based on the \gls{PNP} system of electrodiffusion.
These models use the ion concentrations of all participating species -- next to the electric potential -- as primary variables, i.e.~they are explicitly solved for in the numerical simulations.
In contrast to the work in \cite{qian1989electro} mentioned above, concentrations and electric potential are fully coupled through the Nernst-Planck and Poisson equations, respectively.
These models provide the most fundamental representation of neuronal dynamics among all continuum models, given that magnetic fields can be neglected, which is the case under physiological conditions\footnote{According to \cite{rosenfalck1968intra}, the inclusion of magnetic fields into the calculation makes a difference of about one in $10^9$, as the relevant frequencies in neuronal tissue are small.} \cite{rosenfalck1968intra}.
However, they impose a serious challenge on the implementation, as they have to be solved numerically using problem-specific methods.

In \cite{0901.3914}, the reason for the high computational demand of electrodiffusion models based on the \gls{PNP} equations is discussed: the presence of a thin \emph{Debye layer} close to the membrane -- over which concentrations change significantly -- necessitates a very fine spatial resolution.
A clever approximation is suggested which replaces the Debye layer by a special boundary condition, such that the remaining parts of the domain can be approximated as electroneutral, eliminating the concentrations from the equations.
An extensive explication of the underlying asymptotic studies can be found in \cite{mori2006three}.
An interesting by-product of this remarkable study is the demonstration of the existence of a larger ``intermediate layer'' between Debye layer and electroneutral regions.
In \cite{mori2008ephaptic}, this methodology was applied to study the effect of ephaptic coupling on cardiac action potential propagation in the absence of any synapses or gap junctions.
Many more studies are specifically concerned with ephaptic coupling between neighboring cells, as reviewed in \cite{faber1989electrical,jefferys1995nonsynaptic}.
We will defer the detailed literature review of this particular topic to \cref{chap:multiple_fibers}.

In \cite{lopreore2008computational}, a detailed 3D numerical simulation of electrodiffusion based on the \gls{PNP} equations was carried out for one node of Ranvier, showing the accumulation and depletion of ions close to the membrane, and therefore the invalidity of the electroneutrality approximation close to the membrane, as used in cable equation and \gls{VC} models.
However, the study focused on deviations from the cable equation, not on the extracellular signal, and the membrane thickness was greatly overestimated as a consequence of the coarseness of the spatial discretization dictated by the available computational resources.
This also means the Debye layer was not accounted for.
Although these limitations render quantitative statements difficult, it is the only study we know of dealing with the full set of \gls{PNP} equations in the setting of neuronal excitation.

In other fields, particularly in biophysical studies of ion channels, the application of \gls{PNP} is well-established.
A review is given in \cite{coalson2005poisson}.
Numerical methods for electrodiffusion-reaction equations were analyzed in a comprehensive way in \cite{Lu20106979}, with special regard to surface potentials of biomolecules.

In summary, so far no model exists that explicitly resolves the relevant spatial scales to study neuronal excitation on the detailed level of electrodiffusion. Most importantly, there is no study we know of that actually uses the electrodiffusion approach to model the extracellular potential.

\section{Outline}
In this work we strive to model the detailed evolution of the concentrations of the most relevant ion species and the resulting electric field inside and (particularly) outside the cell during the spread of an action potential along an axonal membrane. 

To this end, the Poisson-Nernst-Planck equations are solved numerically by application of the \gls{FEM}.
To cope with the computational demand, a cylinder symmetry is introduced, which enables us to solve a 3D problem at 2D costs.
We propose an efficient numerical scheme, particularly a suitable computational grid which resolves the multiple spatial scales accurately while still using only a minimal number of unknowns.

The outline of this thesis is as follows: in the following chapter, the relevant theoretical background will be treated. 
\Cref{chap:model_general} discusses the general model, its prominent features and the numerical methods used to solve the system.
A description of the implementation of the numerical algorithm follows in \cref{chap:implementation}.
In \cref{chap:unmyel}, the results for an unmyelinated axon in extracellular fluid are shown.
We try to elucidate the evolution of the extracellular signal by looking at different contributing mechanisms and compare the results with \gls{VC} models.
In \cref{chap:myel}, the modeling approach is extended to also include myelination.
The necessary modifications in the numerical scheme and the emerging simulation results are discussed.
In \cref{chap:multiple_fibers}, we consider neighboring nerve fibers and their ephaptic effects on each other.
\Cref{chap:summary} gives a summary and discusses the results and future directions.

Some material in this work has been previously published in \cite{pods2013electrodiffusion}, parts of which are reproduced and cited in the following without explicit mention. This applies to \cref{sec:intro.related} and large parts of \cref{chap:model_general,chap:unmyel,chap:summary}.

\setchapterpreamble[u]{%
\dictum[H.G. Wells]{Very simple was my explanation, and plausible enough -- as most wrong theories are!}\bigskip}
\chapter{Theory}
This chapter serves as a basic outline of the theoretical background of this work.
It is divided into two parts: a mathematical part concerned mainly with the numerical solution of \glspl{PDE} and a second part that tries to list some fundamental biophysical relations in the context of neuroscience.

\section{Mathematical Aspects of Partial Differential Equations}
This section aims to act as a mini-primer on \glspl{PDE} and its numerical solution by the \gls{FEM}.
We will exclude the treatment of \glspl{ODE} here and refer to the well-established theory \cite{hairer2011solving,hairer2010solving} instead.

After a short look on the types of equations and their most important representatives, we will shortly introduce the basic concept of the finite element method.
By no means is this a complete treatise in the mathematical sense, for this we refer to the excellent script \cite{Bastian2012}, which we will be largely following hereafter.

\subsection{Type Classification}
A general linear second-order \gls{PDE} can -- assuming continuously differentiable coefficients -- be written in the form
\begin{align}
Lu = - \sum_{i,j=1} ^n a_{ij}(x) \partial_{x_j} \partial_{x_i} u  + 
\sum_{i=1}^n b_i(x)  \partial_{x_i} u + c(x) u = f \qquad \text{in $\Omega$} \label{eq:theory.pde_2ndorder}
\end{align}
in the unknown $u$, where $\partial_{x_i}$ is the shorthand notation for the partial derivative $\frac{\partial}{\partial x_i}$ with respect to $x_i$ and $\Omega$ is some domain, usually a subset of the $n$-dimensional Cartesian space $\mathbb{R}^n$. $L$ is called a linear \emph{differential operator}. 

Defining the matrix $(A(x))_{ij} = a_{ij}(x)$ and the vector $b(x) = (b_1(x),\ldots,b_n(x))^T$, \cref{eq:theory.pde_2ndorder} is said to be
\begin{itemize}
  \item \textit{elliptic} in $x$, if all eigenvalues of $A(x)$ are nonzero and have the same sign,
  \item \textit{hyperbolic} in $x$, if all eigenvalues are nonzero, $n-1$ eigenvalues have the same sign
and the remaining eigenvalue has the opposite sign, or
  \item \textit{parabolic} in $x$, if one eigenvalue is zero, the remaining eigenvalues have the same
sign and the $n \times (n+1)$ matrix $(A(x),b(x))$ has full rank.
\end{itemize}
If this classification holds for every $x \in \Omega$, this local property becomes a global one and \cref{eq:theory.pde_2ndorder} is simply called elliptic (parabolic, hyperbolic). 

The names stem from the case $n=2$, where the level set of equal function values $q=\text{const.}$ of the quadratic form $q(x_1,x_2) = a_{11} x_1^2 + 2 a_{12}x_1x_2 + a_{22} x_2^2$ takes either the shape of an ellipse, a parabola, or a hyperbola, depending on the values of the coefficients $a_{ij}$.
Note that this classification is not complete, i.e.~there are also \glspl{PDE} of \emph{mixed type}.

This classification of \glspl{PDE} is useful for their numerical solution.
There does not exist a general theory on the solution of partial differential equations, but for the subclasses, statements can be made about existence and uniqueness of solutions as well as about some of the properties these solutions have.
This can be exploited to develop numerical methods tailored to equations in a certain class.

\subsection{Prototypes}\label{sec:theory.pde_prototypes}
We will now have a look at some typical representatives of the above classes.
In the following, $\Omega\subset\mathbb{R}^n$ is a (spatial) domain, $\Sigma = (0,T]$ is a time interval and $u : \Omega\to\mathbb{R}$ or $u : \Omega\times\Sigma\to\mathbb{R}$ denotes the unknown scalar function.
We will also specify boundary and initial conditions, which will have to fulfill additional criteria (cf.~\cite[chapter 6]{Bastian2012}) in order to obtain a well-posed problem.

\subsubsection{The Poisson Equation}
The Poisson equation is a second-order elliptic \gls{PDE} that arises naturally from many physical problems.
It can for instance be used to calculate the gravitational potential due to certain masses, or the electrostatic potential due to a given charge density. It reads
\begin{align}
\Delta u &= f &&\text{in $\Omega$} \label{eq:theory.pde_prototypes.poisson} \\
u &= g &&\text{on $\Gamma_D$} \nonumber \\
\nabla u\cdot \mathbf{n} &= j &&\text{on $\Gamma_N$} \nonumber 
\end{align}
with \emph{Dirichlet} boundary conditions $g$ and \emph{Neumann} boundary conditions $j$ on the boundary $\partial\Omega = \Gamma_D \cup \Gamma_N, \Gamma_D \cap \Gamma_N = \emptyset$.
Here and in the following, $\mathbf{n}$ denotes the unit outer normal vector.
It is important to note that, since the Poisson equation is not depending on time, it describes an \emph{instantaneous} response or a \emph{stationary} state at which the system will settle given the source term $f$.
If $f=0$, \cref{eq:theory.pde_prototypes.poisson} reduces to the \emph{Laplace equation}.

\subsubsection{The Heat Equation}
The Heat equation is a second-order parabolic \gls{PDE} describing the distribution of a physical quantity (e.g.~heat) as a diffusive process in a certain region over time:
\begin{align}
\partial_t u - \Delta u &= f &&\text{in $\Omega\times\Sigma$} \label{eq:theory.pde_prototypes.heat} \\
u &= g &&\text{on $\Gamma_D\times\Sigma$} \nonumber \\
-\nabla u\cdot \mathbf{n} &= j &&\text{on $\Gamma_N\times\Sigma$} \nonumber \\
u &= u_0 &&\text{on $\Omega\times\{0\}$} \nonumber . 
\end{align}
with Dirichlet values $g$, Neumann values $j$ and the \emph{initial condition} $u_0$.
In contrast to the Poisson equation, the heat equation is \emph{instationary}, as is contains a derivative with respect to time, meaning that it describes a \emph{transient} process.

\subsubsection{The Transport Equation}
The transport or \emph{advection} equation is a first-order hyperbolic \gls{PDE} reading
\begin{align}
\partial_t u + \nabla\cdot (v u) &= f &&\text{in $\Omega\times\Sigma$} \label{eq:theory.pde_prototypes.transport}\\
u &= g &&\text{on $\Gamma_D\times\Sigma$}\nonumber \\
v u\cdot \mathbf{n} &= j &&\text{on $\Gamma_N\times\Sigma$}\nonumber \\
u &= u_0 &&\text{on $\Omega\times\{0\}$} . \nonumber 
\end{align}
with Dirichlet, Neumann and initial conditions as before. It may for instance describe the transport of certain substances in a fluid due to a known velocity field $v$.

\subsubsection{The Convection-Diffusion Equation}
The convection-diffusion equation is a second-order mixed-type \gls{PDE}, as it is a combination of the parabolic heat \cref{eq:theory.pde_prototypes.heat} and the hyperbolic transport \cref{eq:theory.pde_prototypes.transport}. It describes the distribution of the quantity $u$ due to diffusion and advection (convection) over time:
\begin{align}
\partial_t u - \nabla\cdot (D \nabla u) + \nabla\cdot (v u) &= f &&\text{in $\Omega\times\Sigma$} \label{eq:theory.pde_prototypes.convection-diffusion}\\
u &= g &&\text{on $\Gamma_D\times\Sigma$}\nonumber \\
(-D\nabla u + v u) \cdot \mathbf{n} &= j &&\text{on $\Gamma_N\times\Sigma$}\nonumber \\
u &= u_0 &&\text{on $\Omega\times\{0\}$} . \nonumber 
\end{align}
Here, $D$ denotes the \emph{diffusion coefficient}, which might in general be tensorial. In the following, we assume the diffusion to be isotropic, such that $D$ is a scalar value.

The ratio between advective and diffusive forces can be quantified by the \emph{P\'{e}clet number}
\begin{align}
  \text{Pe}_L = \frac{Lv}{D} \label{eq:peclet}
\end{align}
for a characteristic length $L$.

\bigskip

The bored reader at this point might wonder why the previous prototypic \glspl{PDE} were introduced out of any apparent context.
We beg for patience until the governing equations of electrodiffusion are introduced at the end of this chapter.
It will then become clear that the \gls{PNP} system is a combination of the above prototypes consisting of one elliptic and one convection-diffusion-type equation, i.e.~it contains parts of all of the prototypes introduced above.

\section{Numerical Solution of PDEs by the Finite Element Method}\label{sec:theory.fem}
The task of solving \glspl{PDE} like those in \cref{sec:theory.pde_prototypes} is really that of finding a \emph{function} $u(x)$ that satisfies the \gls{PDE} at every point in $\Omega\times\Sigma$.
As there is an (uncountable) infinite number of functions to choose from, this is certainly not an easy task.
When there is no analytical solution at hand, the goal is to find a \emph{numerical} (approximate) solution by reducing the solution candidates to a finite number by \emph{discretization}.
This is the process of finding a finite number of points $x_i \in \Omega, i=0,\ldots,N_S-1$ (in space) and $t_i \in \Sigma, i=0,\ldots,N_T-1$ (in time) on which the solution is to be satisfied.

The common choice for instationary equations is the \emph{method of lines}, where one first discretizes the spatial variables, leaving the time continuous, and afterwards the time variable.
The last step reduces the task to the solution of an \gls{ODE} in only one variable, for which a suitable time-stepping method can be used.
Both discretizations yield a (spatial/temporal) \emph{grid} on which we seek the discrete solution of the continuous problem.
For the space discretization, the \gls{FEM} is the method of choice in the following.

\subsection{Triangulation of the Domain}
The finite element method was designed to find a solution to a given \gls{PDE} in function space.
For this, the spatial domain $\Omega$ is discretized to yield a \emph{triangulation} (also \emph{mesh}, \emph{grid}) $\mathcal{T} = \{t_j : j=0, \ldots ,m-1\}$ consisting of \emph{elements} $t_j$.
In one dimension, the domain $\Omega=(a,b)$ is subdivided into
\begin{align*}
   a = x_0 < x_1 < \ldots < x_m = b
\end{align*}
(not necessarily equidistant). Then for $j=0,\ldots,m-1$, the elements $t_j$ and the \emph{mesh size} $h(t_j)$ are given by
\begin{align*}
   t_j = (x_j,x_{j+1}) \text{ and } h(t_j) = x_{j+1}-x_j \ .
\end{align*}
We use $h = \max_{t \in \mathcal{T}} h(t)$ to denote the maximum mesh size.

In higher dimensions, common choices for the elements are \emph{simplices} (2D: triangles; 3D: tetrahedra) or \emph{cuboids} (2D: rectangles, quadrilaterals; 3D: cubes/cuboids, hexahedra).
Finding a suitable mesh is a problem on its own, as a ``good grid'' depends on the considered equation and its physical properties, as we will see later. 
Here, we will not delve into the details of mesh generation or adaptation and simply assume that a suitable triangulation $\mathcal{T}$ has already been found.

\subsection{The Finite Element Method}
\subsubsection{Weak Form}
Before introducing Galerkin's method, the heart of the finite element method, we need to spend a few words on the requirements for the solution candidate functions $u_h \in V_h$ living on the triangulation $\mathcal{T}$.

Take for example the Poisson \cref{eq:theory.pde_prototypes.poisson}: since it is of second order, a function that satisfies the equation must be at least twice continuously differentiable. 
This imposes a severe restriction on the underlying function space $V_h$, as in practice one might be perfectly content with a piecewise linear solution, which is only once piecewise differentiable.
It is therefore customary to solve the \emph{weak formulation} of the original \gls{PDE}, which is obtained by multiplying the equation by a suitable \emph{test function} (the ``variation'') and integrating over the domain.

For the exemplary \cref{eq:theory.pde_prototypes.poisson} with homogeneous Dirichlet conditions $g = 0 \text{ on } \Gamma_D = \partial\Omega$ ($\Gamma_N = \emptyset$), we choose $v$ to be vanishing at the Dirichlet boundary, yielding
\begin{align}
\int_{\Omega} \Delta u v \,dx = \int_\Omega fv\,dx \ .\label{eq:theory.pde_weakform1}
\end{align}
After integrating by parts, the weak form reads
\begin{align}
\int_{\Omega} \nabla\cdot (\nabla u) v \,dx = - \int_\Omega \nabla u\cdot \nabla v \,dx = 
\int_\Omega fv\,dx \ , \label{eq:theory.pde_weakform2}
\end{align}
where we utilized that the boundary term vanishes due to $v=0$ on $\partial\Omega$.
Defining the \emph{bilinear} and \emph{linear forms}
\begin{align*}
a(u,v) &= -\int_\Omega \nabla u\cdot \nabla v \,dx \qquad \text{and} \qquad
l(v) = \int_\Omega fv\,dx \ ,
\end{align*}
the problem now takes the form 
\begin{equation}\label{eq:theory.pde_bilinear_form}
\text{Find $u\in U$} : \qquad a(u,v) = l(v) \quad \forall v\in V .
\end{equation}
A solution $u$ to \cref{eq:theory.pde_bilinear_form} is called a \emph{weak solution}. 
The weak solution is also a solution of the \emph{strong formulation} \cref{eq:theory.pde_prototypes.poisson} if $u$ is in $C^2(\Omega)$, i.e.~if it is twice continuously differentiable on the given domain (see \cite[chapter 8.1.2]{eriksson1996computational}).
Note that even though we used a concrete example above, every linear \gls{PDE} can be written in the form \cref{eq:theory.pde_bilinear_form}.

We now quickly touch the concept of \emph{linear operators}.
Let $\mathcal{L}(V,\mathbb{R})$ denote the set of all \emph{linear and continuous operators} $A$ from $V$ to $\mathbb{R}$. $V' = \mathcal{L}(V;\mathbb{R})$ is called the \emph{dual space} to $V$.
This way, we obtain an alternative notation of \cref{eq:theory.pde_bilinear_form}, the operator notation:
\begin{equation}\label{eq:theory.pde_operator}
\text{Find $u\in U$}: \qquad A u = l .
\end{equation}
Here, we have used the shortcut notation $Au = A(u)$ and the definition
\begin{equation*}
 \forall u\in U, \forall v\in V : \langle Au,v\rangle_{V',V} = a(u,v) \ ,
\end{equation*} 
so $Au\in V'$ is the continuous linear form that is obtained from $a(u,v)$ by fixing the argument $u$.
The scalar product $\langle \cdot \rangle_{V',V}$ is defined according to the \emph{Riesz Representation Theorem} \cite[Theorem 5.13]{Bastian2012}.

Thus, \cref{eq:theory.pde_bilinear_form} has a unique solution if and only if the corresponding operator $A$ is \emph{invertible}, i.e.~if it is injective and surjective. 
This problem formulation nicely reflects the practical implementation, where $A$ is the matrix, $u$ is the coefficient vector representing the solution, and $l$ is the right-hand side of the linear system to be solved.

In summary, the finite element method relies on the solution of \glspl{PDE} in weak form, as it requires less \emph{regularity} (``numbers of derivatives'') of the candidate functions, allowing for instance the choice of piecewise linear finite element functions for the solution of a second-order problem.
This greater freedom in controlling both accuracy and computational efficiency by the choice of a less regular finite element functions has turned out to be very beneficial in practice.

\subsubsection{Existence and Uniqueness of Weak Solutions}
The weak formulation \cref{eq:theory.pde_bilinear_form} of a given \gls{PDE} obviously has practical advantages over the strong formulation.
But is it also well-posed, i.e.~does it have a unique solution?
To answer this question, we have to become a little more precise and specify the classes of function spaces $V$ we will be using and their different properties.
In the following, we always assume the spaces to be defined on some domain $\Omega$ over the field of real numbers $\mathbb{R}$.
For a formal definition of these important spaces, see \cite[chapter 5]{Bastian2012}.

\begin{itemize}
\item \textbf{Banach space}: A vector space equipped with a \emph{norm} $\lVert \cdot \rVert$ which is \emph{complete} with respect to this norm, i.e.~every Cauchy sequence will converge to an element in the space $V$.
\item \textbf{Hilbert space}: A Banach space which is additionally equipped with a scalar product $\langle u,v \rangle$ and an induced norm $\lVert u \rVert = \sqrt{\langle u,u \rangle}$.
\item \textbf{Lebesque space $L^2$}: A Hilbert space whose elements are required to be square-integrable, i.e. $\int_{\Omega} |u(x)|^2 \, dx < \infty$; here we see the connection to integrals of the form \cref{eq:theory.pde_weakform1}.
This property also allows for the definition of the scalar product $\langle u,v \rangle = \int_{\Omega} u(x)v(x) \, dx$ and the induced $L^2$ norm $\lVert u \rVert = \sqrt{\langle u,u \rangle} = \left( \int_{\Omega} u(x)^2 \, dx \right)^{\frac{1}{2}}$.
\item \textbf{Sobolev space $H^k \subset L^2$}: A subset of $L^2$ which additionally requires the existence of \emph{weak derivatives} up to order $k$; a weak derivative $\partial_x u$ of a function $u$ satisfies $\langle u,\partial_x v \rangle = - \langle \partial u_x,v \rangle$ for every test function $v \in C_0^{\infty}$.
This ensures the existence of integrals of the form of those in \cref{eq:theory.pde_weakform2} obtained by integration by parts.
The scalar product is given by $\langle u,v \rangle = \sum\limits_{0\leq |\alpha|\leq k} \int_\Omega (\partial^\alpha u) (\partial^\alpha v) \, dx $ and the induced norm as before by $\lVert u \rVert = \sqrt{\langle u, u \rangle}$.
One commonly used example is the Sobolev space $H^1$ of functions with weak first order derivatives and scalar product $\langle u,v \rangle = \int_\Omega uv + \nabla u \cdot \nabla v\, dx$ inducing the $H^1$ norm $\lVert u \rVert = \left( \int_\Omega u^2 + \| \nabla u\|^2 \, dx \right)^{\frac{1}{2}}$.
\end{itemize}

Later on, we will see two concrete choices of function spaces for this rather abstract setting: the space of piecewise polynomials on a given simplicial or cuboid mesh, which is also used in the classical \gls{FEM}.

The Lax-Milgram theorem (cf.~\cite[Theorem 5.21]{Bastian2012}) ensures the existence of a unique solution of the weak problem \cref{eq:theory.pde_bilinear_form}.

\begin{theorem}[Lax-Milgram]
Let $V$ be a Hilbert space, $a\in\mathcal{L}(V\times V;\mathbb{R})$
and $l\in V'$. Then Problem \cref{eq:theory.pde_bilinear_form} is well-posed provided
$a$ is \emph{coercive} (or \emph{elliptic}), i.e. it satisfies the condition
\begin{align*}
\exists\alpha>0, \forall u\in V : \qquad a(u,u)\geq \alpha \lVert u \rVert^2 .
\end{align*}
Moreover, the following a-priori estimate holds:
\begin{equation*}
\forall l\in V' : \qquad \|u\|_V \leq \frac{1}{\alpha} \| l \|_{V'} .
\end{equation*}
\end{theorem}

A direct proof can be found in \cite{brenner2008mathematical}.
An alternative is to reduce the Lax-Milgram theorem to the Banach-Nečas-Babuška theorem \cite[Theorem 5.2.1]{Bastian2012}, which is more general and only requires $V$ to be a Banach space, not necessarily a Hilbert space.
The ellipticity of $a$ implies that the corresponding operator $A$ in \cref{eq:theory.pde_operator} is invertible.

\subsubsection{The Continuous Galerkin Method}\label{sec:theory.pde_cg}
To obtain finite-dimensional functions for the representation of solutions, we need a corresponding function space $V_h$.
The subscript $h$ of $V_h$ refers to the mesh size, indicating that it is a finite subspace living on a certain mesh. An appropriate choice of $V_h$ will satisfy
\begin{align*}
   \inf_{u_h \in V_h} \lVert u-u_h \rVert \to 0 \quad \text{as} \quad h \rightarrow 0 \ ,
\end{align*}
such that for small mesh sizes $h$, $u_h$ comes arbitrarily close to the function $u$ living on the original space $V$.
We then pick \emph{trial} or \emph{ansatz functions} which form a \emph{basis} of the underlying function space $V_h$, such that every function $u_h$ can be represented as a linear combination of the ansatz functions.

There are two principal variants for the choice of ansatz functions: in the \gls{cG} method they are defined to be globally continuous, i.e.~there are no jumps at element boundaries.
The \gls{dG} method, which we will not cover here, allows the ansatz functions to be discontinuous at element boundaries, making them especially useful for \glspl{PDE} which themselves have discontinuous solutions, like the transport \cref{eq:theory.pde_prototypes.transport} with an initial step condition.

Both variants need a second function space $W_h$ for the test functions.
In the classical Galerkin method, the spaces for ansatz and test functions are the same, $V_h = W_h$. 
If $V_h \neq W_h$, we obtain the class of \emph{Petrov-Galerkin} methods, which will not be covered here.

Given a mesh $\mathcal{T}$, the abstract \gls{cG} method now reads as follows:
\begin{enumerate}
  \item Obtain the \emph{weak formulation} of the \gls{PDE} by integration and multiplication with test functions $v$.
  \item Choose a suitable function space $V_h$ for the ansatz functions on $\mathcal{T}$, e.g.~the space of piecewise polynomials of degree $k$, $P_k(\mathcal{T})$ / $Q_k(\mathcal{T})$.
  \item With the \emph{bilinear form} $a(u,v)$ and \emph{linear form} $l(v)$ from the weak formulation, state the problem in the finite-dimensional subspace as
\begin{align}
  \text{Find $u_h \in V_h$} : \qquad a(u_h,v) = l(v) \quad \forall v \in V_h. \label{eq:theory.fem_galerkin}
\end{align}
  \item Find a basis of $V_h$
\begin{align*}
   \Phi_h = \{\varphi_1^h, \ldots, \varphi_{N_h}^h \}
\end{align*}
  of size $N_h = \dim V_h$.
  \item As all function $v_h \in V_h$ can be represented as linear combinations of elements of $\Phi_h$, it is sufficient to use these instead of all test functions $v$ in \cref{eq:theory.fem_galerkin}. Utilizing this and inserting the basis representation
\begin{align*}
   u_h = \sum_{j=1}^{N_h} z_j \varphi_j^h
\end{align*}
  yields a linear system of equations
\begin{align*}
& & a(u_h, v) &= l(v) \quad \forall v \in V_h \\
& \Leftrightarrow & a \left( \sum_{j=1}^{N_h} z_j \varphi_j^h, \varphi_j^h \right) &= l(\varphi_i^h) \quad \forall i=1, \ldots, N_h \\
& \Leftrightarrow & \sum_{j=1}^{N_h} z_j a(\varphi_j^h, \varphi_j^h) &= l(\varphi_i^h) \\
& \Leftrightarrow & A z &= b \qquad (A)_{ij} = a(\varphi_j^h, \varphi_i^h), \quad (b)_i = l(\varphi_i^h) \,
\end{align*}
for the coefficient vector $z\in\mathbb{R}^{N_h}$.
\item Solve the linear system by an appropriate linear solver.
\end{enumerate}
Note how the discrete version \cref{eq:theory.fem_galerkin} of the weak problem formulation \cref{eq:theory.pde_bilinear_form} has been reduced to a linear system of equations, representing a discrete version of the operator formulation \cref{eq:theory.pde_operator}.

\subsubsection{Galerkin Orthogonality}\label{theory.fem.galerkin_orthogonality}
We have just seen that Galerkin's method can be used successfully to solve a linear \gls{PDE} in a finite-dimensional subspace.
The question remains: why is this a good strategy? In particular, what does the multiplication with test functions and integration over the domain mean for this problem? 

To understand this, let us again have a look at our Poisson example \cref{eq:theory.pde_prototypes.poisson}.
The residual for this equation reads $R(u) = \Delta u - f$, which is to be minimized by the Galerkin method. 
Rearranging \cref{eq:theory.pde_weakform1} yields
\begin{align}
\int_{\Omega} (\Delta u - f)v\,dx = \int_{\Omega} R(u) v\,dx = 0 \ . \label{eq:theory.pde_galerkin_residual_strong}
\end{align}
Fulfilling this relation with a discrete function $u_h$ for every $v \in V_h$ means that the residual is \emph{orthogonal} to every test function $v$, as can be seen when expressing \cref{eq:theory.pde_galerkin_residual_strong} in terms of the $L^2$ scalar product\footnote{This is completely analogous to the vector space of real numbers, where two vectors are orthogonal if their scalar product vanishes.},
\begin{align}\label{eq:theory.pde_galerkin_residual_scalar_product}
\langle R(u_h), v\rangle = 0 \quad \forall v \in V_h \ .
\end{align}
For the weak solution, \cref{eq:theory.pde_galerkin_residual_strong} takes the slightly different form
\begin{align}
- \int_{\Omega} (\nabla u_h \cdot \nabla v + f v) \,dx = 0 \quad \forall v \in V_h \ ,
\end{align}
or, more generally, for the abstract Galerkin problem \cref{eq:theory.fem_galerkin},
\begin{align}
r(u_h,v) & = a(u_h,v)-l(v) = 0 \quad \forall v \in V_h \ . \label{eq:theory.pde_galerkin_residual}
\end{align}
The approximation error $u-u_h$ therefore fulfills
\begin{align*}
  r(u-u_h,v) &= r(u,v)-r(u_h,v) = a(u,v)-l(v)-\left( a(u_h,v)-l(v) \right)\\
             &= a(u,v)-a(u_h,v) = a(u-u_h,v) = 0 \quad \forall v \in V_h \ .
\end{align*}
This property is called the \emph{Galerkin orthogonality} of the error $u-u_h$ with respect to the bilinear form $a$.
Since, after a choice of the space $V_h$, we are only interested in the errors that are actually \emph{representable} in $V_h$, a solution whose error is orthogonal to all functions in the test space is optimal in this sense.
The orthogonal part of the error can simply not be captured in $V_h$ and therefore we have found the best solution available in $V_h$.

Another view on this problem is related to the $L^2$ norm $\lVert u \rVert = \sqrt{\langle u,u \rangle}$ induced by the scalar product.
By demanding the orthogonality of the residual, Galerkin's method actually calculates the $L^2$ projection $u_h$ of the function $u$ into the finite-dimensional subspace $V_h$, which is known to give the best approximation in the $L^2$ norm \cite[chapter 5.6]{eriksson1996computational}.

\subsubsection{Common Function Spaces}\label{sec:theory.fem.function_spaces}
Now that we are convinced that Galerkin's method gives an optimal solution in a chosen function space $V_h$, we briefly mention two common choices for $V_h$ for simplicial and cuboid grids.

\paragraph*{Pk Finite Elements}
The standard choice of ansatz and test functions are polynomials. If we denote the space of polynomials of degree at most $k$ in $n$ space dimensions as
\begin{align}
   \mathbb{P}^n_k = \{u \in C^\infty(\mathbb{R}^n) : u(x) = \sum_{0 \leq \lvert \alpha \rvert \leq k} c_\alpha x^\alpha \} \ ,
\end{align}
we can use the finite element space
\begin{align}
   P_k(\mathcal{T}) = \{u \in C^0(\overline{\Omega}) : u\rvert_{\overline{t}} \in \mathbb{P}^n_k \ \forall t \in \mathcal{T} \} \ .
\end{align}
on a conforming simplicial mesh (e.g.~triangles, tetrahedra).

\paragraph*{Qk Finite Elements}
For cuboid meshes (rectangles, cubes/hexahedra), the space of polynomials of degree at most $k$ is defined slightly differently as
\begin{align}
   \mathbb{Q}_k^n = \{ u \in C^\infty(\mathbb{R}^n) : u(x) = \sum_{0 \leq \lvert \alpha \rvert_\infty \leq k} c_\alpha x^\alpha \} \label{eq:3.7}
\end{align}
with $\lvert \alpha \rvert_\infty = \max_i \alpha_i, \ \alpha_i \in \mathbb{N}_0$ and $\dim \mathbb{Q}_k^n = (k+1)^n$.
If all cubes in the mesh $\mathcal{T}$ are axi-parallel, the space 
\begin{align}
Q_k(\mathcal{T}) = \{ u \in C^0(\overline{\Omega}) : u\rvert_{\overline{t}} \in \mathbb{Q}_k \ \forall t \in \mathcal{T} \}
\end{align}
can be defined analogous to the simplex meshes.
This finite element space will be used throughout this document with the choice of $k=1$, i.e.~it is a subspace of the Sobolev space $H^1$.
The method of finding a suitable basis for this space is covered in \cite[chapter 7.5]{Bastian2012}.

\subsection{Nonlinear PDEs: Newton's Method}\label{sec:theory.fem.newton}
So far, we have only considered linear \glspl{PDE}. 
But in practice, one often has to deal with nonlinear equations. 
Since the abstract existence theorems (Lax-Milgram, Banach-Nečas-Babuška) only hold for linear \glspl{PDE}, the well-posedness of a nonlinear \gls{PDE} has to be shown by other means, often on a case-by-case basis. 
There does not exist a general theory for the solvability of general \glspl{PDE}, and it is commonly believed unlikely that such a general theory exists.

Assuming a unique solution exists, the generalization of the numerical solution of such problems by the \gls{FEM} is quite straightforward, as the nonlinear operator can be linearized by \emph{Newton's method} and embedded into an iterative procedure.
Here we use the \emph{damped Newton method}, see e.g.~\cite{deuflhard2011newton}.

Given the initial guess $u^0$, compute $r^0 = R(u^0)$. Set $k=0$ and iterate until convergence:
\begin{enumerate}
\item Compute Jacobian matrix $A^k = \nabla R(u^k)$.
\item Solve $A^k z^k = r^k$ with some linear solver.
\item Update $u^{k+1} = u^{k} - \sigma^k z^{k}$, $\sigma^k\in(0,1]$.
\item Compute the new residual $r^{k+1} = R(u^{k+1})$.
\item Set $k = k +1$.
\end{enumerate}
The residual $R(u)$ contains entries $a(u_h,v_h)-l(v_h)$ from our weak form Galerkin approximation in residual form.
Additionally, the Jacobian $\nabla R(u)$, the derivative of the residual with respect to all elements of $u$, is needed.
This can either be obtained by providing an analytical derivative of the operator $L$ beforehand or by using numerical differentiation. 
For the choice of the damping factor $\sigma$, one can for instance use a \emph{line search} strategy that tries to minimize the residual along the search direction $z$.

Note that for a linear operator $L$, $R(u) = Au-b$ and thus $\nabla R(u) = A$. 
In this case, the Newton method will simply solve the linear system and converge in one iteration.

\subsection{Time Discretization}
The theory of \glspl{ODE} \cite{hairer2011solving,hairer2010solving} has produced a vast amount of time-stepping schemes, which is impossible to cover here. 
Instead, we will look at the Euler methods as representatives of the two major classes of methods. 

Suppose the initial value problem to be solved is
\begin{align*}
  y'(t) = f(t,y(t)), && y(t_0) = y_0 \ ,
\end{align*}
where we can think of $y(t)$ as our solution of the \gls{FEM} solution of our \gls{PDE} at a certain time $t$. Suppose now as above the time interval has been discretized at $N_T$ time points $t_i \in \Sigma, i=0,\ldots,N_T-1$ with time intervals $\Delta t_i = t_{i+1} - t_i$, $i=0,\ldots,N_T-2$. 

The \emph{explicit Euler} (or forward Euler) reads
\begin{align}
y_{i+1} = y_i + \Delta t_i f(t_i,y_i), \qquad i=0,\ldots,N_T-2 \ . \label{eq:theory.pde_explicit_euler}
\end{align}
It is a representative of the \emph{explicit} time-stepping methods, as the right-hand side only depends on quantities from the previous iterations.

Its antagonist is the \emph{implicit Euler} (or backward Euler) method
\begin{align}
y_{i+1} = y_i + \Delta t_i f(t_{i+1},y_{i+1}), \qquad i=0,\ldots,N_T-2 \ , \label{eq:theory.pde_implicit_euler}
\end{align}
representing an \emph{implicit} method, as the right-hand side depends on values that are not known yet. 
In general, implicit schemes require the solution of a linear system of equations in each time step and hence tend to be more expensive. 
However, they possess better \emph{stability} properties that in many cases allow the usage of larger time steps $\Delta t_i$.

Both methods in \cref{eq:theory.pde_explicit_euler,eq:theory.pde_implicit_euler} are of \emph{first order accuracy}, meaning that the error of this method scales with $\mathcal{O}(\Delta t_i)$.

\subsection{Linear Solvers}
As seen in the previous sections, the numerical solution of a \gls{PDE} in the end is always reduced to the solution of a linear system of equations.
The variety of methods for this kind of problem is overwhelming, ranging from the direct solution by Gauß elimination with a complexity of $\mathcal{O}(n^3)$ to very sophisticated sparse iterative solvers \cite{saad2003iterative} and parallel multigrid methods \cite{BastianPhD} with an optimal complexity of $\mathcal{O}(n)$. 
A comprehensive list of methods is beyond the scope of this thesis, and since the choice of a suitable solver is highly problem-specific, we will defer this topic to the relevant chapters.

\section{Fundamentals of Biophysics and Neuroscience}
\subsection{Structure and Function of Neurons}
\begin{figure}
\begin{center}%
\centering%
\includegraphics[width=0.75\textwidth]{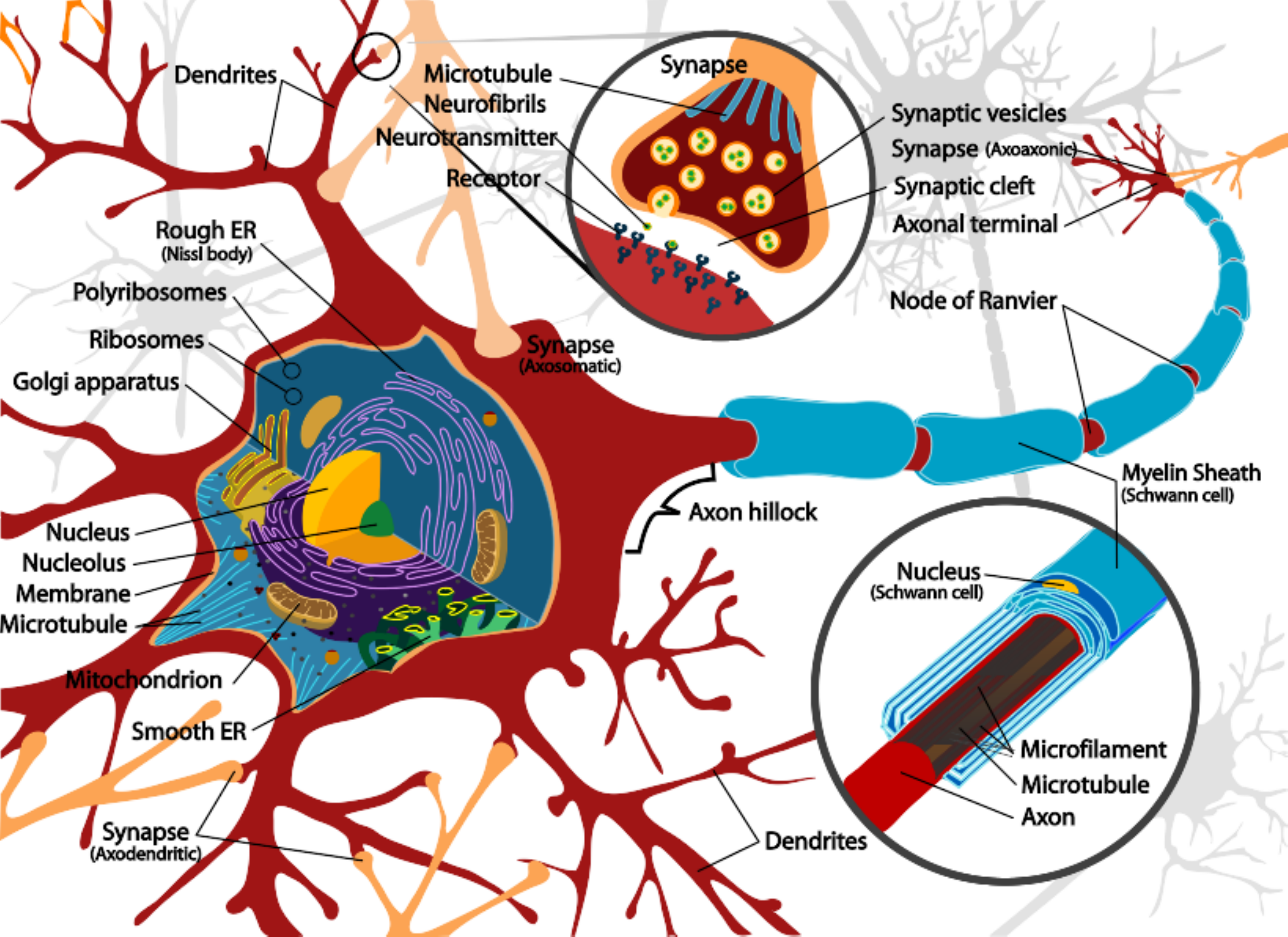}%
\end{center}%
\mycaption[Illustration of a typical neuron]{%
By \href{http://commons.wikimedia.org/wiki/User:LadyofHats}{LadyofHats} [Public domain], \href{http://commons.wikimedia.org/wiki/File\%3AComplete_neuron_cell_diagram_en.svg}{via Wikimedia Commons}}%
\label{fig:theory.neuron_illustration}
\end{figure}

A neuron -- see \cref{fig:theory.neuron_illustration} for an illustration -- can be structurally divided into three parts: The \emph{soma} containing the nucleus as the center of a neuron is the largest part.
Multiple \emph{dendrites} branch off from the soma, creating the \emph{dendritic tree}, which receives its input at \emph{synapses}, the connections to other neurons.
The third part is the \emph{axon}, which is a fiber that is often thinner than the dendrites, but nevertheless can show quite complex branching as well.
A neuron always has one and only one axon.

Apart from this common property, axons can show a wide range of variations among neurons.
They may or may not be \emph{myelinated} as in \cref{fig:theory.neuron_illustration}, i.e.~insulated by a myelin sheath that is provided by a surrounding cell.
In this case, the \emph{nodes of Ranvier} provide periodical segments exposing the underlying axon directly to the \gls{ES}.
The nodes usually contain a high density of ion channels, which provide for an active propagation of a traveling action potential by means of transmembrane currents that keep the \gls{AP} alive.

Between the nodes, at myelinated parts (or \emph{internodes}), the signal propagates passively, but significantly faster than in unmyelinated fibers, since virtually no charge is lost across the myelinated membrane.
The established view is that axons are one-way paths, i.e.~signals are always propagated away from the cell soma towards other cells in an \emph{orthodromic} fashion, although there have been some hints that also \emph{antidromic} propagation in the opposite direction may happen under physiological conditions \cite{bahner2011cellular}. 

\begin{figure}
\begin{center}%
\centering%
\includegraphics[width=0.4\textwidth]{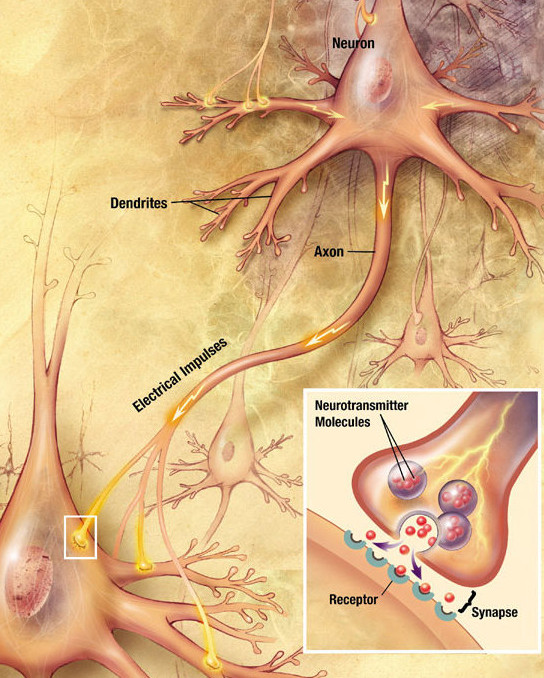}%
\end{center}%
\mycaption[Neuronal signaling mechanisms]{%
By \href{http://commons.wikimedia.org/wiki/user:Looie496}{Looie496} created file, US National Institutes of Health, National Institute on Aging created original [Public domain], via \href{http://commons.wikimedia.org/wiki/File\%3AChemical_synapse_schema_cropped.jpg}{Wikimedia Commons}}%
\label{fig:theory.neuron_signals}
\end{figure}

Figuratively (and keeping with the author's favorite metaphor of electronic devices), one can think of the dendrites as the neuron's antennae collecting input from other neurons, while the axon is the output channel.
Action potentials are generated at the \emph{axon initial segment} close to the soma, and propagate down the axonal arbor, where they will eventually stimulate other neurons (see \cref{fig:theory.neuron_signals}).

Stimulations of this kind, however, do not always have to be \emph{excitatory}, that is driving the target neuron to fire another action potential, but they may also have an \emph{inhibitory} effect, i.e.~preventing the target neuron from firing.
This depends on whether the synapse transmitting the signal between two neurons -- or more specifically, the \emph{neurotransmitters} it uses as a messenger -- has excitatory or inhibitory character.
This is important, since an over-excitation can lead to pathological situations in the brain, one extreme example being epilepsy. The balance of excitation and inhibition is crucial for a functional brain.

The question remains how the elementary unit of information, the action potential, is generated.
As mentioned above, the axon plays the leading role in this process.
It turns out that nature takes advantage of elementary electrodynamics, using ions as charge carriers and the axonal membrane as an electric circuit.
In the following, some basic properties of passive membranes will be touched, before we turn to the active parts of the membrane, the ion channels.

\subsection{Membrane Physics}\label{sec:theory.membrane}
A biological membrane of finite thickness $\dMemb$ and an electric permittivity $\epsMemb$ has some important electric properties with influence on the two electrolytes it separates. 
In the following, we will only consider the case where the electrolytes on both sides of the membrane have the same solvent, water. 
Therefore, the electrolyte permittivity $\epsElec$ is a single value, and the only difference between both sides is the ion concentration. 
For the two electrolytes, the terms \emph{cytosol} and \emph{extracellular space} will be used hereafter, often abbreviated as CY or ES in sub- or superscripts, respectively. 

\begin{figure}
\begin{center}%
\centering%
\includegraphics[width=0.4\textwidth]{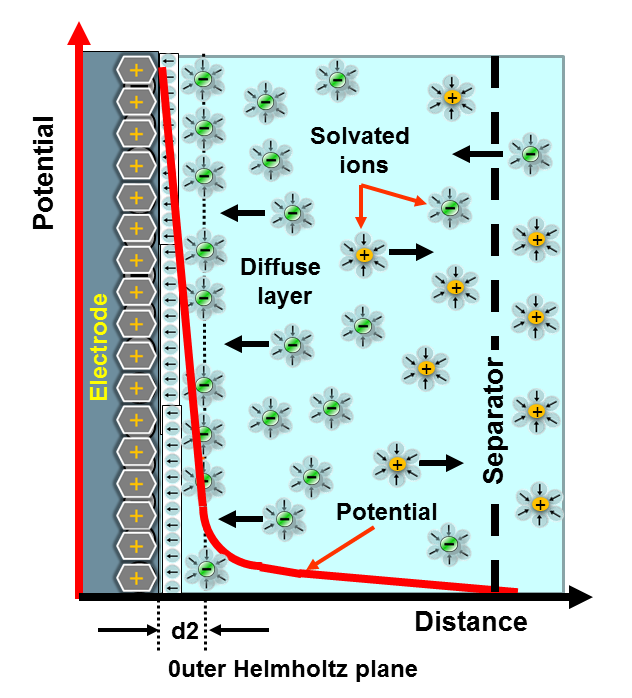}%
\end{center}%
\mycaption[Schematic picture of a membrane and electrolyte]{%
By \href{http://commons.wikimedia.org/wiki/user:Elcap}{Elcap} (Own work) [CC0], via \href{http://commons.wikimedia.org/wiki/File\%3AEDLC-Potentialdistribution.png}{Wikimedia Commons}}%
\label{fig:theory.membrane}
\end{figure}

A schematic depiction of the membrane and one adjacent electrolyte is given in \cref{fig:theory.membrane}. 
We see that the (electrically charged) membrane attracts oppositely charged counterions and repels equally charged co-ions. 
As a consequence, an \gls{EDL} forms: one layer of counterions directly at the membrane, and another layer of co-ions, which is attracted by the counterion layer. 
Several names are linked with the theory of electric double layers. 
Helmholtz \cite{ANDP:ANDP18531650603} was the first to realize that such an \gls{EDL} has the capacity to store electric charges and therefore acts as a capacitor. 
Gouy and Chapman \cite{gouy1910formation,chapman1913diffuse} noted that this capacitance depends on the applied membrane voltage and the ionic concentrations; they were also the first to find that the ion concentrations decrease exponentially with distance from the membrane, which can be described by Maxwell-Boltzmann statistics. 
This added a \emph{diffuse layer} to the Helmholtz layer of fixed charges directly at the membrane interface. 
Further improvements were made by Stern \cite{stern1924theory} and others, resulting in a quite complex theory of different layers and their interactions.

A very good summary of the theoretical background is given in \cite[chapter 12]{lipowsky1995structure}. 
One central equation describing the potential profile in the presence of a membrane is the Poisson-Boltzmann equation.

\subsubsection{Poisson-Boltzmann Equation}
The Poisson-Boltzmann equation
\begin{align}
\nabla\cdot\left(\epsilon\nabla\phi\right) &= -\frac{e^2n^*}{\epsZero \kB T}%
\sum_i z_i n_i^0 \exp(-z_i \phi)\label{eq:pb}
\end{align}
is a special case of the Poisson equation. Here, $\phi$ is the dimensionless relative electric potential energy with respect to the thermal energy ($\phi=eU/ \kB T$ with U given in units of volts; at room temperature, $\phi=1$ corresponds to $U \approx \SI{25}{\milli\volt}$); $\epsilon$ is the relative permittivity, which again may be position-dependent, and $T$ is the temperature of the solvent. 
The remaining constants are the vacuum permittivity $\epsZero$, the Boltzmann constant $\kB$, and the elementary charge $e$. 

In comparison to \cref{eq:theory.pde_prototypes.poisson}, the charge density on the right-hand side has been replaced by a Boltzmann distribution for the ion concentrations at equilibrium
\begin{align}
n_i = n_i^0 \exp(-z_i \phi)\label{eq:pb_conc}
\end{align}
due to a certain potential that has been established at the membrane. Here, $n_i$ denotes the concentration of species $i$ with valence $z_i$ and bulk concentration $n_i^0$.

The equation basically describes how a charged membrane in an ionic solution causes accumulation of counterions, i.e.~ions with opposite charge with respect to the membrane charge. 
This membrane charge could be some surface charge of a biomolecule or, as in this work, the charge due to ions on the other side of the membrane.
In either case, the potential profile will cause a charge density profile with opposite sign, as counterions are attracted and co-ions are repelled from the membrane with a certain charge.
As a consequence, the previously described \gls{EDL} of two oppositely charged regions forms at the boundary between membrane and electrolyte. \Cref{fig:theory.membrane} shows a typical potential profile as described by \cref{eq:pb}.

The Poisson-Boltzmann equation proves useful to test a numerical algorithm, as the calculated equilibrium charge density can be compared with the analytical expression \cref{eq:pb_conc}.

\subsubsection{Debye-Hückel Theory}\label{sec:theory.biophysics.debye-hueckel}
For monovalent solutions and small surface potentials ($< \SI{25}{\milli\volt}$ according to \cite{lipowsky1995structure}), \cref{eq:pb} can be linearized to the Debye-Hückel equation \cite{debye1954theory}
\begin{align}
  \nabla^2 \phi = \frac{1}{\dDebye^2} \phi \ ,
\end{align}
where
\begin{align}\label{eq:debye_length}
\dDebye = \sqrt{\frac{\epsElec \epsZero \kB T}{e^2 2 \IS}}
\end{align}
is called the \emph{Debye length}. This is an important property of an electrolyte which is inversely proportional to the square root of the \emph{ionic strength}
\begin{align}\label{eq:ionic_strength}
  \IS = \sum_i \frac{1}{2} n_i z_i^2 \ .
\end{align}
The Debye length is a characteristic spatial scale over which electrostatic interactions close to the membrane are \emph{screened} (i.e., decaying) exponentially. 
It therefore gives a measure on the distance over which the membrane influences the electric field to a large degree.

For strong electrolytes, $\dDebye$ is small.
However, for electrolytes of lower concentration, the Debye length is larger and has a significant effect in a larger range around the membrane. 
A common term for this vicinity is the \emph{Debye layer}, the region of up to a few Debye lengths. 
Here, we will define the Debye layer to be 10 times the Debye length. 
Outside the Debye layer one can safely assume that concentrations have decayed to their bulk values, such that membrane effects do not play a role.

\subsection{Ion Channels}\label{sec:theory.ion_channels}
It is obvious that a simple solid membrane will not be sufficient to yield some kind of excitation. 
After the previous demonstrations, we know that the Debye layer experiences an accumulation of ions towards the membrane. 
Hence, we can think of the membrane as a \emph{capacitance} separating charges present on the two adjacent electrolytes. 
In order to allow charge carriers to actually cross the membrane, we now introduce \emph{ion channels} in the membrane.

\begin{figure}
\begin{center}%
\centering%
\includestandalone[width=0.2\textwidth]{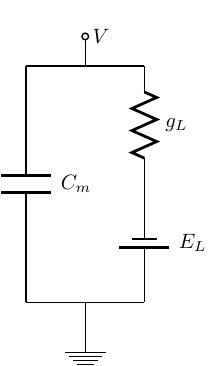}%
\end{center}%
\mycaption[Equivalent circuit for a patch of membrane]{A patch of membrane can be represented by a capacitance $C_m$ in parallel with a series of a leak conductance $g_L$ and a battery $E_L$ representing the (Nernst) reversal potential.}%
\label{fig:theory.membrane_circuit}
\end{figure}

With this, the neuronal membrane can be adequately described by an equivalent circuit as in \cref{fig:theory.membrane_circuit}, consisting of a capacitance in parallel with a conductance and a battery. 
The conductance now represents one out of the multitude of \emph{ion channels} that exist in biological membranes, allowing for a current to establish between intra- and extracellular space. 
The battery here stands for the reversal potential of the ion channel. 
We will now have a look at the most important channel types responsible for the action potential generation.

\subsubsection{Leak Channels}
The simplest kind of ion channels are \emph{leak channels}, which have a constant conductance $\gL$. 
These play an important role for the resting state of a neuron.
The current flowing through channels of this type will depend on the concentration gradient $\nabla n_i$ and the potential gradient $\nabla \phi$ between both sides of the membrane (cf.~\cite{bear2007neuroscience}).
For one ion species, \emph{Nernst's Equation} is very successful in predicting the \emph{reversal potential}
\begin{align}
  E = \frac{RT}{zF} \ln \frac{n^{\text{ES}}}{n^{\text{CY}}} \label{eq:nernst}
\end{align}
of a leak channel that is selective for a certain ion. Here, $R$ denotes the \emph{gas constant} and $F$ the \emph{Faraday constant}.
The \gls{GHK} or simply \emph{Goldman equation} extends this to the case of multiple monovalent species, reading
\begin{align}
  E = \frac{RT}{F} \ln \left( 
\frac{ \sum_i^N P_{M_i^+} [M_i^+]^{\text{ES}} + \sum_j^M P_{A_j^-} [A_j^-]^{\text{CY}} }{
       \sum_i^N P_{M_i^+} [M_i^+]^{\text{CY}} + \sum_j^M P_{A_j^-} [A_j^-]^{\text{ES}} } \right) \ ,
\label{eq:goldman}
\end{align}
with $[M_i^+]$ and $[A_i^-]$ denoting cation and anion concentrations, respectively, and $P_i$ being the relative permeability of the membrane for ion species $i$. $E$ gives the \emph{resting potential} of a neuron with respect to extracellular space. 

The Goldman equation makes some assumptions about the intrinsics of the ion channel: first of all, it assumes a constant electric field (i.e., a linear potential drop) over the membrane; secondly, all ions cross the membrane independently, so there is no interaction between them.
It is clear that we need the (relative) permeabilities of the membrane for all considered ion species in order to predict the resting potential in this way.
In practice, permeabilities are not as easily available as conductances, although the quantities are related.
However, a conductance as a purely electric property can readily be obtained from a direct membrane current measurement by Ohm's law, while this proves more difficult for permeabilities, which is a chemical property of the ions involving the diffusion coefficient and the water-membrane partition coefficient.

When dealing with conductances, one can instead use the \emph{\gls{PCM}} \cite{wyttenbach2008pcm}, which simply uses Ohm's and Kirchhoff's laws applied to an equivalent circuit of the form in \cref{fig:theory.membrane_circuit} with multiple leak conductances in parallel, one for each ion species.
Requiring the net membrane flux to be zero at equilibrium, we get the simple relation
\begin{align}
E = \frac{\sum_i^N g_i E_i}{\sum_i^N g_i} \label{eq:pcm}
\end{align}
with the leak conductances $g_i$ and corresponding reversal potentials $E_i$ as calculated by \cref{eq:nernst} for each ion species separately.
It is easy to see that this reduces to the Nernst \cref{eq:nernst} for a single species.

When dealing with equations that calculate the resting potential of neurons due to the concentrations of certain ions on both sides of the membrane, it is important to note that all of these are based on certain assumptions that might or might not be true in the specific case:
\begin{quote}
  One might ask why the voltage equation is so hard to derive and so closely tied to minute assumptions when the analogous Nernst equation is so simple to obtain [...] and so general.
 The contrast is typical of the difference between equilibrium and nonequilibrium problems.
 The Nernst equation describes a true equilibrium situation and can therefore be derived from thermodynamics as a necessary relation between electrical and ``concentration'' free energies with no reference to structure or mechanism.
 On the other hand, the zero-current voltage equation represents a dissipative steady state. 
[...] Only the sum of charges moving is zero. 
The reversal potential is not a thermodynamic equilibrium potential. 
Such nonequilibrium problems often can make little use of thermodynamics and require empirical relationships closely tied to the structure and mechanism of the flow. 
The assumptions are often simplistic and no more than approximations. 
They are models. \cite[452]{hille2001ionchannels}
\end{quote}\label{quote:theory.ion_channels.equilibrium}
With that in mind, it depends on the underlying model whether \cref{eq:goldman} or \cref{eq:pcm} is adequate to calculate the resting potential $E$.
In practice, it also depends on the availability of parameters, i.e.~whether we think in terms of permeabilities or conductances. 
In the following, we will take the electrical engineer's point of view, and look at the membrane in terms of conductances and equivalent circuits.

Leak channels are also called \emph{passive} channels, in contrast to \emph{active} channels, whose conductance changes in time, depending on the membrane potential or the concentrations of certain compounds.
We will restrict ourselves to the voltage-dependent ones, whose dynamic properties were described in the seminal work by Hodgkin and Huxley.

\subsubsection{The Hodgkin-Huxley System of Membrane Excitation}\label{sec:theory.biophysics.hh}%
\begin{figure}
\begin{center}%
\centering%
\includestandalone[width=0.45\textwidth]{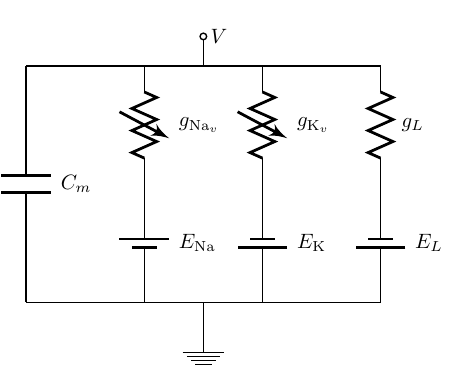}%
\end{center}%
\mycaption[Equivalent circuit for the Hodgkin-Huxley membrane patch model]{It consists of a membrane capacitance $C_m$ in parallel with three branches, each consisting of a series of conductance and battery, representing the ionic current and reversal potentials, respectively. Note that the voltage-gated conductance are dynamic, while the leak conductance does not change over time. Reversal potentials of sodium ($\ENa$) and potassium ($\EK$) have opposite sign.}%
\label{fig:theory.membrane_circuit_HH}
\end{figure}

Hodgkin and Huxley \cite{hodgkin1952quantitative} studied ion channel kinetics in the squid giant axon, a fiber that can be up to \SI{1}{\milli\metre} in diameter, providing a very accessible way for electrophysiologic measurements. 
They considered only three ion channel types: two voltage-dependent channels selective for sodium ($\naIon$) and potassium ($\kIon$), respectively, and one (virtual) leak channel.
This results in the electric circuit of \cref{fig:theory.membrane_circuit_HH} with three different conductances $\gL$, $\gNav$ and $\gKv$ with associated currents $\IL$, $\INa$ and $\IK$.
We largely follow \cite{koch2004biophysics} for the rest of this section.

Using Kirchhoff's current law for a parallel circuit, the equation for the current reads
\begin{align}
C_m \frac{\partial V}{\partial t} = \IL + \INa + \IK + \Iinj \ , \label{eq.hh_currents_sum}
\end{align}
where $V$ is the membrane potential and $C_m$ is the \emph{membrane capacitance} responsible for the capacitive current. $\Iinj$ stands for the injected current or, more generally, any current source contributing to the potential on the intracellular side of the membrane.
The ionic currents can be expanded to
\begin{align}
\IL &= \gL (\EL - V) \\
\INa &= \gNav (\ENa - V) \label{eq.hh_currents} \\
\IK &= \gKv (\EK - V) \ ,
\end{align}
where $E_{i}$ denotes the reversal potential of the ion channel $i$, as can be calculated by \cref{eq:nernst}.
When the channel is permeable to more than one ion species -- as the single leak channel above -- its reversal potential is usually calculated using \cref{eq:goldman} or \cref{eq:pcm}.

The constant leak conductance $\gL$ is the inverse of the \emph{membrane resistance}. 
It is very common to express the electric quantities in terms of unit length or unit area, since the measured values are often only available as \emph{densities}. 
The membrane resistance then becomes a \emph{resistivity} $r_L = R_L \pi r^2$ with the radius of the neurite $r$ in \si{\centi\metre}, yielding the unit \si{\ohm\square\centi\metre}. 
Analogously, the capacitance $c_m$ per unit area in units \si{\farad\per\square\centi\metre} can be used instead of the total capacitance $C_m$.

The conductances above can also be expressed in terms of \emph{conductivities} (commonly in units \si{\milli\siemens\per\square\centi\metre}), and the currents change to \emph{current densities}, which need to be multiplied by the membrane area to get an absolute current.

The remaining active channel conductances $g_i(V, t)$ are defined as
\begin{subequations}
\label{eq:channel_conductances}
\begin{align}\label{eq:theory.hh_conductances}
   g_{\text{Na}_v} &= \bar{g}_{\text{Na}_v} m^3 h \\
   g_{\text{K}_v} &= \bar{g}_{\text{K}_v} n^4 \ ,
\end{align}
\end{subequations}
where $m, h, n$ are time- and voltage-dependent \emph{gating particles} -- taking values from the interval $[0,1]$ -- for the sodium activation, sodium inactivation, and potassium activation, respectively. 
In combination they state which fraction of the maximum conductance $\bar{g}_{i}$ of channel $i$ is open. 
It is notable that the concept of gating particles is a purely theoretical one, i.e.~there is no direct physical equivalent in the chemical structure of an ion channel.

The kinetics of gating particles are given by the \glspl{ODE}
\begin{subequations}
\label{eq:gating_particles}
\begin{align}\label{eq:theory.hh_gating}
\diff{n}{t} &= \alpha_n(V)(1-n) - \beta_n(V)n \\
\diff{m}{t} &= \alpha_m(V)(1-m) - \beta_m(V)m \\
\diff{h}{t} &= \alpha_h(V)(1-h) - \beta_h(V)h
\end{align}
\end{subequations}
with corresponding rate functions
\begin{subequations}
\label{eq:rate_functions}
\begin{align*}
\alpha_n(V) &= c_T 0.01 \vtrap(10-(V-V_{\text{rest}}),10) \\
\beta_n(V) &= c_T 0.125 e^{-(V-V_{\text{rest}}) / 80} \\
\alpha_m(V) &= c_T 0.1 \vtrap(25-(V-V_{\text{rest}}),10) \\
\beta_m(V) &= c_T 4 e^{-(V-V_{\text{rest}}) / 18} \\
\alpha_h(V) &= c_T 0.07 e^{-(V-V_{\text{rest}}) / 20} \\
\beta_h(V) &= c_T \frac{1}{ e^{(30-(V-V_{\text{rest}})) / 10} +1} \ ,
\end{align*}
\end{subequations}
which were slightly adapted to account for temperature dependence by a factor 
\begin{align*}
c_T = 3^{\frac{T-6.3}{10}}
\end{align*}
and zeros in the denominator of rate functions by a function vtrap as used in NEURON \cite{neuron-vtrap}:
\begin{align*}
\vtrap(x,y) =
\begin{cases}
\frac{x}{e^{x/y}-1} & x/y \neq 0 \\
y(1-\frac{x}{2y}) & \text{else} \ . 
\end{cases}
\end{align*}
This completes the description of the \gls{HH} model. 
It has since been modified to include arbitrary ion channel types in addition to the ones present in the original model by defining rate functions, gating particle kinetics and maximum conductances for each added channel, yielding one additional current in \cref{eq.hh_currents_sum}.
Adding dependence on the concentration of a certain ion or molecule is straightforward, given a good model for the time-evolution of concentrations is available. 
Models that follow the basic scheme of the original \gls{HH} are called \emph{\gls{HH}-type models} and are probably the most successful ones in modeling membrane excitation.

\subsection{Cable Equation}
\begin{figure}
\begin{center}%
\centering%
\includestandalone[width=0.8\textwidth]{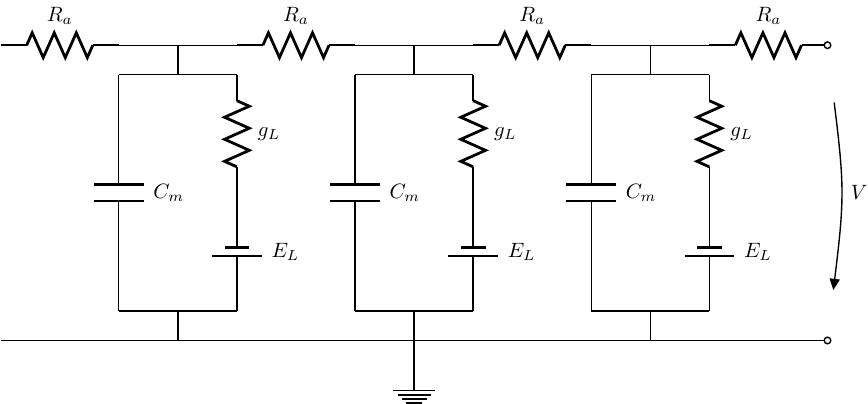}%
\end{center}%
\mycaption[Equivalent circuit for a spatially extended neuronal membrane]{Multiple (in this case passive) membrane patches are connected through axial resistances $R_a$, which in general are position-dependent.}%
\label{fig:theory.membrane_circuit_cable}
\end{figure}

It is important to note that the \gls{HH} model in \cref{sec:theory.biophysics.hh} only represents a single patch of axon, as there is no spatial dependence. 
This corresponds to reducing the neuron to a point in space.
In the real world, however, neurons can show quite complex morphologies, and channel types and densities can vary significantly across different parts of the neuron. 
In search of a mathematical model for this, neuroscientists have rediscovered cable theory -- originally developed to study signal decay in underwater telegraphic cables by Lord Kelvin -- to describe the potential spread in complicated neuronal morphologies (cf.~e.g.~\cite{rall1989cable}). 
The \emph{cable equation} reads
\begin{align}
\frac{1}{R_a(x)} \frac{\partial^2 V}{\partial x^2} = C_m(x) \frac{\partial V}{\partial t} - I_{\text{memb}}(x)
\end{align}
with the axial (cytosol) resistivity $R_a$ and the membrane capacitance $C_m$ (see the schematic equivalent circuit in \cref{fig:theory.membrane_circuit_cable}). 
As in the case of the \gls{HH} model, quantities may be expressed per unit length. 
In any case, $R_a(x)$ and $C_m(x)$ are depending on the position-dependent radius $r(x)$ of the neurite.

This equation is a one-dimensional parabolic diffusion-reaction \gls{PDE} and essentially describes the potential propagation as a transient diffusive process along a cylinder of varying radii, where the signal velocity depends on cytosol resistivity $R_a$ (and implicitly also the fiber diameter $d=2r(x)$). 
The right-hand side corresponds to the \gls{HH} \cref{eq.hh_currents_sum}, where all trans-membrane contributions to the intracellular potential are lumped together into a single term $I_{\text{memb}}(x)$.

The cable equation has been very successful for modeling excitation of spatially extended neurons, and the simulation program NEURON \cite{hines1997neuron} is the de-facto standard among neuroscientists.

\subsection{Extracellular Space}
So far we have only seen models of (parts of) the neuronal membrane and the signal propagation \emph{within} neurons. 
What is missing in these models is the extracellular space. 
This is an important structure, as large amount of experimental data today is obtained by extracellular measurements. 
From the theoretical point of view, it is therefore desirable to have a suitable model of how signals are transmitted in extracellular space.

One more note on the wording. In the literature, the term \gls{LFP} is used with slightly different meanings, apart from the fact that it is poorly named anyway. 
In the most general sense, it describes the potential time course obtained at a single point in extracellular space as a superposition of potential contributions stemming from a number of surrounding neurons. 
In most experimental contexts, however, the term describes an already low-pass filtered signal, which does not contain any fast components like those from action potentials. For the latter, the term \gls{EAP} has found its way into the terminology.
It should be noted that in the following, when the term \gls{LFP} is used, we denote with this the \emph{unfiltered} potential at any point in space, regardless of the number of contributing cells (in the present case, only one or two).
Since we only consider axonal membrane currents following an \gls{AP} as contributions, it should be clear that this is essentially synonymous to an \gls{EAP} in the absence of any synaptic currents or contributions by other cells, in contrast to the common usage.

\subsubsection{Volume Conductor Theory}\label{sec:theory.volume_conductor}
The field of volume conductor theory deals with the propagation of electric and magnetic fields within volume conductors like the brain \cite{plonsey1995volume}. 
It therefore provides a model of the extracellular space, without explicitly representing its complex geometry. 
Instead, it is characterized as a homogeneous medium with an effective conductivity $\sigma$. 
It is based on the quasi-static approximation of Maxwell's equations by neglecting the influence of magnetic fields on the electric field and the existence of any free charges in the medium. 
The electromagnetic part can be neglected by estimating its effect in the relevant frequency range and finding that it is orders of magnitude smaller than the electrostatic contributions. 
When the timescales for the balancing of free charge in the medium are compared with the timescales of neural excitation, the former is found to happen so fast it can be regarded as instantaneous, see \cite{agudelo2013numerical} for an in-depth derivation. 
This results in a Laplace equation
\begin{align}
\Delta \phi &= 0 &&\text{in $\Omega$} \label{eq:volume_conductor} \\
\phi &= g &&\text{on $\Gamma_D$} \nonumber \\
\sigma \nabla \phi\cdot \mathbf{n} &= j &&\text{on $\Gamma_N$} \nonumber 
\end{align}
for the potential due to \emph{current densities} $j$ at the cell boundary and a given conductivity $\sigma$ of the extracellular medium. This equation is dual to the electrostatic equation
\begin{align}
\Delta \phi &= 0 &&\text{in $\Omega$} \label{eq:electrostatic} \\
\phi &= g &&\text{on $\Gamma_D$} \nonumber \\
\epsilon \nabla \phi\cdot \mathbf{n} &= f &&\text{on $\Gamma_N$} \nonumber 
\end{align}
due to a boundary charge density $f$ and an electric permittivity $\epsilon$ of the medium.
Note that both equations have a homogeneous right-hand side by the absence of any free charges in the medium.

Interestingly, \cref{eq:volume_conductor} uses current sources, which are inherently time-dependent. 
It therefore provides a stationary state for the extracellular potential which will establish given the conductivity of the medium. 
Following the above reasoning that this process happens instantaneously, the equation takes the same form as \cref{eq:electrostatic}, predicting an immediate potential response due to a given charge density.

\subsubsection{Line Source Approximation}\label{sec:theory.lsa}
The line source approximation introduced in \cite{holt1999electrical} is an analytical solution to the volume conductor \cref{eq:volume_conductor} under a certain geometry approximation, namely, the finite thickness of the neuron fibers is neglected and collapsed to a line.
It is widely used as an effective model to compute the extracellular potential at any point in space, using only the values of the membrane currents at a finite number of line segments. 

The equation for a point $(r,h)$ due to a single line source $j$ reads
\begin{align}\label{eq:lsa}
   \Phi_j(r,h) = \frac{\rho I_j}{4 \pi \Delta s} \log \left| 
\frac{\sqrt{h^2+r^2}-h}{\sqrt{l^2+r^2}-l} \right| \ ,
\end{align}
where $I_j$ is the \emph{total transmembrane current} of line $j$, $r$ is the radial distance from the line of length $\Delta s$, $h$ is the longitudinal distance from the end
of the line, and $l = \Delta s + h$ is the distance from the start of the line. 
The parameter $\rho = \frac{1}{\sigma}$ describes the resistivity of the extracellular medium. 
The complete extracellular potential $\Phi$ at any point $(r,h)$ of a neuron morphology consisting of a number $M$ of connected lines is then simply given by the superposition of all line source potentials:
\begin{align}\label{eq:lsa_sum}
   \Phi(r,h) = \sum_{j=1}^M \Phi_j(r,h) \ .
\end{align}
This model is very convenient especially when using cable equation models based on a line segment approximation of the original neuron geometry, since all necessary data is readily available and the only free parameter is $\rho$.
Note also that -- in contrast to a numerical solution of \cref{eq:volume_conductor} -- the \gls{LSA} \cref{eq:lsa} does not need to specify any boundary conditions next to the membrane currents, as it implicitly fulfills a potential of zero at infinity.

\subsubsection{The Poisson-Nernst-Planck Equations of Electrodiffusion}\label{pnp_system}%
The most general approach to model a neuron and the surrounding space is to consider the -- in the true sense of the word -- atomic computation unit of the brain: the ion. 
In contrast to the approach of volume conductor theory in \cref{sec:theory.volume_conductor}, free charges are not neglected, but modeled explicitly through ion concentrations.

A suitable model for the movement of ions in a static solvent considers both chemical diffusion (due to a concentration gradient) and electrostatic drift (attraction/repulsion due to a potential gradient), often summarized by the term \emph{electrodiffusion}.

This is described by the Nernst-Planck equation
\begin{subequations}
\begin{align}
\frac{\partial n_i}{\partial t}+\nabla\cdot\mathbf{F}_i &= 0
\intertext{with the ion flux}
\mathbf{F}_i &= -D_i\left(\nabla n_i+z_in_i\nabla\phi\right) \ ,
\end{align}\label{eq:np}
\end{subequations}
where $n_i, \ i=1,\ldots,N$ are defined as relative concentrations (with respect to a scaling concentration $n^*=N_A$, the Avogadro constant) with units \si{\milli\molar} for the $N$ different ion species, $z_i$ (as before) is the valence and $D_i$ the (position-dependent) diffusion coefficient of ion species $i$. 
Together with the Poisson equation for the electric potential
\begin{align}
\nabla\cdot\left(\epsilon\nabla\phi\right) &= -\frac{e^2n^*}{\epsZero\kB T}\sum_i z_in_i \ ,
\label{eq:p}
\end{align}
this constitutes the \gls{PNP} system.

\Cref{eq:np} is of convection-diffusion type and describes the time-dependent change in concentrations due to diffusion and drift through an electrical field, while the elliptic \cref{eq:p} gives the electric potential $\phi$ at any point in space. 
We can see that each \gls{PDE} is linear in its primary variable, but together the system is nonlinear due to the coupling.

Neglecting the finite ion size and representing the ion concentrations as continuous variables in a mean-field approach was found to be valid \cite{lipowsky1995structure}, as the typical size of an ion (about $\SI{1}{\angstrom} = \SI{100}{\pico\metre}$, the extent of the \emph{Stern layer}) is one order of magnitude smaller than the smallest relevant spatial scales (Debye length, about \SI{1}{\nano\metre} for typical ion concentrations in body fluids).

Let us finally spend a few words on the analysis of solutions for the general 3D \gls{PNP} system. 
The majority of publications using the \gls{PNP} system come from the field of semiconductor analysis, although many of them also deal with the electric fields inside ion channels and around biomolecules.
\Cite{Lu20106979} provides a good starting point for this, as it gives a comprehensive overview over previous works involving the \gls{PNP} equations and also touches the topic of its mathematical analysis. 
Existence and stability of solutions has been shown for the steady-state case \cite{jerome1991finite}; solutions are presumed not to be unique and it is assumed that restrictions on the boundary conditions would have to be made for this to hold. 

For the instationary case under closed-system boundary conditions, existence and convergence to stationary solutions for long-time behavior could be shown \cite{biler1994debye}. 
In \cite{eisenberg2007poisson}, the system is analyzed in the presence of permanent charges, explicitly mentioning ion channels as one source. 
In \cite{wu2013global}, existence and uniqueness of global weak solutions to the general drift-diffusion-Poisson system are shown in the presence of an additional reaction term; references therein treat this system under various boundary conditions.

The literature is quite extensive, but we are not aware of a statement on the well-posedness of the general problem \cref{eq:np,eq:p} under arbitrary boundary conditions, especially not for the model at hand -- which will be introduced in the following chapter -- using two (intra- and extracellular) electrolytes separated by a membrane with imposed nonlinear boundary flux conditions by the \gls{HH} ion channel model.

\bigskip\bigskip

After this tour through different neuron models, we have now arrived at the most general continuum model. 
On first sight, this system only describes ion movements in electrolytes. 
We will see in the following chapter how to include the membrane and its ion channels into this framework in order to obtain a very detailed electrodiffusion model of a neuron and its surrounding.

\setchapterpreamble[u]{%
\dictum[Hans-Peter Gail]{Numerik ist ein sehr schmutziges Gesch\"aft.}\bigskip}
\chapter{General Model of an Axon in Extracellular Space}\label{chap:model_general}
The model considered throughout this work will deal exclusively with the axonal part of the neuron. We first mention the governing equations and their boundary conditions, then the numerical methods used to solve those equations in an efficient and accurate way are presented.
This chapter is largely based on an edited version of \cite{pods2013electrodiffusion}.

\section{General Remarks}
In \cref{sec:intro.related}, we mentioned the basic assumptions most cable equations models are based on today.
Half of those can be dropped when using the \gls{PNP} equations of electrodiffusion.

We do not neglect changes in the radial direction (1), nor do we regard the extracellular space as isopotential (2).
Most importantly, we explicitly represent the concentrations as primary variables, which are allowed to change in space and time, in contrast to assumption 6.
Furthermore, we couple intra- and extracellular space, such that the assumption of independence (7) can be discarded as well.

The remaining assumptions are supposed to hold, i.e. the validity of the mean-field approach that allows us to represent all quantities as continuous variables (4), neglecting magnetic fields (3) and the equivalent circuit assumption from the \gls{HH} model (8).
Assumption 5, which states that the \gls{ES} can be regarded as a homogeneous medium, is not a necessary one in our model, as permittivity $\epsilon$ and diffusion coefficients $D_i$ may in principle be position-dependent.
However, we chose these coefficients to be homogeneous and constant in all considered cases.

\section{A Simplified Axon Model in Cylindrical Coordinates}\label{sec:neuron_model.2d}
As the solution of a full 3D system is computationally expensive, we exploit the rotational symmetry of an idealized unbranched axon. 
By representing the computational domain in cylinder coordinates and assuming there is no change in angular direction, it can be reduced to two dimensions.
This enables the calculation of 3D results with a drastically reduced computational complexity. 
For the numerical solution, the rectangular elements of the computational grid will be treated as (hollow) cylinders, and the volumes and integration elements will be calculated accordingly.

\begin{figure}
\begin{center}%
\subfloat{%
\hspace{1cm}%
\includestandalone[width=0.6\textwidth]{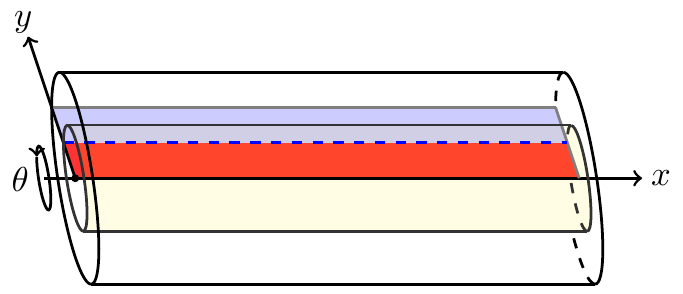}%
}\\
\subfloat{%
\centering%
\includestandalone[width=\textwidth]{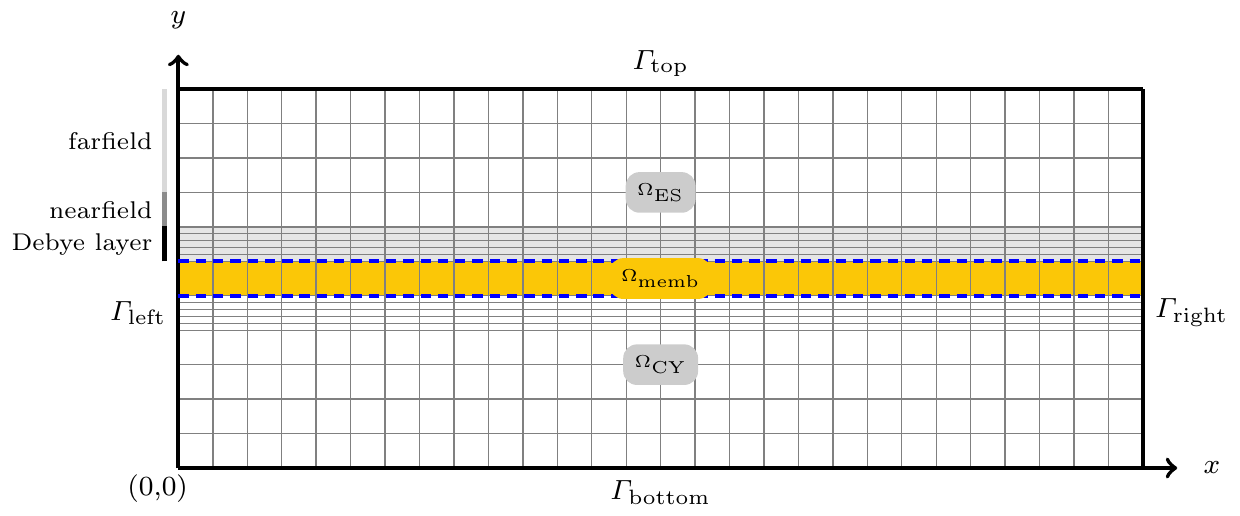}%
}
\end{center}%
\mycaption[Two-dimensional computational domain for the cylinder-symmetric axon model]{The upper part shows the cylinder geometry into which the 2D computational grid is embedded, assuming continuous symmetry in the angular direction.
In the lower part, the domain boundary $\GammaExt$ is represented by solid lines, while the interior (electrolyte-membrane) boundary $\GammaInt$ is plotted with dashed lines.
This divides the domain into three subdomains: the (non-connected) electrolyte domain $\OmegaElec = \OmegaCytosol \cup \OmegaExtra$ consisting of two subdomains and the separating membrane subdomain $\OmegaMemb$.
The Debye layer of $\OmegaExtra$ close to the membrane is highlighted in gray, followed by the nearfield and farfield parts.
The lower boundary represents the inner-cell symmetry axis.
Note that this scheme is not to scale, as the actual mesh sizes in $y$-direction differ by several orders of magnitude between Debye layer and farfield, making the grid very anisotropic.}%
\label{fig:domain_2D}
\end{figure}

In the upper part of \cref{fig:domain_2D}, the cylinder geometry is shown with the two-dimensional subset highlighted, constituting the effective computational domain.
The $x$-axis represents the domain's symmetry axis in \emph{axial} direction (usually denoted $h$ or $z$ in cylinder coordinates), eliminating the angular coordinate $\theta$ from the equations.
The $y$-axis is usually denoted by $r$ or $\rho$ in cylinder coordinates and stands for the \emph{radial} direction. 
For historical reasons\footnote{Previous versions of this model were in 1D and plain 2D, with the coordinate axes named accordingly.} we will stick to the variables $x$ and $y$ for the coordinate axes in the following.

The domain consists of three partitions: cytosol, membrane, and extracellular space.
Cytosol and extracellular space are electrolytes, yielding the electrolyte domain $\OmegaElec = \OmegaCytosol \cup \OmegaExtra$. 
It may contain an arbitrary number $N$ of concentrations of different ion species.
Here, however, we will restrict ourselves to the minimal set of $N=3$ monovalent species sodium ($\naIon$), potassium ($\kIon$), and chloride ($\clIon$).
Sodium and potassium are needed for the ion channel dynamics triggering an action potential; chloride is a representative of the anions needed for electroneutrality in the bulk solution and does not cross the membrane in this model.

The membrane domain $\OmegaMemb$ separates the two parts of the electrolyte domain, therefore $\OmegaElec$ is not connected. $\OmegaElec$ and $\OmegaMemb$ form a partition of the computational domain: $\bar{\Omega} = \bar{\Omega}_{\text{elec}} \cup \bar{\Omega}_{\text{memb}}$.

\section{PNP Equations and Boundary Conditions}\label{sec:model.boundary_conditions}
This setup necessitates the introduction of additional boundary conditions on the membrane-electrolyte-interface $\GammaInt$ -- next to the obligatory boundary conditions on the domain boundary $\GammaExt$ -- such that the set of all boundary points is given by $\varGamma = \GammaInt \cup \GammaExt$.
To properly define the boundary conditions in this setup in a general way, we first need to introduce a little bit of notation for the description of the boundaries and their corresponding boundary condition types.

As can be seen in \cref{fig:domain_2D}, the exterior boundary $\GammaExt = \GammaExtBottom \cup \GammaExtLeft \cup \GammaExtRight \cup \GammaExtTop$ consists of four parts, and the interior boundary $\GammaInt = \GammaIntCytosol \cup \GammaIntExtra$ of two non-connected parts at either side of the membrane.

To increase the notational complexity even more, each of these boundary subsets may again be a partition of two subsets for Dirichlet and Neumann conditions, denoted schematically by subscripts $\varGamma_{\cdot,D}$ and $\varGamma_{\cdot,N}$, respectively. 
Additionally, the boundary condition type may be different for each equation of the \gls{PNP} system. 
In this context, however, it will not be necessary to distinguish between individual concentrations; is is sufficient to have one set of boundaries for each equation type, denoted by the addition of superscripts $\varGamma_{\cdot,\cdot}^{\text{P}}$ and $\varGamma_{\cdot,\cdot}^{\text{NP}}$ for Poisson and Nernst-Planck equation, respectively.

In summary, we arrive at the final notation of a boundary in the schematic form $\varGamma_{L,T}^{(E)}$, where $L$ specifies the \emph{location} of the boundary, $T$ the boundary condition \emph{type} (\emph{D}irichlet or \emph{N}eumann), and $E$ the \emph{equation} for which it is defined (\emph{P}oisson or \emph{N}ernst-\emph{P}lanck).

With this notation of the boundaries, we now have the Nernst-Planck \cref{eq:np} defined on $\OmegaElec$ with boundary conditions
\begin{subequations}\label{eq:np_bc}
\begin{align}
\mathbf{F}_i \cdot \mathbf{n} &= j_i^{\text{(NP)}} \quad \text{on } \GammaExtN^{\text{(NP)}} \cup \GammaInt^{\text{(NP)}} \label{eq:np_bc_N}\\
n_i &= g_i^{\text{(NP)}} \quad \text{on } \GammaExtD^{\text{(NP)}} %
\label{eq:np_bc_D}
\end{align}
\end{subequations}
and the Poisson \cref{eq:p} defined on the whole domain $\Omega$ with boundary conditions
\begin{subequations}\label{eq:p_bc}
\begin{align}
\epsilon\nabla\phi \cdot \mathbf{n} &= j^{\text{(P)}} \quad \text{on } \GammaExtN^{\text{(P)}} \label{eq:p_bc_N} \\%
\phi &= g^{\text{(P)}} \quad \text{on } \GammaExtD^{\text{(P)}} \ ,\label{eq:p_bc_D}%
\end{align}
\end{subequations}
where $\mathbf{n}$ denotes the unit outer normal vector.
For consistency, of course, $\GammaExtN^{(\cdot)} \cup \GammaExtD^{(\cdot)} = \GammaExt$ holds for each equation. 
From \cref{eq:np_bc} we see that the internal concentration boundary is always a Neumann boundary, as we want to describe membrane \emph{fluxes}, which are equivalent to the more commonly used term of membrane \emph{currents} by a constant factor $ezn^*$. 
For the external boundaries, any combination of boundary conditions is possible in principle. 
Nevertheless, as the solution of the Poisson \cref{eq:p} is only defined up to a constant, we need at least one point with a Dirichlet condition in order to have a well-posed problem with a unique solution.

One more word about the domains of \cref{eq:np} and \cref{eq:p}, which are obviously different:
the Poisson equation is defined (and continuous) on the whole domain, while the Nernst-Planck equation is only defined on the electrolyte subdomain. 
This means that, for simplicity, we assume the membrane to be free of charge carriers. (Fixed) membrane surface charges could easily be added as an additional source term $f_{\text{P}}$ in the Poisson equation, but those are not considered here. 

For the Nernst-Planck equation, the precise locations of \cref{eq:np_bc,eq:p_bc} may be problem-dependent, but in most considered cases they will be the same. Therefore, we give the definition that holds for the majority of setups hereafter; if boundary conditions are defined differently, this will be explicitly stated in the following. The general scheme reads
\newcommand{\FixedSize}[1]{\makebox[2.7cm][l]{\ensuremath{#1}}}%
\begin{align*}
   j_i^{\text{(NP)}} &= 
\begin{cases}
  f_i^{\text{memb}}(n_i, \phi, t) \quad & 
\text{on } \GammaInt^{\text{(NP)}}\\
  0 \quad & \text{on } \GammaExtN^{\text{(NP)}} = \GammaExt \setminus \GammaExtTop
\end{cases}\\
   g_i^{\text{(NP)}} &= \FixedSize{n_i^0} \quad \text{on } \GammaExtD^{\text{(NP)}} = \GammaExtTop\\
   j^{\text{(P)}} &= \FixedSize{0} \quad \text{on } \GammaExtN^{\text{(P)}} = \GammaExt \setminus \GammaExtTop\\
   g^{\text{(P)}} &= \FixedSize{0} \quad \text{on } \GammaExtD^{\text{(P)}} = \GammaExtTop \ ,
\end{align*}
where we call the constant $n_i^0$ the \emph{bulk concentration} of species $i$. We can see that in this general scheme the locations of Neumann and Dirichlet boundary conditions match for Poisson and Nernst-Planck equations, respectively.

The membrane flux function $f_i^{\text{memb}}$ is the heart of this model and will be defined in the next section. The Dirichlet boundary conditions for the concentrations $g_i^{\text{(NP)}}$ on the upper exterior boundary model an infinite reservoir for each ion species. The potential is clamped to zero at the upper extracellular boundary by means of $g^{\text{(P)}}$, which introduces an error equal to the value of the real potential value $\hat{\phi}(y_{\text{max}})$ calculated for an unrestricted domain, where the potential is 0 for $y \rightarrow \infty$. 

For a point charge on the membrane, this error would correspond to an absolute shift at each point in the domain. 
As the potential of a point charge falls off as $\frac{1}{r}$ in 3D, an increase of the domain size by a factor 100 will reduce the error by a factor $\frac{1}{100}$.
For a finite line charge (like in the present case of an active membrane), this is only true if the radial distance is large compared to the length of the line charge.
Otherwise, the potential drop is a logarithmic function of the membrane distance (cf.~\gls{LSA}, e.g.~in \cite{gold2006origin} and \cref{eq:lsa}). 
In any case, increasing the domain size in $y$-direction will reduce the error introduced by the upper Dirichlet boundary. 
A sufficiently large domain size of $y_{\text{max}} = \SI{10}{\milli\metre}$ was chosen to account for this.

At the lower boundary, which represents the intracellular symmetry axis, the potential gradient and ion fluxes are vanishing by the definition of $j_i^{\text{(NP)}}$ and $j^{\text{(P)}}$, such that no boundary artifacts are introduced.

\section{Derivation of the Membrane Flux}\label{sec:neuron_model.membrane_flux}
The most important part for the boundary conditions -- and, as we will see later on, the dynamics of the system as a whole -- are the membrane currents. 
To represent those, we use the \gls{HH} system from \cref{sec:theory.biophysics.hh} in a slightly modified form: the leak channel was split into two separate leak channels for $\naIon$ and $\kIon$, respectively.
 The corresponding conductances now read
\begin{align}
\gNa &= \gNav + \gNaL\\
\gK &= \gKv + \gKL \\
\gCl &= 0 \ ,
\end{align}
where the voltage-dependent parts $\gNav$ and $\gKv$ of the total conductances $\gNa$ and $\gK$ are given in \cref{eq:theory.hh_conductances} as before.
The leak parts $\gNaL$ and $\gKL$ add up to the total leak conductance of the original \gls{HH} model,
\begin{align*}
\gNaL + \gKL = \gL \ .
\end{align*}
The reason for splitting up the leak channel becomes apparent when we look at the Goldman \cref{eq:goldman}: here, different permeabilities of the membrane for different ion species determine the membrane potential, given the intra- and extracellular bulk concentrations.
When we now think in terms of \gls{HH}-type ion channels and, hence, in terms of conductances instead of permeabilities, the \gls{PCM} \cref{eq:pcm} is the right choice for the membrane potential, using \emph{relative leak conductances} $\frac{\gNaL}{\gL}$ and $\frac{\gKL}{\gL}$.

Whether we take the view of permeabilities (Goldman) or conductances (\gls{PCM}), it is the \emph{relative} quantities of channels that matter.
We need a means to express these relative values in the model, since a single channel -- as in the original \gls{HH} model -- is not sufficient if we include concentrations explicitly.

For the definition of the membrane flux, each point $\mathbf{x} \in \GammaIntCytosol$ on the cytosol-membrane interface is associated with a point $\mu(\mathbf{x}) \in \GammaIntExtra$ on the opposite membrane-extracellular interface by a map $\mu(\mathbf{x}) = \mathbf{x} + d_{\text{memb}} \cdot \mathbf{n_{\text{CY}}}$, where $d_{\text{memb}}$ is the membrane thickness and $\mathbf{n_{\text{CY}}}$ is the unit outer normal at $\mathbf{x}$ pointing in the direction from cytosol to membrane.
The values of potential and concentrations evaluated at these points are called $\phi^{\text{CY}} = \phi(\mathbf{x})$, $n_i^{\text{CY}} = n_i(\mathbf{x})$, $\phi^{\text{ES}} = \phi(\mathbf{\mu(x)})$ and $n_i^{\text{ES}} = n_i(\mathbf{\mu(x)})$.

We now take the single Hodgkin-Huxley membrane current 
\begin{align*}
I_i^{\text{memb}} = g_i(\phi, t) (\membPot - E)
\end{align*}
for the membrane potential $\membPot = \phi^{\text{CY}} - \phi^{\text{ES}}$ and replace the constant battery $E$ by a variable concentration-dependent reversal potential calculated from the Nernst \cref{eq:nernst}. After adding the necessary scaling factors to bring $f_i^{\text{memb}} = I_i^{\text{memb}}/ ezn^*$ to SI units \si{\mole\per\square\metre\per\second}, we obtain the concentration-, potential- and time-dependent \emph{membrane flux} $f_i^{\text{memb}}$ of species $i \in \{\text{Na}, \text{K}\}$,
\begin{align}
f_i^{\text{memb}}(\mathbf{x}) = f_i^{\text{memb}}(\mu(\mathbf{x}))
= g_i(\phi, t) \frac{k T}{e^2 z^2 n^{\ast}} \left( z \membPot
+ \ln \frac{n_i^{\text{ES}}}{n_i^{\text{CY}}}
\right) \ . \label{eq:fmemb_2D}
\end{align}
Note that two opposite points on the membrane interface are identified with each other here, i.e.~the membrane thickness is essentially neglected.
This is a compulsory assumption with respect to the underlying membrane model, which was replaced by an equivalent circuit by the \gls{HH} model in \cref{fig:theory.membrane_circuit_HH}, where the spatial extent of the membrane is inherently not represented. 

But this is also reasonable from the physical point of view, as the membrane thickness (in the range of a few \si{\nano\metre}) is so small that any delay from an ion crossing the membrane can be neglected and considered instantaneous in view of the governing \gls{AP} time scales in the range of milliseconds. 
Of course, the Poisson equation is not affected by this.
The correct potential fall-off over the membrane is respected, as its spatial extent is represented explicitly.

This interior boundary condition fits nicely into our framework, as it unifies the potential-dependent \gls{HH} system with the concentration-dependent Nernst equation to arrive at an expression that represents all the features of the potential- and concentration-dependent ion flux $\mathbf{F}_i$ of the \gls{PNP} system.
An equilibrium state of this model is expected to satisfy the Nernst \cref{eq:nernst} (for a single ion species) or the \gls{PCM} \cref{eq:pcm} (for multiple species) for the potential as well as the concentration distribution \cref{eq:pb_conc} predicted by the Poisson-Boltzmann \cref{eq:pb}. 
This will be tested in \cref{chap:unmyel}.

\section{Numerical Methods}\label{sec:model.numerical_methods}
\subsection{Weak Form and Discretization}
For the numerical solution of the \gls{PNP} system as defined in \cref{pnp_system} and with boundary conditions given in \cref{sec:model.boundary_conditions}, we use the finite element method introduced in \cref{sec:theory.fem} with piecewise linear, globally continuous $Q_1(\mathcal{T})$ nodal basis functions on an axi-parallel rectangular grid, as defined in \cref{sec:theory.fem.function_spaces}. 

More specifically, the computational grid $\mathcal{T}$ will always be \emph{tensor grid}, i.e.~it can be written as the Cartesian product of the two coordinate vectors, $\mathcal{T} = X \times Y$.
This allows for several optimizations in the numerical code, which turn out to be beneficial for the computational efficiency.

For application of the \gls{FEM}, we obtain the weak formulation in residual formulation for each equation $i$, $i=1,\ldots,N+1$ of the \gls{PNP} system by multiplication with test functions $v_i$, integrating over the respective domain and applying integration by parts.
We assume the $v_i$ have been chosen to fulfill the respective Dirichlet boundary conditions $g_i^{\text{(NP)}}$ and $g^{\text{(P)}}$.

For the Nernst-Planck equation, the temporal part
\begin{subequations}
\begin{align}
R_{\text{NP,T}} &= \int_{\OmegaElec} \frac{\partial n_i}{\partial t} v_i \, dx \ , \qquad i=1,\ldots,N \label{eq:residual_np_time}
\end{align}
and the spatial part
\begin{align}
R_{\text{NP,S}} &= \int_{\OmegaElec} D_i \left( \nabla n_i + z_i n_i \nabla \phi \right) 
\cdot \nabla v_i \, dx \nonumber \\
& + \int_{\Gamma_{\text{N}}} j_{\text{NP}} v_i \, ds \ , \qquad i=1,\ldots,N \label{eq:residual_np_space}
\end{align}
are combined yielding
\begin{align}
R_{\text{NP}} &= R_{\text{NP,T}} + R_{\text{NP,S}} \ . \label{eq:residual_np}
\end{align}
\end{subequations}
For the Poisson equation, we get
\begin{align}
R_{\text{P}} &= \int_{\Omega} - \epsilon \nabla \phi \cdot \nabla v_{N+1} 
+ \left( \frac{e^2 n^\ast}{\epsilon_0 k T} \sum_i z_i n_i \right) v_{N+1} \, dx \nonumber \\
& + \int_{\Gamma_{\text{N}}} j_{\text{P}} v_{N+1} \, ds \label{eq:residual_p}
\end{align}
and therefore the residual for the full system reads
\begin{align}
R &= \binom{R_{\text{NP}}}{R_{\text{P}}} \label{eq:residual_full} \ .
\end{align}
If we now denote the unknown function by $u = \left(n_1, \ldots,n_N, \phi \right)^T$ and the vector of test functions by $v = \left(v_1, \ldots,v_{N+1} \right)^T$, the finite element problem according to \cref{theory.fem.galerkin_orthogonality} can be written as
\begin{align}
  \text{Find $u\in Q_1^{N+1}(\mathcal{T})$} : \qquad R = R(u,v) = 0 \quad \forall v\in Q_1^{N+1}(\mathcal{T}) \ .
\end{align}

By applying the method of lines, each equation is discretized in space first by representing the unknown functions $n_i$ and $\phi$ as well as the test functions $v_i$ by $Q_1(\mathcal{T})$ nodal basis functions on the tensor grid $\mathcal{T}$, and then in time using the implicit Euler time-stepping scheme from \cref{eq:theory.pde_implicit_euler}.

\subsubsection{Space Discretization}\label{sec:model.numerical_methods.space_discretization}
As suggested in \cref{fig:domain_2D}, the grid is refined toward the membrane in $y$-direction.
This is essential in order to resolve the Debye length, the characteristic length scale over which the electrolyte ion concentrations deviate significantly from their bulk values close to the membrane (see \cref{sec:theory.biophysics.debye-hueckel}).

In $x$-direction, the grid is allowed to be much coarser, as there is no Debye layer to resolve.
This results in a very \emph{anisotropic} grid, especially at grid cells close to the membrane with ratios of up to $\frac{h_x}{h_y} = \num{200000}$ between mesh sizes in $x$- and $y$-direction.

The cylinder symmetry introduces another subtle difficulty for the numerical treatment of the system.
Since the cell volumes increase super-linearly in positive $y$-direction (roughly with $y\ dy$), the entries of the full residual $R$ in \cref{eq:residual_full} differ by several orders of magnitude ($10^9$ for a domain size of \SI{10}{\milli\metre}) solely by the presence of volume integrals in the weak form of the equations. This imposes a severe difficulty for the linear solver.

A \emph{threshold volume scaling} strategy is applied to account for this: at a certain distance from the membrane, a reference volume $V_{\text{ref}}$ is calculated.
All entries of the residual from an unknown at node $i$, where the corresponding volume $V_i$ is greater than $V_{\text{ref}}$, are scaled by a factor $V_{\text{ref}} / V_i$.
Here, $V_i$ is defined as the minimum volume of all adjacent cells of node $i$. 
So we are essentially compensating for the large cell volumes by scaling down those residual entries that stem from cells with a volume larger than the threshold volume $V_{\text{ref}}$.
By choosing a certain threshold volume $V_{\text{ref}}$, we can exclude the cytosol, membrane and Debye layer cells close to the membrane from this scaling procedure. 
These cells are comparably small in volume, but contribute large entries to the residual because of the exhibited gradients close to the membrane.
Therefore, scaling these residual entries would increase those already large entries, which is not desired. 
The threshold volume is chosen such that only residual contributions from large volume elements in more distant extracellular space are scaled down.

Mathematically, this  corresponds to multiplying a diagonal matrix from the left to the linear system, meaning that the \emph{same} linear system is solved in each Newton iteration. 
This scaling greatly improves the convergence properties of the Newton algorithm for this
cylinder geometry setup.

\subsubsection{Time Discretization}\label{sec:model.numerical_methods.time_discretization}
The choice of the time-stepping scheme strongly depends on the P\'{e}clet number (cf.~\cref{eq:peclet}), or, in the discrete case, the maximum \emph{cell P\'{e}clet number} of the Nernst-Planck \cref{eq:np},
\begin{align*}
  \text{Pe}_{h,\max} = \max_{\substack{t_k \in \mathcal{T}\\i \in 1,\ldots,N}} \left\| \frac{\mathbf{h}_k \cdot \boldsymbol{\omega}_i}{D_i} \right\|_{\infty} \label{eq:cell_peclet}
\end{align*}
with mesh size vector $\mathbf{h}_k = {h_x \choose h_y}$ and velocity vector $\boldsymbol{\omega}_i = D_i z_i \nabla \phi$ evaluated on each cell $t_k$, respectively.

A value of $\text{Pe}_h > 1$ means the advective (hyperbolic) part of \cref{eq:np} is dominating, otherwise the equation is diffusion-dominated and it is more parabolic.
Usually, for $\text{Pe}_h > 2$, an explicit time-stepping scheme is used to account for the advection-dominance limiting the maximally possible time step size. 
In the remaining cases, an implicit scheme will yield a stable solution and enable the use of larger time steps due to the diffusion dominance.

We see that the scalar diffusion coefficient $D_i$ cancels out in the above equation, such that, for monovalent ions, the condition $\text{Pe}_{h,\max} < 1$ is satisfied if the potential does not vary by more than $1$ (corresponding to about \SI{25}{\milli\volt}) over the extent of one grid cell.

For the present model with parameters in the physiological range and a mesh resolving the Debye layer, the potential gradient is always sufficiently fine resolved. 
Thus, the maximum grid P\'{e}clet number was always significantly smaller than 1, and the implicit Euler scheme was chosen.
Since the system is very susceptible to numerical oscillations, the choice of an implicit scheme provides an additional benefit.
Implicit schemes tend to smooth out these unphysical oscillations over time, while they may be amplified in explicit schemes.

The \gls{HH} system as the driving force of an \gls{AP} shows a great variability of the time scales on which the system dynamics act. During an \gls{AP}, membrane fluxes and potential differences are large and change rapidly, so a small time step is needed to capture the dynamics during this period. On the other hand, potential differences during inter-spike intervals are small and so are the magnitudes of ion fluxes, allowing for the use of a larger time step. 
Therefore, an adaptive time-stepping strategy is used to speed up the simulation by controlling the time step $\Delta t$ depending on the dynamics of the system.

The time step is bounded by minimum and maximum values of $\dtMin = \SI{0.05}{\micro\second}$ and $\dtMax$, respectively. 
The value of $\dtMax$ depends on the problem and will range between \SIrange{10}{50}{\micro\second}. During an action potential (membrane potential $\membPot > \SI{-50}{\milli\volt}$) or when an external stimulation is present, the maximum time step is additionally limited to $\dtMaxAP = \SI{10}{\micro\second}$.

Apart from these fixed bounds, the change of the time step depends on the number of Newton iterations $it_k$ needed to complete the
previous time step $k$:
\begin{align}
   \dt_{k+1} = \begin{cases}
      \dt_k \times 1.1 & it_k < \itMax \wedge it_k \leq it_{k-1} \\
      \dt_k / 1.2 & it_k > \itMin \\
      \dt_k       & \text{else} \ .
   \end{cases}
\end{align}
The upper and lower iteration bounds for adjusting the time step, $\itMax$ and $\itMin$, depend on the problem.

\subsection{Solving the PDE System}\label{sec:model.numerical_methods.newton}
As seen in \cref{pnp_system}, each equation \cref{eq:np} and \cref{eq:p} alone is linear in its unknowns. 
One could therefore use an operator-splitting approach and alternately solve the equations by a linear solver until convergence, in each time step. 
However, it could be observed that a very small time step (in the order of nanoseconds) is necessary to solve the system this way.

Since the nonlinearity of the whole system results from the coupling of both equations, it seems reasonable to represent this crucial feature also in the numerical method. 
The system is therefore solved fully-coupled using Newton's method, which requires the solution of one single large linear system in each iteration.

As described in \cref{sec:neuron_model.membrane_flux}, the dynamic channel conductances from the \gls{HH} scheme are needed to calculate the total membrane flux $f_i^{\text{memb}}$.
Hence, for each channel type $i$, one additional \gls{ODE} per gating particle has to be solved in each time step. 
This is done using a simple implicit Euler step.
The membrane flux is calculated \emph{once} at the beginning of each time step and kept fixed for the whole Newton iteration.
This way, the unknowns from the \gls{HH} scheme do not enter the full system matrix for the \gls{PNP} system, which avoids convergence issues arising from changing boundary conditions.
However, this introduces a small splitting error of the order of $\mathcal{O}(\Delta t)$.

The time step size is adapted in a rather conservative manner (cf.~\cref{sec:model.numerical_methods.time_discretization}).
Still, the Newton iteration might not converge for a chosen time step value $\dt$ for various reasons. 
A restart mechanism was implemented to account for this: if the solver does not converge for a given time step $\dt$, the procedure is repeated for the halved time step $\frac{\dt}{2}$.
A maximum number of three restarts is allowed, otherwise the simulation will be terminated.

A significantly higher stability was observed for the fully-coupled approach using Newton's method, leading to possible time steps of the order of tens of microseconds, as indicated by the values of time step thresholds from the previous section.

\subsubsection{Linear Solvers}
For small systems, SuperLU \cite{superlu99} was used, for larger systems with more than \num{50000} unknowns, a \gls{BiCGStab} iterative solver, preconditioned by an \gls{ILU} decomposition, turned out to be faster while maintaining the same accuracy. A combination of a restarted \gls{GMRes} method in combination with an \gls{AMG} preconditioner proved to be very robust and efficient when solving the system in parallel. 
Both \gls{ILU} and \gls{AMG} preconditioners are capable of coping with the grid anisotropy introduced by the spatial discretization in \cref{sec:model.numerical_methods.space_discretization}, and the usage of state-of-the art iterative linear solvers like \gls{BiCGStab} and \gls{GMRes} is crucial in order to attain a reasonable computation time for the problem at hand.

\subsection{Validation of the Numerical Algorithm}
To test the numerical algorithm from \cref{sec:model.numerical_methods} and to validate the implementation as a \Dune module, which will be covered in the next chapter, a number of simulations have been run and compared to recently found unsteady analytical solutions of the \gls{PNP} equations for the case of a single electrolyte domain \cite{schoenke2012}.

These solutions provide a useful test suite, as they allow for tests of the numerical methods with different boundary conditions and under different physical conditions. In particular, test cases for both diffusion- and advection-dominated conditions are presented. 

For the rest of this section, we use a normalized version of the \gls{PNP} system, where the temperature $T$ in \cref{eq:p} is chosen such that the constant $-\frac{e^2n^*}{\epsZero\kB T}$ on the right-hand side of the Poisson equation is equal to $1$. 
Similarly, the diffusion coefficient is fixed at $D=\SI{1}{\square\metre\per\second}$. 
As a consequence, the potential and concentrations are scale-free and it does not make much sense to express them in common physical units.
Here, we consider them as dimensionless.
The interval $[-5 \ 5]$ is used as the computational domain.

We now give a short overview over the three one-dimensional solutions\footnote{Since in 1D, the \gls{PNP} can be transformed to the \emph{Burgers equation}, each test case was named after a different burger, for no scientific reason.} and provide numerical results that demonstrate the proper convergence behavior. 

\begin{figure}%
\centering%
\subfloat[Potential]{\includegraphics[width=0.45\textwidth]%
{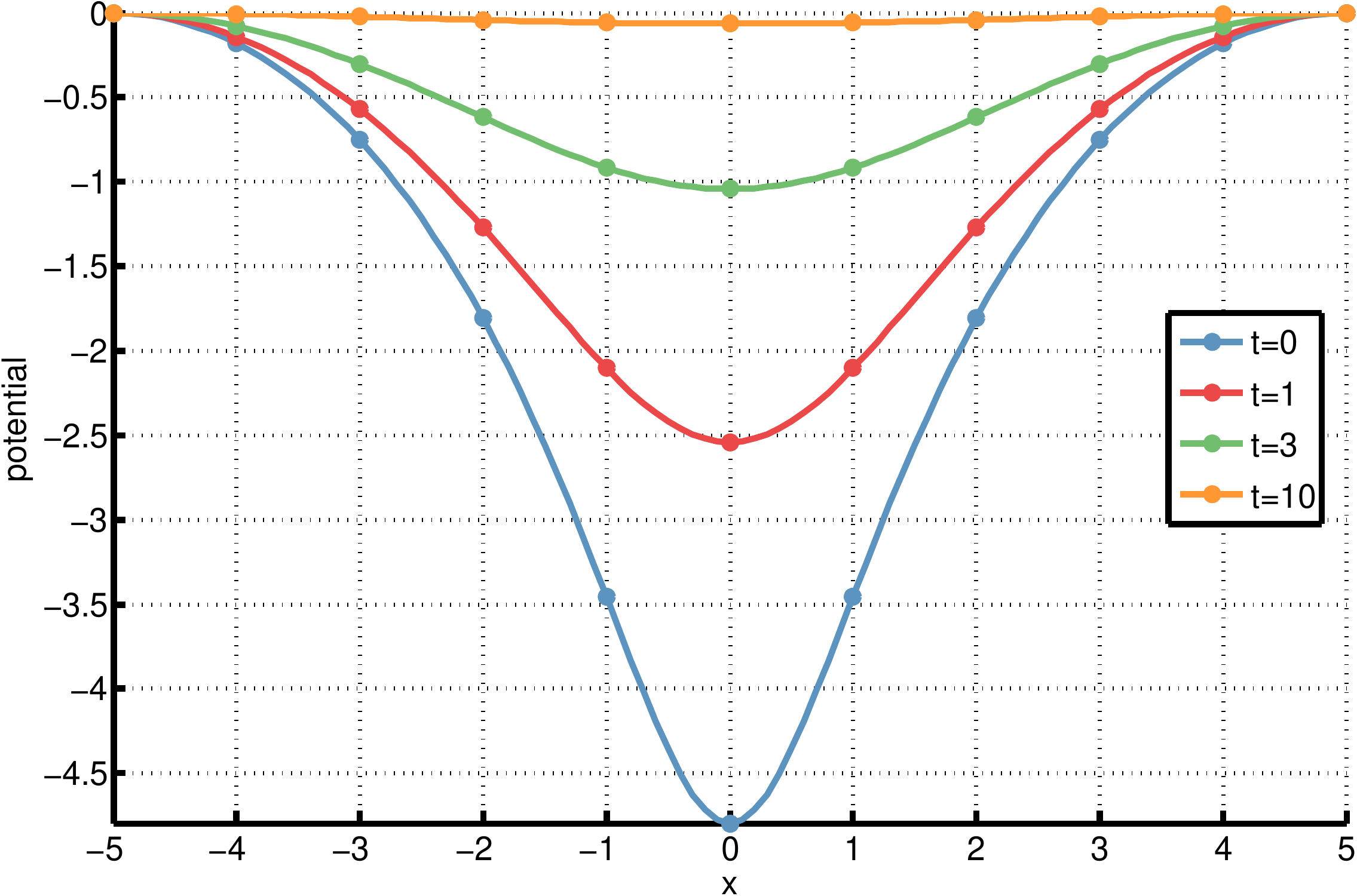}\label{fig:model_general.convergence_hamburger_solution_pot}}%
\subfloat[Concentration]{\includegraphics[width=0.45\textwidth]%
{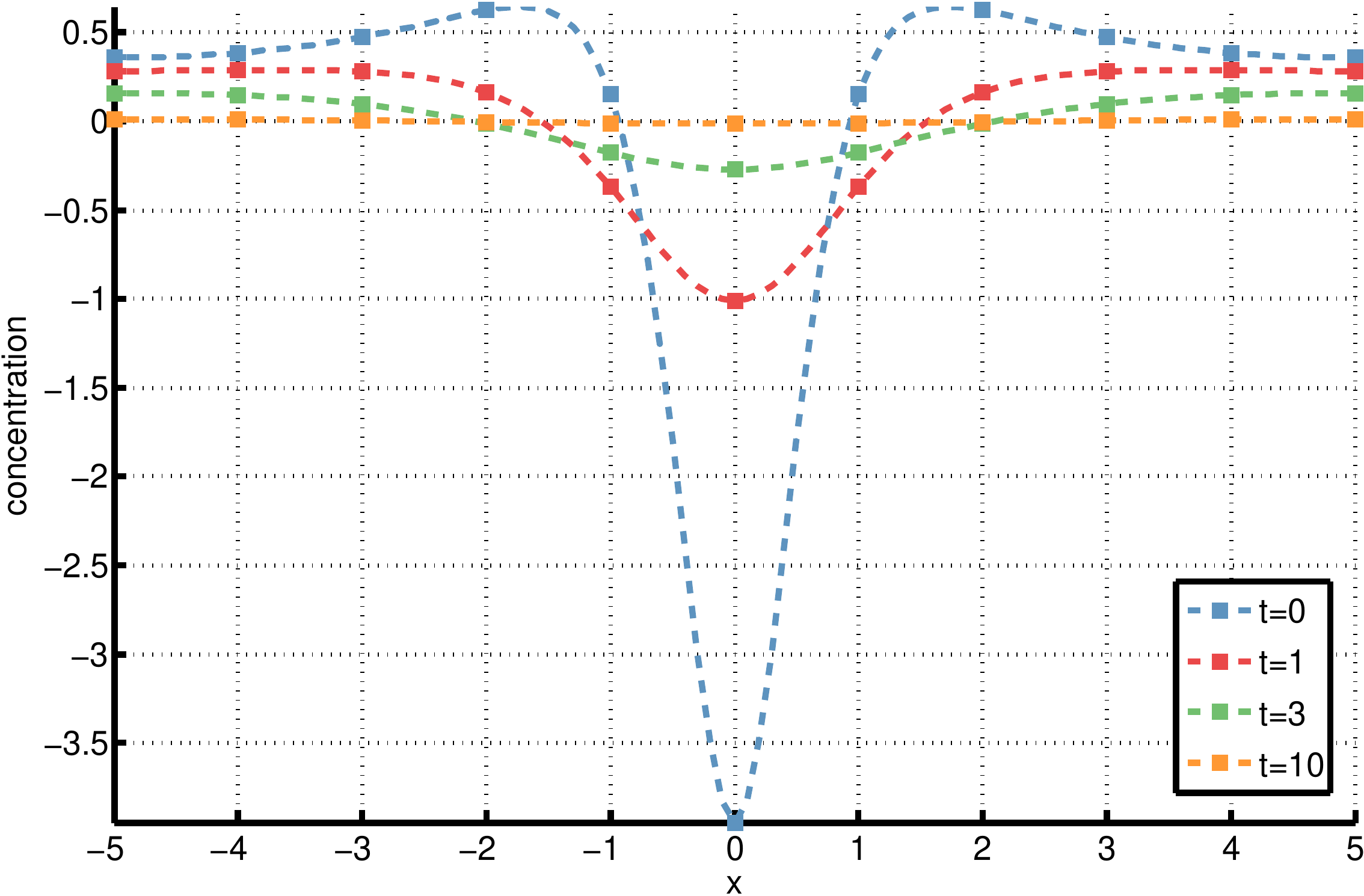}\label{fig:model_general.convergence_hamburger_solution_con}}%
\mycaption[Solution for hamburger example at different time points]{}%
\label{fig:model_general.convergence_hamburger_solution}%
\end{figure}

\paragraph*{Closed system (\textit{hamburger})}
This solution describes an insulated system, i.e. it has zero boundary flux. The initial condition is smeared out over time, and, as such, this solution represents a diffusion-dominated case. We depict the solution for the potential and a single monovalent ion species (say, sodium) in \cref{fig:model_general.convergence_hamburger_solution}. 

\begin{figure}%
\centering%
\subfloat[Error to analytical solution]{\includegraphics[width=0.45\textwidth]%
{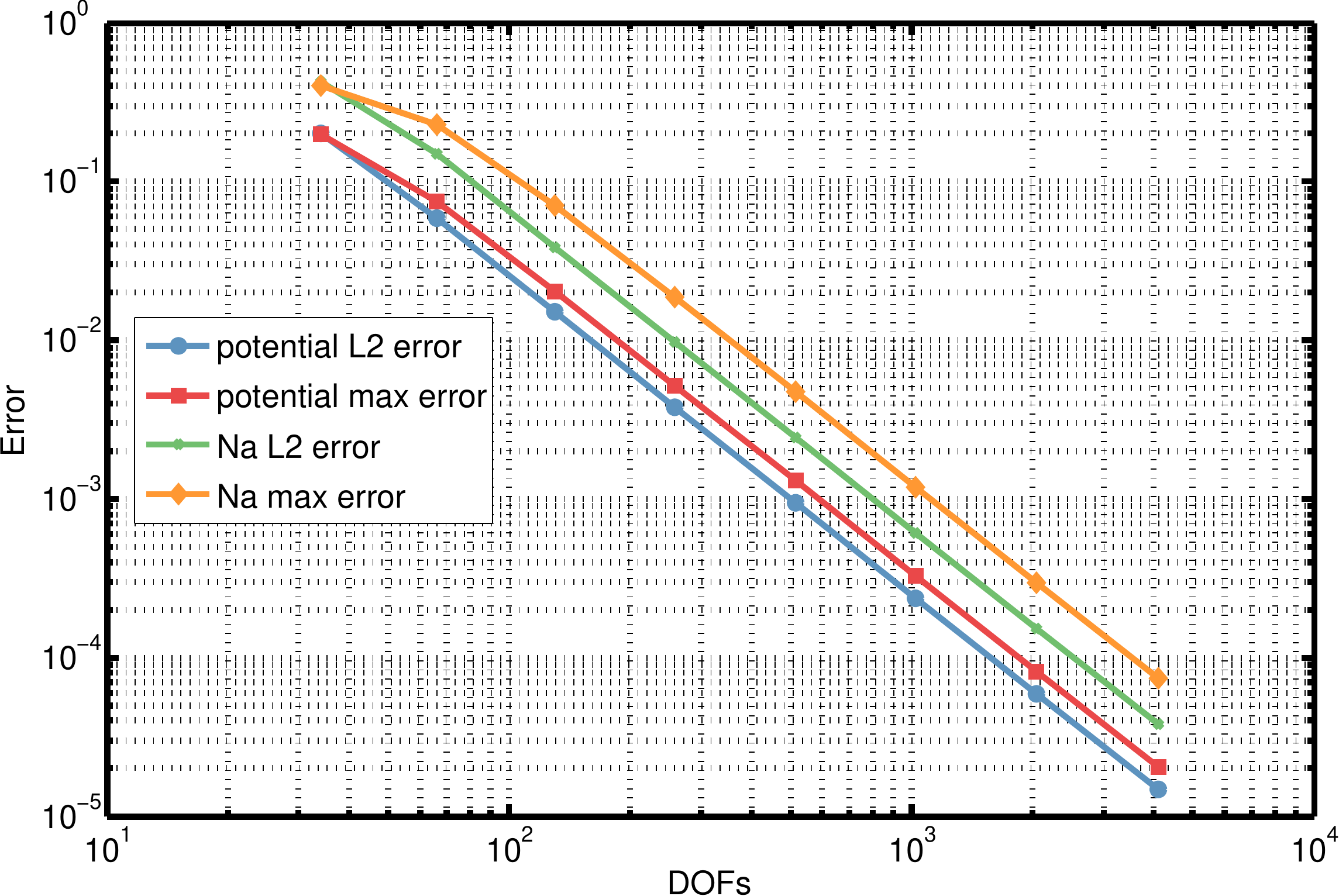}\label{fig:model_general.convergence_hamburger_conv_error}}%
\subfloat[Order of convergence]{\includegraphics[width=0.45\textwidth]%
{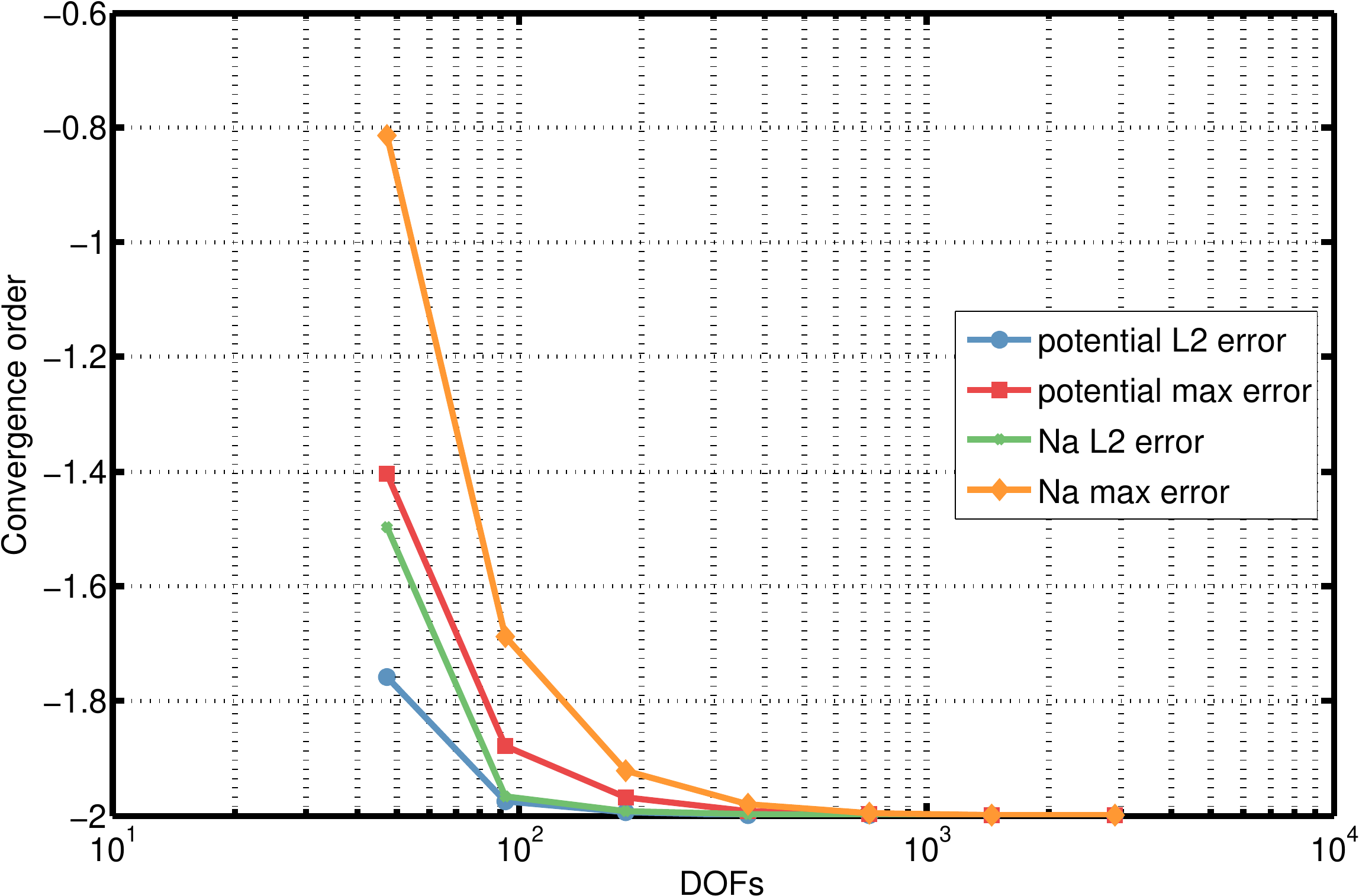}\label{fig:model_general.convergence_hamburger_conv_order}}%
\mycaption[Hamburger convergence behavior in space]{}%
\label{fig:model_general.convergence_hamburger_conv}%
\end{figure}

In order to assess the convergence behavior in space and time separately, we first chose a very small time step $\dt = 10^{-9}$ and carried out 1000 time steps. 
At $\tEnd = 10^{-6}$, the errors in $L^2$ and $L^\infty$ norms were calculated with respect to the analytical solution for different mesh sizes $h=2^{-4}$ to $2^{-11}$, which can be seen in \cref{fig:model_general.convergence_hamburger_conv_error}. 
Due to the very small time steps, time discretization errors should be negligible compared to the spatial errors. 
This is confirmed when looking at the convergence order in \cref{fig:model_general.convergence_hamburger_conv_order}, which shows the expected order of $2$ for $Q_1$ finite elements as soon as the grid is fine enough to accurately resolve the initial condition.

\begin{figure}%
\centering%
\subfloat[Error to analytical solution]{\includegraphics[width=0.45\textwidth]%
{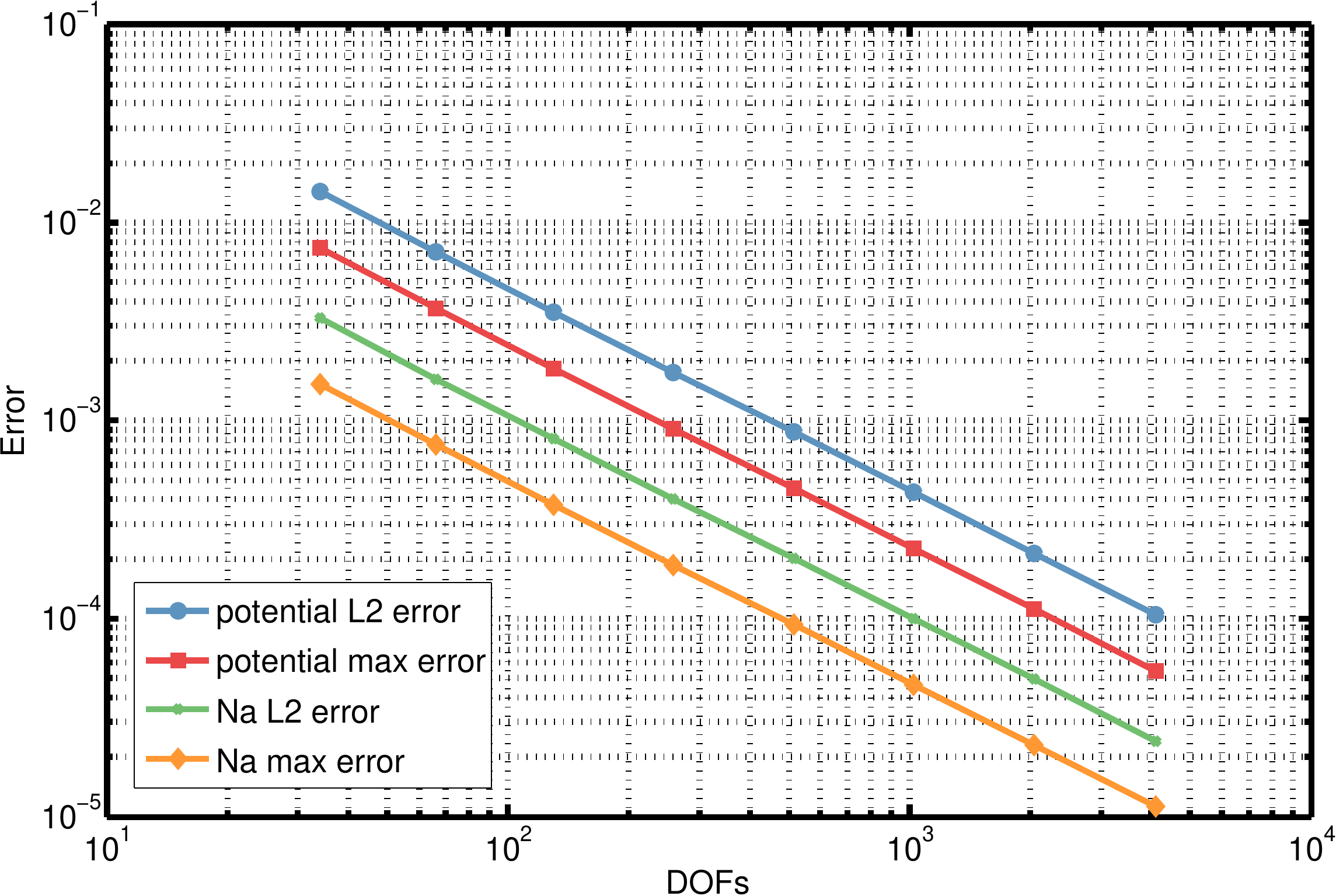}\label{fig:model_general.convergence_hamburger_time_conv_error}}%
\subfloat[Order of convergence]{\includegraphics[width=0.45\textwidth]%
{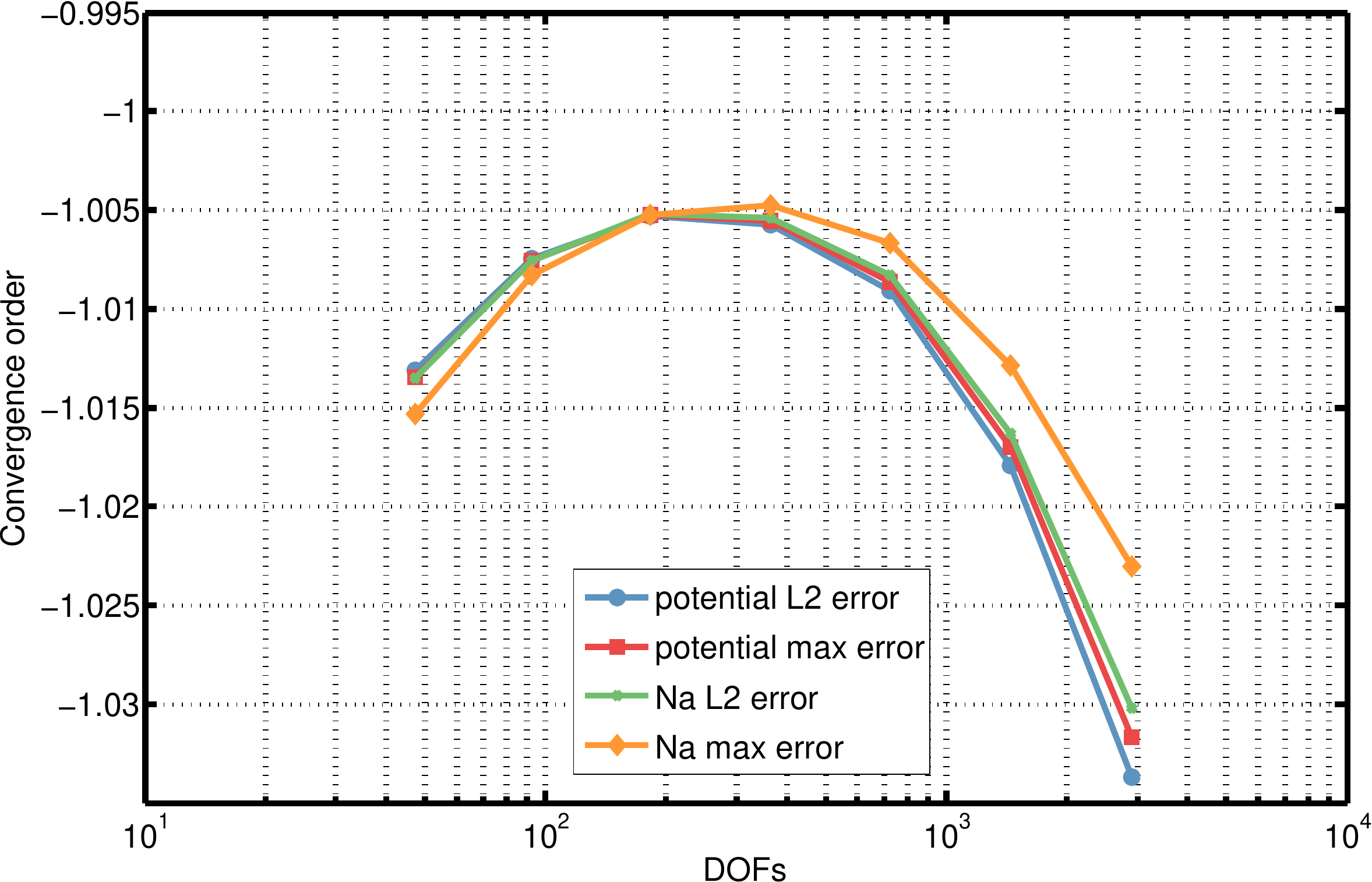}\label{fig:model_general.convergence_hamburger_time_conv_order}}%
\mycaption[Hamburger convergence behavior in time]{}%
\label{fig:model_general.convergence_hamburger_time_conv}%
\end{figure}

Then, we chose a fine spatial resolution of $2^9$ cells and assessed the temporal errors by varying the time step $\dt$ from $1$ down to $2^{-10}$. 
At $\tEnd=10$, errors were calculated as before, this time assuming spatial errors to be negligible, see \cref{fig:model_general.convergence_hamburger_time_conv_error}. 
Again, we find the expected convergence order of 1 for the first order implicit Euler scheme in \cref{fig:model_general.convergence_hamburger_time_conv_order}.

\begin{figure}%
\centering%
\subfloat[Potential]{\includegraphics[width=0.45\textwidth]%
{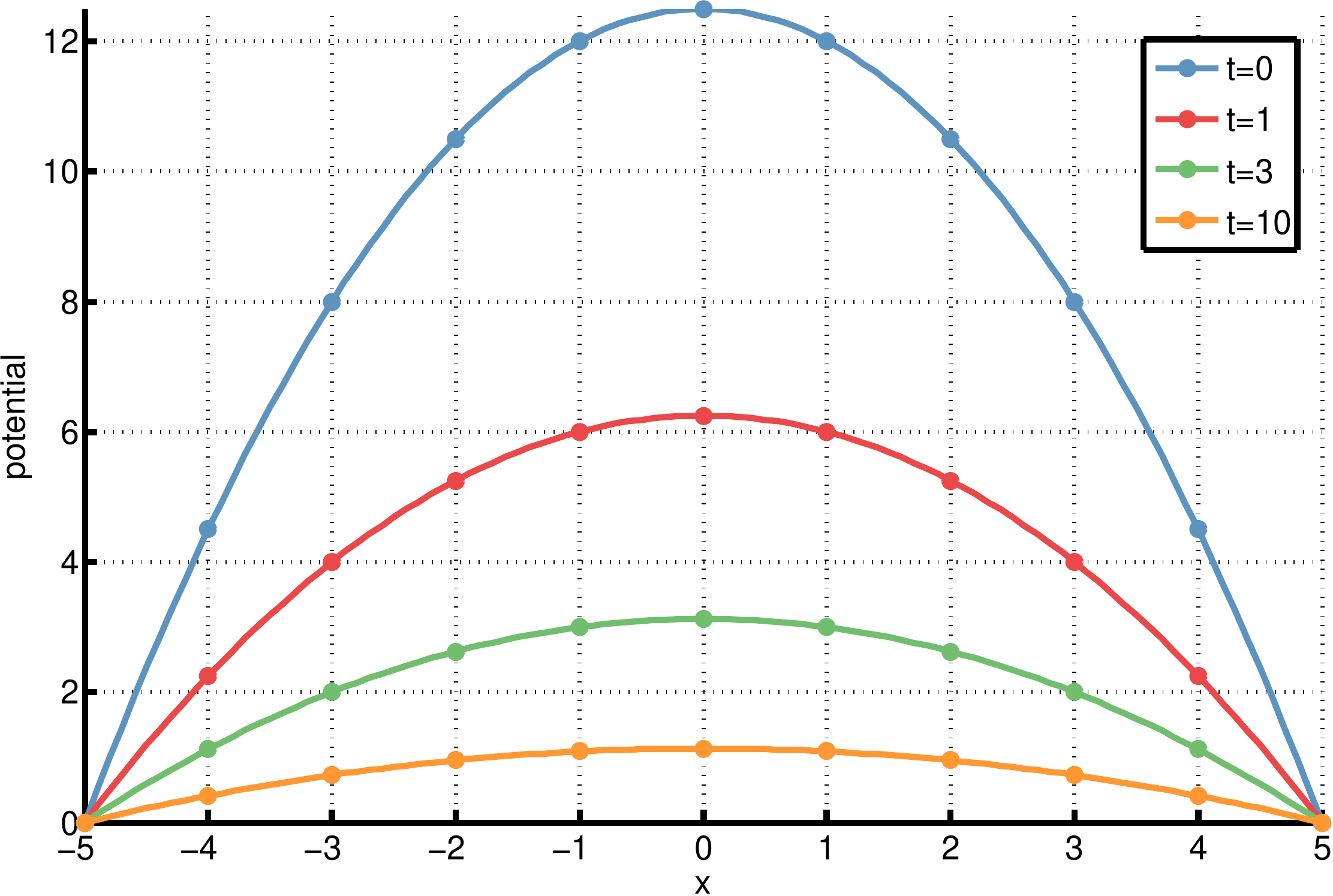}\label{fig:model_general.convergence_cheeseburger_solution_pot}}%
\subfloat[Concentration]{\includegraphics[width=0.45\textwidth]%
{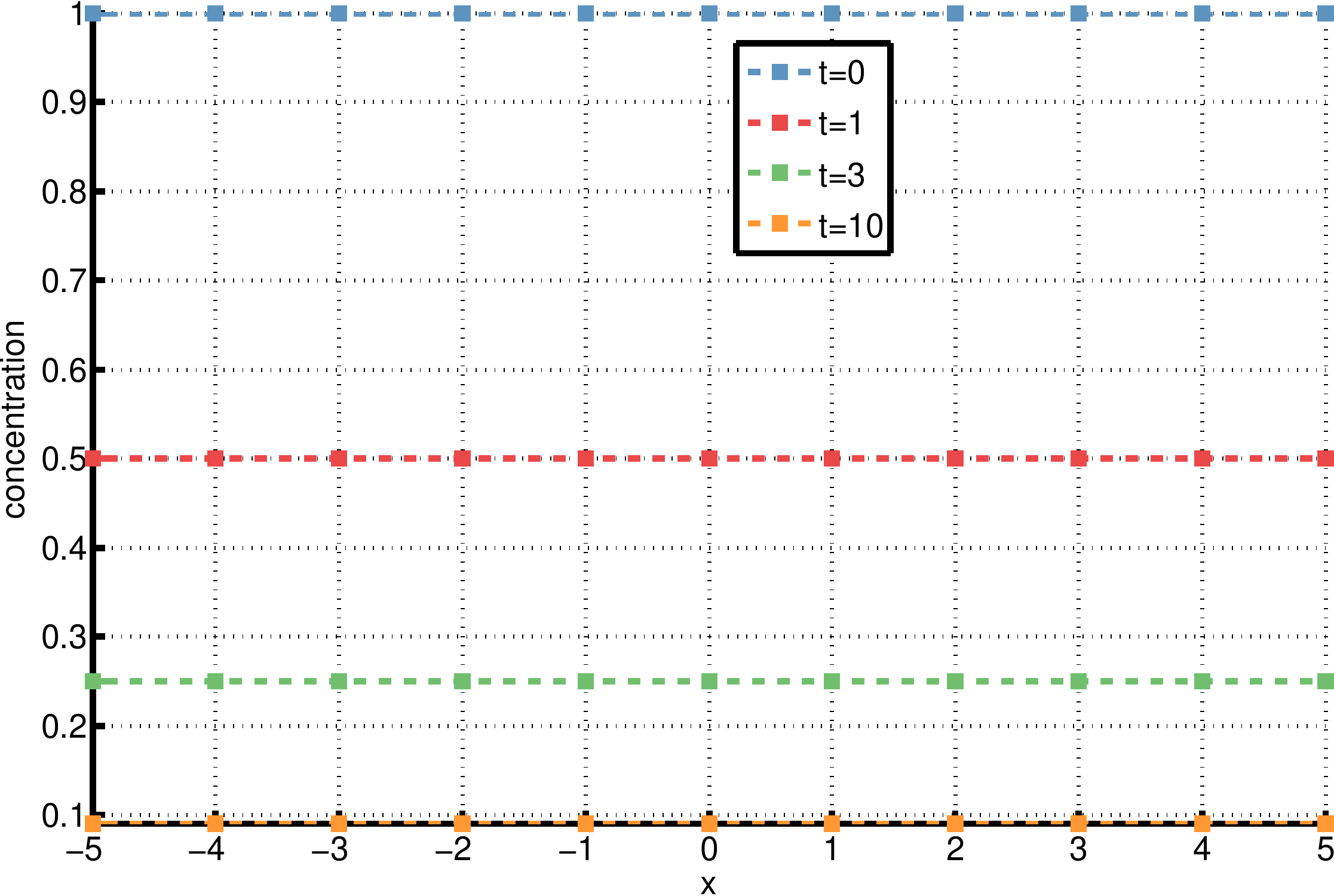}\label{fig:model_general.convergence_cheeseburger_solution_con}}%
\mycaption[Solution for cheeseburger example at different time points]{}%
\label{fig:model_general.convergence_cheeseburger_solution}%
\end{figure}

\paragraph*{Spatially homogeneous outflow (\textit{cheeseburger})} The second example describes a system with a spatially uniform concentration and an outflow boundary condition.
Over time, the concentration leaks out of the domain boundary due to a symmetric outward flow centered at $x=0$. This results in a successive decrease of the spatially constant concentration, as visible in \cref{fig:model_general.convergence_cheeseburger_solution}.
This example tests the correctness of the numerical algorithm regarding non-zero time-dependent Neumann boundary conditions.

\begin{figure}%
\centering%
\subfloat[Error to analytical solution]{\includegraphics[width=0.45\textwidth]%
{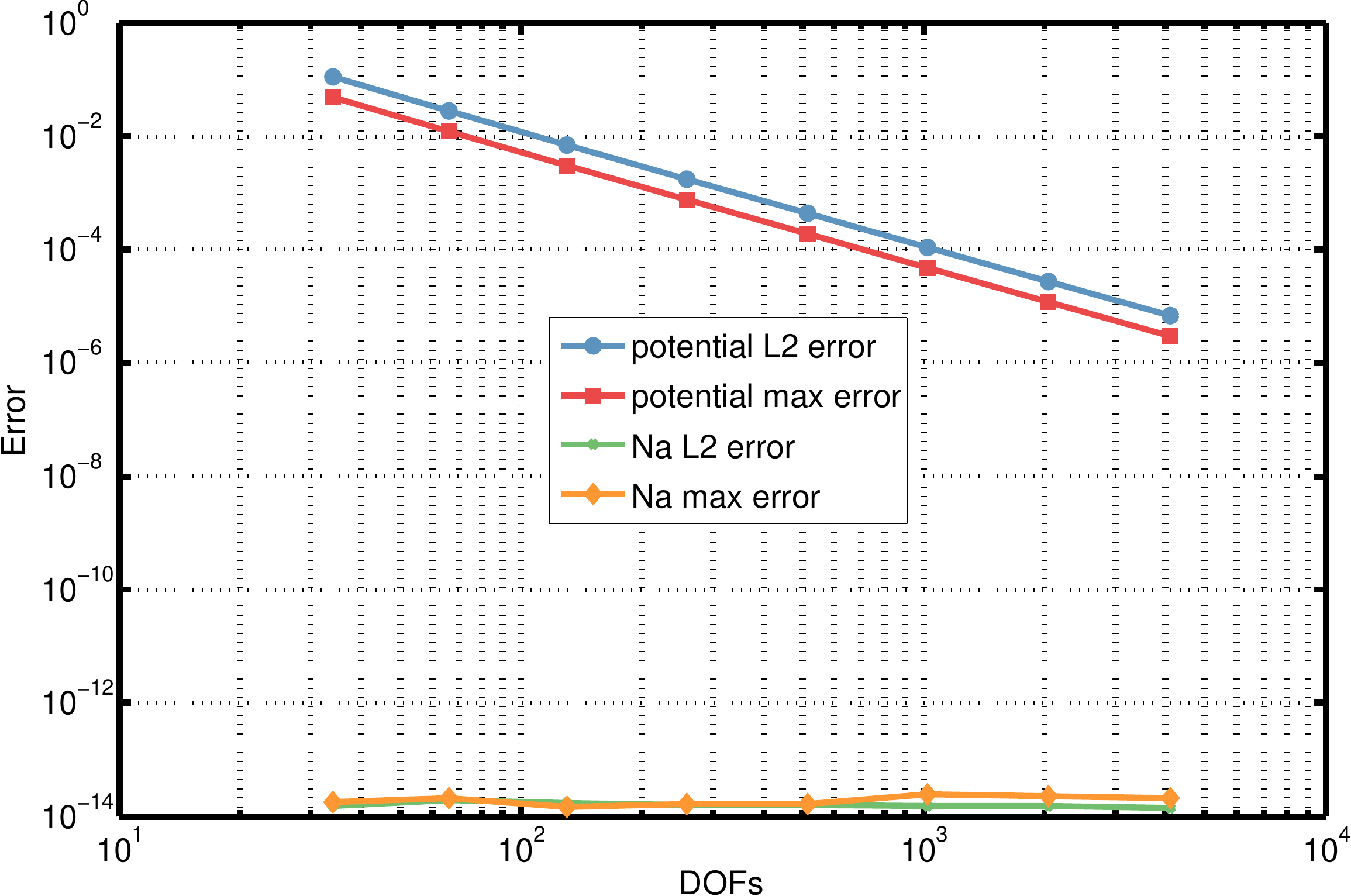}\label{fig:model_general.convergence_cheeseburger_conv_error}}%
\subfloat[Order of convergence]{\includegraphics[width=0.45\textwidth]%
{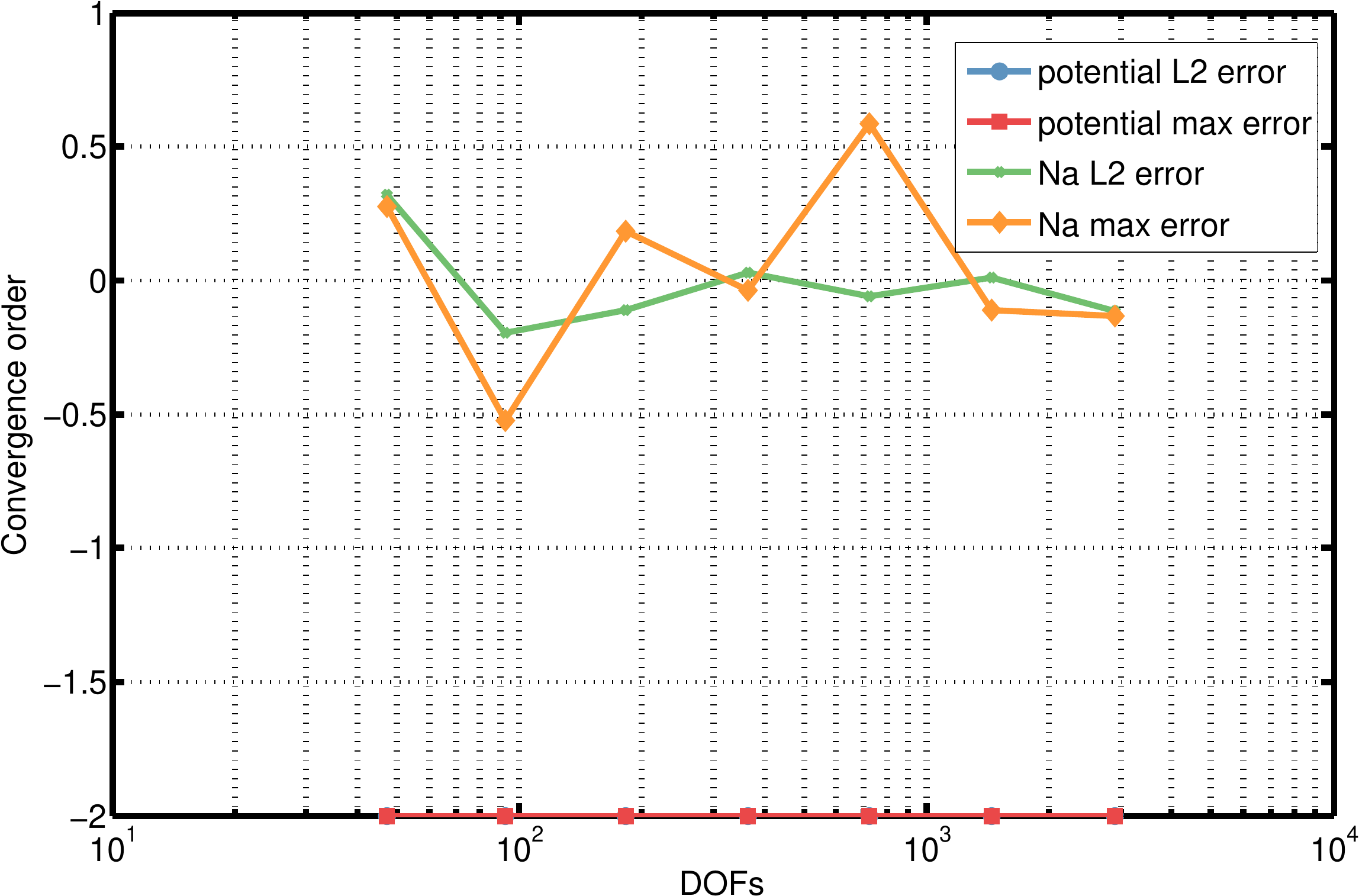}\label{fig:model_general.convergence_cheeseburger_conv_order}}%
\mycaption[Cheeseburger convergence behavior for spatial discretization]{}%
\label{fig:model_general.convergence_cheeseburger_conv}%
\end{figure}

\begin{figure}%
\centering%
\subfloat[Error to analytical solution]{\includegraphics[width=0.45\textwidth]%
{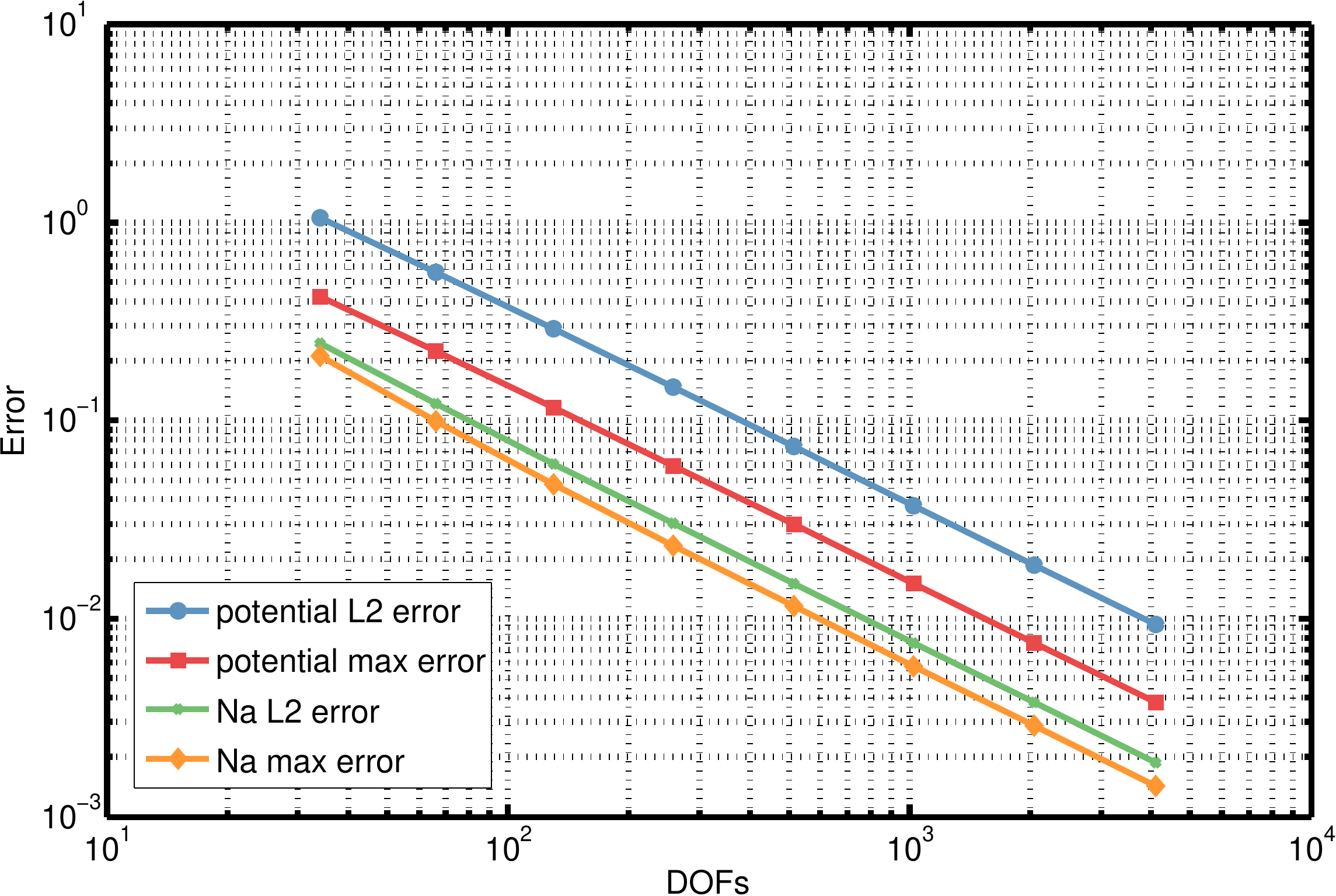}\label{fig:model_general.convergence_cheeseburger_time_conv_error}}%
\subfloat[Order of convergence]{\includegraphics[width=0.45\textwidth]%
{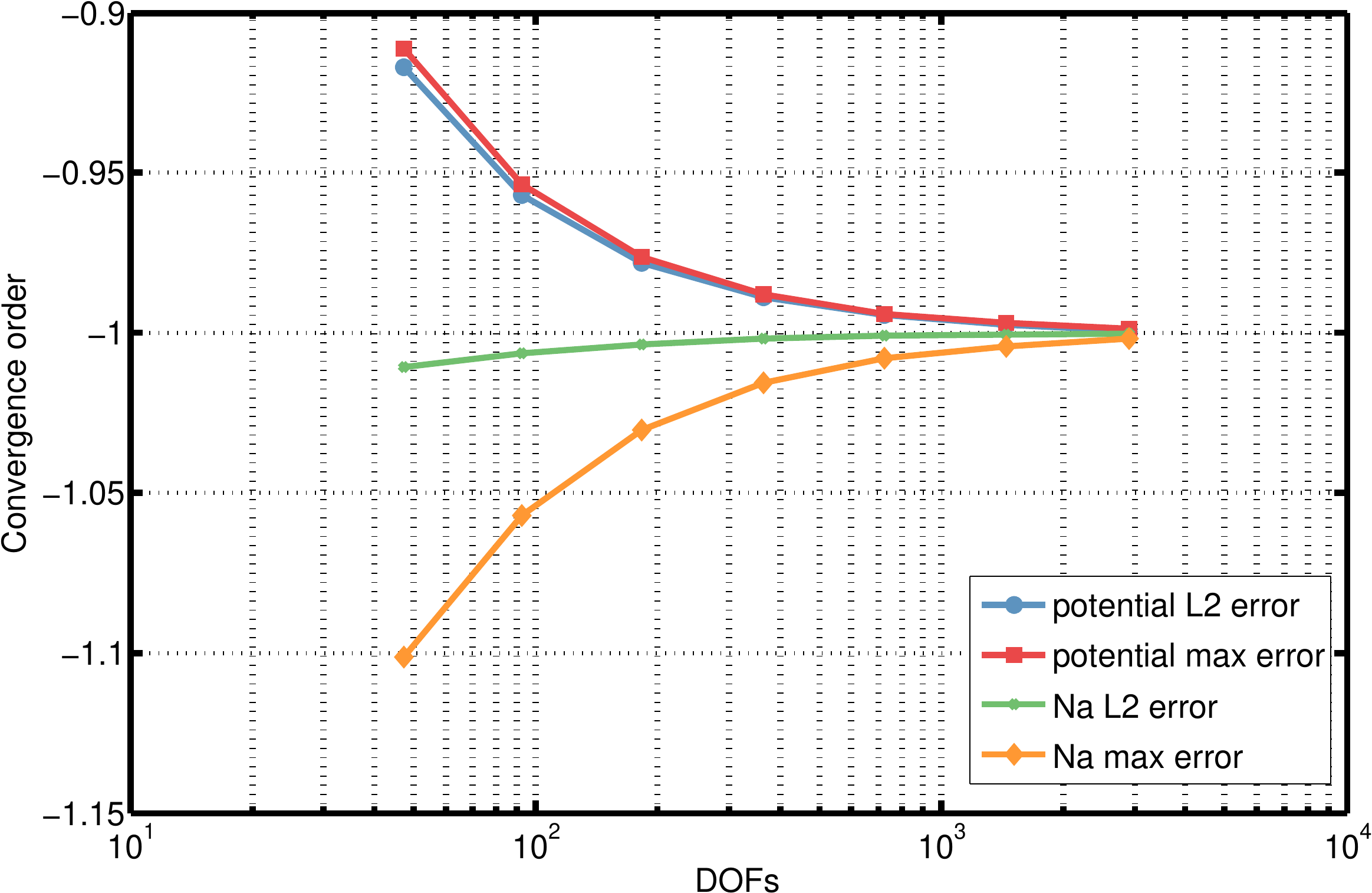}\label{fig:model_general.convergence_cheeseburger_time_conv_order}}%
\mycaption[Cheeseburger convergence behavior in time]{}%
\label{fig:model_general.convergence_cheeseburger_time_conv}%
\end{figure}

Convergence in space for the potential is second-order, but this does not hold for the concentrations in \cref{fig:model_general.convergence_cheeseburger_time_conv_order}. 
This result is intuitively clear respecting the concentration is spatially homogeneous.
An increase in grid resolution can not be expected to yield an improvement for a function which can already be represented accurately with only a single cell. 
We confirm this by noting that the concentration error in \cref{fig:model_general.convergence_cheeseburger_conv_error} is already close to machine precision at the coarsest resolution. Therefore, the strange behavior in the convergence order plot \cref{fig:model_general.convergence_cheeseburger_conv_order} is uncovered to be numerical noise.

The behavior in time is not affected by this and shows first-order convergence in \cref{fig:model_general.convergence_cheeseburger_time_conv_order} as soon as the time step is sufficiently small. Interestingly, the concentrations initially show a slightly super-linear convergence speed.

\begin{figure}%
\centering%
\subfloat[Potential]{\includegraphics[width=0.45\textwidth]%
{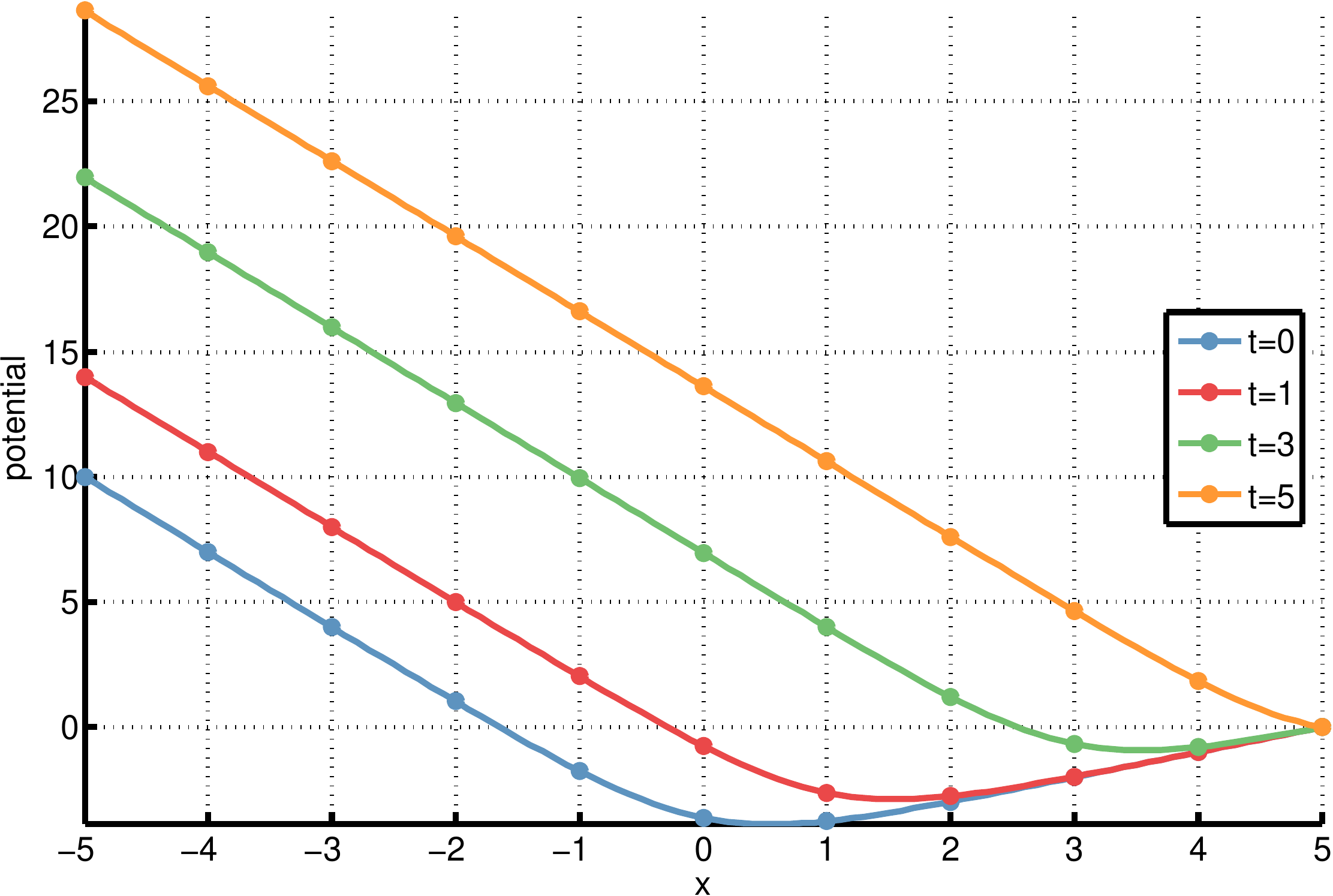}\label{fig:model_general.convergence_bigmac_solution_pot}}%
\subfloat[Concentration]{\includegraphics[width=0.45\textwidth]%
{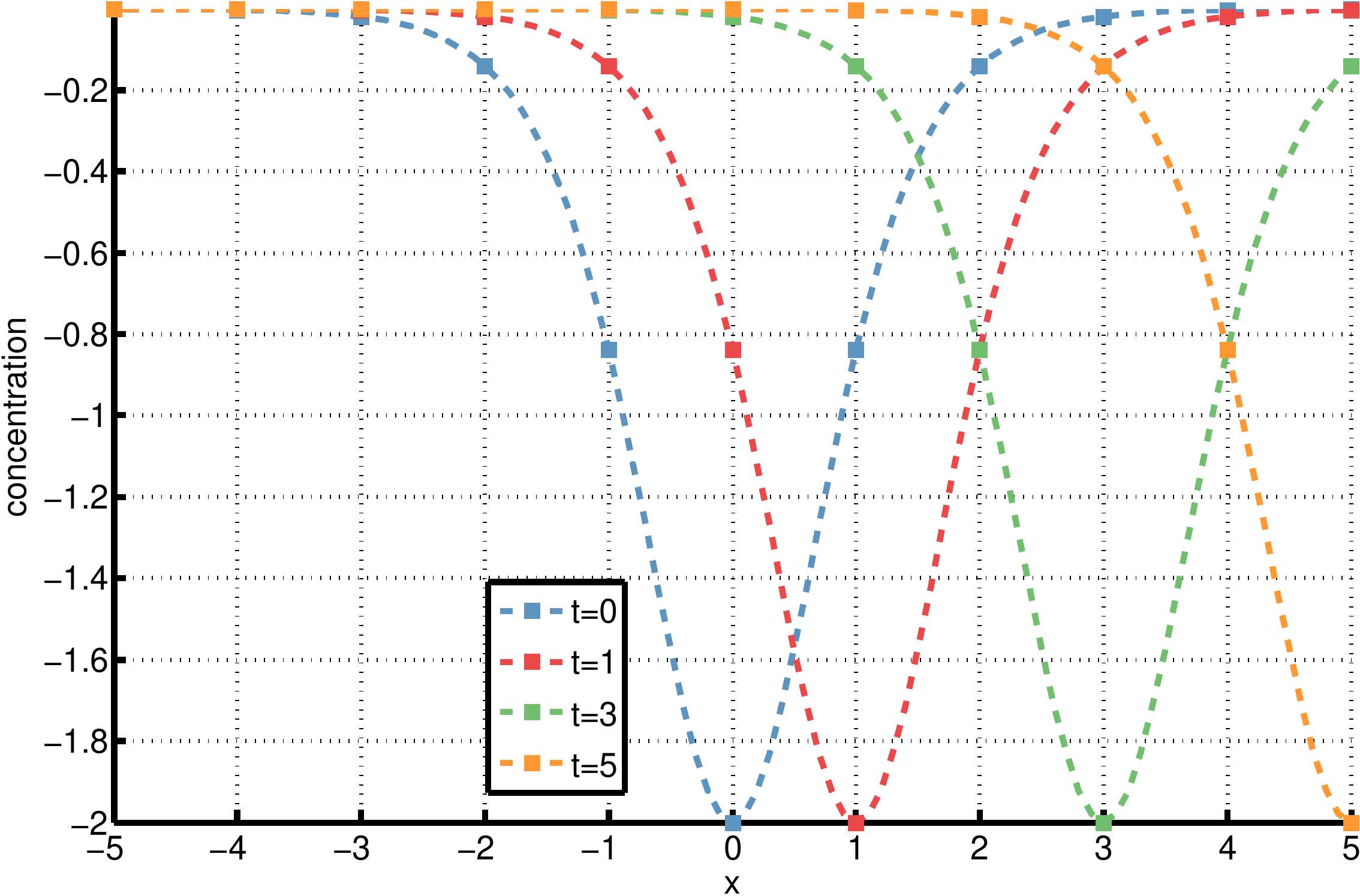}\label{fig:model_general.convergence_bigmac_solution_con}}%
\mycaption[Solution for bigmac example at different time points]{}%
\label{fig:model_general.convergence_bigmac_solution}%
\end{figure}

\paragraph*{Pure advection (\textit{bicmac})}
The last solution is purely advective, therefore the ion flux $\mathbf{F} = -D\left(\nabla n_i+z_in_i\nabla\phi\right)$ is reduced to $\mathbf{F} = -D_i z_i \nabla\phi n_i$ and the drift velocity $v = D_i z_i \nabla\phi$ is replaced by a constant advection velocity $v=1$. 
As a result, the initial concentration profile is simply transported to the right in \cref{fig:model_general.convergence_bigmac_solution}. 
This test case especially checks the numerical algorithm's ability to cope with a drift-dominated system.

\begin{figure}%
\centering%
\subfloat[Error to analytical solution]{\includegraphics[width=0.45\textwidth]%
{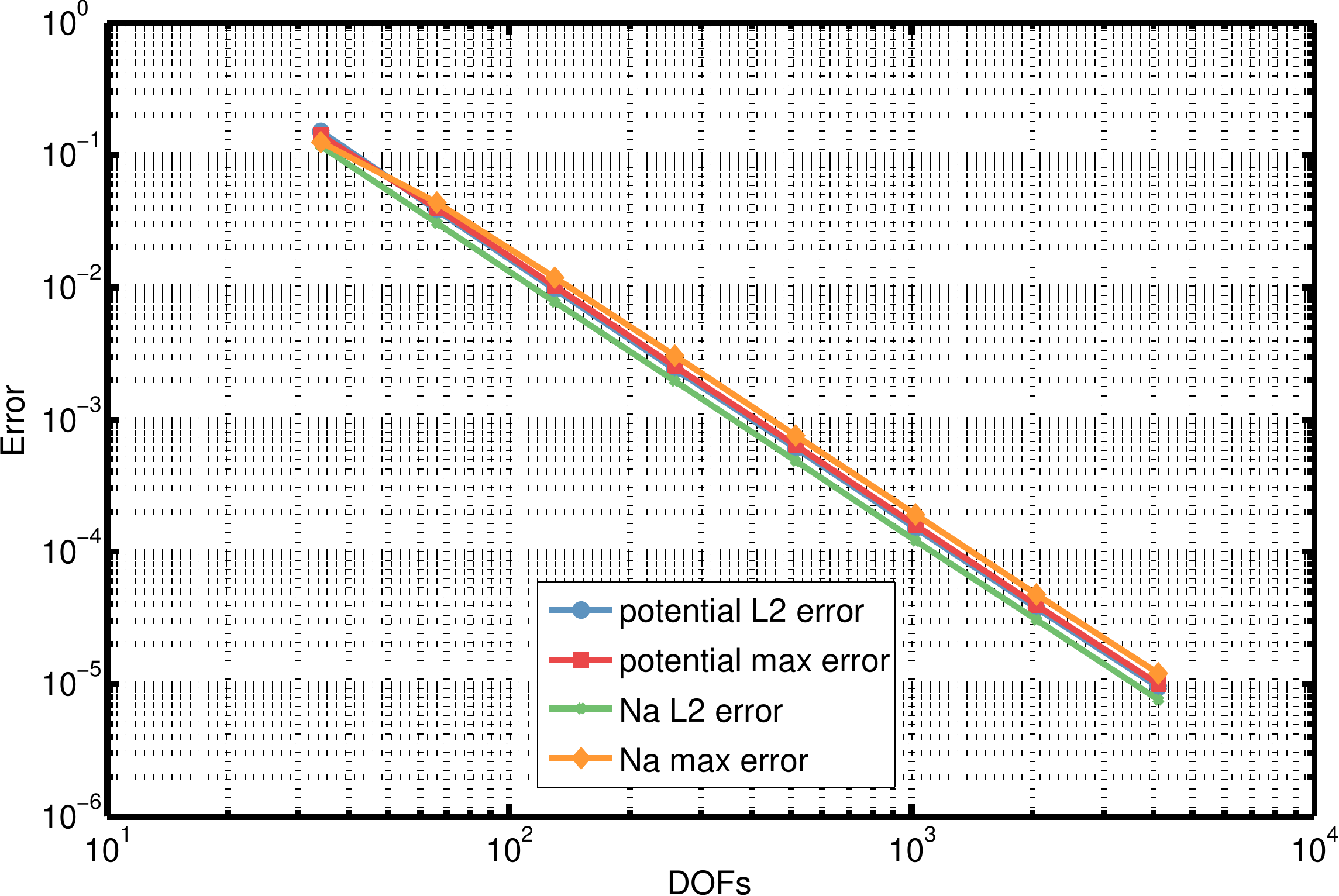}\label{fig:model_general.convergence_bigmac_conv_error}}%
\subfloat[Order of convergence]{\includegraphics[width=0.45\textwidth]%
{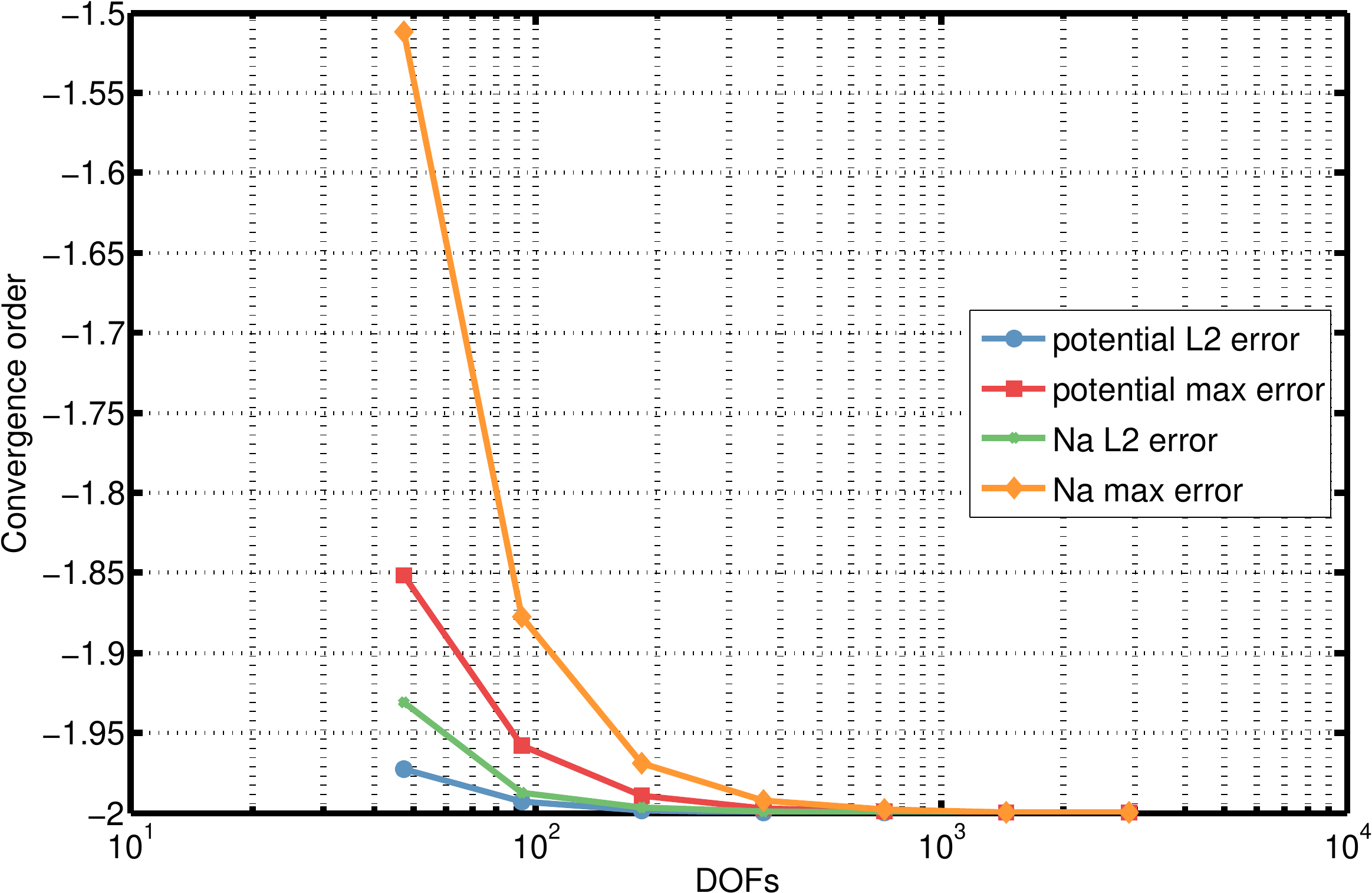}\label{fig:model_general.convergence_bigmac_conv_order}}%
\mycaption[Bigmac convergence behavior for spatial discretization]{}%
\label{fig:model_general.convergence_bigmac_conv}%
\end{figure}

\begin{figure}%
\centering%
\subfloat[Error to analytical solution]{\includegraphics[width=0.45\textwidth]%
{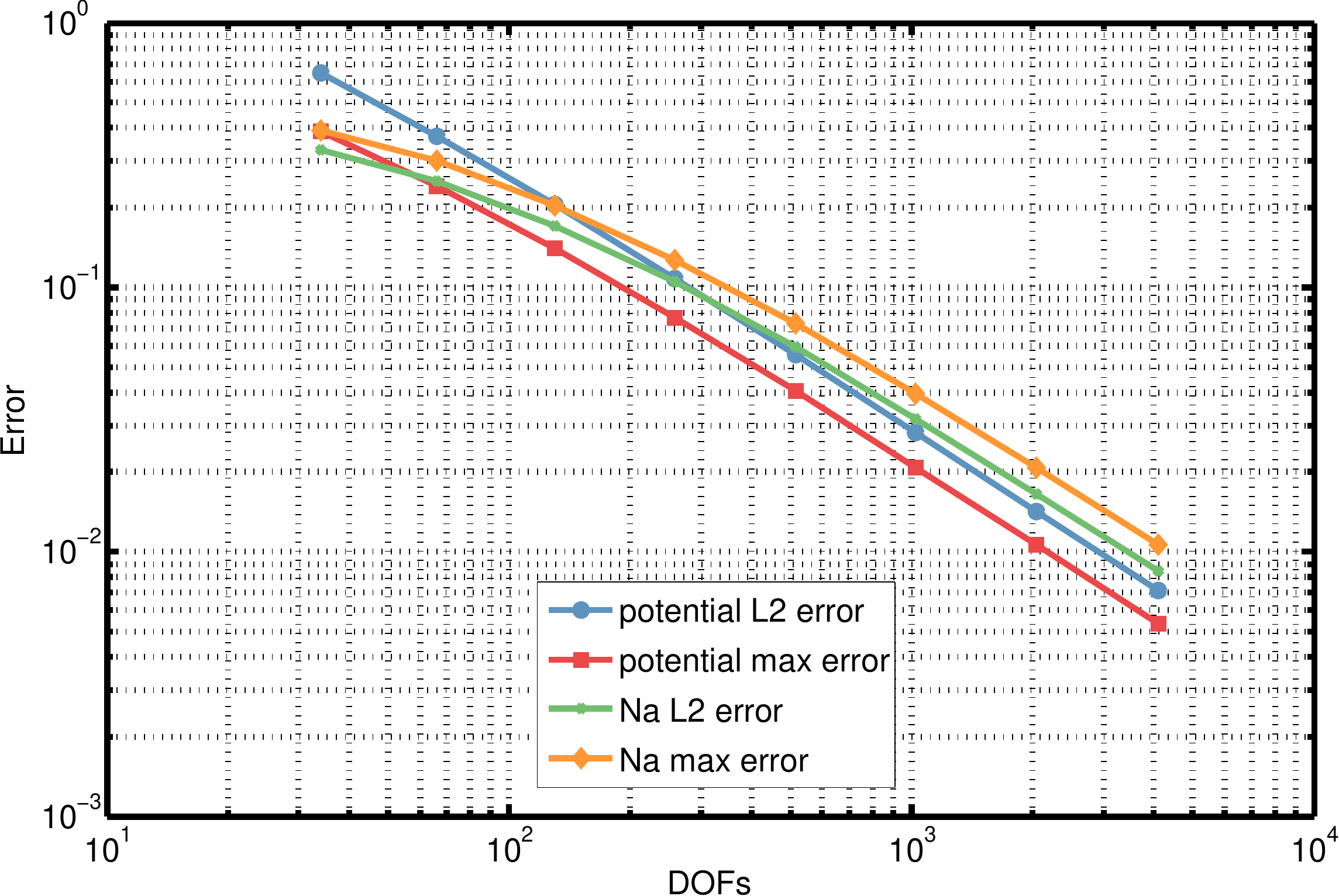}\label{fig:model_general.convergence_bigmac_time_conv_error}}%
\subfloat[Order of convergence]{\includegraphics[width=0.45\textwidth]%
{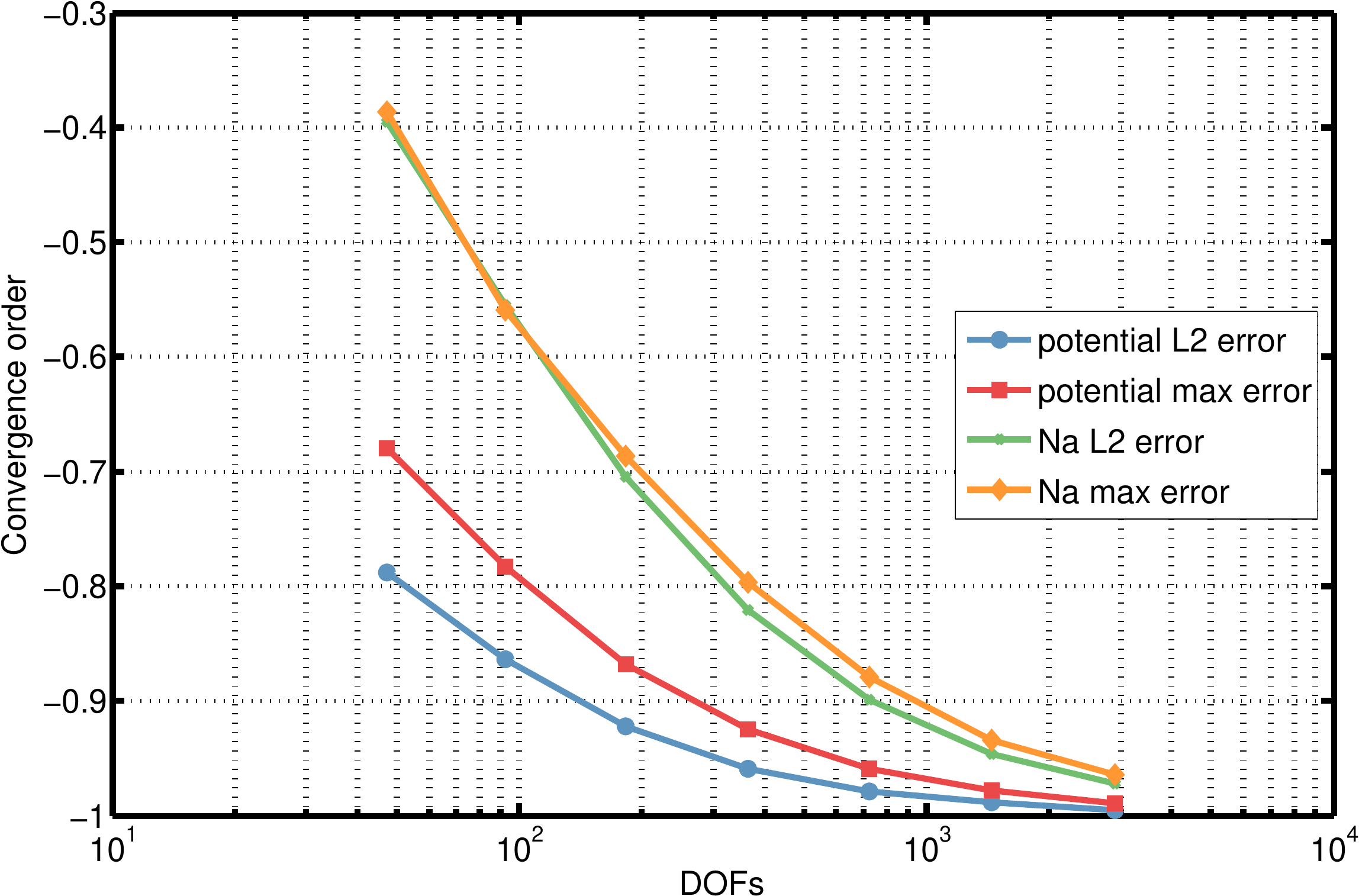}\label{fig:model_general.convergence_bigmac_time_conv_order}}%
\mycaption[Bigmac convergence behavior in time]{}%
\label{fig:model_general.convergence_bigmac_time_conv}%
\end{figure}

This example is not as reluctant as the previous one and we see the full convergence order for both potential and concentrations, i.e.~second order in space (\cref{fig:model_general.convergence_bigmac_conv_order}) and first order in time (\cref{fig:model_general.convergence_bigmac_time_conv_order}).
In comparison to the previous examples, the final convergence order is reached only for relatively fine discretization levels. 
The maximum concentration error is especially susceptible to this at coarser mesh sizes, since in these cases the transported concentration peak can not be captured exactly by the grid. 
We also see quite a delay in the time convergence at large time steps, which can be attributed to the smoothing character of the implicit Euler method.

\bigskip

Summarizing the validation by analytical solutions, we obtain the expected order of convergence in space and time for all test cases. \Citeauthor{schoenke2012} \cite{schoenke2012} also describes a methodology to construct solutions in higher dimensions from the 1D solutions, but we refrain from carrying out extensive convergence analyses within the scope of this work.
\setchapterpreamble[u]{%
\dictum[gcc 4.5]{Warning: ISO C++ says that these are ambiguous, even though the worst conversion for the first is better than the worst conversion for the second.}\bigskip}
\chapter{Implementation}\label{chap:implementation}
This chapter strives to give a minimal overview over the implementation of the numerical algorithm from \cref{sec:model.numerical_methods}.

\section{The \Dune Framework}
The implementation was done in C++ using \Dune (Distributed and Unified Numerics Environment) \cite{dune-web-page}, a framework for the grid-based solution of \glspl{PDE}.
It consists of several modules, the heart of which is \lstinline!dune-grid! containing the abstract grid interface \cite{dunegridpaperII:08} based on a rigorous mathematical description of hierarchical grids \cite{dunegridpaperI:08}.

Existing  grid implementations (or \emph{grid managers}, in  \Dune jargon), can be plugged into the framework by e.g.~an adapter fulfilling the abstract interface.
This is made possible by the extensive usage of C++ template techniques \cite{veldhuizen2000techniques}, which allows for the inclusion of arbitrary implementations without a big performance loss.
This also summarizes the main design principles in \Dune: \emph{flexibility} with regard to the reusability of software components and -- at the same time -- \emph{efficiency} by removing the interface overhead at compile-time using generic programming techniques.
The key here is to replace conventional inheritance (\emph{dynamic polymorphism}) by \emph{static polymorphism}, where the complete inheritance hierarchy is known at compile-time, thereby eliminating the runtime overhead of e.g.~function table lookups necessary in dynamic polymorphism.

One main advantage of this approach is that one is not restricted to a certain grid implementation.
There exist quite a number of grid managers with different features, but none of these is universal.
Most grid managers are rather specialized towards a certain class of algorithms.
\Dune makes it possible to simply exchange the underlying grid without having to rewrite the code for the numerical solution, as it is based on the abstract interface.
 This has proven to be very handy in practice, as virtually any application can be solved within a single framework.

\Dune modules are classified into two groups: The \emph{core modules} providing the basic functionality, and additional modules which extend the functionality of the core modules or implement a specific application.
The core modules are
\begin{itemize}
  \item \lstinline!dune-common!: classes used by all \Dune modules, including data structures for dense vectors and matrices as well as the program \lstinline!dunecontrol! providing the build system logic.
  \item \lstinline!dune-geometry!: provides geometric information of the grid cells (included in the more general term of an \emph{entity} in \Dune) based on generic \emph{reference elements}, their mapping into the global space and quadrature rules for integration.
  \item \lstinline!dune-grid!: contains the abstract grid interface and a small number of grid implementations as well as adapters for external grid managers.
  \item \lstinline!dune-istl!: the \gls{ISTL} contains a number of iterative linear solvers and preconditioners that were designed specifically with respect to parallel efficiency.
  \item \lstinline!dune-localfunctions!: this module defines functions living on the reference elements which can be used to assemble global finite element functions.
\end{itemize}
In addition, the following modules were used for the implementation:
\begin{itemize}
  \item \lstinline!dune-pdelab!: PDELab \cite{pdelabalgoritmy} is a discretization module which allows the user to specify a \emph{local operator} living on a single grid cell. The functionality of PDELab then allows the generic assembly of the global matrix by means of a \emph{grid operator} and its (sequential or parallel) solution by arbitrary combination of preconditioners and solvers from \lstinline!dune-istl!. It also contains a Newton implementation.
  \item \lstinline!dune-multidomaingrid!: this module provides a \lstinline[breaklines=true]!Dune::MultidomainGrid! metagrid on top of a \Dune grid which allows for the definition of arbitrary subdomains, useful for multi-physics applications \cite{dune-multidomaingrid}.
  \item \lstinline!dune-multidomain!: this is an add-on module for PDELab which, in conjunction with \lstinline!dune-multidomaingrid!, allows for the definition of different local operators on different subdomains of the \lstinline!Dune::MultidomainGrid! \cite{dune-multidomain}. The operators can be coupled in a very flexible way, integrating nicely into the PDELab framework.
\end{itemize}
The joint functionality of these modules is used in the application module \lstinline!dune-ax1!, whose main components will be described in the following section.
 
\section{The \Dune Module \texttt{dune-ax1}}\label{sec:implementation.dune-ax1}
The module \lstinline!dune-ax1! contains all the code that was used to implement the numerical solution of the model from \cref{sec:neuron_model.2d}. The main actors in this module are listed below.

\subsection{Directory Structure}
The directory structure of the \lstinline!dune-ax1! module looks as follows:
\dirtree{%
.1 dune-ax1.
.2 src. 
.2 dune. 
.3 ax1. 
.4 acme0. 
.4 acme1. 
.4 acme1MD. 
.4 acme2. 
.4 acme2\_cyl. 
.5 common. 
.5 configurations. 
.6 default. 
.6 ES.
.6 laplace.
.6 step. 
.5 operator. 
.4 channels. 
.3 common. 
}
The \lstinline!src! directory contains the application drivers, while the \lstinline!dune/ax1! folder contains a number of subfolders which demonstrate the historical evolution of the application.
The folders \lstinline!acme0! to \lstinline!acme2! represent the different evolution stages from a simple 1D model without a membrane on a \lstinline!Dune::OneDGrid! to a 2D model with a membrane using \lstinline!Dune::MultidomainGrid!.
The reason for keeping the older stages was that initially the Subversion \cite{subversion} version control system was used, which has its difficulties when using different development branches.
After the switch to Git \cite{git}, keeping different versions in different branches has become quite easy, but the older stages are still there as a Subversion legacy.
The most recent version of the application resides in \lstinline!acme2_cyl!.
The folder \lstinline!channels! contains classes representing a variety of \gls{HH}-type ion channels that can be vectorized to a single \lstinline!ChannelSet!.
Additional classes used by every application can be found in \lstinline!common!.
The next section will describe the main classes in \lstinline!acme2_cyl! and \lstinline!common!.

\subsection{Main Components}\label{sec:implementation.components}
\setcounter{secnumdepth}{3}
In the following, the main components responsible for the implementation are listed. Yet, this section does not only serve to list the purpose of each of these components; we will also highlight some of the concepts used, both with respect to programming techniques and to numerical subtleties.

\paragraph{\texorpdfstring{\lstinline!acme2_cyl_par.cc!}{acme2\_cyl\_par.cc}} This is the only \lstinline!.cc! file in the application, all the others are header files containing (mostly templated) classes.
This is due to the fact that, when using C++ templates, the usual separation between header and source files is no longer possible \cite[10]{josuttis1999c++}.
The name contains the acronym ``acme'', which stands for ``active membrane''\footnote{Intentionally, it is also a reference to the ACME (``A company that manufactures everything'') company from the TV series Looney Tunes.}. The other building blocks of the name refer to the grid dimension (``2''), the cylindrical coordinate system (``cyl''), and the parallel solution of the equations (``par'').

\lstinline!acme2_cyl_par.cc! as the application driver contains the \lstinline!main()!-routine, which reads command line arguments and config file parameters, calls the grid generation procedure in \lstinline!Ax1GridGenerator!, and sets up the \lstinline!Dune::Multidomain\-Grid!.
Afterwards, it calls \lstinline!Acme2CylFactory!, which instantiates the central data class \lstinline!Acme2CylPhysics! and hands all the objects over to \lstinline!Acme2CylSetup!.

\paragraph{\texorpdfstring{\lstinline!Ax1GridGenerator!}{Ax1GridGenerator}} This class has a static method \lstinline!generateTensorGrid()!, which fills two \lstinline!CylinderGridVector!s with coordinates for the $x$- and $y$-direction.
The vector class \lstinline!CylinderGridVector! is a modified version of a class by Dominic Kempf \cite{kempf2011geothermie}, which contains several methods for conveniently adding successive coordinates according to different strategies, e.g.~equidistant spacing, linearly increasing spacing, spacing increasing according to a geometric series, and many more.
These methods are particularly useful in the present case of strongly varying mesh sizes in $y$-direction (cf.~\cref{sec:model.numerical_methods.space_discretization}), as they allow for a smooth transition of grid spacing between regions with very small and rather large mesh sizes.
The method \lstinline!generateTensorGrid()! calls these functions according to the minimum/maximum grid spacings specified in the config file in order to set up the tensor grid vectors $X$ and $Y$. 
The generated vectors have the desired properties of a very fine resolution in $y$-direction at the membrane, and sufficiently coarse spacings away from the membrane and in $x$-direction, in order to minimize the number of grid points.

After generating the grid coordinate vectors, the actual grid hierarchy is set up.
In particular, this hierarchy consists of a \lstinline!Dune::YaspGrid! in a \lstinline[breaklines=true]!Dune::\-Geometry\-Grid! metagrid, which itself is wrapped by a \lstinline!Dune::MultidomainGrid!.
The reason for this stacked metagrid hierarchy is the following: the structured \lstinline!Dune::YaspGrid! is very efficient and it allows for an arbitrary parallel domain decomposition; however, it is an equidistant grid in each coordinate direction.
To make this a tensor grid, the functionality of \lstinline!Dune::GeometryGrid! is used, which enables to specify a coordinate function mapping the (equidistant) base grid nodes to those provided by the tensor product $X \times Y$.
Finally, \lstinline!Dune::MultidomainGrid! provides the possibility to define (even non-connected) subdomains and interior boundaries, which is just what we need for the multi-domain problem at hand.
Each level in the grid hierarchy is paid for by an additional computational overhead.
In performance tests, the runtime for a program that excessively called functions on the grid geometry -- the most basic functions that are called in the innermost loops of the local operator, and therefore the most time-consuming operations introduced by the nested grid hierarchy -- increased by approximately 30\% when using \lstinline!Dune::GeometryGrid!.
We consider this acceptable with respect to the functionality provided.
A similar overhead is expected for the \lstinline!Dune::MultidomainGrid!.

\paragraph{\texorpdfstring{\lstinline!Acme2CylFactory!}{Acme2CylFactory}} By implementing the (static polymorphism version of the) \emph{factory pattern} \cite{gamma1994design}, \lstinline!Acme2CylFactory! creates an instance of an \lstinline[breaklines=true]!Acme2\-Cyl\-Physics! template class.
Most of the template parameters of \lstinline!Acme2CylPhysics! are specified by a static configuration class residing in the subdirectory \lstinline[breaklines=true]!acme2_\-cyl/\-configurations!, containing compile-time constants and the class types for initial and boundary conditions.
The resulting object of \lstinline!Acme2CylPhysics! knows about the classes for initial and boundary conditions as well as all relevant model parameters for the desired simulation setup.
The configurations classes, on the contrary, are only used as ``read-and-forget'' classes by \lstinline!Acme2CylFactory!, i.e.~their sole purpose is to provide read-only data for the instantiation of the physics class.

The reason for this approach is the restriction imposed by using generic programming, where all of the types used as template parameters must be known at compile-time.
Therefore, for each of the configurations, there exists one corresponding class.
At compile-time, the code for each of these configurations is generated. This increases compilation time, but for a reasonably low number of configurations, this overhead is acceptable, as it adds the benefit that a certain simulation setup can then be chosen at runtime, by means of a config file parameter.

\paragraph{\texorpdfstring{\lstinline!Acme2CylPhysics!}{Acme2CylPhysics}} Once created, this class contains all relevant model parameters.
It is the central data class in the application and provides additional methods for extracting data attached to certain grid cells and, most importantly, to both membrane interfaces.
The values at the two opposite sides of the membrane are necessary for the calculation of the trans-membrane flux $f_i^{\text{memb}}$, as could be seen in \vref{sec:neuron_model.membrane_flux}.
For this purpose, \lstinline!Acme2CylPhysics! internally holds a map that identifies two opposite \lstinline!Dune::IntersectionIterator!s with each other, corresponding to the map $\mu(\mathbf{x})$ from \cref{sec:neuron_model.membrane_flux}.
For all the other data attached to grid entities, \lstinline!std::vector!s are used in combination with \lstinline!Dune::IndexSet!s, which provide the corresponding index used in the vectors for a given grid element.

Since grid data and some meta information about the grid and its elements is needed virtually anywhere in the application, almost every one of the following classes hold a reference to the physics object.
It is a bit unsatisfactory from the software design point of view to have such a central data class that is used everywhere in the program, but it is the only way if we do not want to sacrifice efficiency for a cleaner design.

\paragraph{\texorpdfstring{\lstinline!Acme2CylGeometrySwitch!}{Acme2CylGeometrySwitch}} This \lstinline!struct! was added when extending the simulator from Cartesian 2D to cylinder coordinates.
It seemed advantageous to be able to use both coordinate systems without having to maintain two different code bases or suffer from performance impairments, which could be achieved by using \glspl{TMP} \cite{veldhuizen2000techniques}, a technique that utilizes template specialization in order to implement conditional behavior depending on the type of an object.
This is best exemplified when looking at the code:
\begin{lstlisting}
struct Acme2CylGeometrySwitch
{
    template<typename GEO, bool useCylinderCoords = USE_CYLINDER_COORDINATES>
    struct GeometrySwitch
    {
      typedef GEO type;
    };

    template<typename GEO>
    struct GeometrySwitch<GEO,true>
    {
      typedef Acme2CylGeometry<GEO> type;
    };
  
    //...
}
\end{lstlisting}
We see that the procedure is actually quite simple.
\lstinline!Acme2CylGeometrySwitch! contains a template \lstinline!struct GeometrySwitch! with two template parameters: the type of the original (2D, Cartesian) \lstinline!Dune::Geometry! class of the current grid, and a boolean flag that evaluates to \lstinline!true! when the cylinder coordinate system is to be used. 
In the simplest case, the flag \lstinline!useCylinderCoords! is \lstinline!false! and thus the nested typedef \lstinline!type! of \lstinline!GeometrySwitch! will evaluate to the template parameter \lstinline!GEO!.
A second version of \lstinline!GeometrySwitch! is partially specialized on the second template parameter.
The compiler will give preference to this specialized version when the boolean flag is \lstinline!true!, and the nested typedef \lstinline!type! will evaluate to \lstinline!Acme2CylGeometry<GEO>! instead of \lstinline!GEO!.
It is immediately clear that the flag has to be a compile-time constant in order for this to work.
The default value \lstinline!USE_CYLINDER_COORDINATES! can be set in a global header \lstinline!constants.hh! for convenience.
The functionality of \lstinline!Acme2CylGeometry! is described next.

\paragraph{\texorpdfstring{\lstinline!Acme2CylGeometry!}{Acme2CylGeometry}} This class can be seen as a wrapper class around the original geometry class providing the transformation to cylinder coordinates.
But technically, it works differently, since it is designed as a \emph{mixin}:
\begin{lstlisting}
template<typename Geometry>
class Acme2CylGeometry : public Geometry
{
  //...
};
\end{lstlisting}
The mixin pattern describes the method of deriving from a class which itself is given as a template parameter.
It can be seen as the static polymorphism version of the famous Gang of Four \emph{decorator} pattern \cite{gamma1994design}.
Consequently, an \lstinline!Acme2CylGeometry! object is not a wrapper around the original \lstinline!Geometry!, it \emph{is} a \lstinline!Geometry! object and therefore inherits all of its methods.
This way, overhead is only added for those member functions that need to be adapted. 
Furthermore, it eliminates the overhead of additional lookups that would arise when using dynamic polymorphism, and it allows the compiler to perform better optimizations, as the father class type is known at compile-time.
In conjunction with the switch \lstinline!Acme2CylGeometrySwitch!, this can be implemented as follows:
\begin{lstlisting}[breaklines=true]
// Get geometry object from a given entity iterator
// 'eit'
const GeometryOrig& geoOrig = eit->geometry();
// Use switch to choose original or cylinder geometry 
// type, depending on compile-time flag
typedef typename Acme2\-Cyl\-Geometry\-Switch::\-GeometrySwitch<GeometryOrig,useCylinderCoords>::\-type Geometry;
// 'geo' is now either a plain 2D or a cylinder geometry
const Geometry& geo(geoOrig);
\end{lstlisting}
When the flag \lstinline!useCylinderCoords! is \lstinline!true!, the 2D geometry \lstinline!geoOrig! will be replaced by a new \lstinline[breaklines=true]!Acme2\-Cyl\-Geometry! object \lstinline!geo!, which is copy-constructed from \lstinline!geoOrig!.

When it is \lstinline!false!, the typedef \lstinline[breaklines=true]!Acme2\-Cyl\-Geometry\-Switch::\-GeometrySwitch\-<GeometryOrig>::\-type! will evaluate to the original type \lstinline!GeometryOrig!, and the only overhead is the creation of a single const reference.
With this setup we can switch between both coordinate systems simply by setting a single boolean compile-time flag.
Finally, we can also create two different executables for each case from the same code, which allows for direct comparison between 2D and 3D results.

\paragraph{\texorpdfstring{\lstinline!Acme2CylSetup!}{Acme2CylSetup}} The \lstinline!setup()! method of this class essentially plugs together all the different \Dune components, particularly the local operators.
Following the residual definition in \cref{sec:model.numerical_methods}, one operator for the assembly of each residual in \cref{eq:residual_np_time,eq:residual_np_space,eq:residual_p} has to be defined. 

In order to couple Nernst-Planck \cref{eq:np} and Poisson \cref{eq:p}, however, the residuals for the spatial part of the subdomain $\OmegaElec$ are treated together by a single operator \lstinline[breaklines=true]!Acme2CylOperatorFullyCoupled!, assembling the combined residual $R_{\OmegaElec} = \left( R_{\text{NP,S}}, R_{\text{P,}\OmegaElec} \right)^T$.
Here, $R_{\text{P,}\OmegaElec}$ denotes those entries of the full Poisson residual $R_{\text{P}}$ from elements belonging to the $\OmegaElec$ subdomain.
The class \lstinline[breaklines=true]!Acme2CylOperatorFullyCoupled! is a modification of the \lstinline[breaklines=true]!Dune::PDELab::ConvectionDiffusionFEM! local operator from \lstinline!dune-pdelab!.
The modification involves the treatment of all $N+1$ variables ($N$ concentrations, 1 potential) in a single operator.

Using a single operator of convection-diffusion type is possible because both Nernst-Planck and Poisson equations can be written in the form of a convection-diffusion \cref{eq:theory.pde_prototypes.convection-diffusion}.
The two parameter classes \lstinline!NernstPlanckParameters! and \lstinline!PoissonParameters! are used to insert the correct coefficients for \cref{eq:np} and \cref{eq:p}, respectively, into the convection-diffusion equation of form \cref{eq:theory.pde_prototypes.convection-diffusion}.

The remaining entries $R_{\text{P,}\OmegaMemb}$ of the full Poisson residual are assembled by a separate operator taking only contributions
from the subdomain $\OmegaMemb$ into account.
For this, the existing class \lstinline!Dune::PDELab::ConvectionDiffusionFEM! is used, again in conjunction with a \lstinline!PoissonParameters! class.

A third operator \lstinline!NernstPlanckTimeLocalOperator! handles the temporal part of the residual, $R_{\text{NP,T}}$.
A combined instationary operator \lstinline!Dune::\-PDELab::\-OneStep\-GridOperator! is obtained automatically by the functionality of \lstinline|dune-\-multidomain|.
This grid operator is handed over to the \lstinline!Dune::PDELab::Newton! class, and all solver and infrastructure objects are forwarded to the class \lstinline[breaklines=true]!Acme2\-Cyl\-Simulation!.

\paragraph{\texorpdfstring{\lstinline!Acme2CylSimulation!}{Acme2CylSimulation}} The simulation class is responsible for carrying out the time loop for the instationary problem.
In each iteration, the new boundary conditions are calculated, especially the membrane flux, which is done in the helper class \lstinline!MembraneFluxGridFunction!.
Then the grid operator is used to assemble the full Newton matrix and residual, which is then solved for in each iteration according to \cref{sec:theory.fem.newton}.

By a suitable choice of template parameters for the class \lstinline[breaklines=true]!Dune::\-PDELab::\-Multi\-Domain::\-Multi\-Domain\-Grid\-Function\-Space!, the ordering of \glspl{DOF} in the vector and matrix data structures can be specified.
In conjunction with additional template parameters in \lstinline[breaklines=true]!Dune::\-PDELab::\-ISTL\-Vector\-Backend!, \glspl{DOF} can be \emph{blocked} together.
In this case, all unknowns belonging to a certain node are blocked together, yielding a matrix with $(N+1) \times (N+1)$ blocks that shows a block-tridiagonal pattern.

Note that this is only possible when using a single membrane element layer, since then the number of unknowns at each grid vertex is the same, i.e.~$N$ concentration and one potential variable. 
This is the precondition for using the vertex-blocking strategy, as the \Dune matrices currently only allow block matrices of equal size in the large system matrix.
For $n > 1$ membrane elements, there are isolated potential variables at grid vertices inside the membrane, rendering the blocking strategy impossible and prohibiting the solution of large systems, as the linear solver would not converge anymore.

\begin{figure}%
\centering%
\subfloat[Lexicographic order]{\includegraphics[width=0.32\textwidth]%
{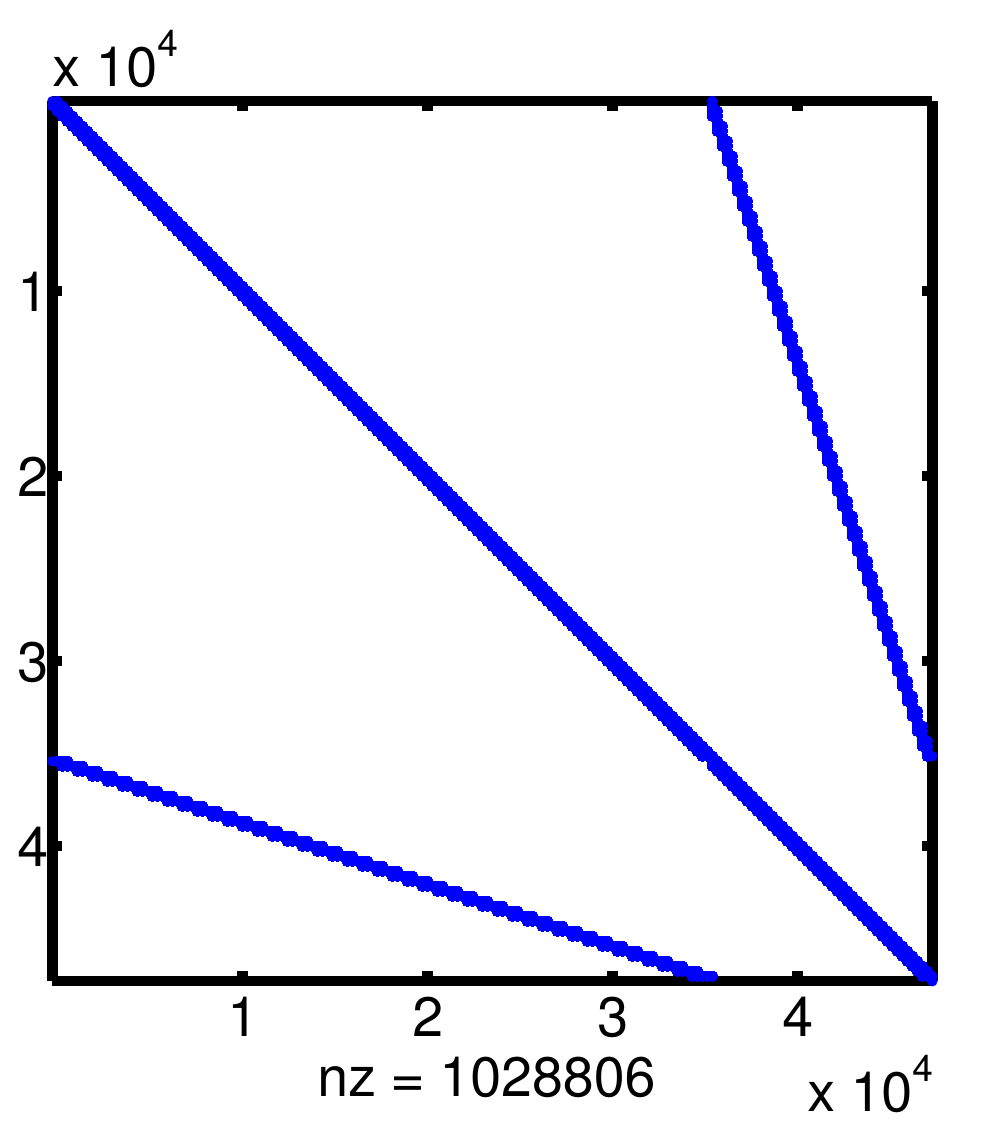}\label{fig:impl.matrix_bad}}\ %
\subfloat[Vertex-blocked order]{\includegraphics[width=0.32\textwidth]%
{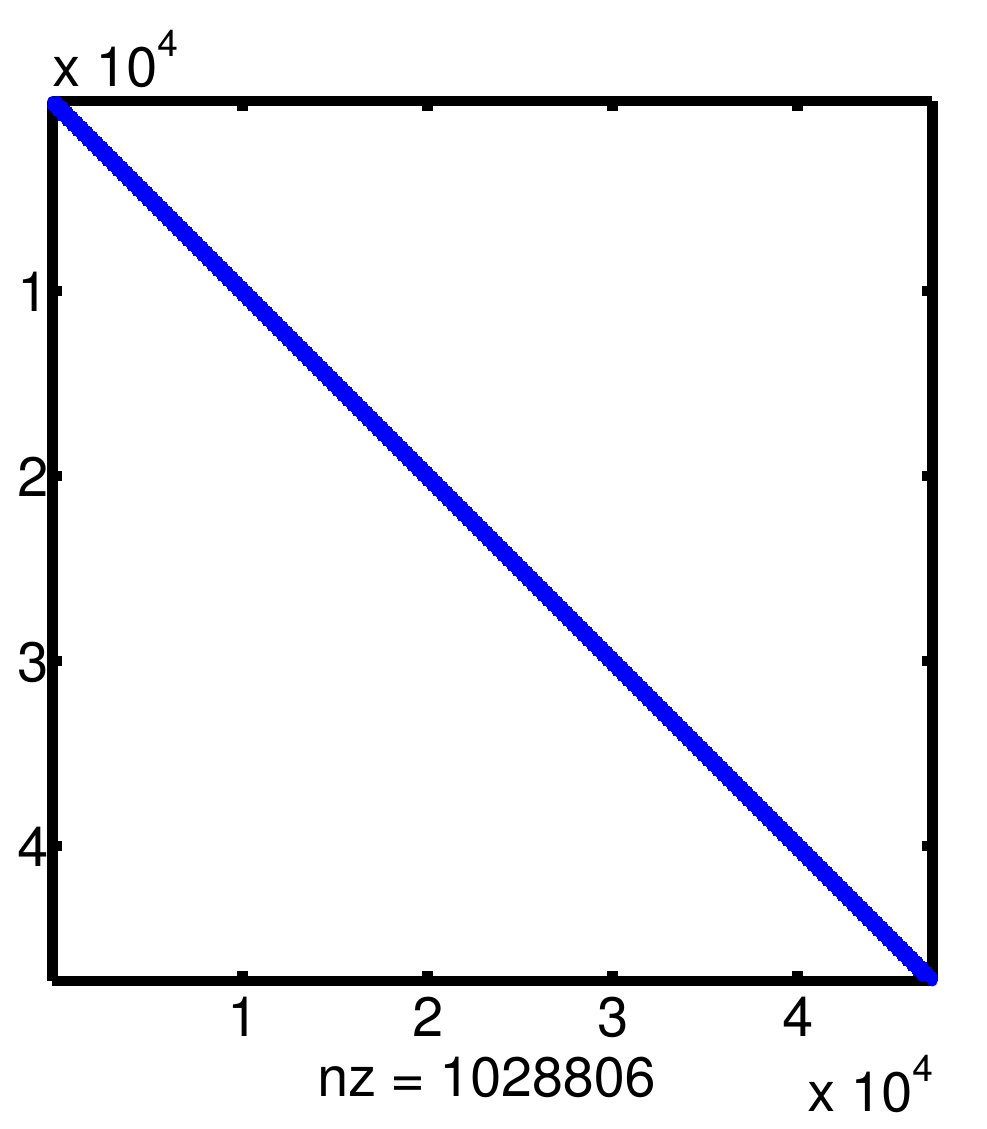}\label{fig:impl.matrix_good}}\ %
\subfloat[Zoom into (b)]{\raisebox{0.5em}{\includegraphics[width=0.332\textwidth]%
{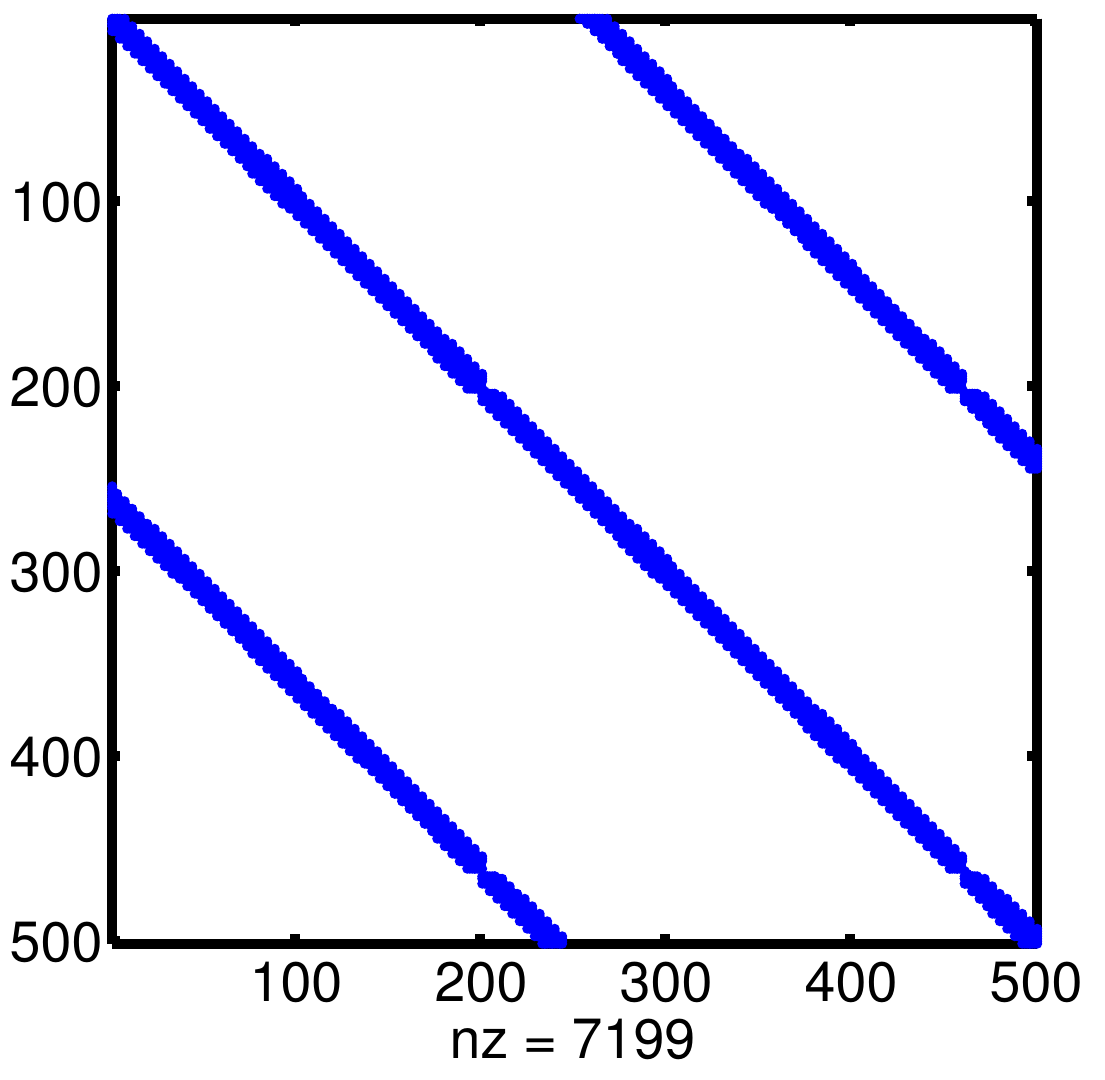}}\label{fig:impl.matrix_good_zoom}}%
\mycaption[Comparison of matrix structure depending on DOF ordering]{The left panel shows a lexicographic ordering of unknowns with an unfavorable sparsity pattern. Using a vertex-blocked ordering (\textit{center}) yields an advantageous diagonal pattern, which reveals to consist of multiple block-diagonal matrices when zooming in (\textit{right}).}%
\label{fig:impl.matrix}%
\end{figure}

The matrix pattern is illustrated in \cref{fig:impl.matrix}: in comparison to a naive lexicographic order of \glspl{DOF} as in \cref{fig:impl.matrix_bad}, the vertex-blocked order in \cref{fig:impl.matrix_good} shows a much more advantageous diagonal structure, which reveals to consist of three block-diagonals in the zoom-in \cref{fig:impl.matrix_good_zoom}.

This structure is very beneficial for the linear solver performance, as both \gls{ISTL} implementations of the \gls{ILU} and \gls{AMG} preconditioners make great use of the block structure.
For the iterative solver itself, i.e.~\gls{BiCGStab} or \gls{GMRes}, the block-diagonal pattern is absolutely crucial, since otherwise it would not converge for setups with a large number of unknowns.

The termination criterion of the \lstinline!Dune::PDELab::Newton! implementation depends on two values: the absolute norm of the residual, the \emph{defect} $\|r^k\|$ in iteration $k$, denoted by \lstinline!absLimit!, and the relative norm with respect to the initial residual $\frac{\|r^k\|}{\|r^0\|}$, the \lstinline!reduction!.
Tolerances for both errors can be provided, and convergence is acknowledged as soon as one of the error tolerances is satisfied.

Usually, prescribing a certain reduction is desired, as the absolute value of the defect depends on the mesh resolution.
But in certain situations, providing an absolute limit is necessary, for example when the desired reduction can not be reached in every time step due to changing dynamics (and hence the condition of the problem).
One example for such a case is when a system reaches steady-state. Then, the initial defect will usually be small in each time step and using the reduction from previous time steps with higher initial defects would be too restrictive, as the initial defect now is already close to the reachable limit.
These error tolerances have to be found individually for each problem, and they critically determine convergence and accuracy properties of the numerical solution.

After each solve of the \gls{PNP} system, the simulation class triggers the output of solution vectors and diagnostic data in \lstinline!Acme2CylOutput!.

\paragraph{\texorpdfstring{\lstinline!Acme2CylOutput!}{Acme2CylOutput}} The output class essentially only writes out all relevant simulation data to files.
However, the generic implementation is intrinsically quite complicated, as there is a plethora of PDELab grid functions living on different parts of the grid and on different function spaces.
Therefore, not only do the grid function classes have different domain and field types, they also require different output strategies.
For example, the membrane potential is a single value that is only defined on membrane interfaces, while the concentrations are aggregated into a vector of size $N$ that lives on electrolyte elements only.
This problem is solved again using \glspl{TMP} \cite{veldhuizen2000techniques}.
The \gls{TMP} \lstinline!OutputStrategy! switches between the different output methods, based on the type of the grid function to be written. 

Two file formats are used for the output: Gnuplot \cite{gnuplot} and HDF5 \cite{hdf5}.
For the gnuplot output, a custom output class \lstinline!GnuplotTools2D! writes ASCII files that can be examined on-the-fly while the simulation is running. 
The adapter class \lstinline!HDF5Tools! internally uses the HDF5 libraries to write the data to the complex binary \lstinline!.h5! file format.
In production runs, the gnuplot output is used only for the output of small diagnostic data, while the large solution vectors are written to the more storage-efficient HDF5 format.

\setcounter{secnumdepth}{4}

\setchapterpreamble[u]{%
\dictum[Michail Kalashnikov]{Things that are complex are not useful, things that are useful are simple.}\bigskip}
\chapter{Model of an Unmyelinated Axon}\label{chap:unmyel}
After the definition of the model in \cref{chap:model_general} and its implementation in \cref{chap:implementation}, we are now ready to look at the results for an unmyelinated, homogeneous axon in extracellular fluid.
Again, this chapter is based largely on an edited version of \cite{pods2013electrodiffusion}.

\section{Simulation Setup}\label{sec:results.parameters}
In the following, we consider a square computational domain of size \SI{10}{\milli\metre} $\times$ \SI{10}{\milli\metre}, where the axon extends from $y=0$ to $\yMemb=\SI{500}{\nano\metre}$.
Note that $\yMemb$ represents the axon radius due to cylinder symmetry. The membrane thickness was chosen to be $\dMemb = \SI{5}{\nano\metre}$.

The choice of grid parameters for the main setup used for most of the results in this chapter can be found in \cref{tab:sim_params_unmyel}.
Cases with a different choice of parameters will be explicitly mentioned.
In $x$-direction, a uniform spacing of $h_x$ is used.
The minimum Debye length for the given intra- and extracellular concentrations is about \SI{0.9}{\nano\metre}, so a minimum grid spacing of $\hyMin = \SI{0.5}{\nano\metre}$ was chosen at the membrane in $y$-direction to account for this.
For the rest of the $y$-direction, a mixture of geometrically increasing and equidistant mesh widths was used: starting from the membrane, the grid spacing is smoothly increased up to a maximum of $\hyMax = \SI{100}{\micro\metre}$.
The large difference between these lengths underlines the multi-scale character of this model, resulting in a maximum anisotropy of $\frac{h_x}{h_y}= \num{200000}$.

The diffusion coefficients $D_i$ were chosen to be the diffusivity in water for each ion species.
The relative permittivity $\epsilon$ was 80 in the electrolytes and 2 on the membrane, in accordance with \cite{Lu20106979}.
The temperature was fixed at $T = \SI{6.3}{\celsius}$ as in the original \gls{HH} model \cite{hodgkin1952quantitative}.

\newpage
\begin{longtabu} to \textwidth {X[0.5] S s p{0.48\textwidth}}
\caption[Simulation parameters for the unmyelinated axon model]{\textbf{Simulation parameters for the unmyelinated axon model}.}\label{tab:sim_params_unmyel}\\
\toprule
  Parameter & {Value} & {Unit} & Description \\
\midrule
\endfirsthead
\toprule
  Parameter & {Value} & {Unit} & Description \\
\midrule
\multicolumn{4}{c}{\sffamily NUMERICS}\\
\midrule
\endhead
  \multicolumn{4}{c}{\sffamily GRID}\\
\midrule
  $\xMax$ & 10 & \milli\metre & Domain size ($x$-direction) \\
  $\yMax$ & 10 & \milli\metre & Domain size ($y$-direction) \\
  $\yMemb$ & 500 & \nano\metre & Radius of the axon \\
  $\dMemb$ & 5 & \nano\metre & Membrane thickness \\
  $h_x$ & 100 & \micro\metre & Mesh size in $x$-direction\\ 
  $\hyMin$ & 0.5 & \nano\metre & Minimum mesh size in $y$-direction (Debye layer)\\
  $\hyMax$ & 100 & \micro\metre & Maximum mesh size in $y$-direction \\
  \#\glspl{DOF} & 73124 && Total number of unknowns \\
\midrule
  \multicolumn{4}{c}{\sffamily PHYSICS}\\
\midrule
  $\epsMemb$ & 2 && Membrane permittivity \\
  $\epsElec$ & 80 && Electrolyte permittivity \\[0.5em]
  $\naCytosol$ & 12 & \milli\molar & Intracellular $\naIon$ bulk concentration \\
  $\kCytosol$ & 125 & \milli\molar & Intracellular $\kIon$ bulk concentration \\
  $\clCytosol$ & 137 & \milli\molar & Intracellular $\clIon$ bulk concentration \\
  $\naExtra$ & 100 & \milli\molar & Extracellular $\naIon$ bulk concentration \\
  $\kExtra$ & 4 & \milli\molar & Extracellular $\kIon$ bulk concentration \\
  $\clExtra$ & 104 & \milli\molar & Extracellular $\clIon$ bulk concentration \\
  $\DNa$ & 1.33e-9 & \square\metre\per\second & $\naIon$ diffusivity \\
  $\DK$ & 1.96e-9 & \square\metre\per\second & $\kIon$ diffusivity \\
  $\DCl$ & 2.03e-9 & \square\metre\per\second & $\clIon$ diffusivity \\
  $\gNav$ & 120 & \milli\siemens\per\centi\metre\squared & Conductance of the voltage-gated Na channel \\
  $\gKv$ & 36 & \milli\siemens\per\centi\metre\squared & Conductance of the voltage-gated K channel \\
  $\gL$ & 0.5 & \milli\siemens\per\centi\metre\squared & Total leak conductance\\
\midrule
 \multicolumn{4}{c}{\sffamily NUMERICS}\\
\midrule
 \lstinline!reduction! & 1e-5 && Newton reduction \\
 \lstinline!absLimit! & 1e-5 && Newton absolute limit \\
 $\tEnd$ & 20 & \milli\second & Simulated time \\
 $\dtMin$ & 0.05 & \micro\second & Minimum time step \\
 $\dtMax$ & 50 & \micro\second & Maximum time step \\
 $\dtMaxAP$ & 10 & \micro\second & Maximum time step during \gls{AP} \\
 $\itMin$ & 10 && Newton iteration threshold for time step increase \\
 $\itMax$ & 30 && Newton iteration threshold for time step decrease \\
\bottomrule
\end{longtabu}

\subsection{Parallelization}\label{sec:unmyel.parallelization}
For the simulation parameters summarized in \cref{tab:sim_params_unmyel}, the problem consists of \num{73124} unknowns per time step.
When simulating until $\tEnd = \SI{20}{\milli\second}$, we obtain an average time step size of \SI{13.6}{\micro\second}, resulting in a total computation time of about \SI{19}{\hour} for the sequential solution on a single processor. 

While this is not intractably long, it is certainly not suitable for rapid prototyping.
Therefore, it seemed beneficial to parallelize the algorithm, also considering that we planned to add myelin to the model, which would supposedly require a much higher number of \glspl{DOF}.

\begin{figure}
\begin{center}%
\centering%
\includestandalone[width=\textwidth]{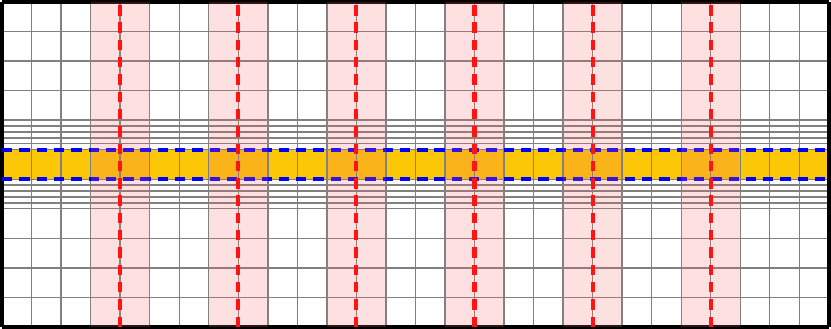}%
\end{center}%
\mycaption[Partition of the unmyelinated axon computational domain for the parallel case]{The computational domain with its three subdomains is
shown as before; additionally, processor boundaries (\textit{vertical dashed lines}) and overlap elements (\text{shaded}) are shown, in this case exemplary for $p=7$ processors. Note that this method gives optimal control over the load balancing and ensures that membrane interfaces never coincide with processor boundaries.}%
\label{fig:unmyelinated_domain_partitioning}
\end{figure}

In an effort to minimize problems due to an improper domain decomposition, the grid was chosen to be partitioned only in $x$-direction.
This way, the membrane will only be cut vertically, as suggested in \cref{fig:unmyelinated_domain_partitioning}, where the processor boundaries are marked by vertical dashed lines.
This partition ensures that the two electrolyte subdomains do not get separated from their associated patch of membrane, which would cause problems implementation-wise for the calculation of membrane fluxes.

It also prevents numerical problems, as membrane and Debye layers should be handled on a single processor to cope with the grid anisotropy.
An overlap of one cell is used at processor boundaries, marked by the shaded area in \cref{fig:unmyelinated_domain_partitioning}.

To assess the performance of the parallelization, we ran simulations on the same problem with different processor counts (\emph{strong scaling}).
\Cref{tab:unmyel.parallel_times} shows the timings for different processor counts.
For $p=10$ processors the total computation time is reduced to about \SI{2.5}{\hour}, yielding a speedup of $7.58$.

These results are quite good, considering that, according to the developers\footnote{i.e.~\Dune developers Steffen M\"uthing for \lstinline[breaklines=true]!Dune::MultidomainGrid! and Christian Engwer for \lstinline[breaklines=true]!Dune::GeometryGrid! (personal communication)}, the stacked metagrid hierarchy described in \cref{sec:implementation.components} can not be expected to scale perfectly in parallel.
In addition, we have to consider that even in the parallel runs, quite a large amount of debug output is written to disk and standard output sequentially by the root node, which further impairs the scaling performance.

Since the goal of the parallelization was not getting perfect scaling properties, but rather achieving a reasonable speedup in computation time, we were content with these results and no further attempt was made to specifically optimize the algorithm to yield a better parallel performance.

\begin{table}
\mycaption[Simulation timings for the parallel solution using different processor counts $p$]{
For each processor count $p$, the total computation time, the needed number of time steps, and the average
solution time per time step (full Newton iteration) are shown together with the resulting speedup
with respect to the sequential problem.}\label{tab:unmyel.parallel_times}
\centering
\begin{tabu} to 0.8\textwidth {@{} p{0.5cm} *4{X[l]} @{}}
\toprule
$p$ & Total comp. time [s]& \# time steps & avg. time / time step [s] & speedup\\
\midrule
1 & 69042.3 & 1469 & 47.00 &\\
2 & 35556.1 & 1469 & 24.20 & 1.94\\
4 & 18721.3 & 1469 & 12.74 & 3.69\\
10 & 9101.12 & 1469 & 6.20 & 7.58\\
\bottomrule
\end{tabu}
\end{table}

\subsection{Linear Solver and Numerical Performance}
The number of \num{73124} \glspl{DOF} is a small one within the context of \gls{HPC}, and this would allow for the usage of a direct \gls{LS} like SuperLU on each processor, even in the sequential case.
Nevertheless, we chose to use an iterative solver in perspective of solving larger problems.
For the present problem, an overlapping \gls{BiCGStab} solver preconditioned by an \gls{ILU}0 decomposition proved to be a good choice, as visible in the solver statistics \cref{tab:unmyel.solver_statistics}.

Only one or two linear solver iteration per Newton iteration were needed on average, indicating that the \gls{ILU}0 preconditioner is very effective.
This is not surprising, since we deliberately chose to group together all unknowns at a certain grid vertex in order to create dense $4 \times 4$ blocks.
This arrangement allows the \gls{ILU}0 preconditioner to invert the blocks on the diagonal \emph{exactly}, which essentially captures the nonlinearity in the \gls{PNP} system, the coupling between Nernst-Planck and Poisson equations.

\begin{table}
\mycaption[Solver statistics per time step for the parallel solution using different processor counts $p$]{All values are averages over all time steps. The total time includes matrix and residual assembly as well as the actual solution time by Newton's method. The solver time includes both \gls{ILU} decomposition and \gls{LS} time, the \gls{LS} time only the actual \gls{BiCGStab} solve. The number of linear solver iterations per Newton iteration stays at a very low level for all processor counts.}\label{tab:unmyel.solver_statistics}
\centering
\begin{tabu} to \textwidth {@{} p{0.3cm} X[0.4,l] *3{X[0.8,l]} X[1,l] X[0.4,l] X[1,l] X[0.7,l]@{}}
\toprule
$p$ & {speed\-up} & \centering total time & \centering assembler time [s] & solver time [s] & \centering \gls{LS} time [s] &\centering \gls{LS} it. & \centering \gls{LS} time / it. [s] & \centering Newton it.\\
\midrule
1 && 44.9519 & 38.0977 & 0.30445 & 0.085314 & 1.22 & 0.070337 & 1.9993\\
2 & 1.94 & 23.1669 & 19.4495 & 0.34784 & 0.137967 & 2.3 & 0.059875 & 1.9993\\
4 & 3.69 & 12.1984 & 10.2244 & 0.22298 & 0.095068 & 2.44 & 0.043907 & 1.9993\\
10 & 7.55 & 5.9506 & 4.8945 & 0.27284 & 0.128984 & 2.8 & 0.046578 & 1.9993\\
\bottomrule
\end{tabu}
\end{table}

\section{Simulation Results}
The remaining part of this chapter serves to show several simulation results for the case of an unmyelinated axon.
The most interesting case of an axon firing an action potential and demonstrating the various processes involved in this state of excitation, however, requires a valid initial state, which is the resting state of the membrane.

Since the concentration distribution at resting potential as well as the precise potential profile are unknown, a transient \emph{equilibration} simulation has to be carried out for each simulation setup with a change in the computational grid, boundary conditions, or bulk concentration values $n_i^0$.
The process of obtaining a physically consistent equilibrium state is described first, followed by the actual action potential simulations.

\subsection{Equilibrium States} \label{sec:unmyel.results.equilibrium}
As quoted in \vref{quote:theory.ion_channels.equilibrium}, the resting state of the neuronal membrane is not a true equilibrium state in the thermodynamical sense, as only the \emph{sum} of fluxes goes to zero.
Nevertheless, we will use the terms \emph{resting state} and \emph{equilibrium state} synonymously, which is also common practice in the literature.
The reader should keep in mind that indeed the more general concept of a \emph{flux equilibrium} is meant by this.

The procedure of obtaining the equilibrium state is as follows: the model from \cref{sec:neuron_model.2d} is initialized by setting the ion concentrations within one electrolyte domain uniformly to their intra- and extracellular bulk values $n_i^0$ (see \cref{tab:sim_params_unmyel}).
Then the leak channels are opened and the evolution of membrane currents is simulated using a fixed time step $\dt = \SI{10}{\micro\second}$ until the sum of trans-membrane fluxes is sufficiently close to zero, which, for most setups, was the case after a simulated time of \SI{1}{\second}.
It is important to note that voltage-dependent channels are kept closed during the whole equilibration to avoid premature \gls{AP} firing.

The bulk concentrations were chosen such that each electrolyte initially is electroneutral, i.e.~the net charge is zero, which is a reasonable assumption both physically and biologically with respect to energy minimization principles.

In the following, the generated equilibrium states are depicted by plotting the relevant values along a line perpendicular to the membrane, i.e.~parallel to the $y$-axis.
This is possible because the $x$-components of the ion fluxes during equilibration are negligible, therefore the solution only changes notably in $y$-direction. 
This fact allows for a fast equilibration procedure: the equilibrium state is obtained for a grid with only one element in $x$-direction, and the hereby obtained values are interpolated onto the fine grid when starting the actual \gls{AP} simulations.
This way, the equilibrium state can be generated within a few minutes instead of hours.

When selectively opening only the leak channel for one ion species, the equilibrium membrane potential is expected to be equal to the corresponding ionic reversal potential, as predicted by \cref{eq:nernst}.
The first two rows in \cref{tab:leak_conductances} show the calculated equilibrium potentials which indeed match the value calculated by Nernst's equation.

When opening both Na and K leak channels, the equilibrium membrane potential will reach a value between the two channels reversal potentials, depending on the ratio of Na and K leak conductances.
The relative leak conductances that result in a resting potential of about \SI{-65}{\milli\volt} can be found in the third row of \cref{tab:leak_conductances}.
The resting potential exactly matches the value predicted by the \gls{PCM} \cref{eq:pcm}.
The total leak conductance (\SI{0.5}{\milli\siemens\per\square\centi\metre}) was always kept constant.

\begin{table}
\mycaption[Relative leak conductances and resulting equilibrium membrane potentials]{}\label{tab:leak_conductances}
\centering
\noindent \begin{tabu} to \textwidth {@{} S @{ $\times$ } S @{ } s[table-alignment=left]%
S @{ $\times$ } S @{ } s[table-alignment=left ] %
l @{\qquad\qquad} S[table-alignment=right] @{ } s[table-alignment=left] @{}}%
\toprule%
\multicolumn{6}{c}{\small Leak conductances} & \multicolumn{3}{c}{\small Equilibrium membrane potential}\\
\multicolumn{3}{c}{$\gNaL$} & \multicolumn{3}{c}{$\gKL$} & \multicolumn{3}{c}{$\membPot$}\\
\midrule
1.0 & 0.5 & \milli\siemens\per\centi\metre\squared 
& 0.0 & 0.5 & \milli\siemens\per\centi\metre\squared && 50.62 & \si{\milli\volt}\\
\addlinespace[1em]
0.0 & 0.5 & \milli\siemens\per\centi\metre\squared 
& 1.0 & 0.5 & \milli\siemens\per\centi\metre\squared && -82.18 & \si{\milli\volt}\\
\addlinespace[1em]
0.13 & 0.5 & \milli\siemens\per\centi\metre\squared
& 0.87 & 0.5 & \milli\siemens\per\centi\metre\squared && -64.92 & \si{\milli\volt}\\
\bottomrule
\end{tabu}
\end{table}

\begin{figure}%
\centering%
\includegraphics[width=\textwidth]{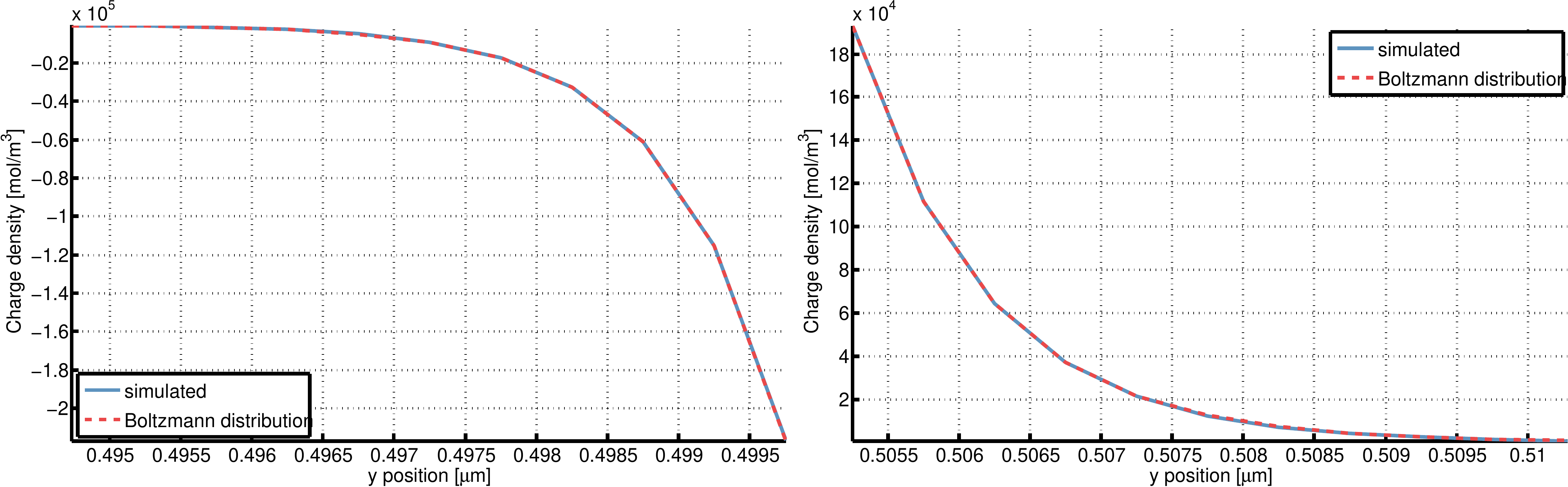}%
\mycaption[Equilibrium charge density]{The equilibrium charge density resulting from contributions of all three ion species (\textit{solid lines}), compared with a Boltzmann distribution (\textit{dashed lines}). Only the range close to the membrane is depicted, where the charge density profile undergoes its greatest change.}%
\label{fig:equi_conc}%
\end{figure}

\begin{figure}%
\centering%
\includegraphics[width=0.8\textwidth]{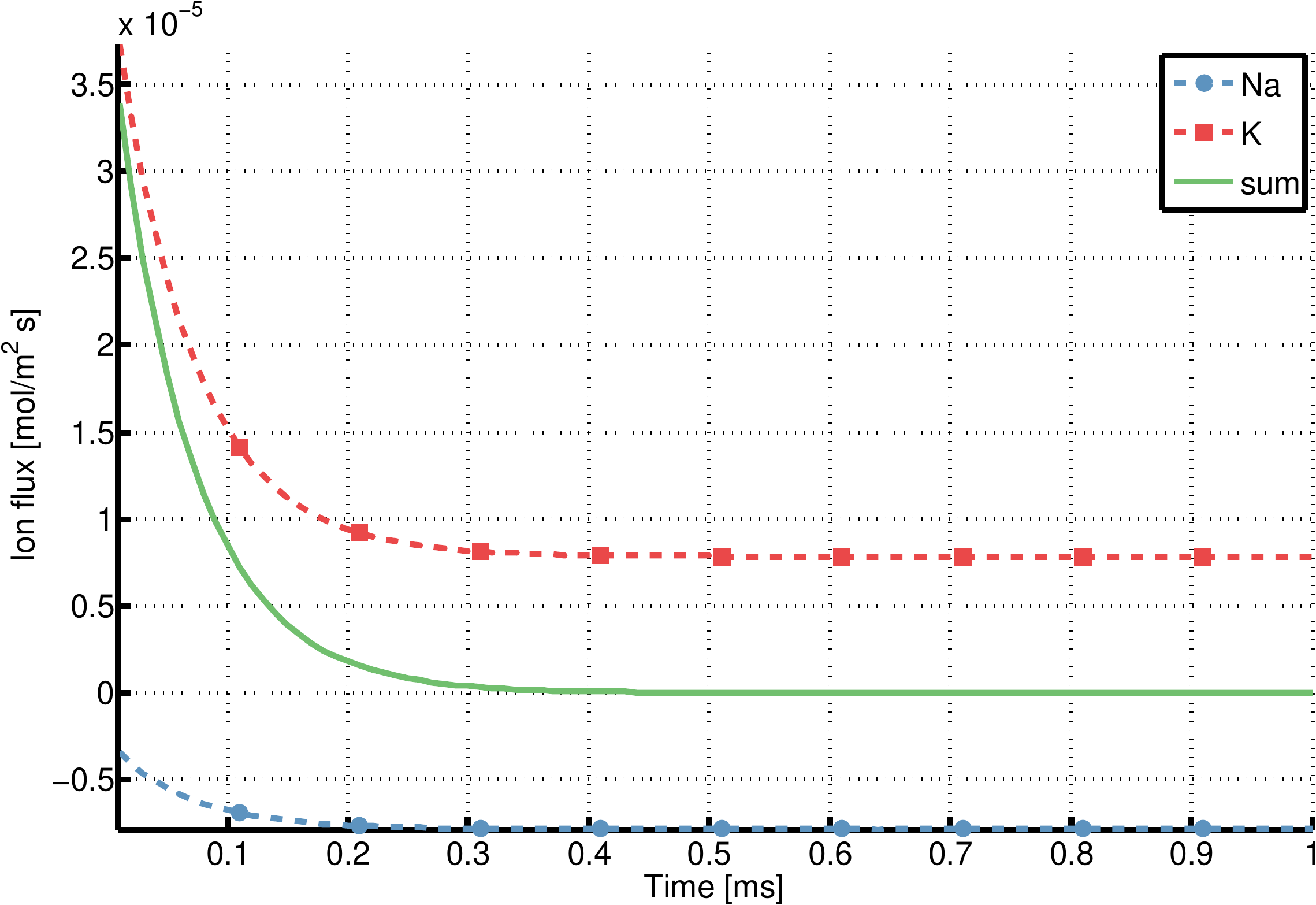}%
\mycaption[Equilibration of membrane fluxes]{Na (\textit{lower line}), K (\textit{upper line}), and summed membrane fluxes (\textit{middle line}) are shown for the equilibration phase. The sum of fluxes vanishes at equilibrium.}%
\label{fig:equi_memb_flux}%
\end{figure}%

\Cref{fig:equi_conc} shows the intra- and extracellular charge density profile at equilibrium. As predicted by the Poisson-Boltzmann \cref{eq:pb}, both electrolytes adjust their concentrations to follow a Boltzmann distribution towards the membrane.
\Cref{fig:equi_memb_flux} shows the evolution of membrane fluxes during the equilibration phase. The sum of inward- and outward-directed fluxes tends to zero, marking the neuron's resting state.

\subsection{Action Potential}
The action potential simulations are carried out by loading the equilibrium state generated as described in the previous section.
Then the gating particles $p$ of voltage-gated channels are initialized to their steady state 
\begin{align*}
  p_{\infty}(\vRest) = \frac{\alpha_p (\vRest)}{\alpha_p (\vRest) + \beta_p (\vRest)}
\end{align*}
with respect to the calculated membrane resting potential $\vRest$, as in \cite[146]{koch2004biophysics}.
This is also the default initialization procedure in NEURON \cite[chapter 8.3]{hines1997neuron}.
As opening the voltage-gated conductances changes the ratio of conductances between ion species, the leak channel ratio is corrected such that the ratio of each ion species' sum of conductances with respect to the total membrane conductance stays constant, corresponding to the values chosen for the equilibration procedure (cf.~\cref{tab:leak_conductances}).
This ensures that the membrane potential $\membPot$ does not drift off from the generated resting potential $\vRest$ after opening the voltage-gated channels.

Here we acknowledge again that only \emph{relative} conductances matter for the membrane potential, as can be seen both from the Goldman \cref{eq:goldman} and the \gls{PCM} \cref{eq:pcm}, i.e. the general validity of the equilibrium state is maintained by this modification.
However, the recalculation of leak conductances involves rounding errors, which have a small, but observable effect on the membrane potential in the sub-millivolt range.
The system is allowed to settle with respect to the changed channel conductances for a low number of time steps to account for this.

To evoke an action potential, a sodium rectangle pulse is injected into the cell by adding a fixed amount of sodium at the stimulation site located near the left domain boundary at $\textbf{x}_\text{stim} = (\SI{150}{\micro\metre},\SI{0}{\micro\metre})$ for \SI{2}{\milli\second}.
The pulse had a value of \SI{0.965}{\nano\ampere} in this setup.

The membrane is depolarized close to the stimulation site and, after reaching threshold, an action potential is generated due to the ion channel kinetics from \cref{sec:theory.biophysics.hh}.
The potential wave travels along the axon, opening more channels along the way and keeping the action potential alive, resulting in a wave traveling at constant velocity.
The conductance velocity depends on the time constants of the ion channel kinetics, but also on the intra- and extracellular ion diffusion coefficients, and has a value of about \SI{0.93}{\metre\per\second} for this setup.

\subsubsection{Intracellular Potential}
\begin{figure}%
\centering%
\includegraphics[width=\textwidth]{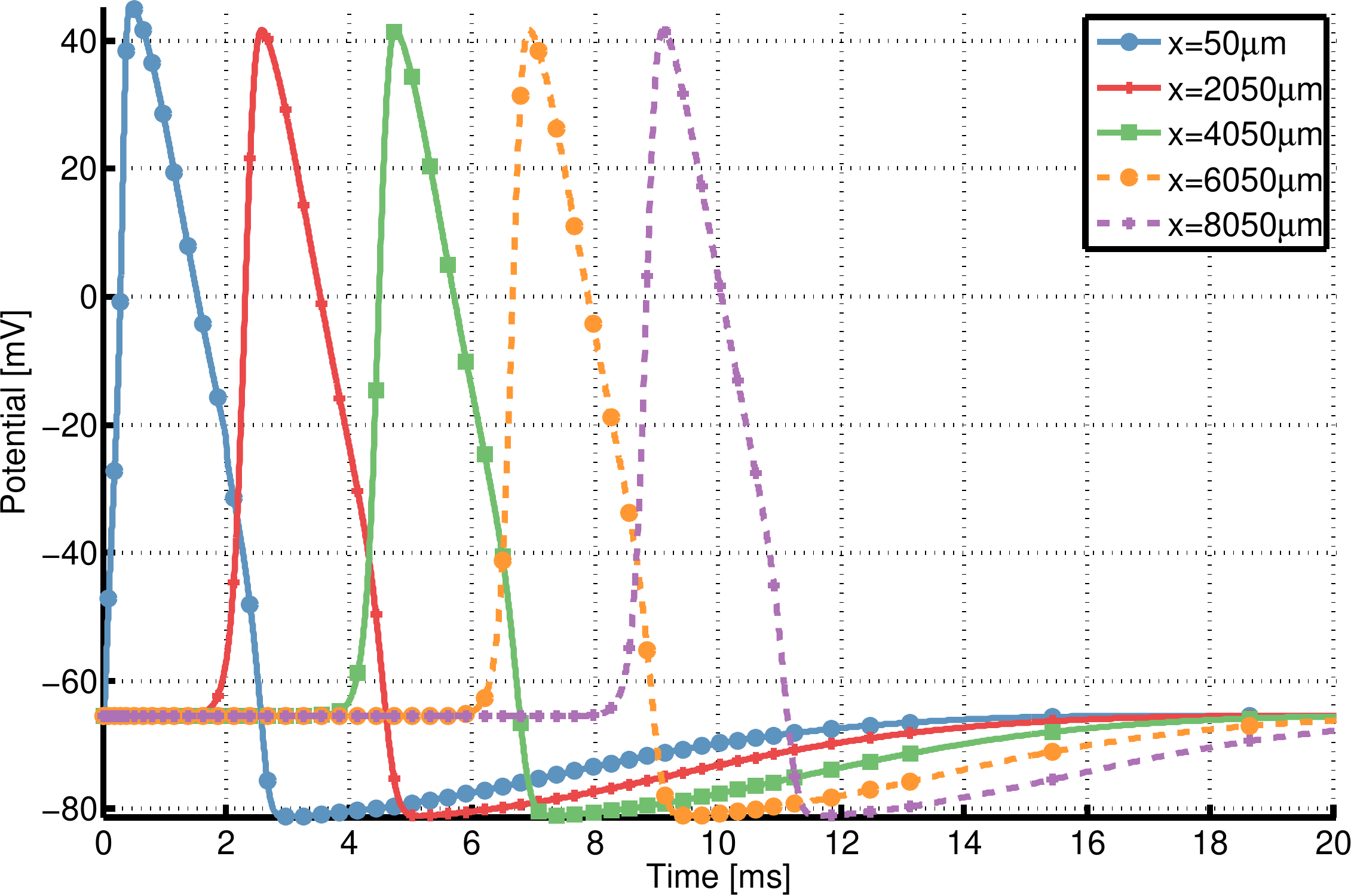}%
\mycaption[Action potentials evoked at equidistant positions along the axon]{The leftmost
AP has a higher amplitude than the others due to the vicinity to the stimulus site. The following
curves at equidistant positions along the axon are identical and show constant onset delay, 
indicating a wave traveling with constant speed.}%
\label{fig:ap}%
\end{figure}

\Cref{fig:ap} shows the potential time courses at different $x$-positions along the axon.
The $y$-position is about \SI{488}{\nano\metre}, but this does not have a large impact, as the intracellular potential is fairly constant outside the Debye layer in this direction.
In a first approximation, it is also equal to the membrane potential, since the extracellular potential is much smaller. 
The first \gls{AP} has a higher amplitude than the following ones, caused by the proximity to the stimulation site.
Also, switching off the stimulus is reflected by an artifact in the repolarization phase of the first \gls{AP} at $t=\SI{2}{\milli\second}$.

\subsubsection{Membrane Flux}\label{sec:unmyel.memb_flux}
\begin{figure}%
\centering%
\includegraphics[width=0.9\textwidth]{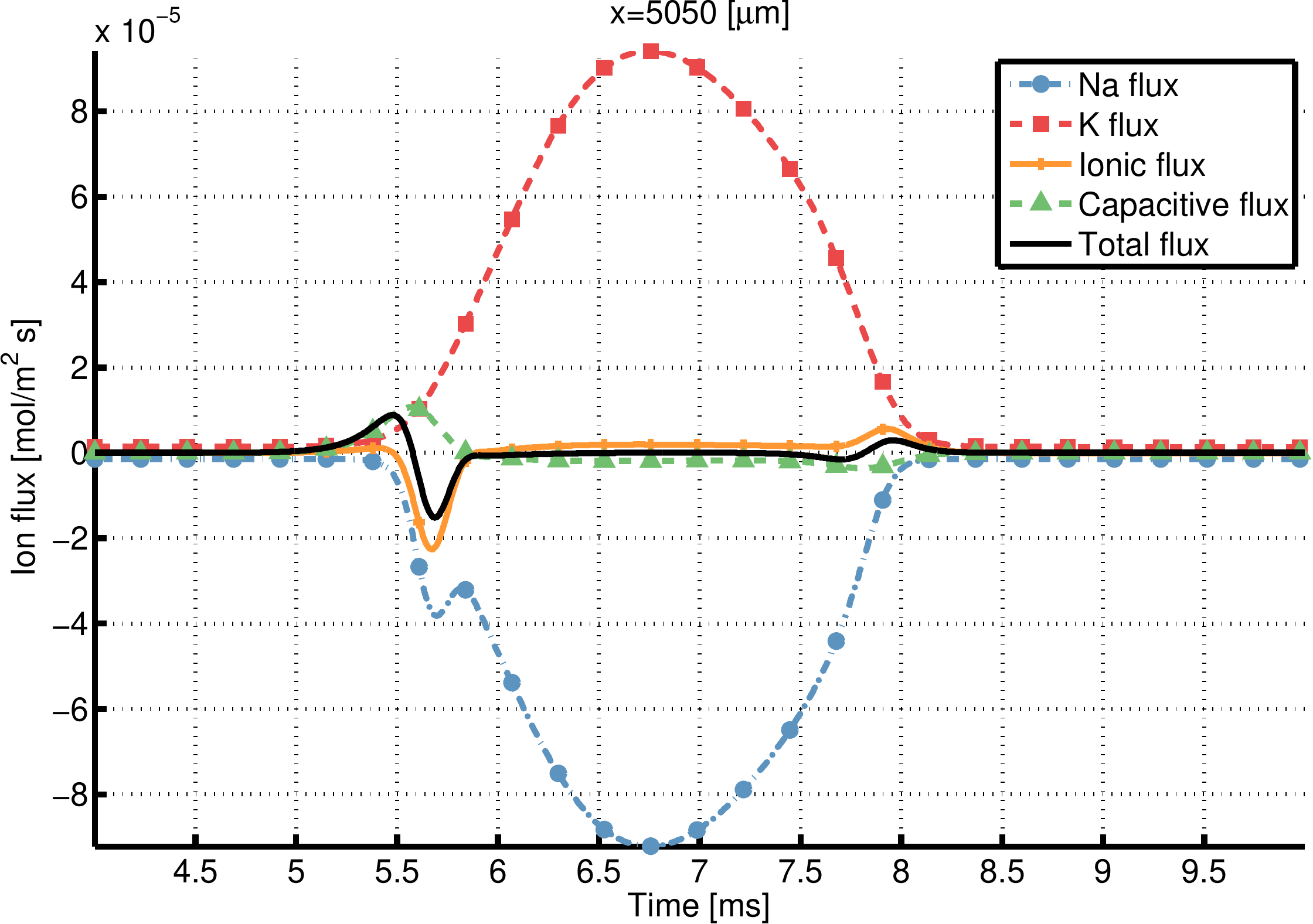}%
\mycaption[Different components of the membrane flux at fixed point on the membrane]{These are shown at the same
$x$-coordinate as the potential curves from \cref{fig:lfp_distance}. The total flux consists of
capacitive and ionic flux, which itself is the sum of Na and K channel fluxes.}%
\label{fig:ap_memb_flux}%
\end{figure}

The total membrane flux consists of two main components: an ionic and a capacitive component, see \cref{fig:ap_memb_flux}. 
The ionic flux itself is the sum of sodium and potassium fluxes through the membrane's (active and passive) ion channels. 
The other component is a consequence of the electric properties of the membrane. 
As seen in \cref{sec:theory.ion_channels}, the membrane separates charges and therefore acts as a capacitor. 
The capacitive flux is a consequence of the virtual current caused by charge redistribution at both sides of the membrane. 
In cable equation models, this current has to be specified explicitly by setting a membrane capacitance parameter, while in the electrodiffusion model, it is implicitly contained in the \gls{PNP} equations through the concentrations contained in the adjacent electrolytes. 
It can be calculated by using the textbook formula for the capacitive current
\begin{align}\label{eq:capacitive_current}
  \Icap = C \diff{U}{t} = C \frac{k T}{e} \diff{\membPot}{t},
\end{align}
where $U$ is the potential across the capacitor, in this case the membrane potential $\membPot$ brought to units of Volts, and $C$ is the membrane capacitance.

In the following, $C$ will be expressed as the capacitance per unit area, as it is also the common choice in electrophysiology literature. It can be calculated by the formula for a cylinder capacitor
\begin{align}\label{eq:memb_capacity_cylinder}
  C = 2 \pi \epsilon_0 \epsMemb \frac{l}{A \ln\frac{r_2}{r_1}} \approx \SI{0.35e-2}{\farad\per\square\metre}
\end{align}
with $l$ the length, $A$ the surface area of the membrane patch, and $r_1 = \yMemb$ and $r_2 = \yMemb + \dMemb$ marking the opposite membrane boundaries, respectively. The value turned out to be the same when using the formula for a parallel-plate capacitor
\begin{align}\label{eq:memb_capacity_parallel}
  C = \frac{\epsilon_0 \epsMemb}{\dMemb}\ ,
\end{align}
suggesting that the membrane can be regarded as a parallel capacitor in a first approximation, as the membrane thickness is small compared to the axon diameter. 
In summary, we have
\begin{align*}
  \fcap = \frac{1}{e N_A A} \Icap = \frac{1}{e N_A A} \frac{\epsilon_0 \epsMemb}{\dMemb} \frac{k T}{e} \diff{\membPot}{t}
\end{align*}
as the third trans-membrane flux next to the ionic fluxes $\fNa$ and $\fK$ defined in \vref{sec:neuron_model.membrane_flux}.

\begin{figure}%
\centering%
\subfloat[Ionic flux]{\includegraphics[width=0.45\textwidth]%
{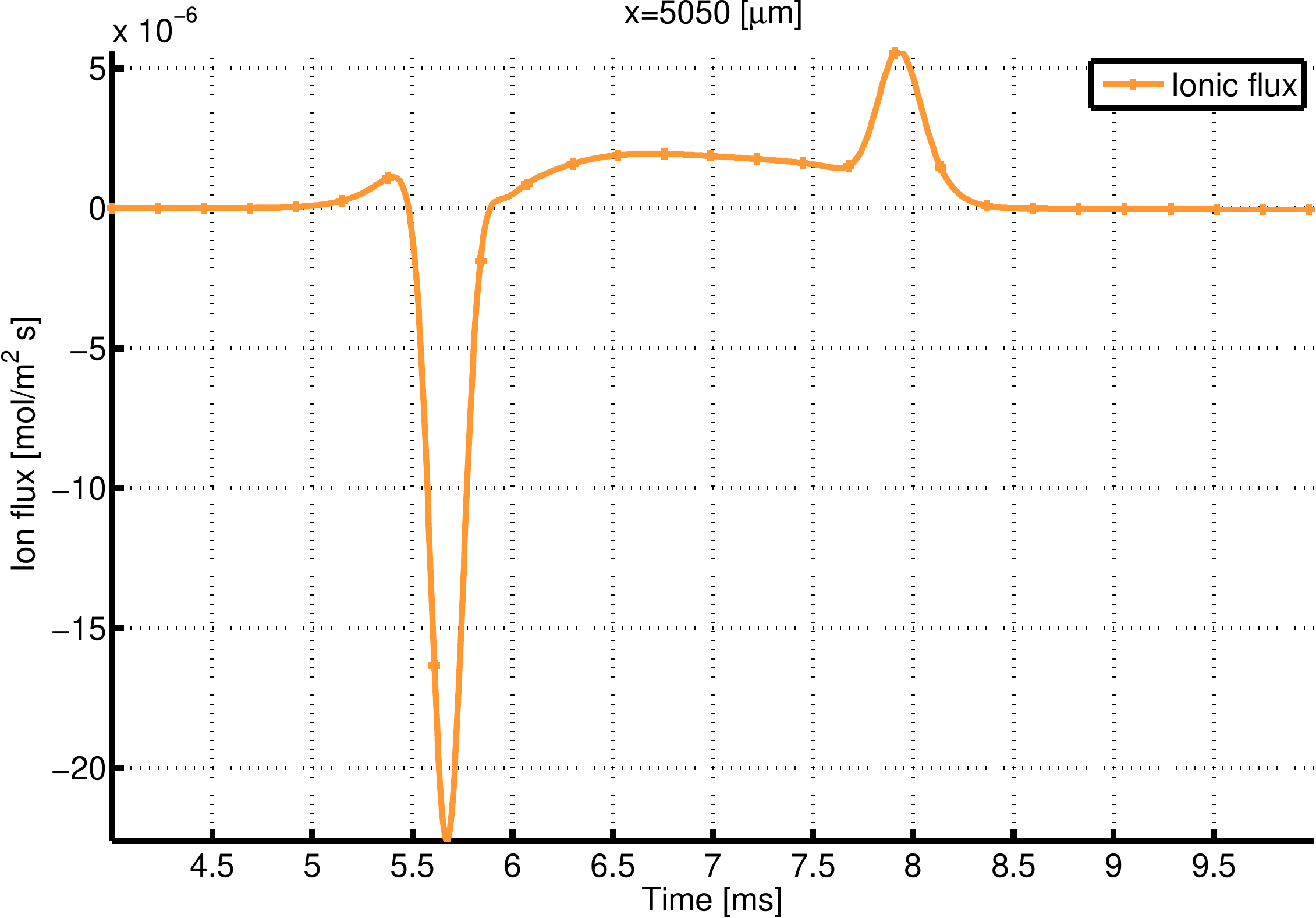}\label{fig:ap_memb_flux_ionic}}%
\subfloat[Capacitive flux]{\includegraphics[width=0.45\textwidth]%
{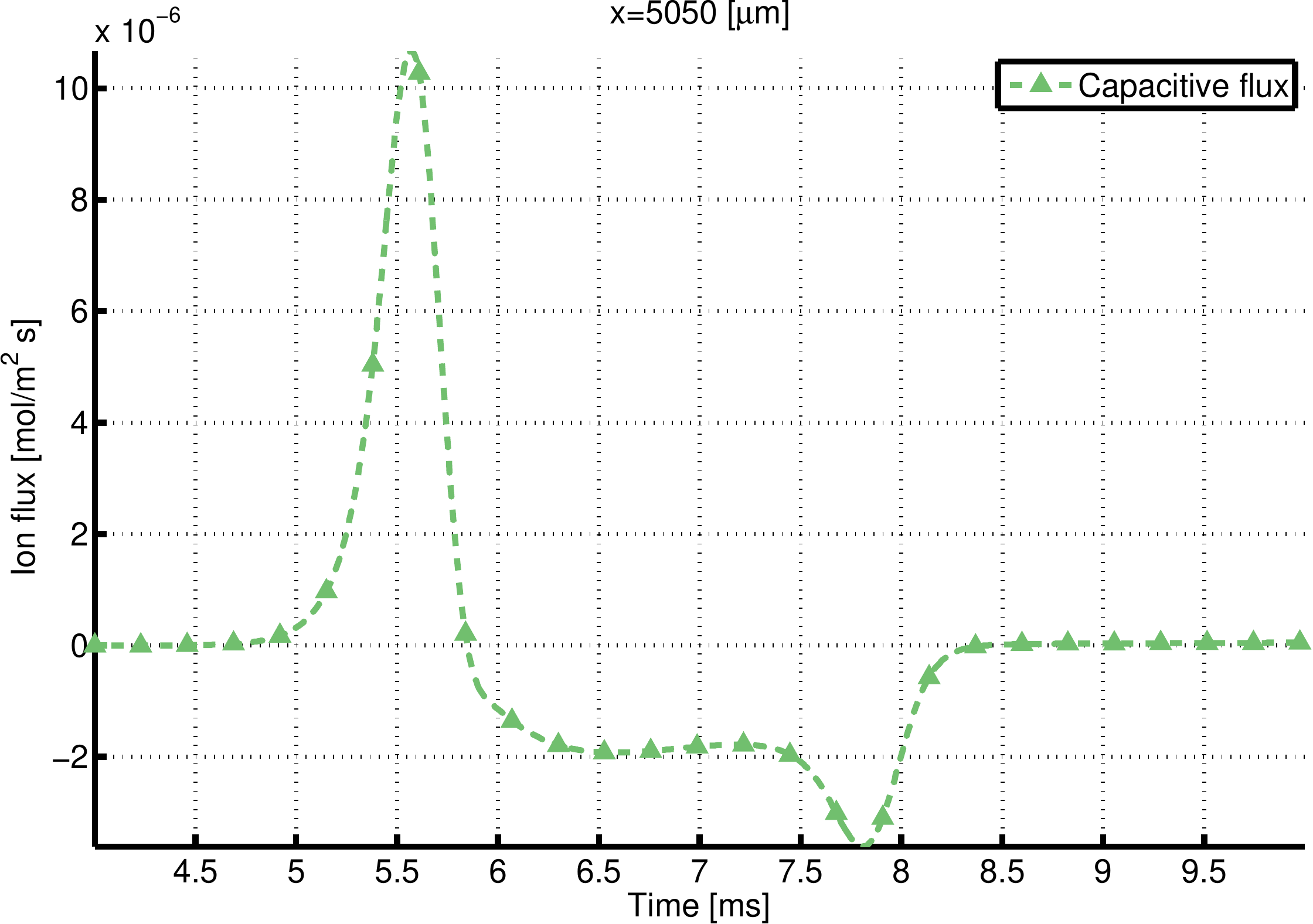}\label{fig:ap_memb_flux_cap}}%
\mycaption[Ionic and capacitive membrane flux at fixed point on the membrane]{%
Shown are the respective components of the total membrane flux from \cref{fig:ap_memb_flux}. 
The ionic flux is dominated by the sodium current during the depolarization phase, and later by
the potassium current during repolarization; therefore it has a ``down-up'' shape. On the contrary, 
the capacitive flux is proportional to the time derivative of the membrane potential, therefore 
it goes up before it goes down. Together, these two components form the ``up-down-up'' shape
of the total membrane flux in \cref{fig:ap_memb_flux}.}%
\label{fig:ap_memb_flux_components}%
\end{figure}

The total ionic flux follows the sodium flux in the rising phase of an \gls{AP} and is later antagonized by the potassium flux, resulting in a ``down-up'' shape in \cref{fig:ap_memb_flux_ionic}.
One interesting detail here is the small peak at the rear end of the ionic flux. This comes out of the standard \gls{HH} model directly, because the sodium current declines faster than the potassium current. This feature has been referred to as a ``gratuitous bump'' in \cite[307]{cole1968membranes} and can be interpreted as an artifact of the original \gls{HH} model, although later studies showed that such a structure can be observed under the influence of certain drugs in experiments.

\begin{figure}%
\centering%
\includegraphics[width=0.65\textwidth]{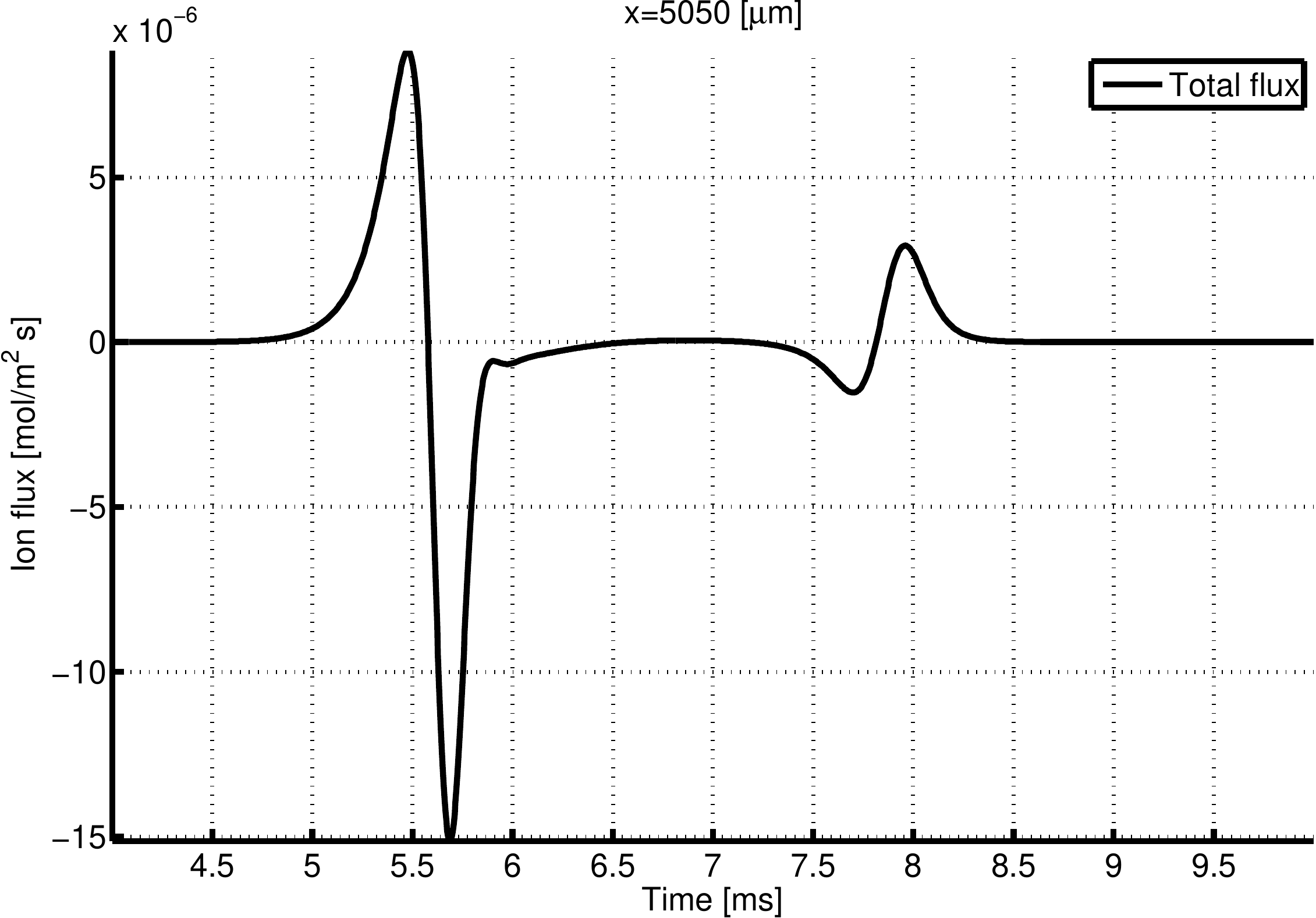}%
\mycaption[Total membrane flux at fixed point on the membrane]{This is the same total flux as in
\cref{fig:ap_memb_flux}, shown separately.}%
\label{fig:ap_memb_flux_total}%
\end{figure}

In contrast, the capacitive flux is proportional to the time-derivative of the membrane potential and therefore shows an opposed behavior in \cref{fig:ap_memb_flux_cap}.
The sum of both components results in the total membrane flux in \cref{fig:ap_memb_flux_total}, following roughly a triphasic ``up-down-up'' shape.

\subsubsection{Near- and Farfield Extracellular Potential}
\begin{figure}%
\centering%
\includegraphics[width=0.7\textwidth]{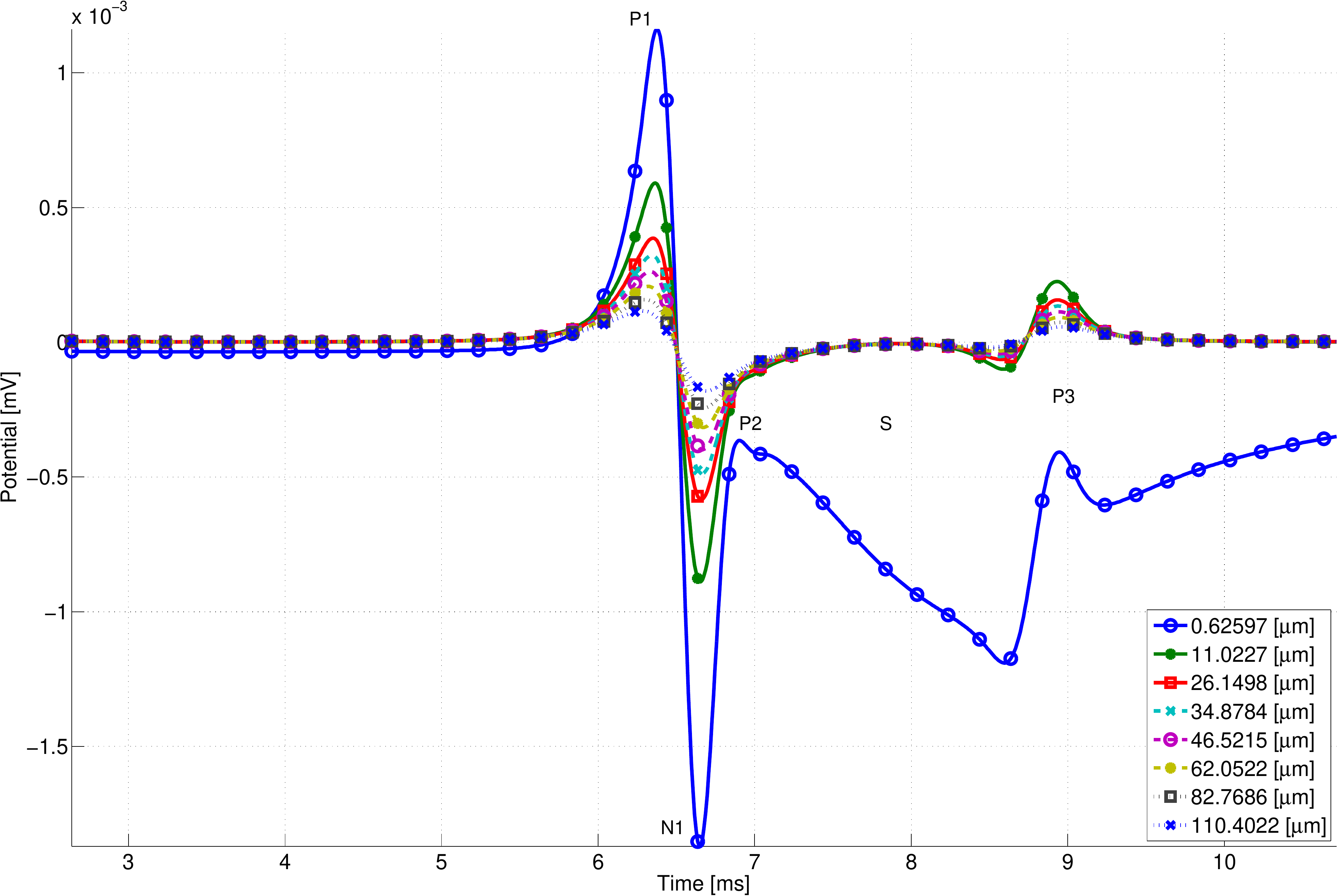}%
\mycaption[Extracellular potential at various distances from the membrane]{The potential time courses for a fixed $x$-coordinate $x=\SI{5.05}{\milli\metre}$ are shown for several different $y$ positions.
While the characteristic triphasic shape is maintained, the amplitude spans several orders of magnitude as the signal is strongly attenuated with distance.}%
\label{fig:lfp_distance}%
\end{figure}

We will now focus on the time evolution of \gls{EAP} signals at any point in the extracellular domain.
In \cref{fig:lfp_distance}, the potential time courses are plotted for the same $x$-coordinate at increasing distances from the membrane.
Some major features can be identified from these curves: a first positive peak (P1) followed by a larger negative peak (N1), then a (very) small second positive peak (P2) with a subsequent longer phase of slowly varying potential with negative curvature (S), and a last peak (P3).
This characteristic ``up-down-up'' shape is maintained at various distances from the membrane.
The potential time course generally shows similarities with the total membrane flux at the same $x$-coordinate (\cref{fig:ap_memb_flux}), suggesting that the membrane currents are the main contributors to the \gls{EAP}.

The \gls{EAP} of the point closest to the membrane, however, shows deviations from the general pattern, notably in the rear part P2 -- S -- P3.
Consequently, the second peak P2 does not look like a peak, but more like a kink at a distance of only a few micrometers away from the membrane, because the following part S shifts from a negative edge towards a more or less constant, slightly negatively curved bow.
We will see later on that this can be attributed to electrostatic forces from the membrane and the resulting concentration redistributions influencing the nearfield potential.

Since the action potential is a traveling wave, we can alternatively look at snapshots of the extracellular potential and concentrations at a fixed point in time.
These profiles simply move through space with a constant (known) velocity, such that the complete information about the potential and the \gls{EAP} dynamics at any point in space can be gained from these plots.
As the signal moves in positive $x$-direction for all following snapshots, we read from ``right to left'' when assessing the behavior of the profile in time, as opposed to the time course in \cref{fig:lfp_distance}.

\begin{figure}%
\centering%
\includegraphics[width=0.7\textwidth]{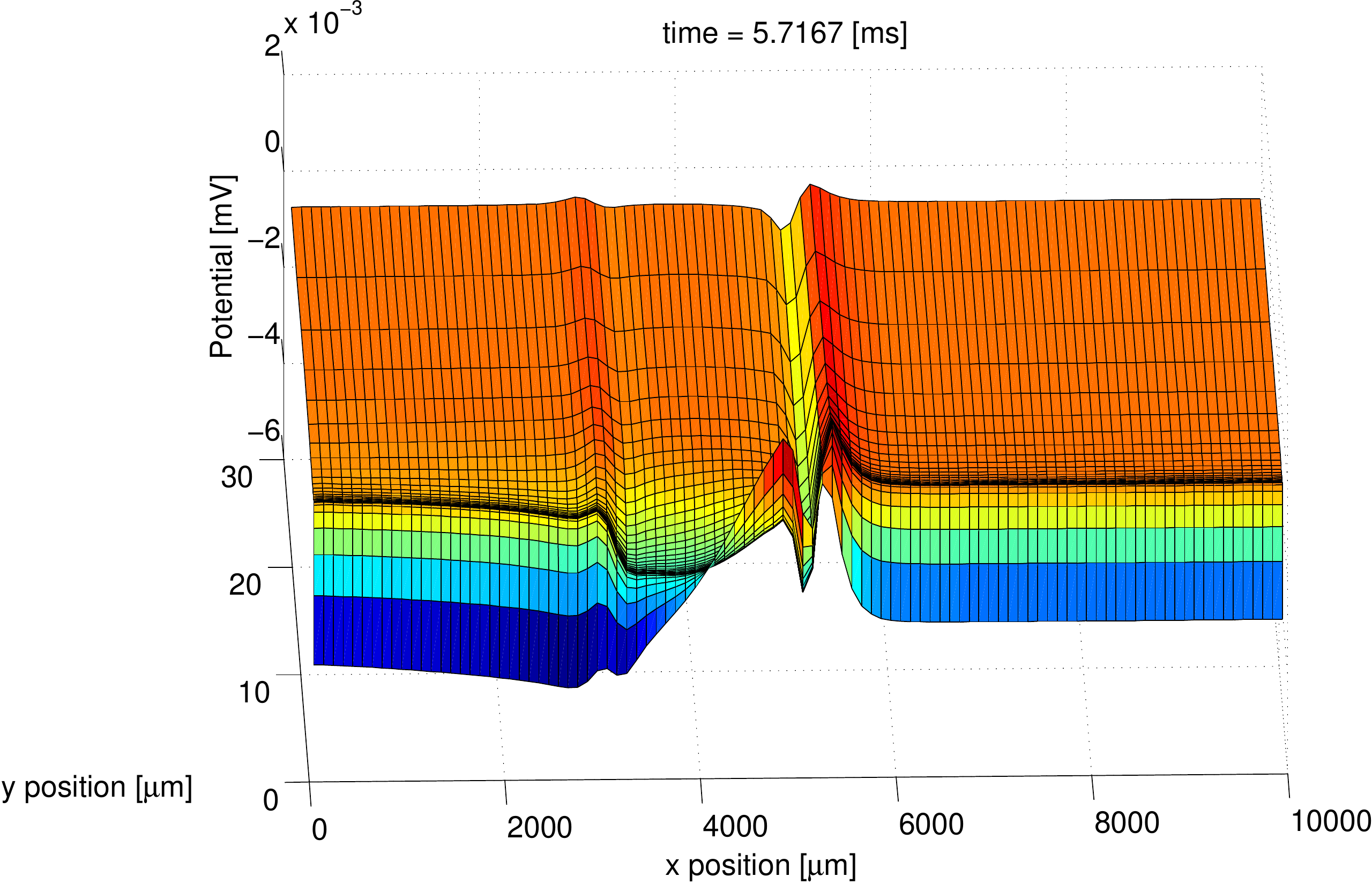}%
\mycaption[Snapshot of the nearfield extracellular potential profile]{The plot shows the potential values for a stripe of the extracellular domain $\OmegaExtra$ just above the  extracellular Debye layer (i.e., above the gray area from \cref{fig:domain_2D}), at a fixed time.
It shows a more complex structure compared to the Debye layer.}
\label{fig:lfp_nearfield_snapshot}%
\end{figure}

As expected, the nearfield potential profile (a few nanometers to about \SI{10}{\micro\metre} from the membrane) in \cref{fig:lfp_nearfield_snapshot} shows the same pattern as the total membrane flux in \cref{fig:ap_memb_flux_total} and the \gls{EAP} in \cref{fig:lfp_distance}.
The profile begins to the right with a rise in the potential (corresponding to P1) followed by a sharp drop (N1) and another rise (P2).
After this first phase, the potential has a longer phase of low variation (S) until another, less pronounced peak is observed at the rear end of the traveling action potential (P3).

The distance between the end of P1 and the beginning of P3 at the beginning of the nearfield region is a good measure for the timescale of the extracellular field we term the ``\gls{EAP} valley length'', the region with negative potential values.
It gives a characteristic length scale for the range of simultaneous ion channel activity along the axon.
In this simulation, it is about \SI{2000}{\micro\metre} at the beginning of the nearfield region. 
Of course, the \gls{EAP} valley length is largely determined by the \gls{AP} velocity.

\begin{figure}%
\centering%
\includegraphics[width=0.8\textwidth]{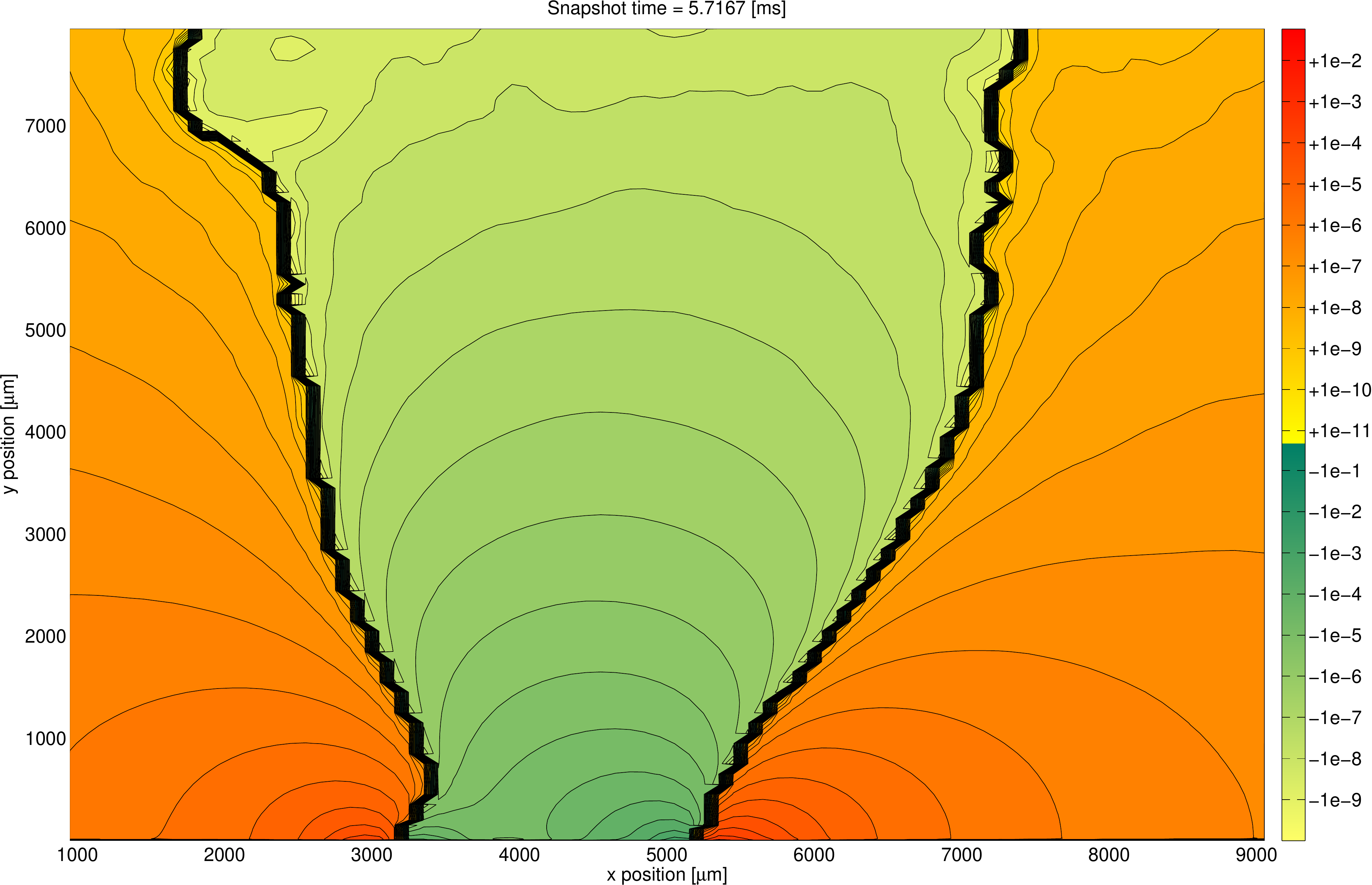}%
\mycaption[Snapshot of the whole extracellular potential]{A contour plot of the log-scaled extracellular potential %
at a fixed time point. The domain was cut at the left, right, and upper boundary to 
exclude artifacts introduced by the boundary where the potential is almost zero and
switches signs due to small numerical errors.}%
\label{fig:lfp_contour}%
\end{figure}

\Cref{fig:lfp_contour} shows the potential for a large part of the extracellular space on a logarithmic scale, demonstrating that the nearfield potential profile essentially continues into the farfield, albeit attenuating quickly with distance.
The details of the shape that could be previously seen in \cref{fig:lfp_distance,fig:lfp_nearfield_snapshot} have been smoothed out in this depiction by the logarithmic scaling, but the general pattern of the \gls{EAP} stays the same: a positive upwind domain (P1) just in front of the opening channels, followed by a negative middle region (N1, S), and then again a positive rear domain (P3).
The \gls{EAP} valley length (the diameter of the green area) increases notably with distance, which we account to the diffusive character of the system. 

\subsubsection{Debye Layer Extracellular Potential}
\begin{figure}%
\centering%
\subfloat[Debye layer potential]{\includegraphics[width=0.5\textwidth]%
{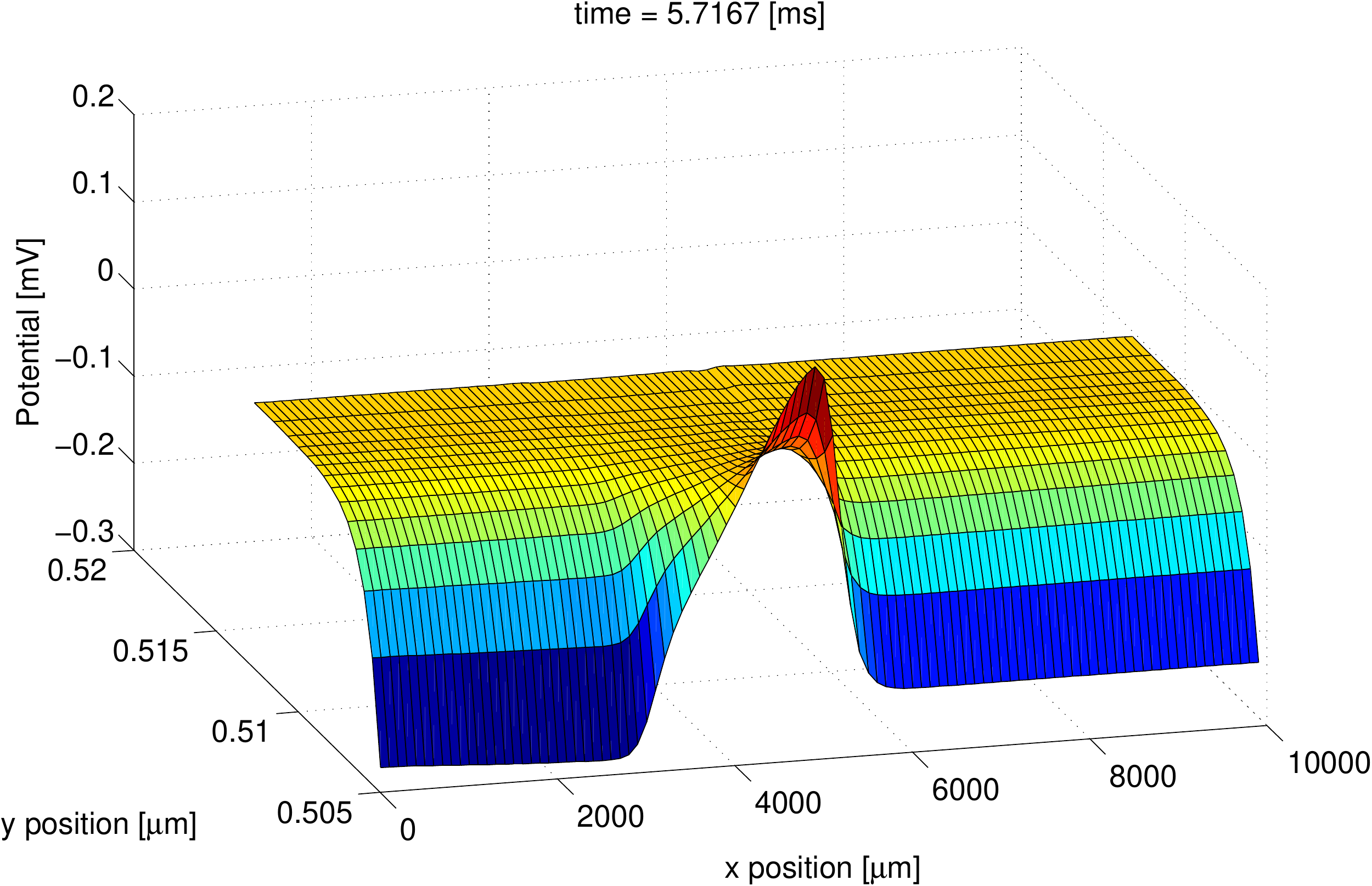}\label{fig:lfp_debye_snapshot1}}%
\subfloat[Intracellular potential]{\includegraphics[width=0.5\textwidth]%
{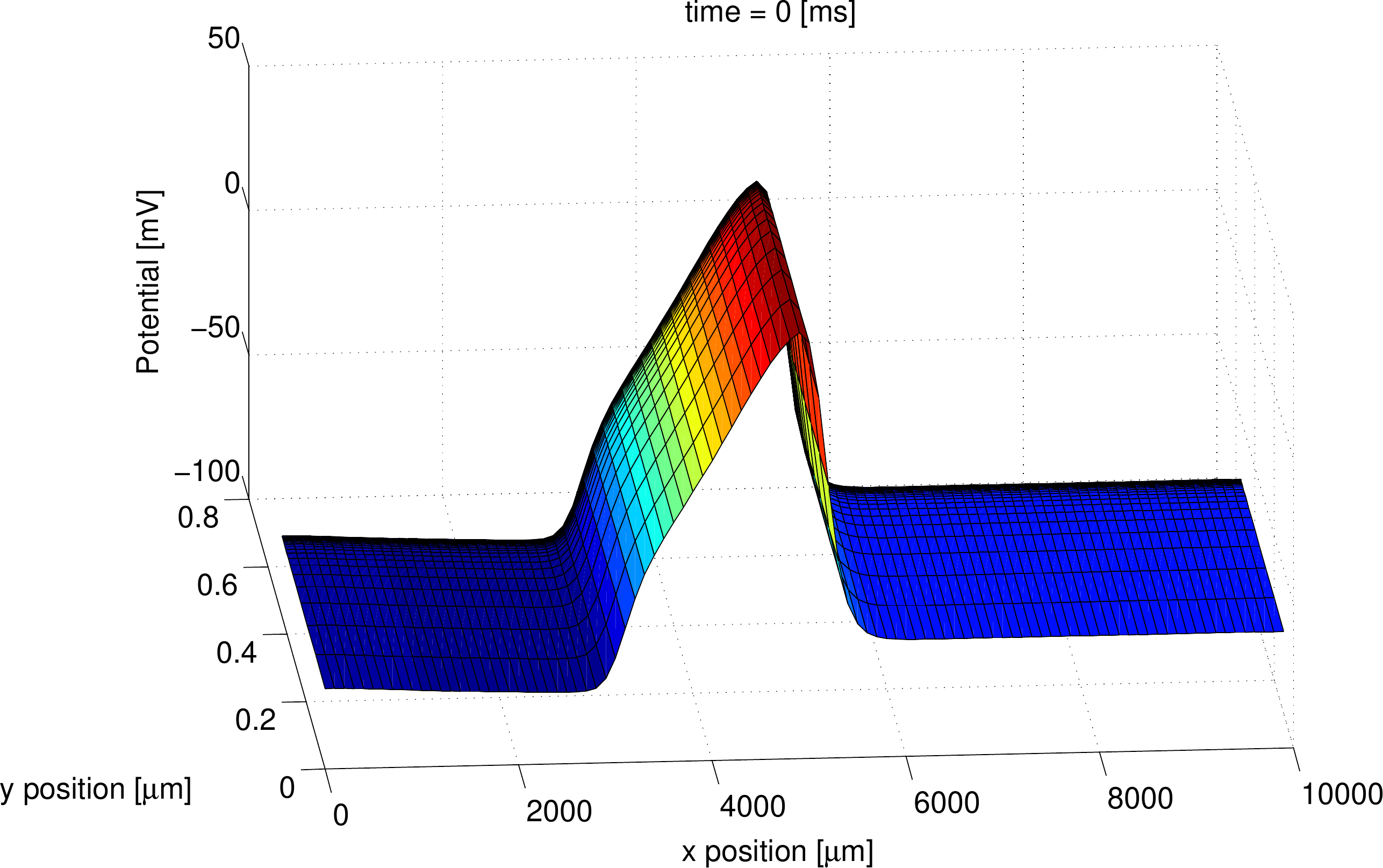}\label{fig:lfp_debye_snapshot2}}%
\mycaption[Snapshot of the Debye layer extracellular potential profile]{In the left part, %
the potential %
values for a narrow stripe of the extracellular domain $\OmegaExtra$ %
(corresponding to the gray area from \cref{fig:domain_2D}) %
just above the membrane are plotted at a fixed time. %
This turns out to be the intracellular action potential (right) propagating over the membrane, %
with a significantly reduced amplitude.}%
\label{fig:lfp_debye_snapshot}%
\end{figure}

The spatial structure of the extracellular potential profile within the Debye layer, i.e.~at most a few nanometers from the membrane, shows a different picture, see \cref{fig:lfp_debye_snapshot1}.
It is almost exclusively dominated by the intracellular potential (cf.~\cref{fig:lfp_debye_snapshot2}), meaning that the intracellular potential spreads across the membrane into the extracellular space, if with a greatly reduced amplitude.
We call this potential the ``echo'' of the action potential. 

The potential damping over the membrane will now be quantified.
In \cite[chapter 12]{lipowsky1995structure}, a parameter $s$ is used to estimate the electrical coupling between intra- and extracellular electrolytes.
It is defined as
\begin{align}\label{eq:s_coupling}
s = \frac{\dDebye}{\dMemb} \frac{\epsMemb}{\epsElec}
\end{align}
in our usual notation of lengths and permittivities, and with $\dDebye$ being the Debye length of the extracellular electrolyte.
For $s=1$, the electrolytes are fully coupled, for $s=0$ fully decoupled, meaning that the extracellular side does not see anything of the intracellular potential ($\phiOut = 0$).
In the present setup, $s \approx 0.0045$.
This is small, but not at all negligible, since extracellular potentials are generally orders of magnitude smaller than intracellular ones. As a result, a small $s$ might still yield a potential that is comparable in amplitude to the extracellular potential, in absolute numbers.
A value of $s=0.0045$ describes two weakly coupled electrolytes.

Moreover, we found that $s$ is not only a vague estimation for the coupling, it is indeed the \emph{exact} constant of proportionality between intra- and extracellular potential at the membrane interface:
\begin{align}\label{eq:pot_in_out}
\phiOut = s \phiIn \ .
\end{align}
We found this relation to be very accurate over the whole time course of a simulation for an arbitrary $x$-coordinate along the membrane, implying that the Debye layer profile in \cref{fig:lfp_debye_snapshot1} is just the constant $s$ times its intracellular counterpart in \cref{fig:lfp_debye_snapshot2}.
As a result, the membrane potential can be expressed in terms of only one of these two potentials:
\begin{align*}
\membPot = \phiIn - \phiOut = \phiIn - s \phiIn &= (1-s) \phiIn \\
&= \frac{1-s}{s} \phiOut \ . 
\end{align*}
This means we have found an analytical expression for the relation of the electric potentials at opposite membrane boundaries, provided only a few physical parameters of membrane and electrolyte are known.
It remains to be shown under which conditions this relation is valid.
In extreme cases -- where the bulk electrolyte ceases to exist, as the bulk concentrations are not constant anymore (see also \cref{chap:multiple_fibers}) -- we found that \cref{eq:pot_in_out} does not hold anymore.
Under most physiological conditions, however, it might be a valid relationship that would enable the calculation of the extracellular membrane interface potential from the intracellular one and vice versa.

Looking at the definition of $s$, another interesting connection can be made.
An extracellular solution with higher ionic strength in the sense of \cref{eq:ionic_strength} implies, according to \cref{eq:debye_length}, a smaller Debye length, which, in turn, means faster screening of concentrations.
As a result, also the constant $s$ is decreased, lowering the electrical coupling with the intracellular electrolyte.
Essentially, this corresponds to a local increase in permittivity.
It can be understood by recognizing that the concentrations always counteract the membrane potential -- as counterions are attracted and co-ions repelled -- such that a stronger electrolyte with a higher concentration of counterions will be more successful in ``swallowing'' the membrane charge and therefore screening the potential.

The question of how much of the intracellular potential is ``felt'' on the extracellular side is hence not only a property of the membrane, but to a large degree also of the extracellular electrolyte itself.
This fact raises the question whether the assumption of fully independent intra- and extracellular electrolytes -- which some models rely on, see below -- can be made in general.
We now see that the degree of dependence heavily relies on the magnitude of concentrations, which might vary significantly 
among different brain areas.

\subsubsection{Transition Interval from Debye Layer to Nearfield}
We have seen that nearfield/farfield and Debye layer potential time courses behave quite differently. While the former is dominated by membrane currents, the latter is driven by the intracellular potential, damped by a constant factor $s$.
To further investigate the change of \gls{EAP} shape during the transition between Debye layer and nearfield, we look at extracellular time courses much in the same way as in \cref{fig:lfp_distance}, but this time only for positions located between Debye layer and nearfield, in \cref{fig:lfp_debye}. The transition between the \gls{AP} echo shape very close to the membrane (\cref{fig:lfp_debye_1,fig:lfp_debye_2})
and the more intricate shape following the membrane flux (\crefrange{fig:lfp_debye_6}{fig:lfp_debye_9}) can
be observed to happen within the range of only a few nanometers, as depicted in~\crefrange{fig:lfp_debye_3}{fig:lfp_debye_5}.

\begin{figure}%
\centering%
\foreach \n in {1,...,3}{%
\subfloat[]{\includegraphics[width=0.33\textwidth]%
{img/matlab_temp/lfp_increasing_distance_debye_\n-crop.pdf}\label{fig:lfp_debye_\n}}%
}\\
\foreach \n in {4,...,6}{%
\subfloat[]{\includegraphics[width=0.33\textwidth]%
{img/matlab_temp/lfp_increasing_distance_debye_\n-crop.pdf}\label{fig:lfp_debye_\n}}%
}\\
\foreach \n in {7,...,9}{%
\subfloat[]{\includegraphics[width=0.33\textwidth]%
{img/matlab_temp/lfp_increasing_distance_debye_\n-crop.pdf}\label{fig:lfp_debye_\n}}%
}\\
\mycaption[Extracellular potential time courses over Debye layer]{%
For a selected number of points in the 
Debye layer, the extracellular time courses show a transition from the \gls{AP} echo dominated shape
directly at the membrane (panels a-b), via an intermediate shape (panels c-e) towards the 
characteristic up-down-up shape (panels f-i), which is also found at more distant positions,
see \cref{fig:lfp_distance}}%
\label{fig:lfp_debye}%
\end{figure}

In the course of this transition, the single peak from \cref{fig:lfp_debye_snapshot1} we termed the \gls{AP} echo is divided into two parts P1 and P2 by an interrupting negative peak N1.
Comparing again with \cref{fig:ap_memb_flux}, we see that N1 results from the negative peak present in the membrane flux, which is the consequence of opening voltage-gated sodium channels and the following massive depletion of sodium ions.
Roughly speaking, the activation of sodium channels and the resulting negative peak N1 splits up the single positive peak from the AP echo into two peaks P1 and P2.
Looking closely, we can also see that the interrupting negativity N1 time-shifts from left to right between \cref{fig:lfp_debye_3} and \cref{fig:lfp_debye_6}, shifting also the relation in magnitude between P1 and P2.
Furthermore, the rear part of the \gls{AP} is superimposed by the membrane flux components S and P3.

\begin{figure}%
\centering%
\subfloat[normal scale]{\includegraphics[width=0.5\textwidth]%
{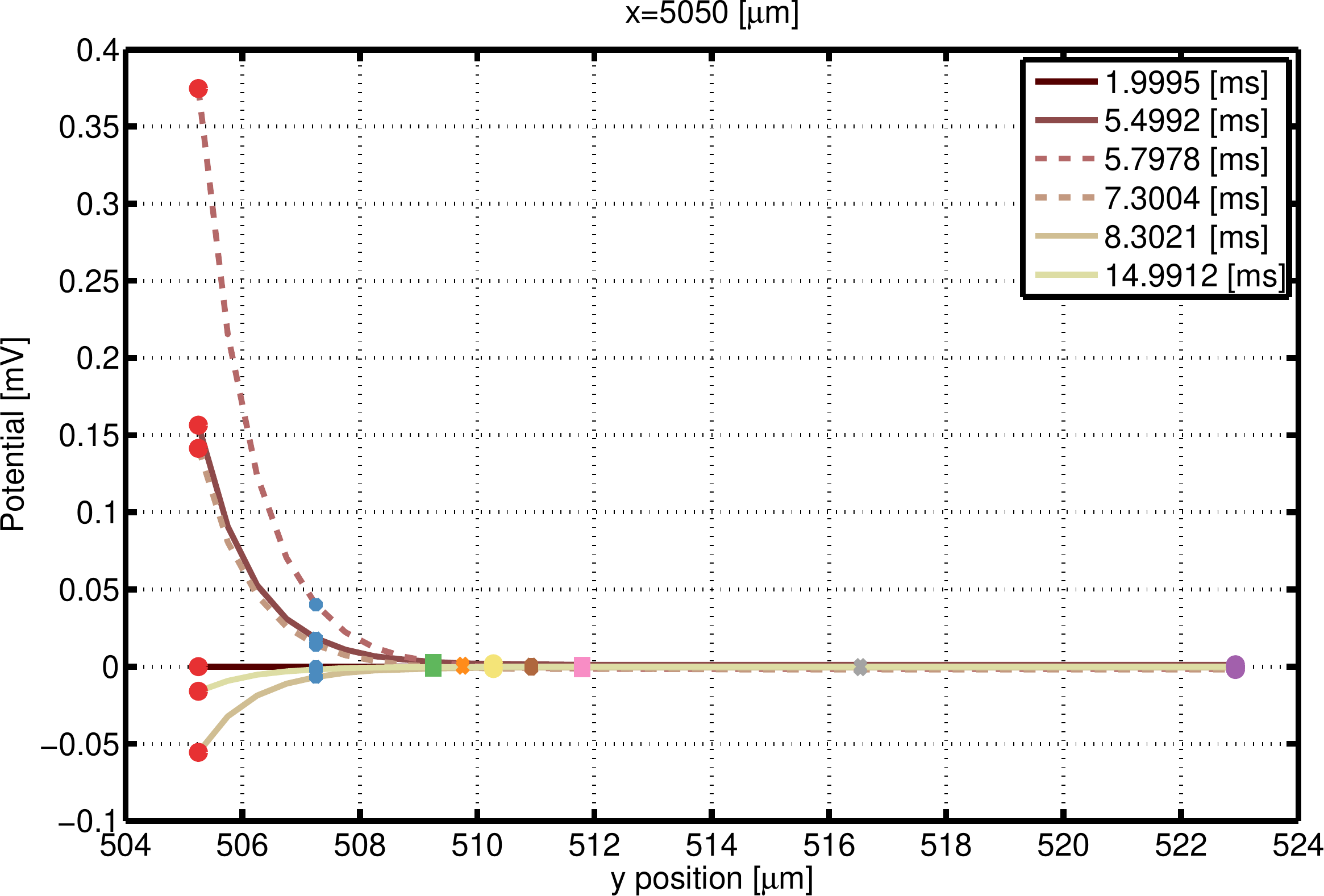}\label{fig:lfp_debye_pot_normal}}%
\subfloat[log scale]{\includegraphics[width=0.5\textwidth]%
{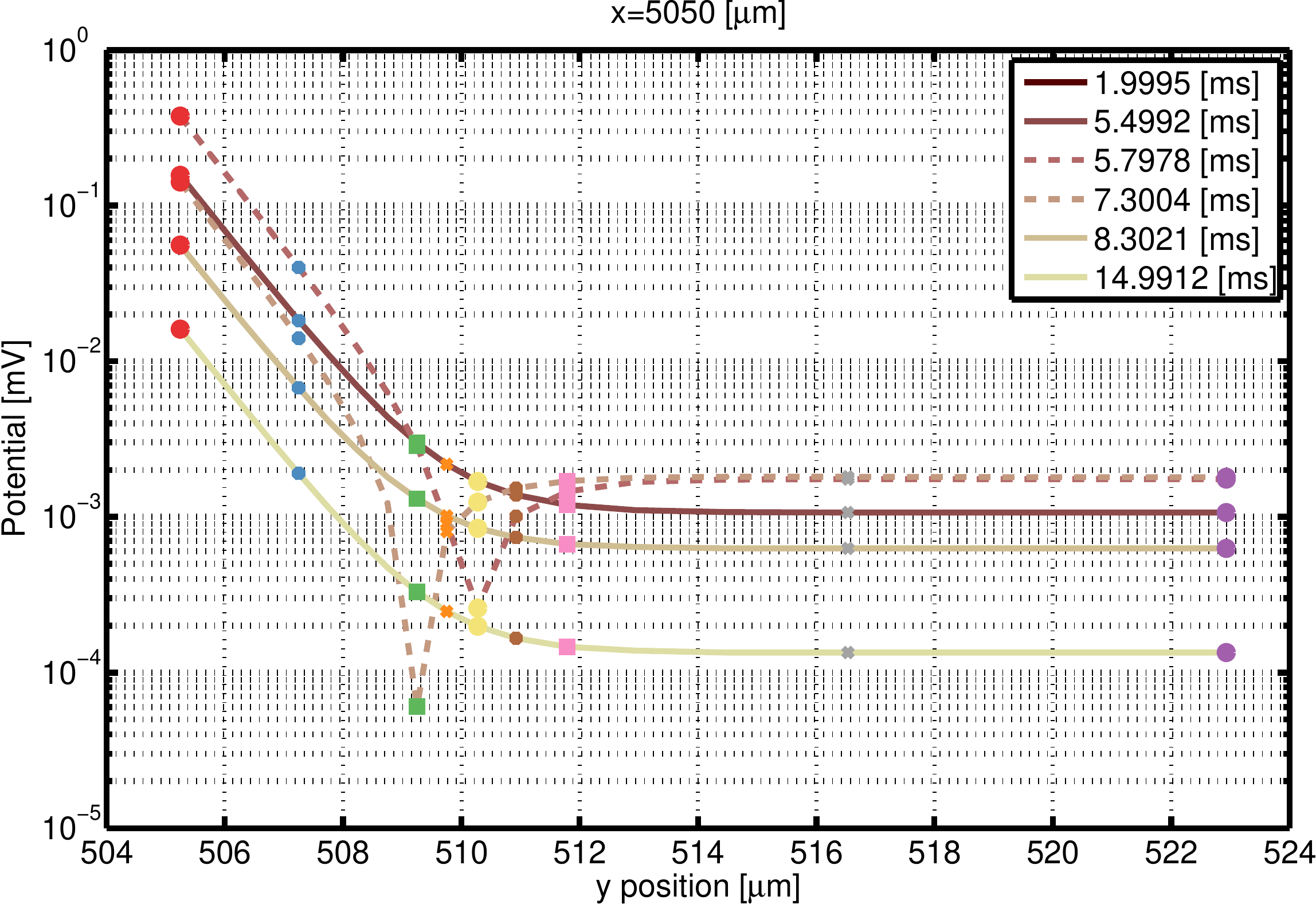}\label{fig:lfp_debye_pot_log}}%
\mycaption[Potential profile over Debye layer at various time points]{%
The values are relative to baseline at $t=\SI{2}{\milli\second}$ in each point. 
The left panel shows the potential profile
close to the membrane, which is attenuated quickly with distance, independent of its value at the 
extracellular membrane interface at different times. The right panel shows the log-scaled absolute
values of the same data, demonstrating the exponential screening over the Debye layer. 
Since the potential courses are not directly comparable due to the change in shape over the Debye layer, the log-scaled plot shows some irregularities in this region.}%
\label{fig:lfp_debye_pot}%
\end{figure}

This interesting behavior can be further elucidated by looking at the large concentration and potential gradients present in this range, illustrated for the potential in \cref{fig:lfp_debye_pot}. 
It shows the spatial profile of the \gls{EAP} time courses from \cref{fig:lfp_debye} -- normalized to baseline at $t=\SI{2}{\milli\second}$ in each point -- at distinct time points, which are indicated as vertical lines in \cref{fig:lfp_debye}.
The markers in \cref{fig:lfp_debye_pot}, on the other hand, correspond to the respective positions of \crefrange{fig:lfp_debye_1}{fig:lfp_debye_9}.

\begin{figure}%
\centering%
\includegraphics[width=0.7\textwidth]{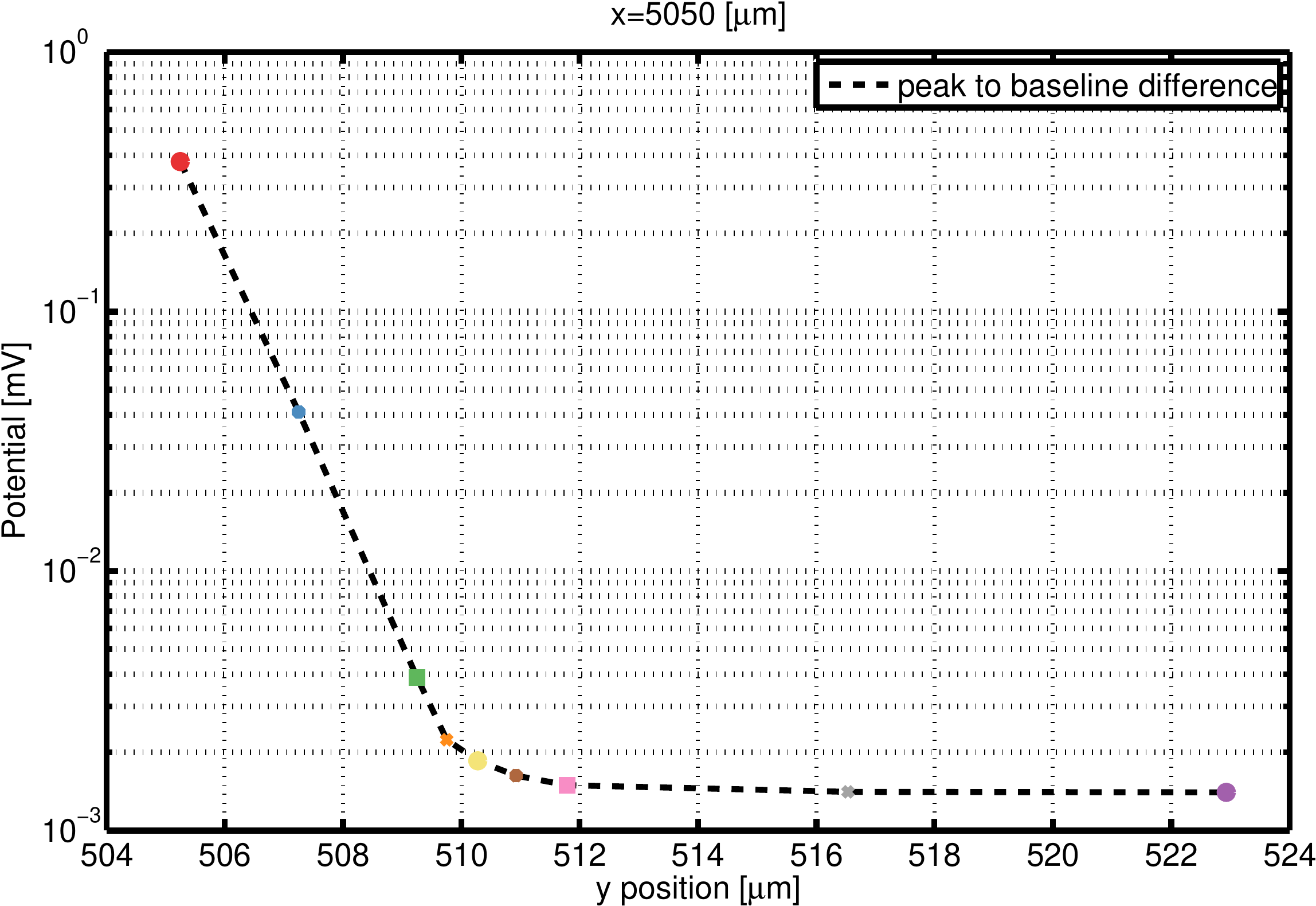}%
\mycaption[Decay of peak-to-baseline in EAP over Debye layer]{%
For the \glspl{EAP} in \cref{fig:lfp_debye}, the difference between peak and baseline potential 
(value at $t=\SI{2}{\milli\second}$) is plotted on a log-scale, showing an abrupt switch in the fall-off rate.}
\label{fig:lfp_debye_pot_peak2baseline_decay_log}%
\end{figure}

We can see a rapid fall-off in the potential profile with increasing distance from the membrane in the normal scale \cref{fig:lfp_debye_pot_normal}, but the log-scaled \cref{fig:lfp_debye_pot_log} shows a switch in fall-off behavior.
Note that there are some irregularities in the transition interval where the \gls{EAP} shape changes.
From this, we can read that the potential time courses are not directly comparable at the chosen time points.
For example, peak P1 in \cref{fig:lfp_debye_3} is smaller than P2, whereas in \cref{fig:lfp_debye_4} the opposite is the case, due to the time-shift of N1 over distance.
A better approach to define the potential decay is via its maximum value in time.
This is illustrated in \cref{fig:lfp_debye_pot_peak2baseline_decay_log}, which shows the peak values of the potential time courses from \cref{fig:lfp_debye} (again relative to baseline). The peak for each point possibly occurs at a different point in time.

\begin{figure}%
\centering%
\subfloat[normal scale]{\includegraphics[width=0.5\textwidth]%
{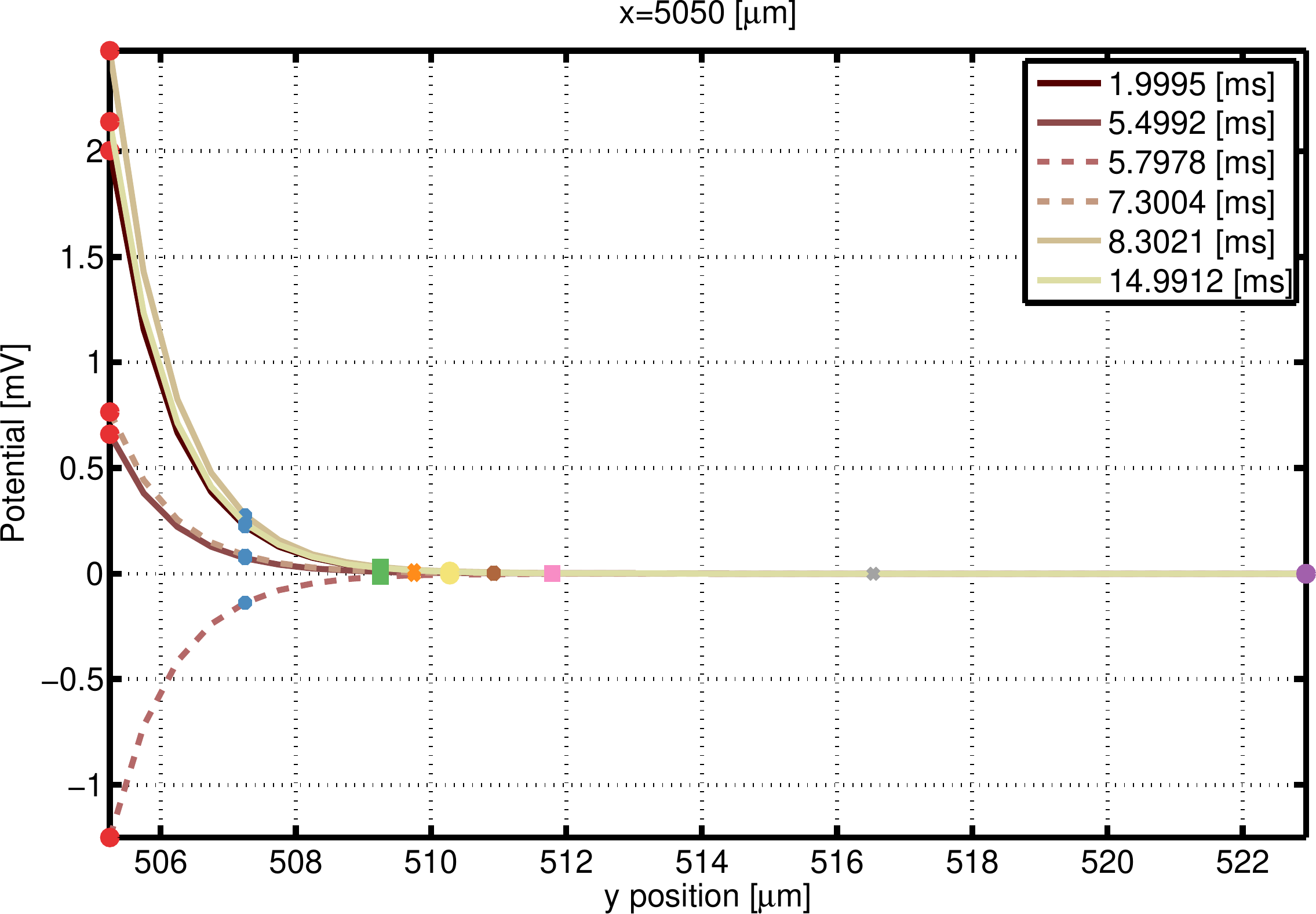}\label{fig:lfp_debye_cd_normal}}%
\subfloat[log scale]{\includegraphics[width=0.5\textwidth]%
{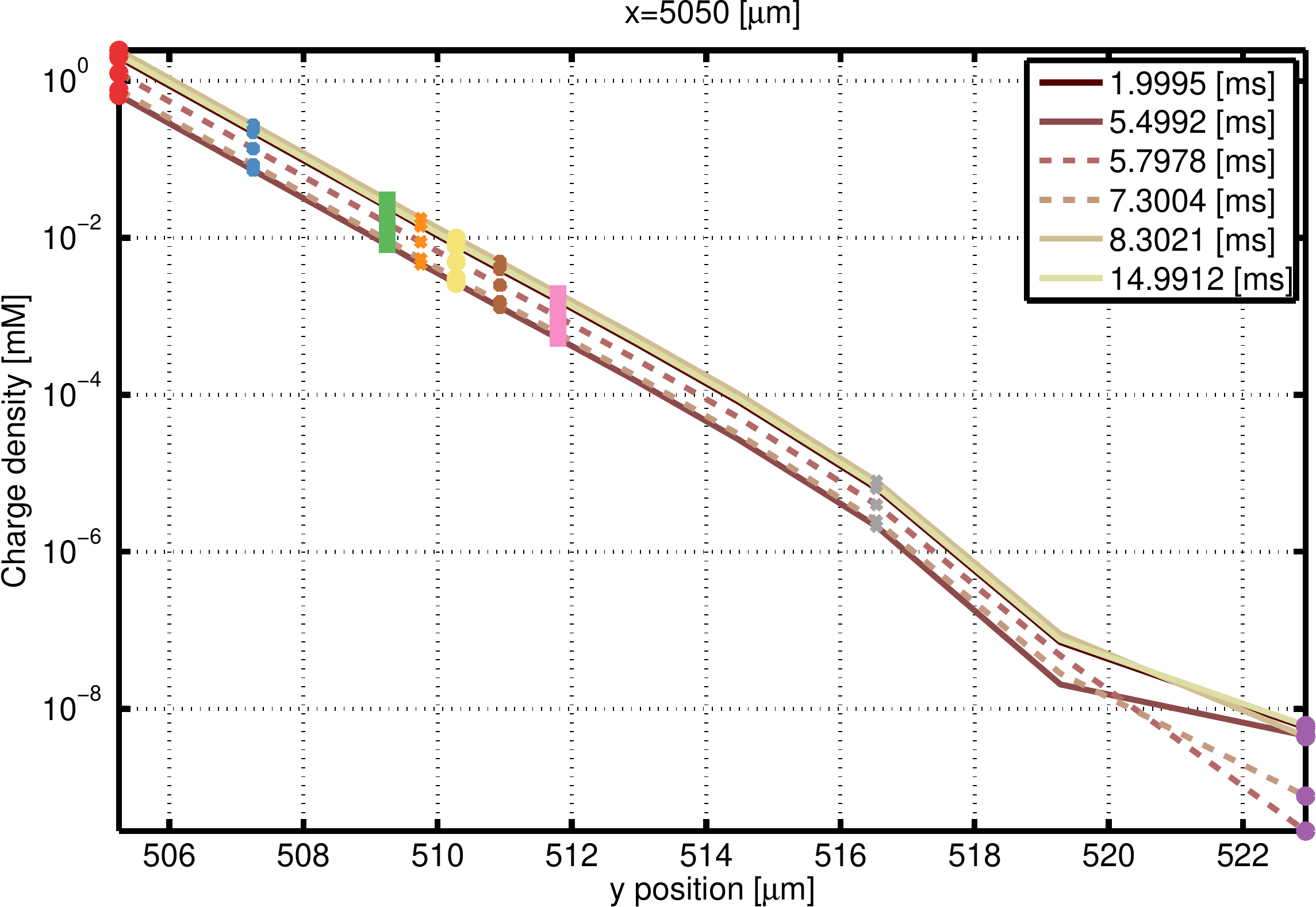}\label{fig:lfp_debye_cd_log}}%
\mycaption[Charge density profile over Debye layer at various time points]{%
As in \cref{fig:lfp_debye_pot}, the left panel is in normal scale, the right panel are the log-scaled absolute values. 
The kinks between $y=\SI{518}{\nano\metre}$ and $y=\SI{520}{\nano\metre}$ are a consequence of the charge density switching signs, as 
it is already close to zero. In both panels, the plotted markers correspond to the positions of the time courses 
in \cref{fig:lfp_debye}.}%
\label{fig:lfp_debye_cd}%
\end{figure}

The analog is shown for the charge density in \cref{fig:lfp_debye_cd}.
As before, we see a rapid fall-off in normal scale plot (\cref{fig:lfp_debye_cd_normal}), which can be shown to be exponential with an appropriate logarithmic scaling (\cref{fig:lfp_debye_cd_log}).

In contrast to the charge density, a significant change in fall-off behavior is found in the potential when traversing the Debye layer.
Over the first \SI{5}{\nano\metre} from the membrane, it drops exponentially together with the charge density.
Then the rapid decay changes to a much slower one, where it is driven primarily by membrane currents whose influence does not drop that drastically with distance.
Thus, instead of dropping towards zero as the concentrations do, the potential only drops exponentially until it has reached a value in the same order as the membrane current contributions.

We conclude that close to the membrane, the influence of the intracellular potential is so large that all contributions by membrane fluxes are hidden in the dominating \gls{AP} echo.
In this small range close to the membrane, the \gls{PNP} system seems to be described very well by the stationary Poisson-Boltzmann \cref{eq:pb}, letting the concentrations undergo an exponential decay towards their bulk values.
As the right-hand side of \cref{eq:pb} is an exponential function, the potential -- after integrating twice -- must also follow an exponential function and decay rapidly towards the ``bulk potential'', which is zero for an electroneutral bulk solution.
However, this bulk potential is never reached, as by traversing through the Debye layer, the influence of membrane fluxes increases, and the Poisson-Boltzmann equation does not provide a suitable representation anymore.

\subsubsection{Extracellular Concentrations}
\begin{figure}%
\centering%
\subfloat[Na concentration]{\includegraphics[width=0.5\textwidth]%
{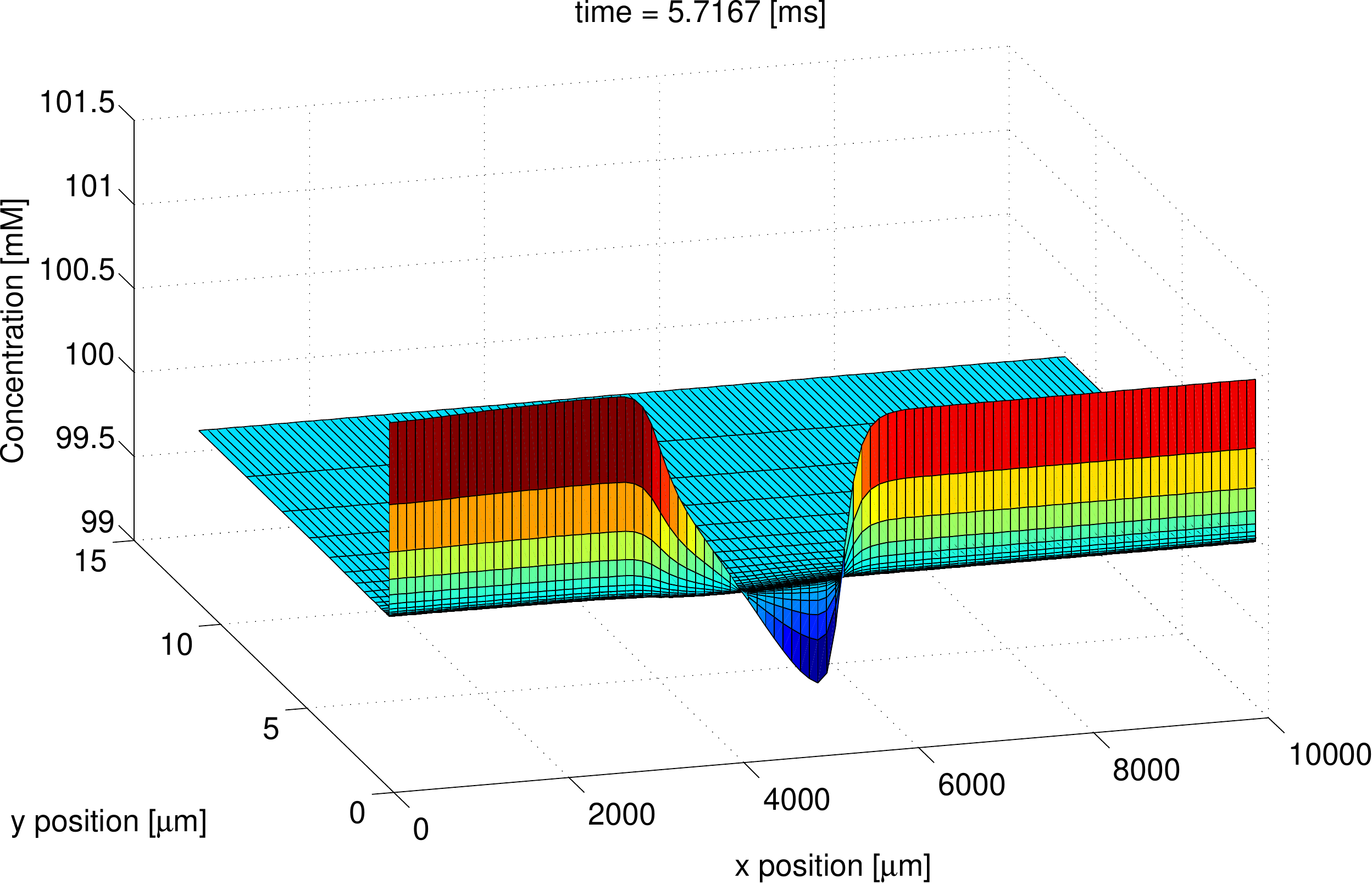}\label{fig:snapshot_conc_na}}%
\subfloat[K concentration]{\includegraphics[width=0.5\textwidth]%
{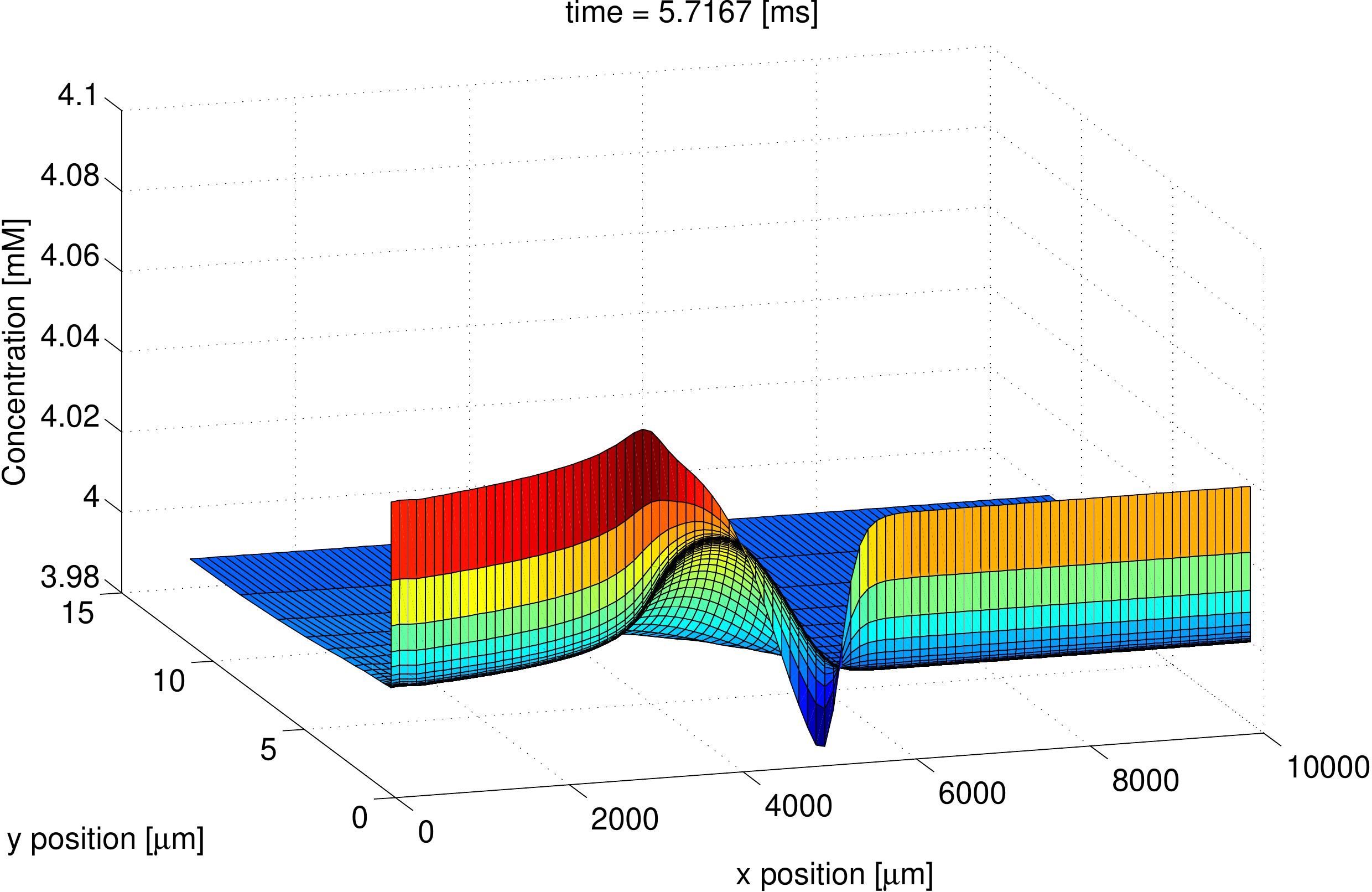}\label{fig:snapshot_conc_k}}\\%
\subfloat[Cl concentration]{\includegraphics[width=0.5\textwidth]%
{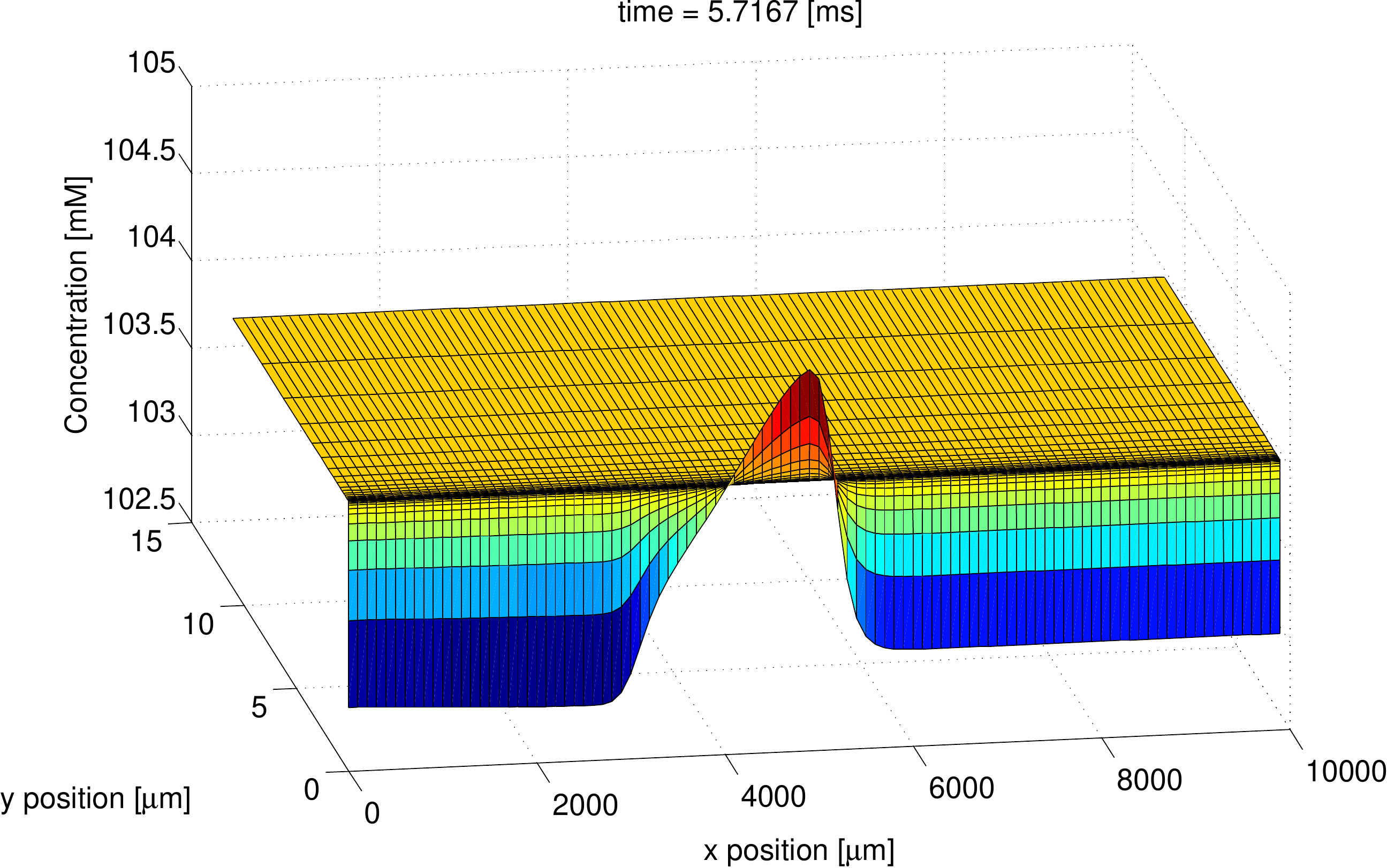}\label{fig:snapshot_conc_cl}}
\subfloat[(Scaled) charge density]{\includegraphics[width=0.5\textwidth]%
{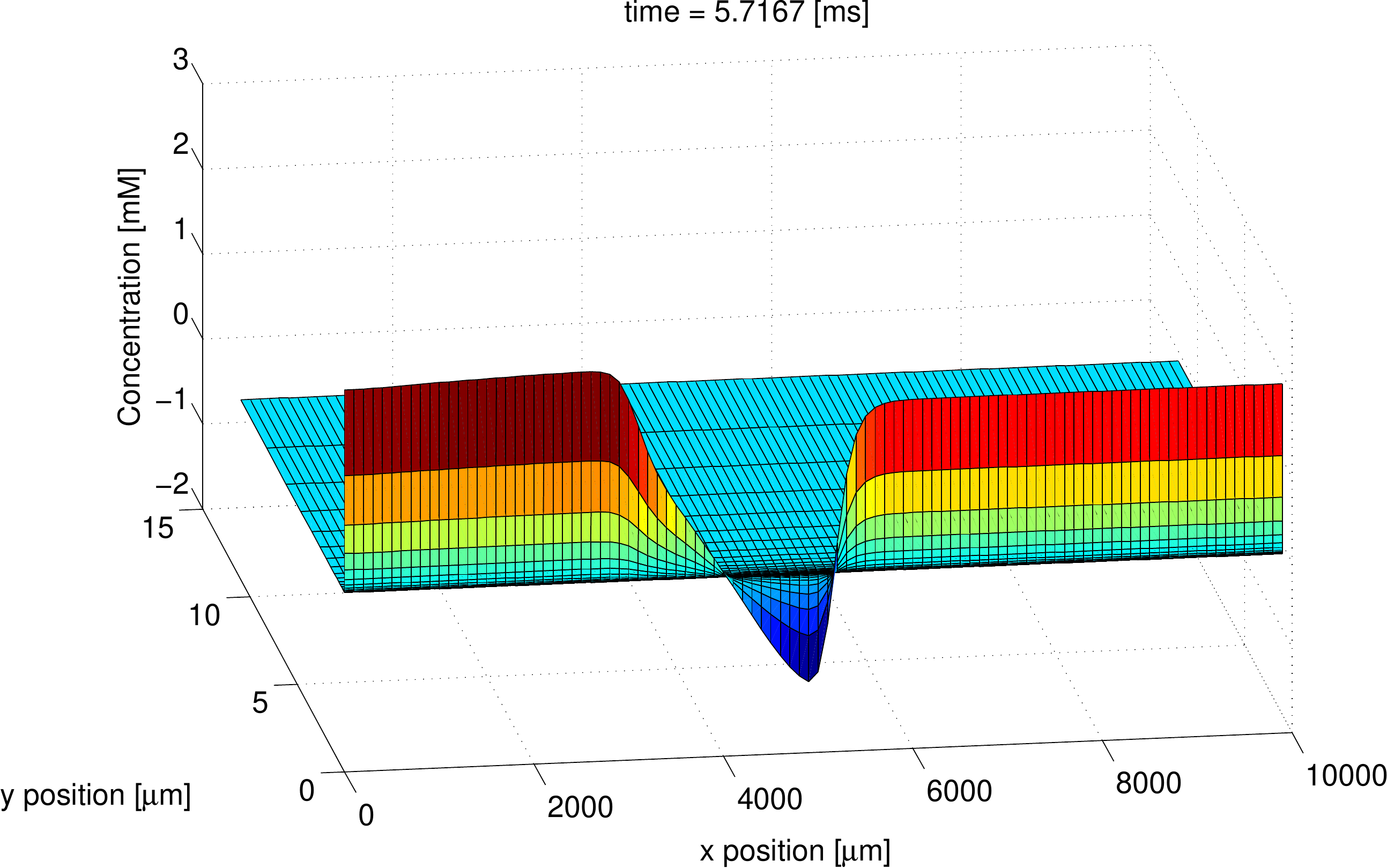}\label{fig:snapshot_conc_cd}}\\%
\mycaption[Snapshot of the Debye layer and nearfield concentration profiles]{The concentration profiles of all three ion species and the charge density
are plotted up to a distance of about \SI{10}{\micro\metre} from the membrane. The
strong concentration gradients close to the membrane demonstrate the influence of
the Debye layer. The charge density can be seen to be dominated by the sodium
concentration.}%
\label{fig:snapshot_conc}%
\end{figure}

In analog to the nearfield and Debye layer potential in \cref{fig:lfp_debye_snapshot,fig:lfp_nearfield_snapshot}, we can also look at concentration snapshots, see \cref{fig:snapshot_conc}.
We stated above that the charge density (\cref{fig:snapshot_conc_cd}) is always oriented opposed to the potential, as counterions are attracted to and co-ions repelled from the membrane, consistent with the Poisson-Boltzmann \cref{eq:pb}.
In \cref{fig:lfp_debye_cd_log}, we also saw that the charge density drops exponentially over the Debye layer at any given time, satisfying the equilibrium Boltzmann distribution on the right-hand side of \cref{eq:pb}.
This indicates that in the Debye layer, the stationary Poisson-Boltzmann equation is valid also for the timescales present in this instationary case. The membrane transshipment happens so fast it can be considered instantaneous.

As expected from the charge density profile in \cref{fig:lfp_debye_cd}, the concentrations fall off very quickly.
Close to the membrane, sodium (\cref{fig:snapshot_conc_na}) as the main counterion (at equilibrium, the membrane is negatively charged from the extracellular point of view) shows an inverted \gls{AP} shape, since it is repelled by the incoming (positive) action potential.
Chloride as a co-ion, on the other hand, follows the \gls{AP} shape due to its negative charge (\cref{fig:snapshot_conc_cl}).
A slightly different picture shows for potassium in \cref{fig:snapshot_conc_k}, whose shape shows similarities to the sodium profile, albeit somewhat distorted.
Since the extracellular potassium bulk concentration is two orders of magnitude lower than those of sodium and chloride, the previous reasoning of membrane currents being hidden in the steep concentrations gradients does not fully apply here.
In contrast to the other two ion species, a small variation by membrane currents in absolute numbers has a larger effect due to its lower bulk concentration, so that the influence of membrane currents can be clearly seen in the potassium concentrations, especially the increase after opening of $\text{K}_v$ channels to the left of the profile.
The total charge density in \cref{fig:snapshot_conc_cd} follows the counterions, mainly the higher concentrated sodium ions.

\section{Comparison with Other Models}\label{sec:unmyel.comparison}
In order to compare the electrodiffusion results with existing models for the extracellular potential, a 10-fold finer $x$-resolution of $h_x = \SI{10}{\micro\metre}$ was used to ensure an accurate resolution of membrane dynamics, especially membrane potential and channel currents.
A finer $y$-resolution of $\hyMax = \SI{10}{\micro\metre}$ did not yield a notable difference, so it was kept at $\hyMax = \SI{100}{\micro\metre}$.

\subsection{Line Source Approximation}\label{sec:unmyel.lsa}
The \gls{LSA} model as an analytical solution of \vref{eq:volume_conductor} has been introduced in \cref{sec:theory.lsa}. It can be used to calculate the extracellular potential as a superposition of current contributions of a finite number of line segments, as produced e.g.~by a NEURON simulation.
But also the extracellular field computed by the electrodiffusion model can easily be compared with the \gls{LSA} results, as the membrane currents can be calculated directly as the sum of ionic currents -- known from the membrane flux boundary conditions -- and a capacitive current\footnote{Note that unfortunately, the capacitive current $I_C$ was omitted in \cite{pods2013electrodiffusion}. A correction is available under \url{http://dx.doi.org/10.1016/j.bpj.2013.12.026}. This work already contains the corrected version.}
\begin{align}\label{eq:lsa_currents}
   I_j &= I_{\text{ionic}}(x) + I_C(x)\\
       &= e N_A \left( \sum_{i=1}^N f_i^{\text{memb}}(x) + f_C(x) \right) \ ,
\end{align}
where the capacitive flux $f_C$ was calculated as described in \cref{sec:unmyel.memb_flux}, and $x$ is the center of the membrane-extracellular interface corresponding to line $j$ in \gls{LSA} notation. 

The resistivity $\rho$ was chosen manually in such a way that the positive peaks P1 of electrodiffusion and \gls{LSA} approximately matched for moderately large distances from the membrane ($> \SI{5}{\micro\metre}$ in this case). 
This resulted in a value of $\rho =  \SI{72}{\ohm\centi\metre}$, corresponding to a conductivity of $\sigma = \SI{1.39}{\siemens\per\metre}$, which is in good agreement with \cite{baumann1997electrical} and 20-27.5\% lower than in \cite{gabriel1996dielectric}. 
This is plausible, respecting that measurements in \cite{baumann1997electrical,gabriel1996dielectric} were carried out at room/body temperature, while the data in this work was generated at a lower temperature of $T=\SI{6.3}{\celsius}$, causing a lower conductivity.
It should also be noted that this value matches well with the conductivity of \SIrange{60}{70}{\ohm\centi\metre} for \emph{Ringer's fluid}, an electrolytic solution that is isotonic to body fluids like blood or cerebrospinal fluid \cite{faber1989electrical}.

\begin{figure}%
\centering%
\subfloat[]{\includegraphics[width=0.33\textwidth]%
{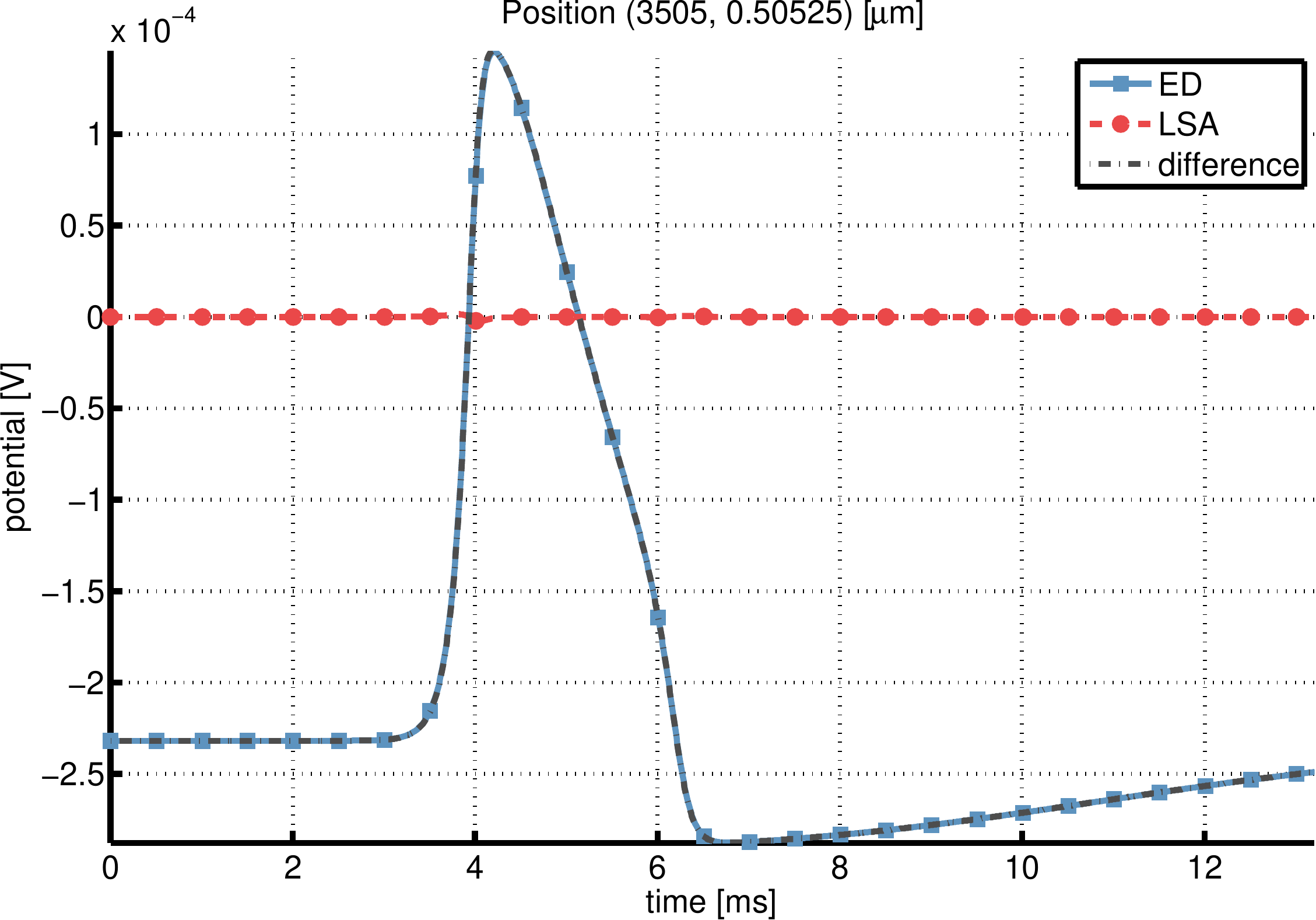}\label{fig:lsa_ed1}}%
\subfloat[]{\includegraphics[width=0.33\textwidth]%
{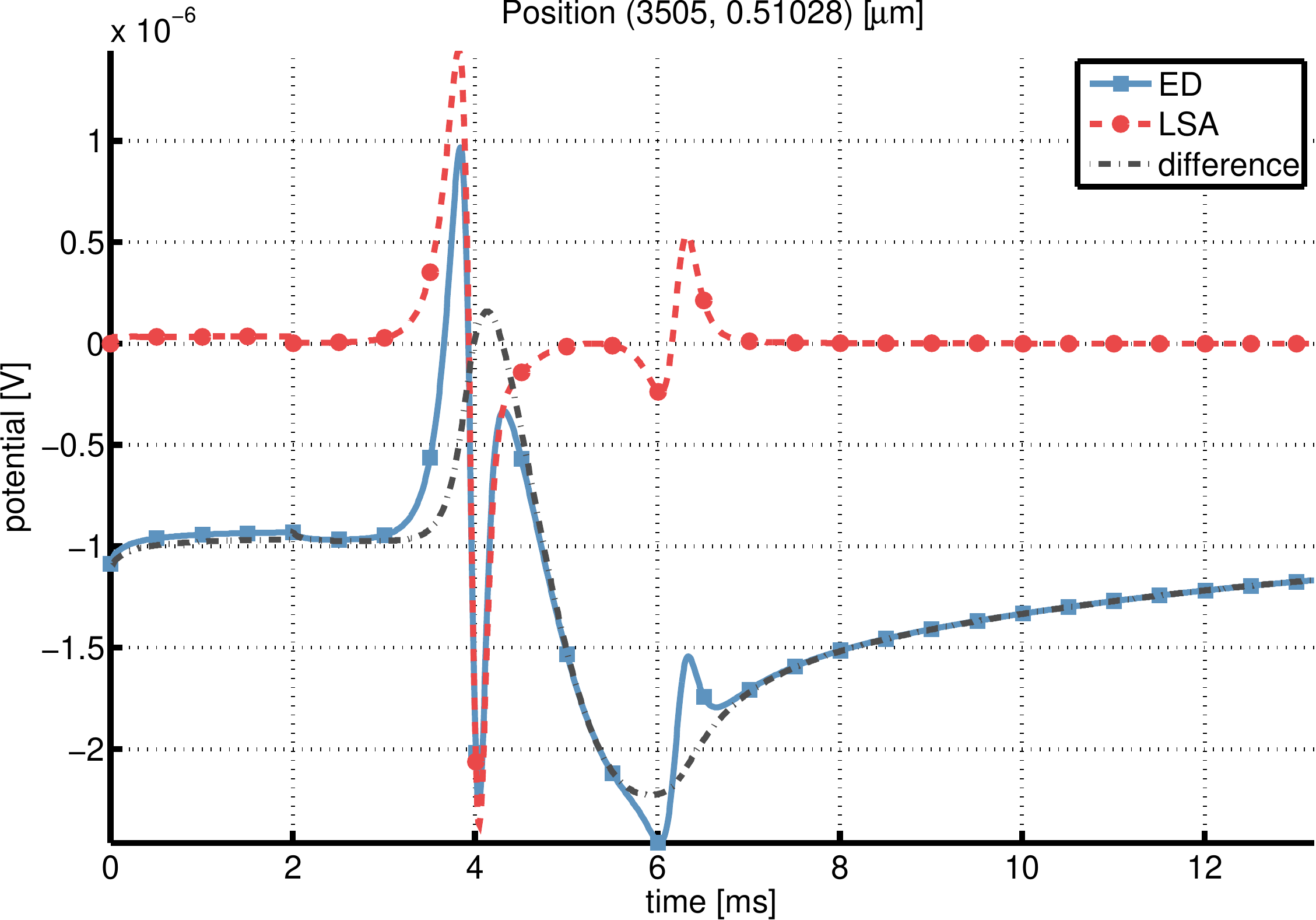}\label{fig:lsa_ed2}}%
\subfloat[]{\includegraphics[width=0.33\textwidth]%
{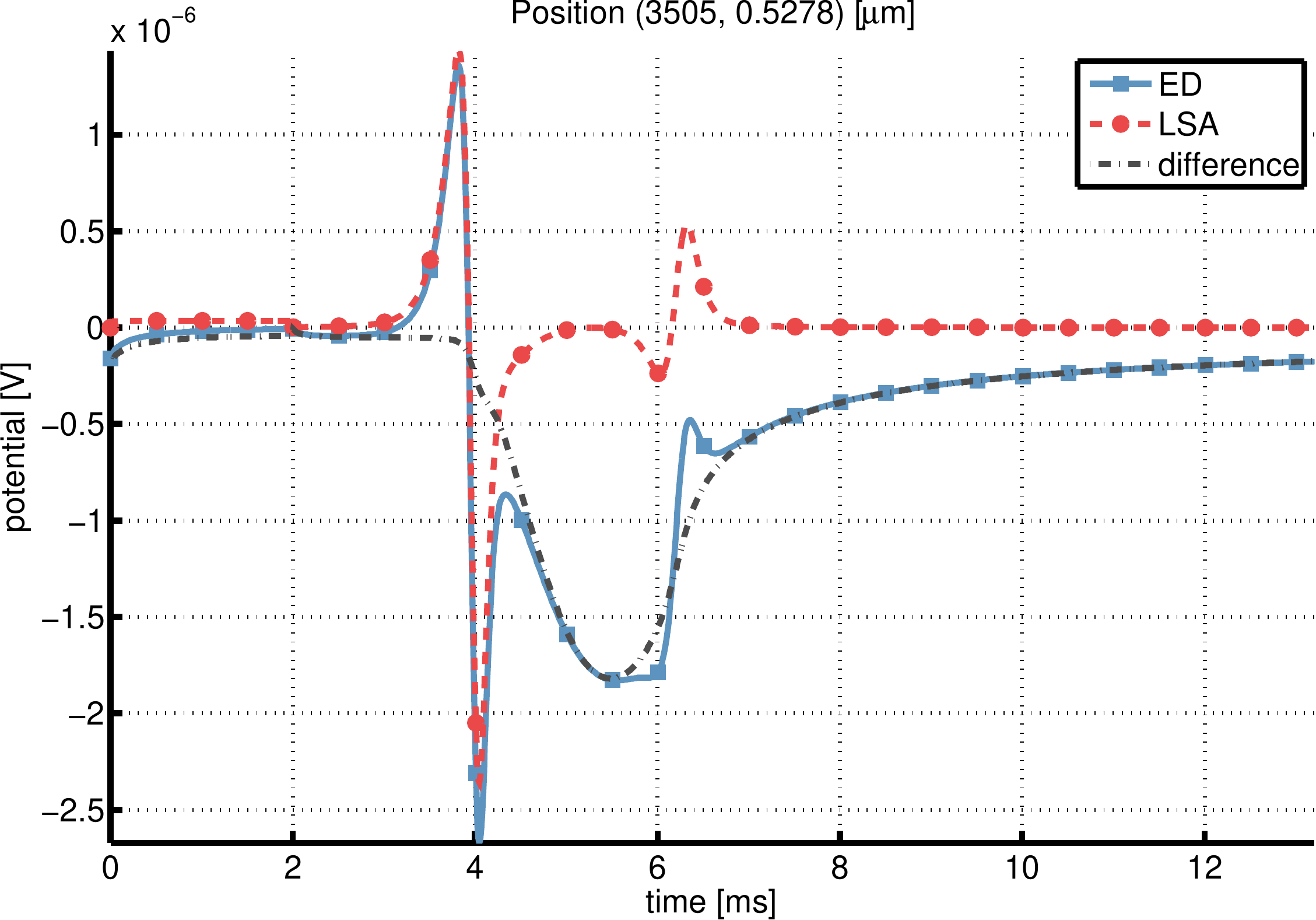}\label{fig:lsa_ed3}}\\%
\subfloat[]{\includegraphics[width=0.33\textwidth]%
{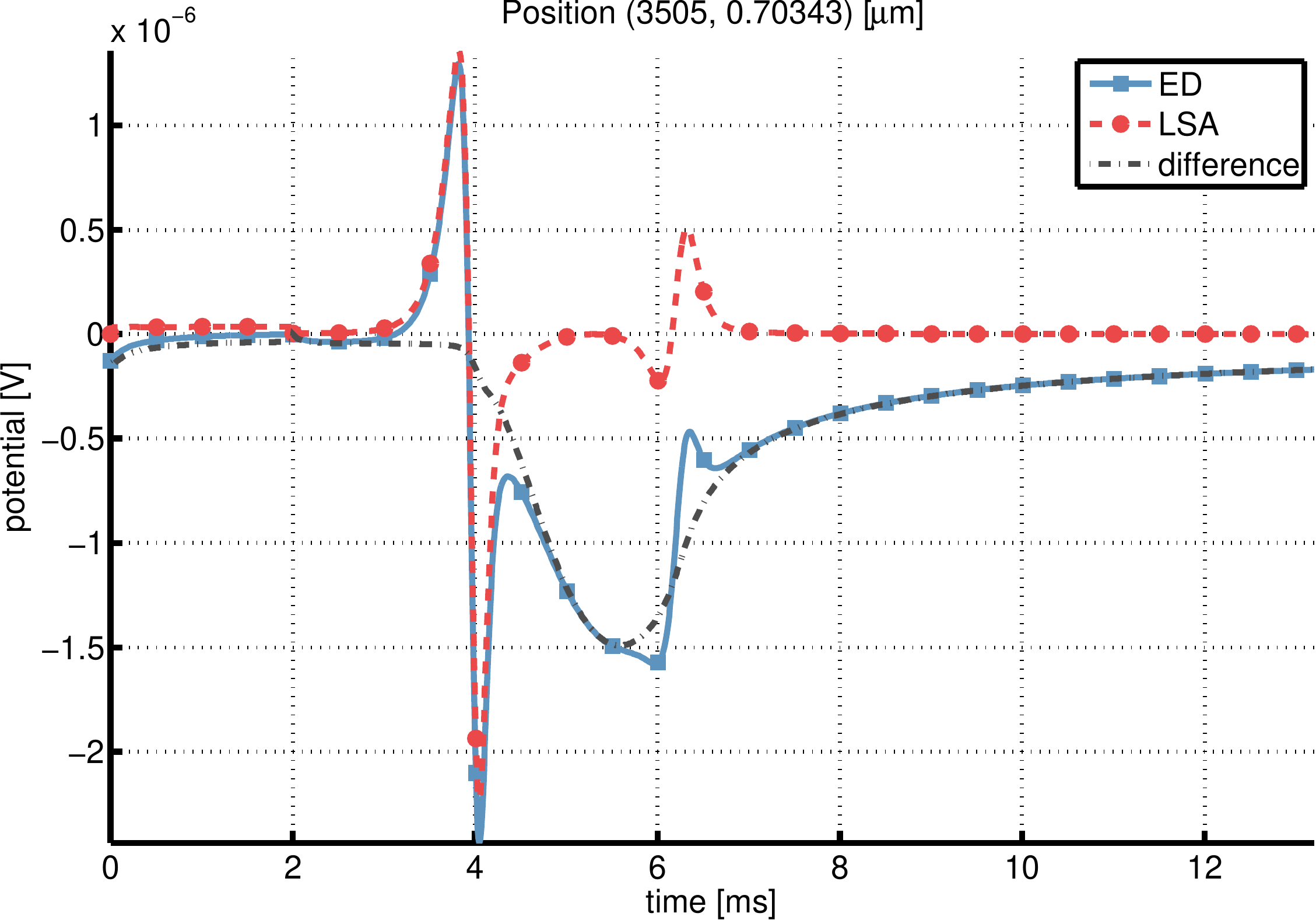}\label{fig:lsa_ed4}}%
\subfloat[]{\includegraphics[width=0.33\textwidth]%
{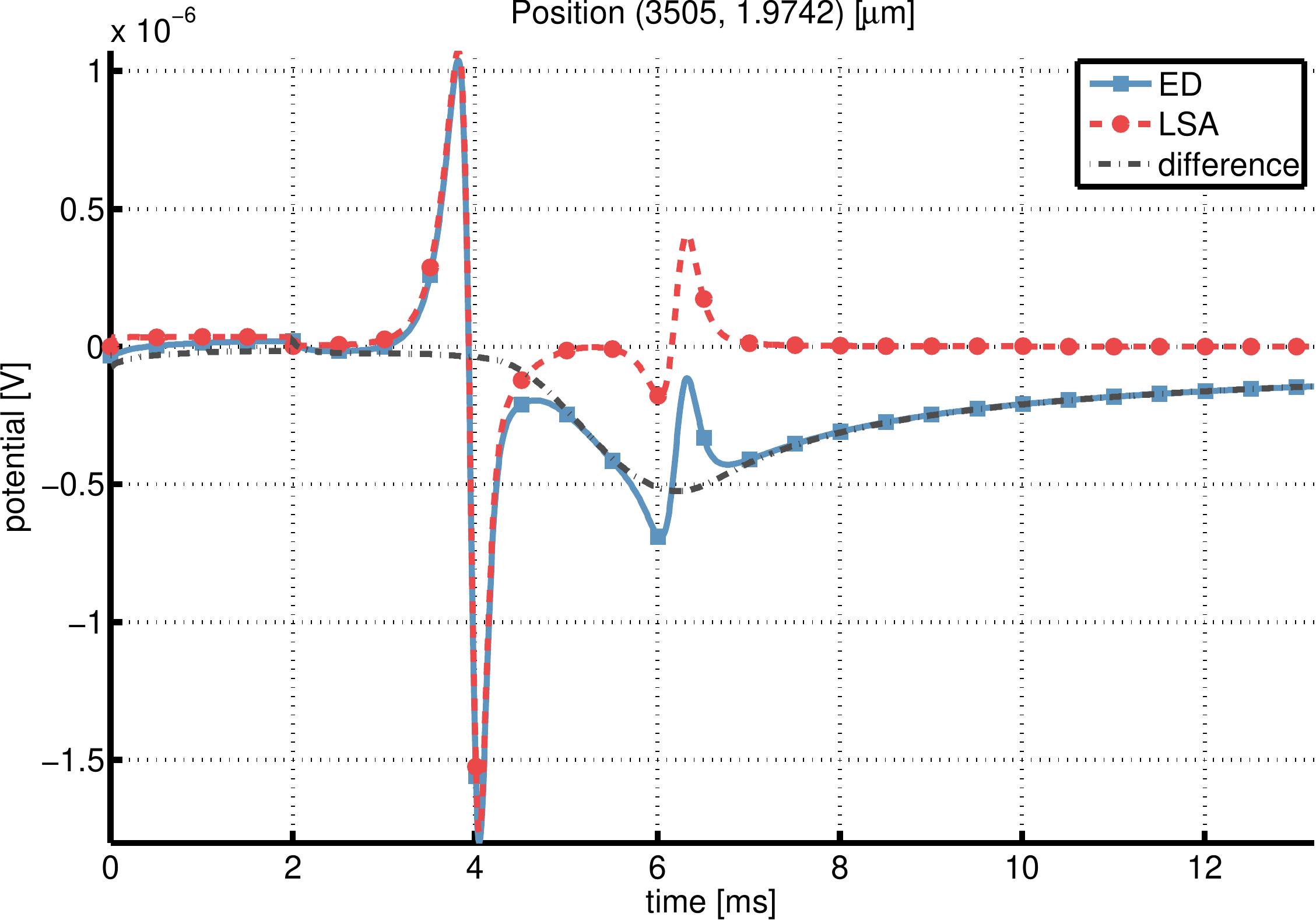}\label{fig:lsa_ed5}}%
\subfloat[]{\includegraphics[width=0.33\textwidth]%
{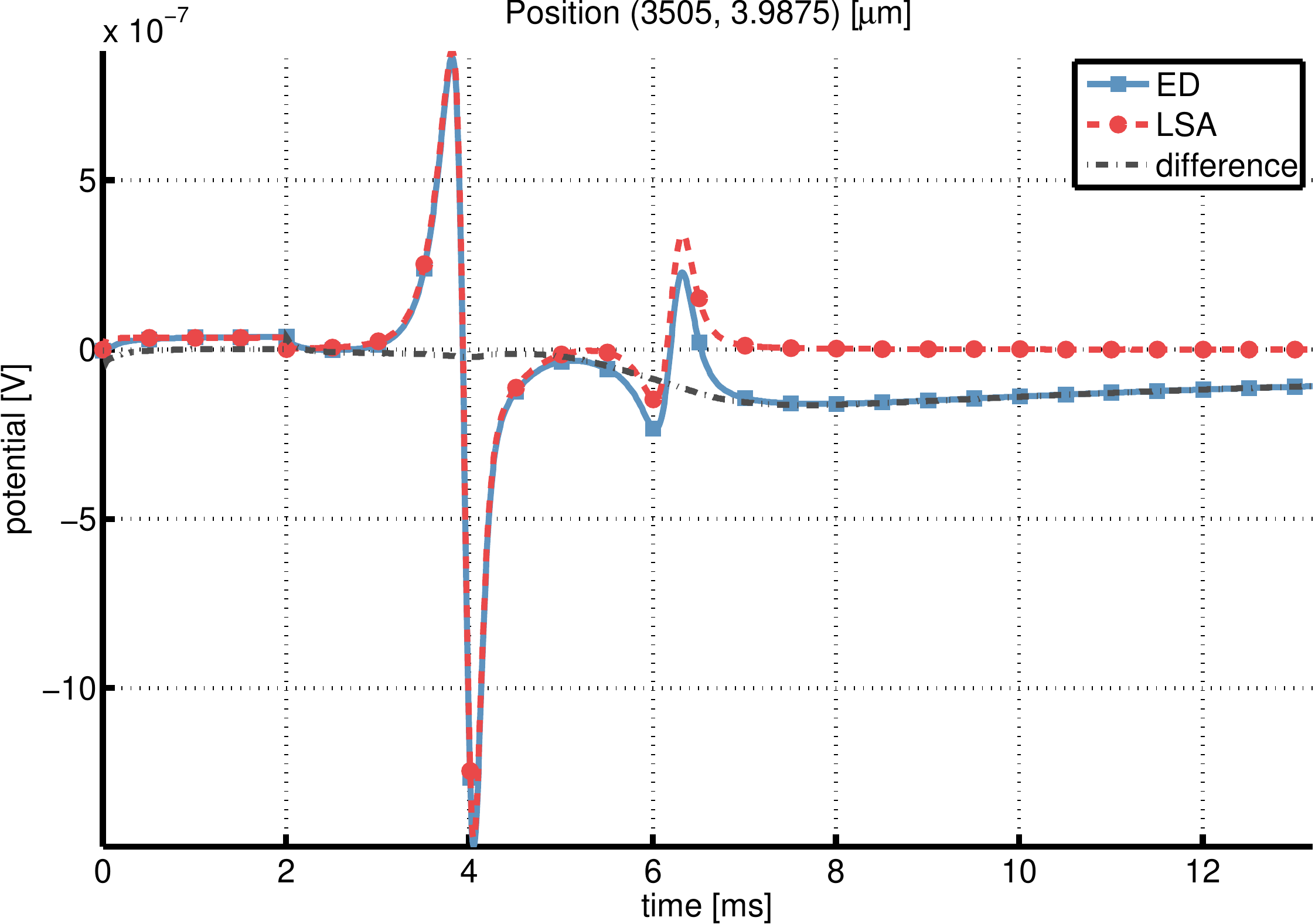}\label{fig:lsa_ed6}}\\%
\subfloat[]{\includegraphics[width=0.33\textwidth]%
{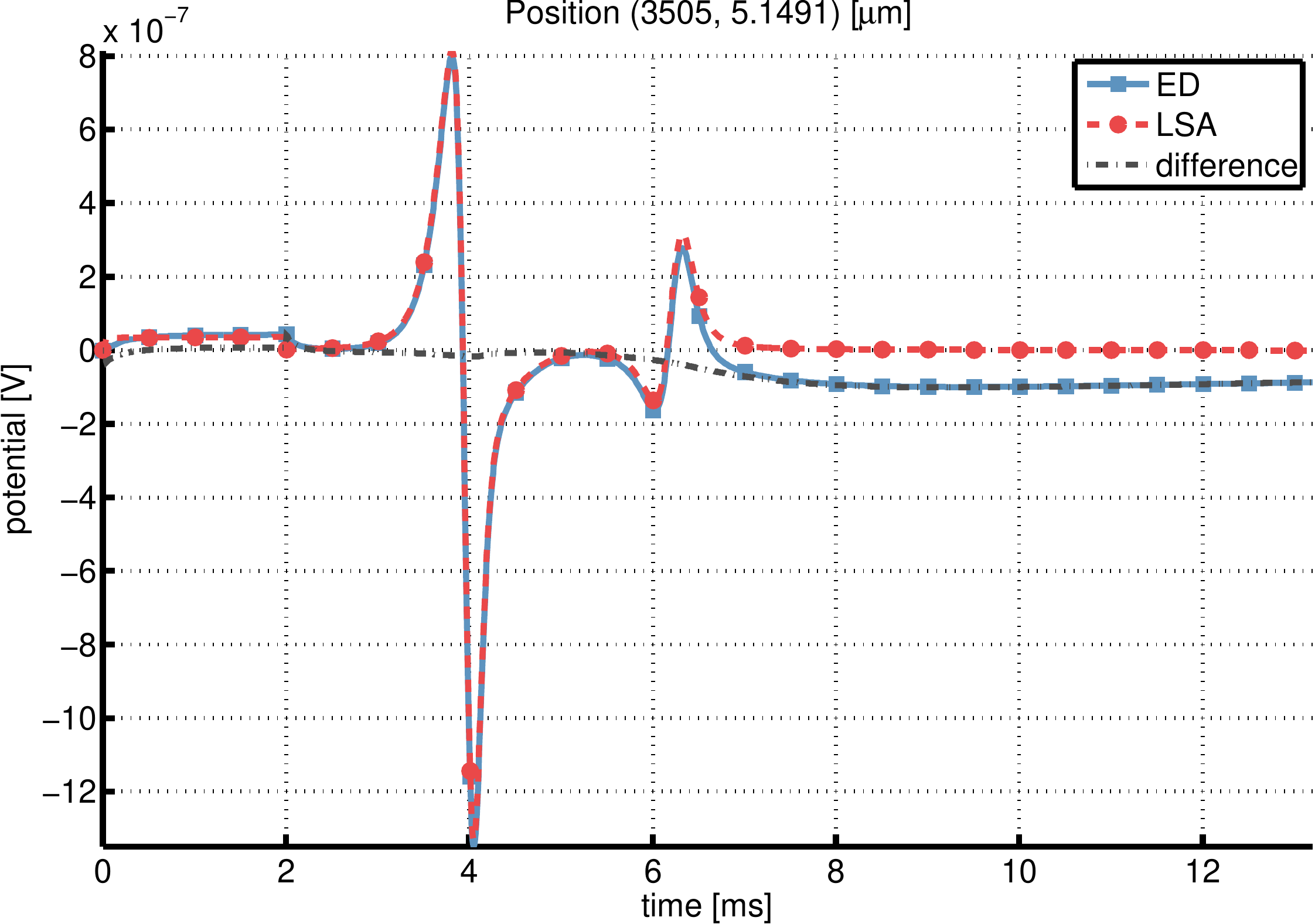}\label{fig:lsa_ed7}}%
\subfloat[]{\includegraphics[width=0.33\textwidth]%
{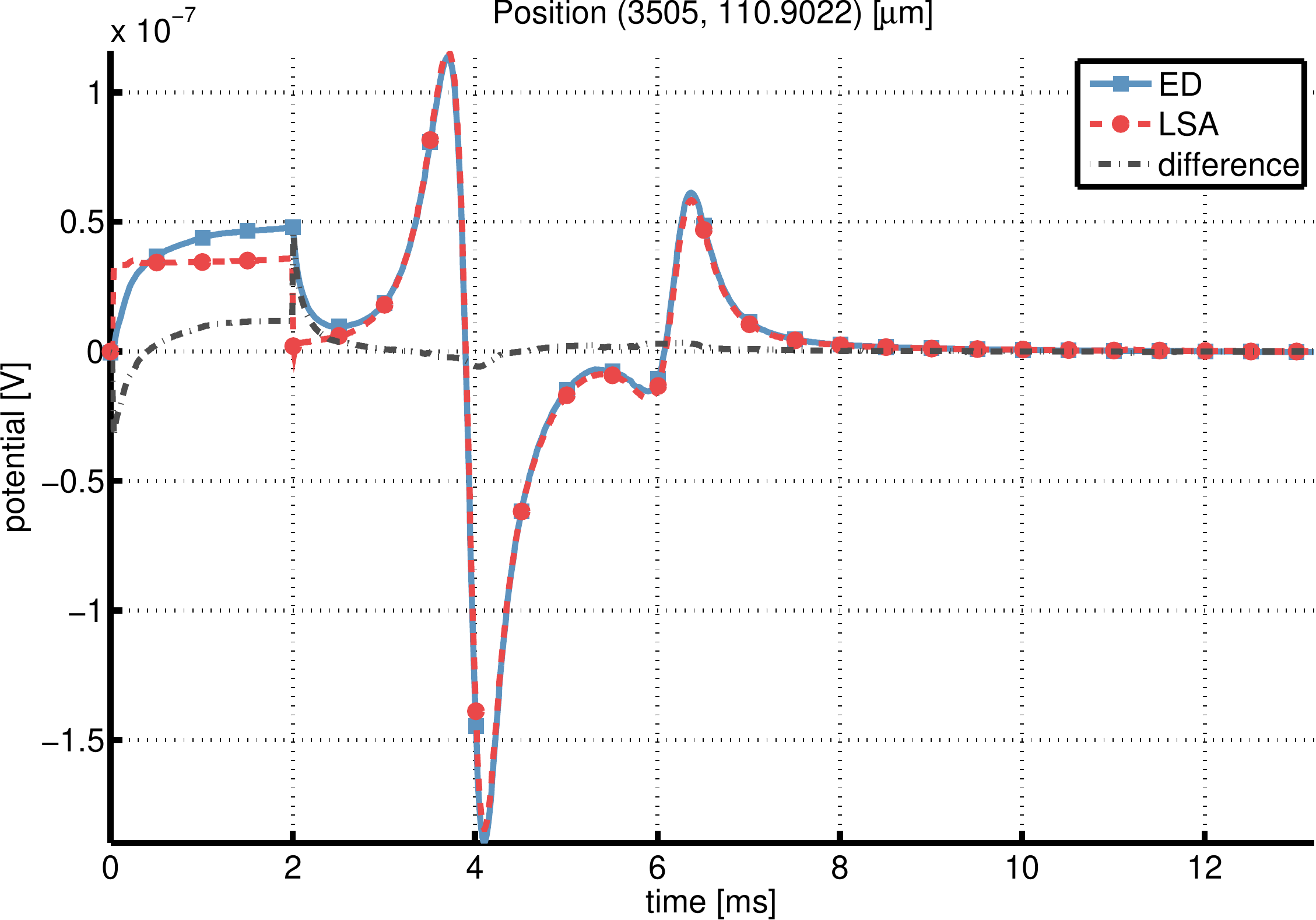}\label{fig:lsa_ed8}}%
\subfloat[]{\includegraphics[width=0.33\textwidth]%
{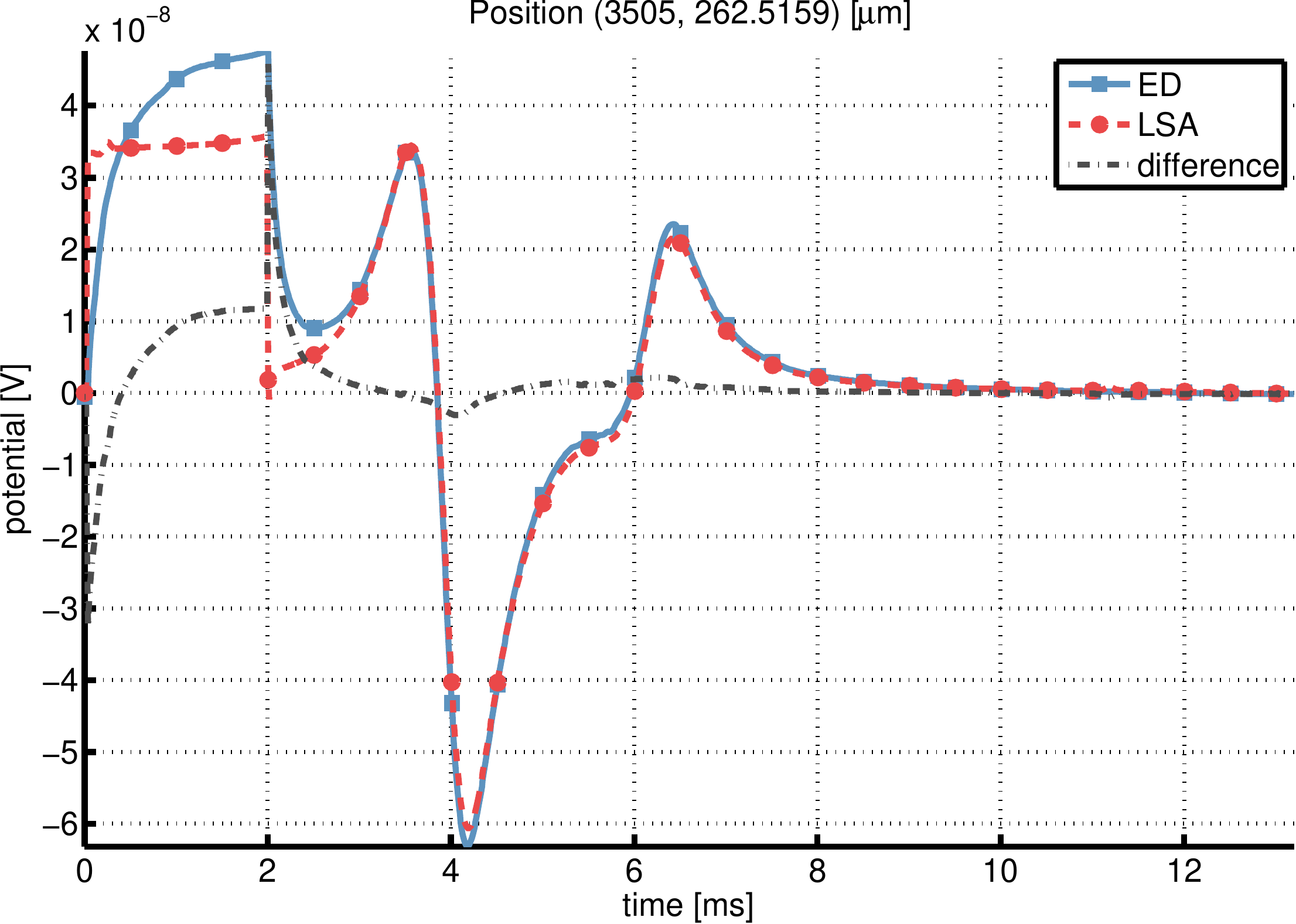}\label{fig:lsa_ed9}}%
\mycaption[Comparison of ED and LSA potentials]{%
The time courses of extracellular potentials calculated by electrodiffusion (\gls{ED}, \textit{solid lines}) and line source approximation 
(\gls{LSA}, \textit{dashed lines}) models are compared at different distances from the membrane (a-i) for a fixed $x$-coordinate. 
Additionally, the difference (\textit{dash-dotted lines}) between the two curves is shown. The
different behavior for the part between N1 and P3 is apparent at lower distances, but for larger distances, a good agreement
can be found.}%
\label{fig:lsa_ed}%
\end{figure}

\Cref{fig:lsa_ed} shows the comparison of potential time courses for \gls{LSA} and \gls{ED} models at various distances from the membrane.
Additionally, the difference between the two curves is plotted.

One can see that both signals agree very well for larger distances from the membrane ($> \SI{5}{\micro\metre}$). 
For distances below a few microns, however, the deviations are clearly visible.
Especially the latter part of the signal after N1 shows qualitative differences.
A special case is the potential at a point directly adjacent to the membrane in \cref{fig:lsa_ed1}, which is exclusively dominated by the \gls{AP} echo in the electrodiffusion simulation, while \gls{LSA} predicts a much smaller amplitude.
Continuing across the Debye layer and into the nearfield regime, both signals show better matches for increasing distances, where the \gls{EAP} calculated by the electrodiffusion model tends to have smaller amplitudes and a tail (P2 -- S -- P3) that is shifted downwards.
A notable difference is also the potential answer to the stimulus artifact, which happens instantaneously in the form of a rectangle pulse in \gls{LSA}, while the gradual electrodiffusion answer is much smoother in comparison.

\subsubsection{Separate Inspection of EAP Contributions}\label{sec:unmyel.lsa.separate}
To further investigate the reason for the nearfield differences between \gls{ED} and \gls{LSA} results, an additional simulation was carried out to eliminate the contribution of ionic currents.
To this end, an ``echo simulation'' was set up with the same parameters as before, but with all membrane channels closed, resulting in zero membrane fluxes. 
The lower potential Neumann-0 boundary condition was replaced by a time-dependent Dirichlet boundary, generated separately by an action potential simulation as before (the ``source simulation'').
This resulted in a setup where the contributions of ionic membrane currents are eliminated, enabling the assessment of the membrane potential and induced capacitive currents alone.
This can, in turn, be compared to the \gls{LSA} model with only capacitive currents, allowing for the evaluation of differences of capacitive contributions separately.

A notable difference between these simulations and the one considered before is the usage of discrete Dirichlet values from the source simulation as a boundary condition in the passive membrane solution. 
An interpolation of these discrete values generated kinks in the lower potential boundary condition, which made the system very susceptible to numerical oscillations. 
The kinks got picked up and amplified quickly in the numerical solution, especially at larger distances from the membrane, probably due to the threshold volume scaling.
When using a finer space grid -- as mentioned above, with mesh parameters $\hyMax = \SI{100}{\micro\metre}$ and $\hxMax = \SI{10}{\micro\metre}$ -- together with a finer time grid, with upper time step values of $\dtMax = \SI{10}{\micro\second}$ and $\dtMaxAP = \SI{1}{\micro\second}$ for the source simulation, and additionally using the same space and time grid in the passive membrane setup, the oscillations could be successfully eliminated.

\begin{figure}%
\centering%
\subfloat[]{\includegraphics[width=0.33\textwidth]%
{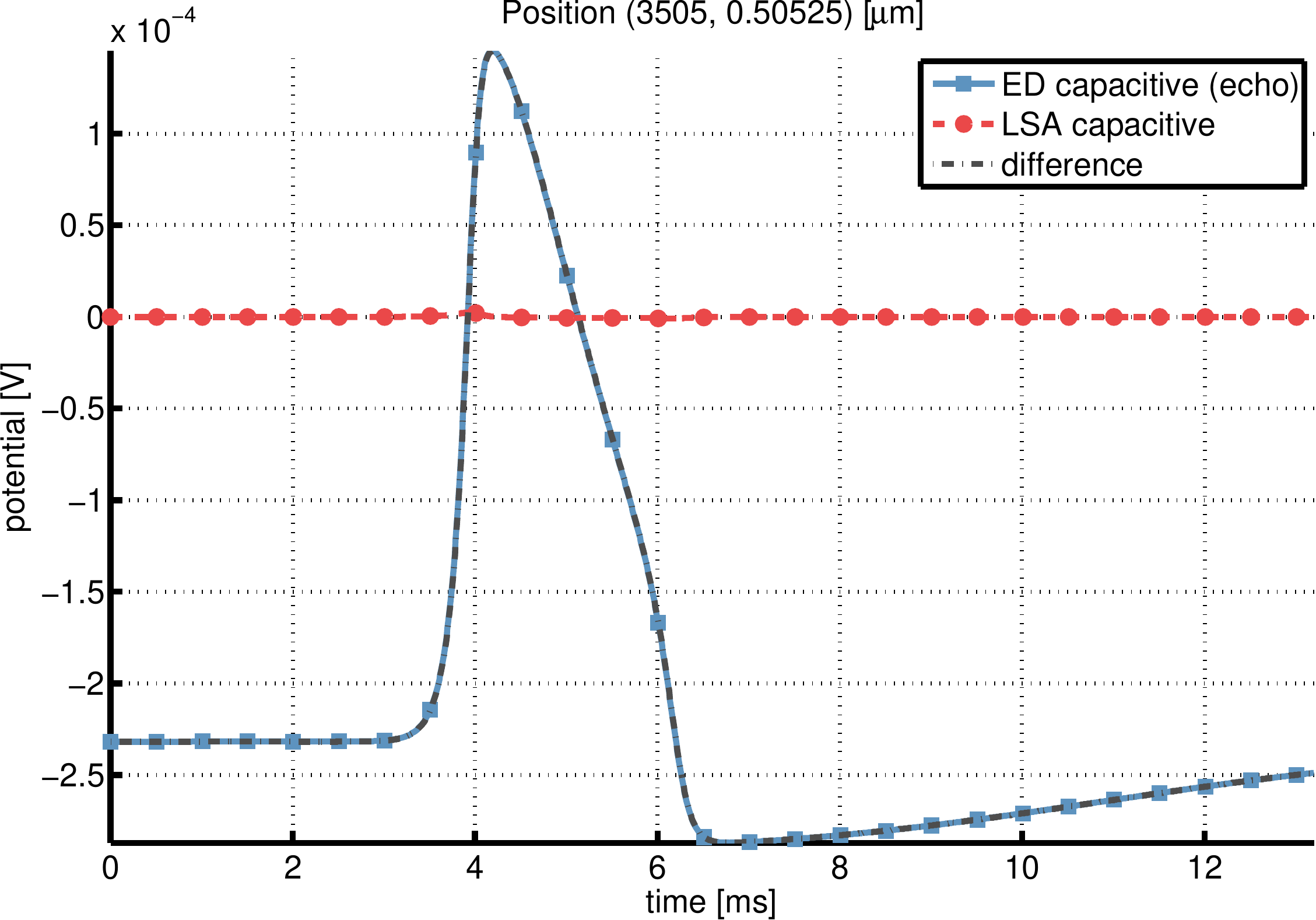}\label{fig:ap_echo-lsa_ed_capacitive1}}%
\subfloat[]{\includegraphics[width=0.33\textwidth]%
{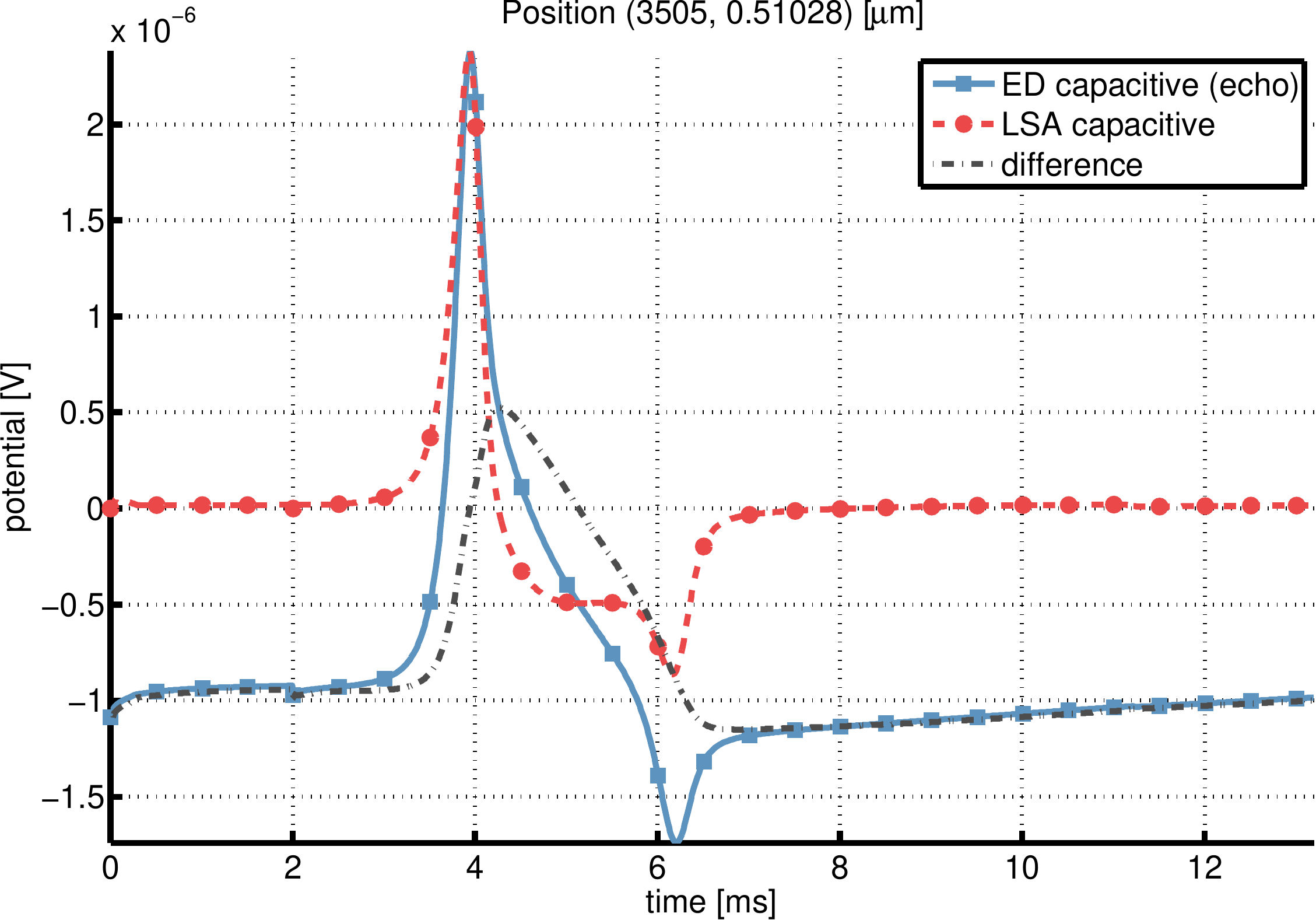}\label{fig:ap_echo-lsa_ed_capacitive2}}%
\subfloat[]{\includegraphics[width=0.33\textwidth]%
{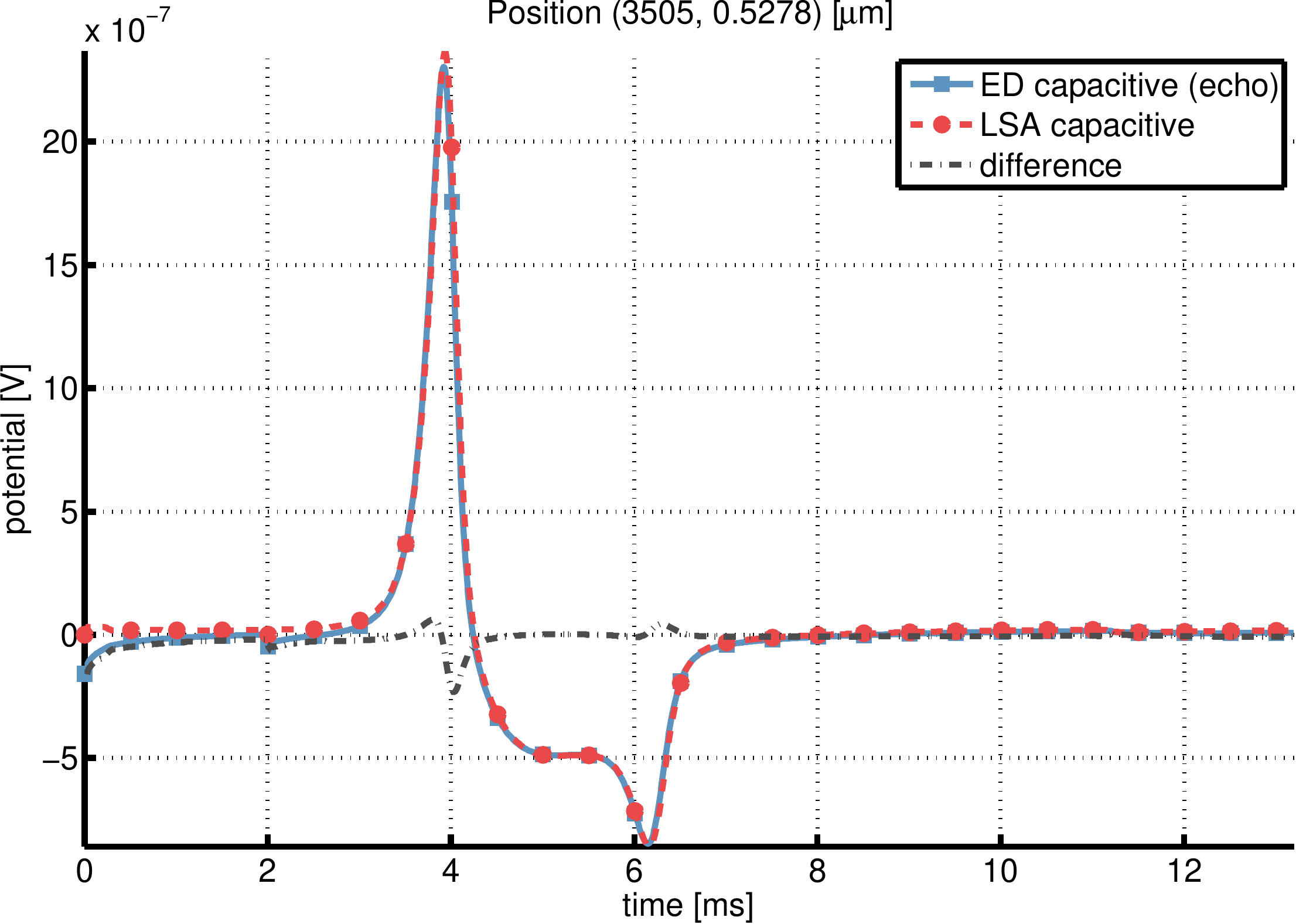}\label{fig:ap_echo-lsa_ed_capacitive3}}\\%
\subfloat[]{\includegraphics[width=0.33\textwidth]%
{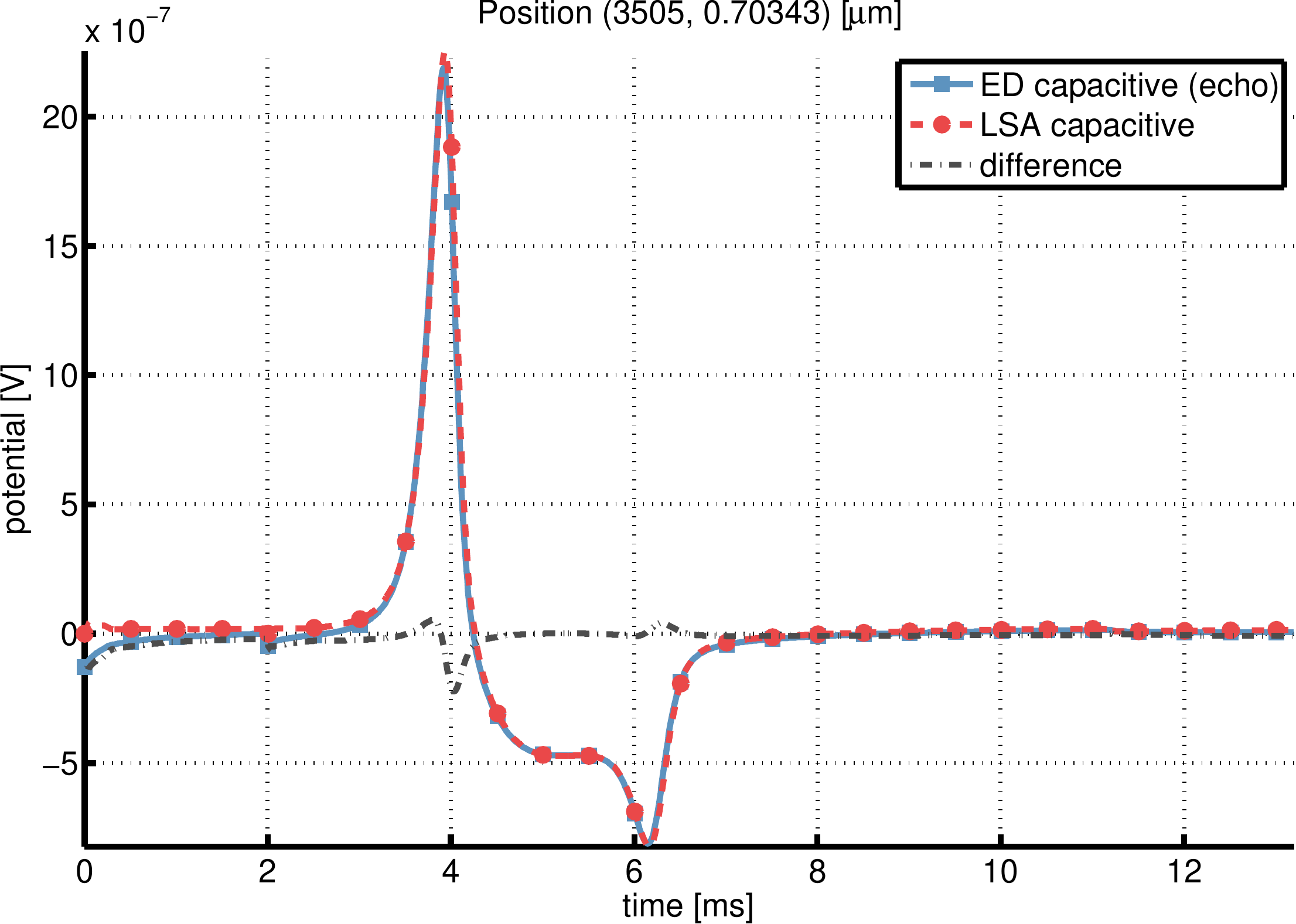}\label{fig:ap_echo-lsa_ed_capacitive4}}%
\subfloat[]{\includegraphics[width=0.33\textwidth]%
{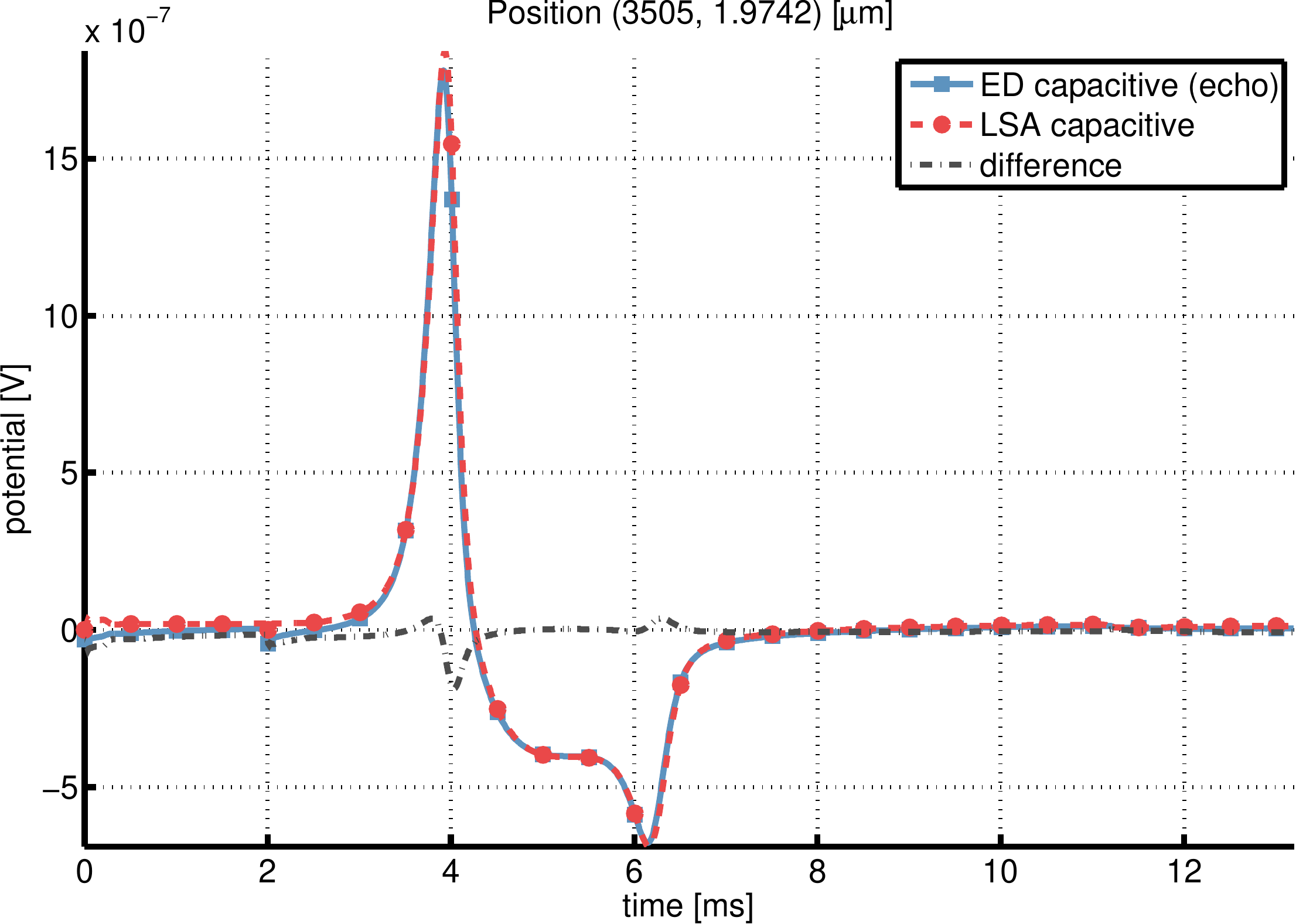}\label{fig:ap_echo-lsa_ed_capacitive5}}%
\subfloat[]{\includegraphics[width=0.33\textwidth]%
{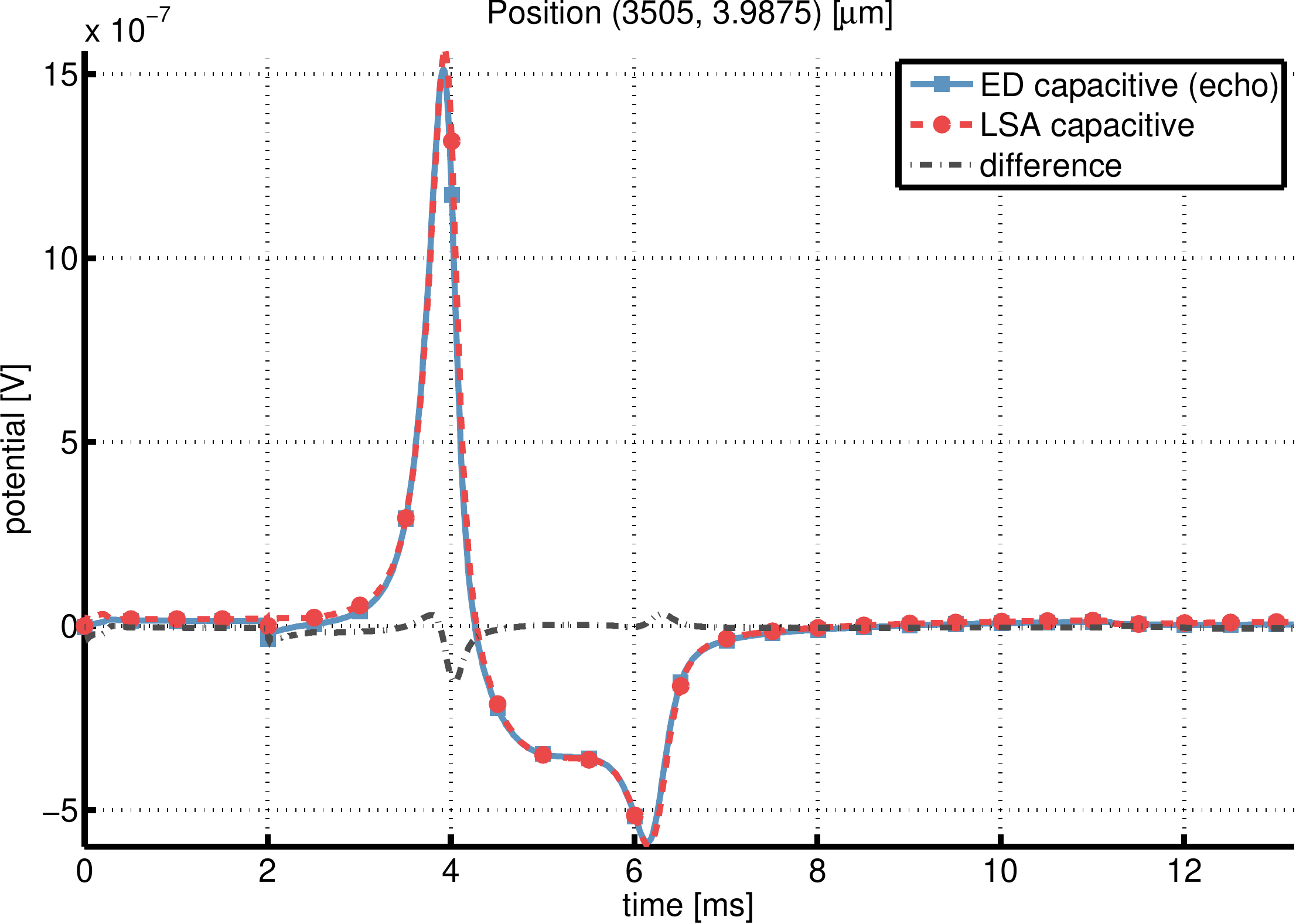}\label{fig:ap_echo-lsa_ed_capacitive6}}%
\mycaption[Comparison of the capacitive component of the EAP for ED and LSA]{This shows only the capacitive component of the full \gls{EAP} in \cref{fig:lsa_ed} for \gls{LSA} (\textit{dashed lines}) and \gls{ED} (\textit{solid lines}) model, as generated by the echo simulation.}%
\label{fig:ap_echo-lsa_ed_capacitive}%
\end{figure}

\Cref{fig:ap_echo-lsa_ed_capacitive} shows the \gls{EAP} time courses at the same points as before for the full \gls{ED} simulation in \cref{fig:lfp_distance}, in comparison with \gls{LSA} calculated using only the capacitive flux $I = I_C$.
As expected, Debye layer potentials deviate completely, as the \gls{AP} echo simulation still captures strong Debye layer gradients.
But interestingly, the match outside the Debye layer is remarkably well.
Notable differences can be made out during the on- and offset of the stimulation artifact until $t=\SI{2}{\milli\second}$.

\begin{figure}%
\centering%
\subfloat[]{\includegraphics[width=0.33\textwidth]%
{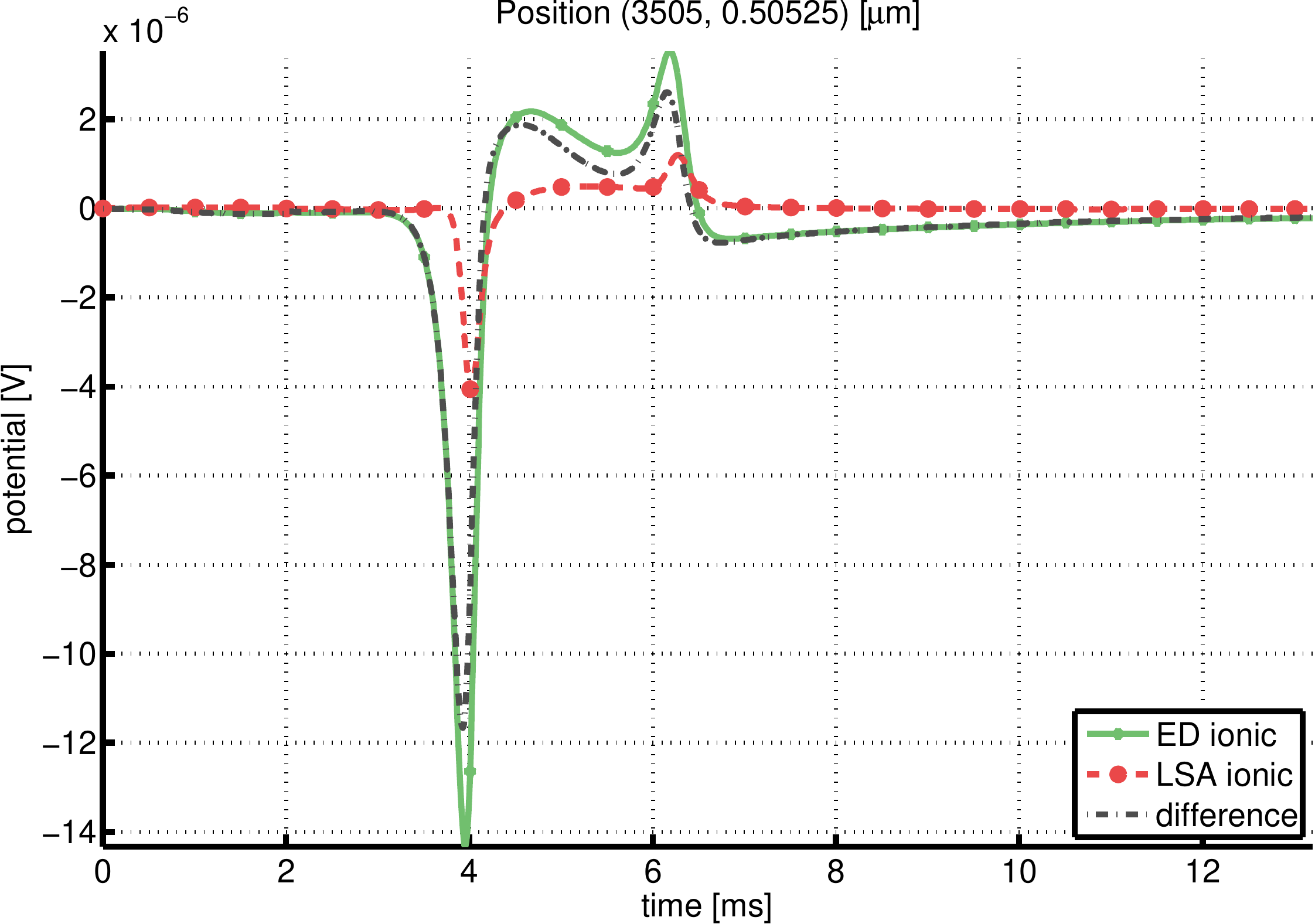}\label{fig:ap_echo-lsa_ed_ionic1}}%
\subfloat[]{\includegraphics[width=0.33\textwidth]%
{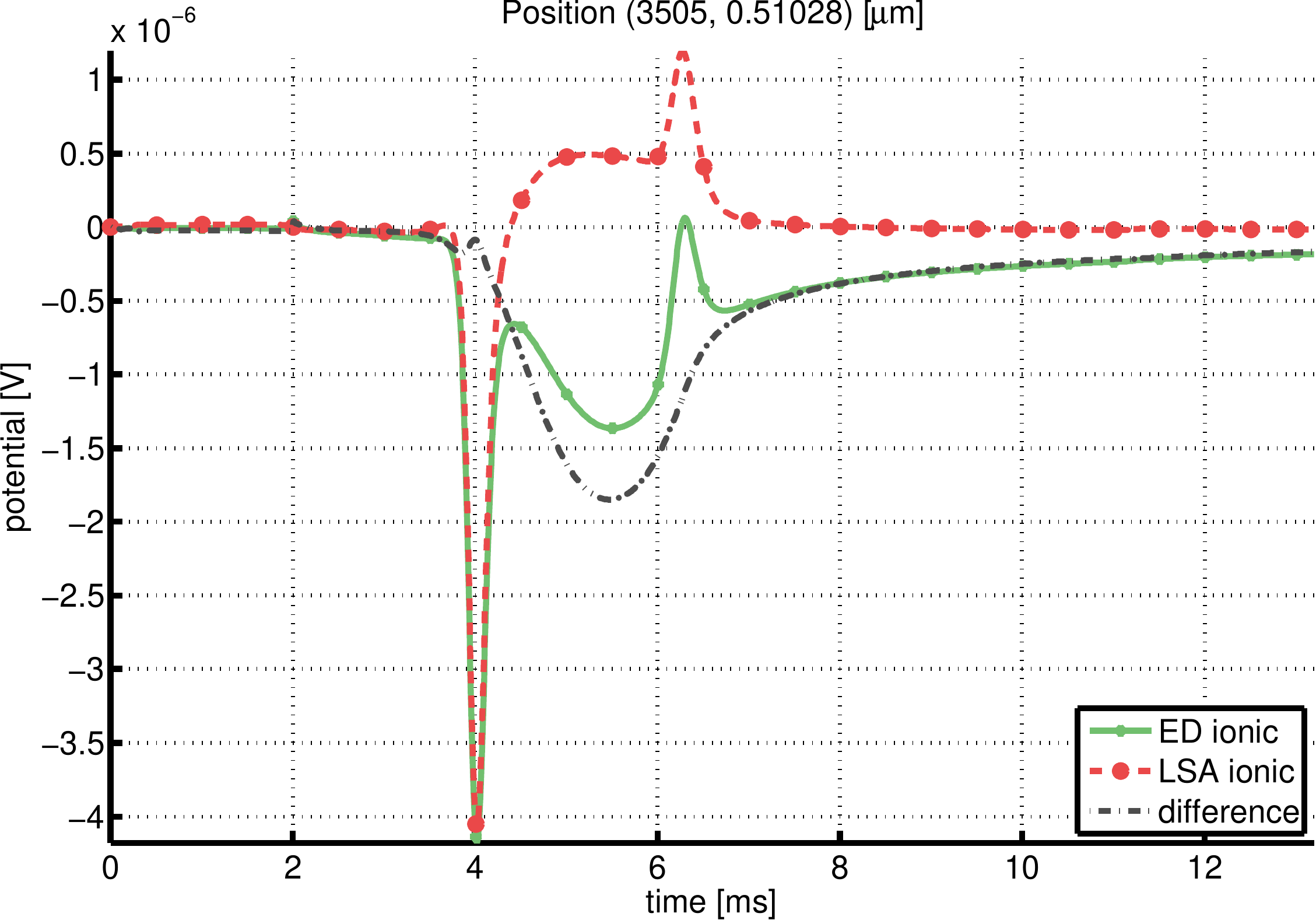}\label{fig:ap_echo-lsa_ed_ionic2}}%
\subfloat[]{\includegraphics[width=0.33\textwidth]%
{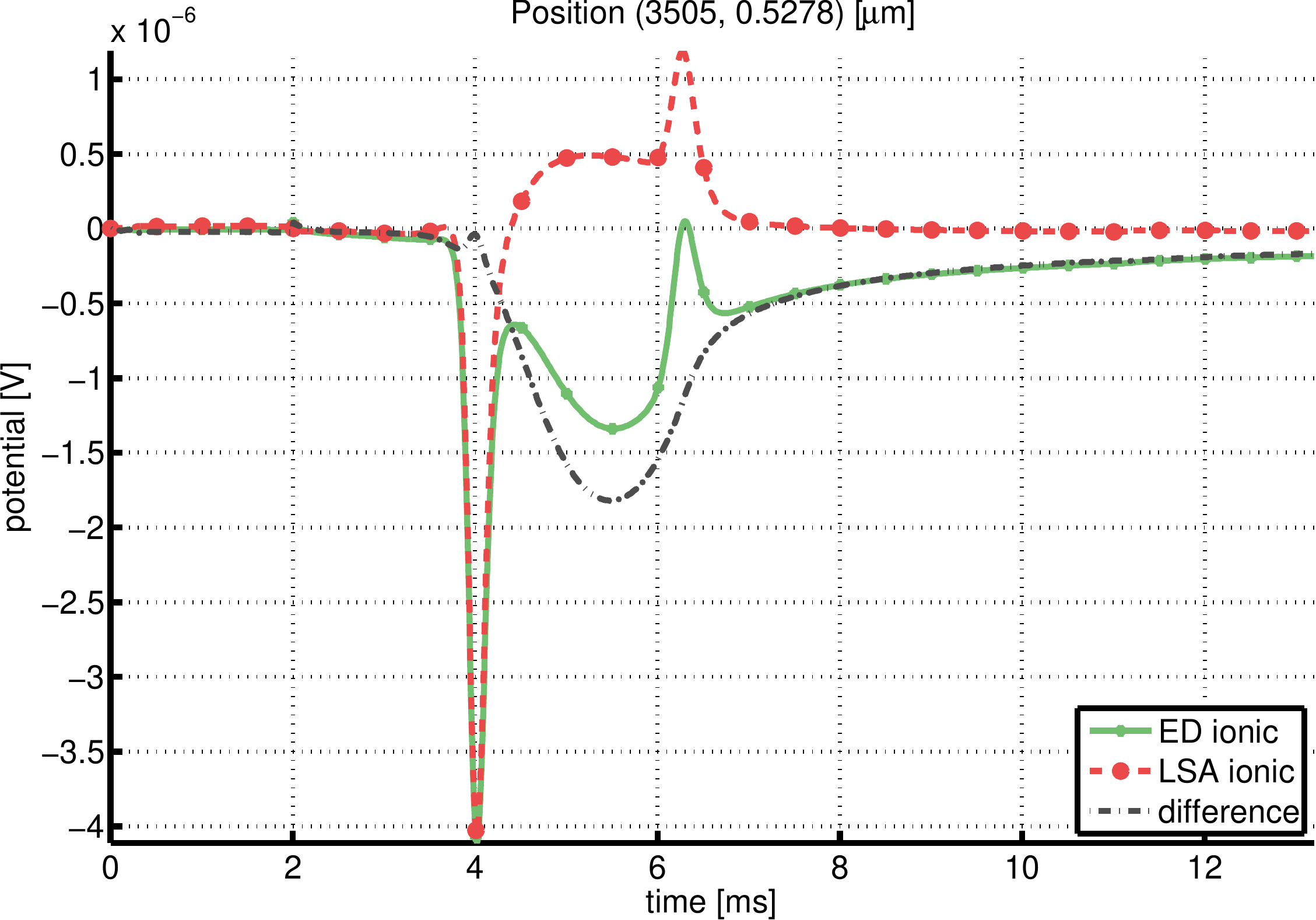}\label{fig:ap_echo-lsa_ed_ionic3}}\\%
\subfloat[]{\includegraphics[width=0.33\textwidth]%
{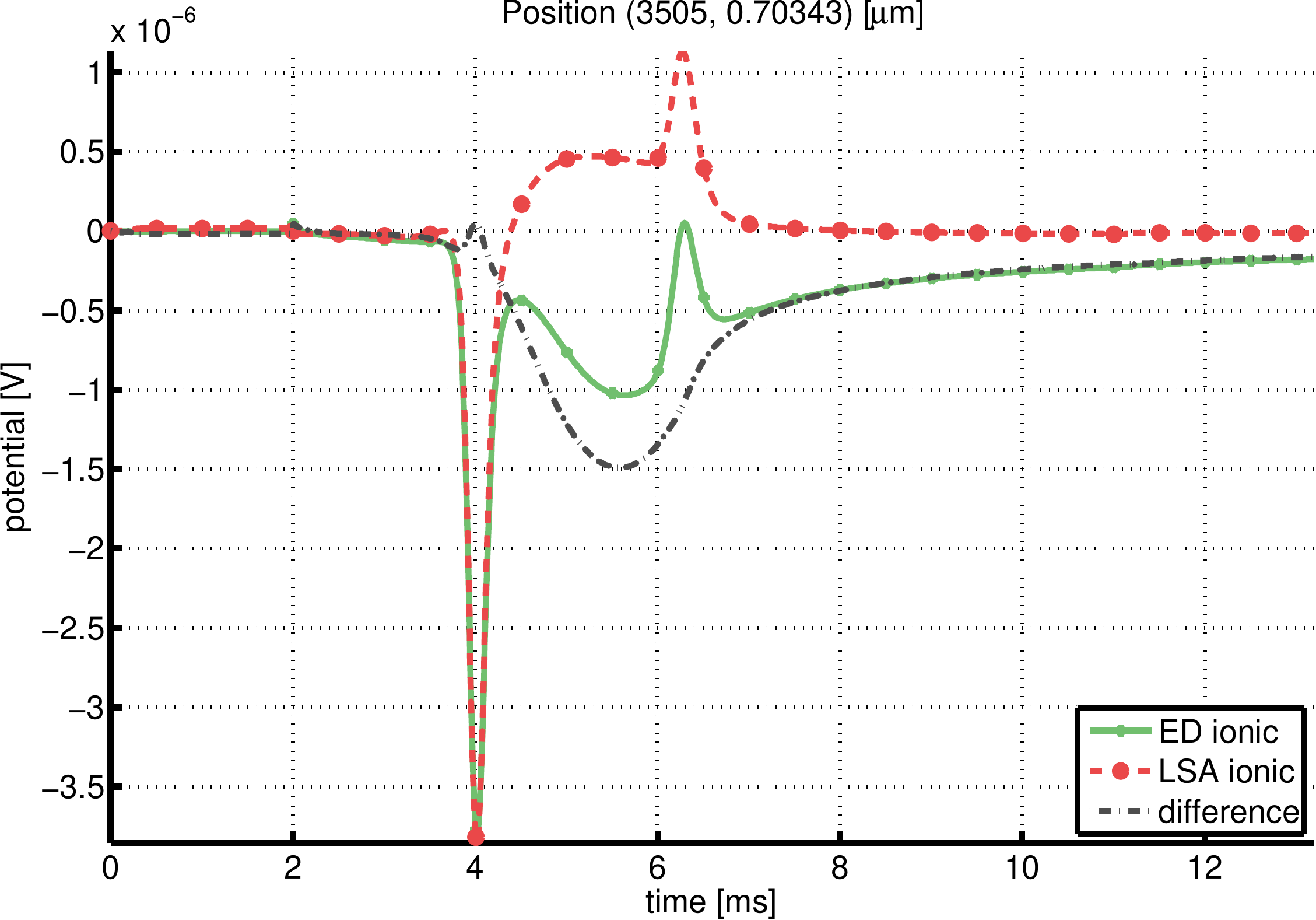}\label{fig:ap_echo-lsa_ed_ionic4}}%
\subfloat[]{\includegraphics[width=0.33\textwidth]%
{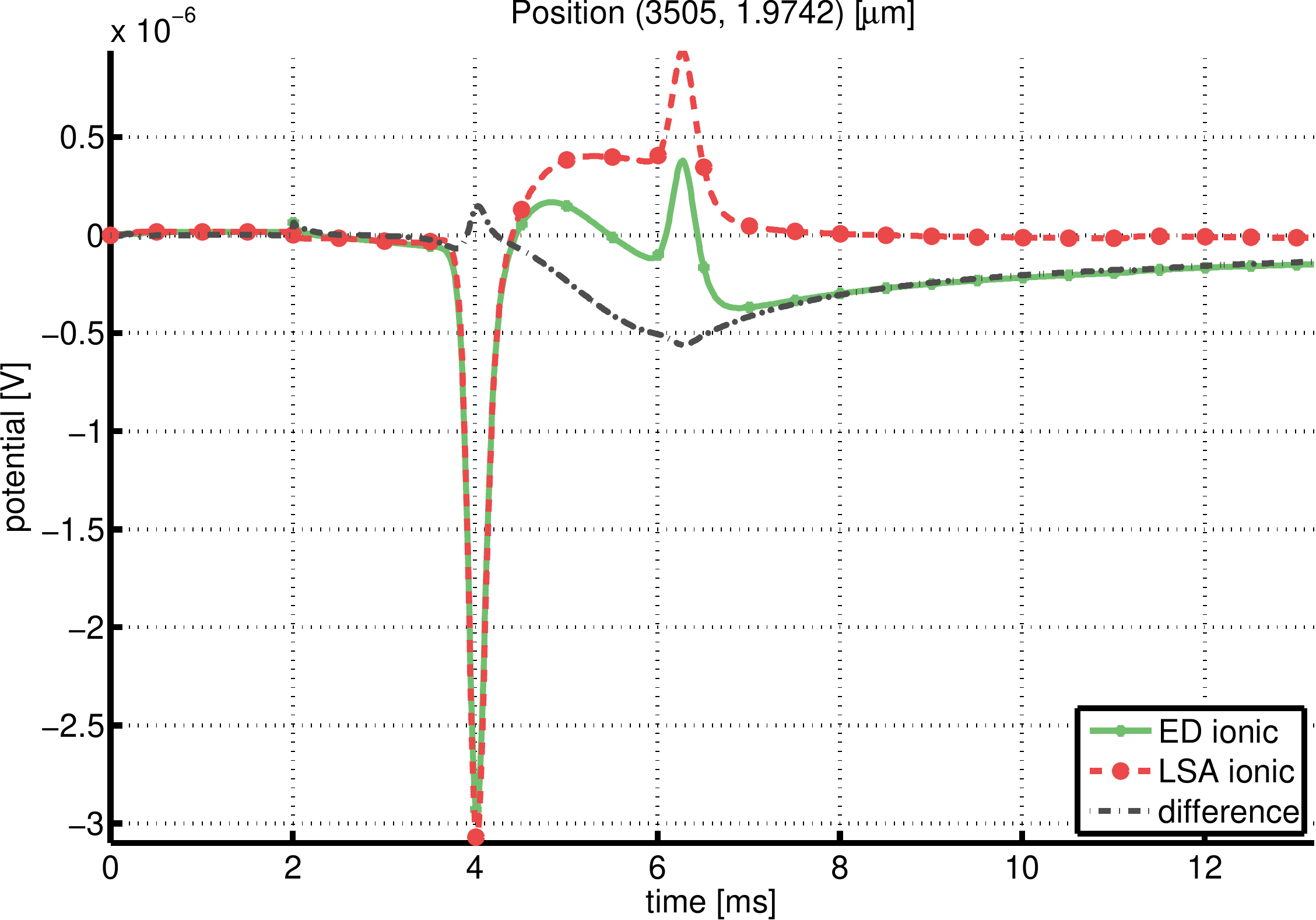}\label{fig:ap_echo-lsa_ed_ionic5}}%
\subfloat[]{\includegraphics[width=0.33\textwidth]%
{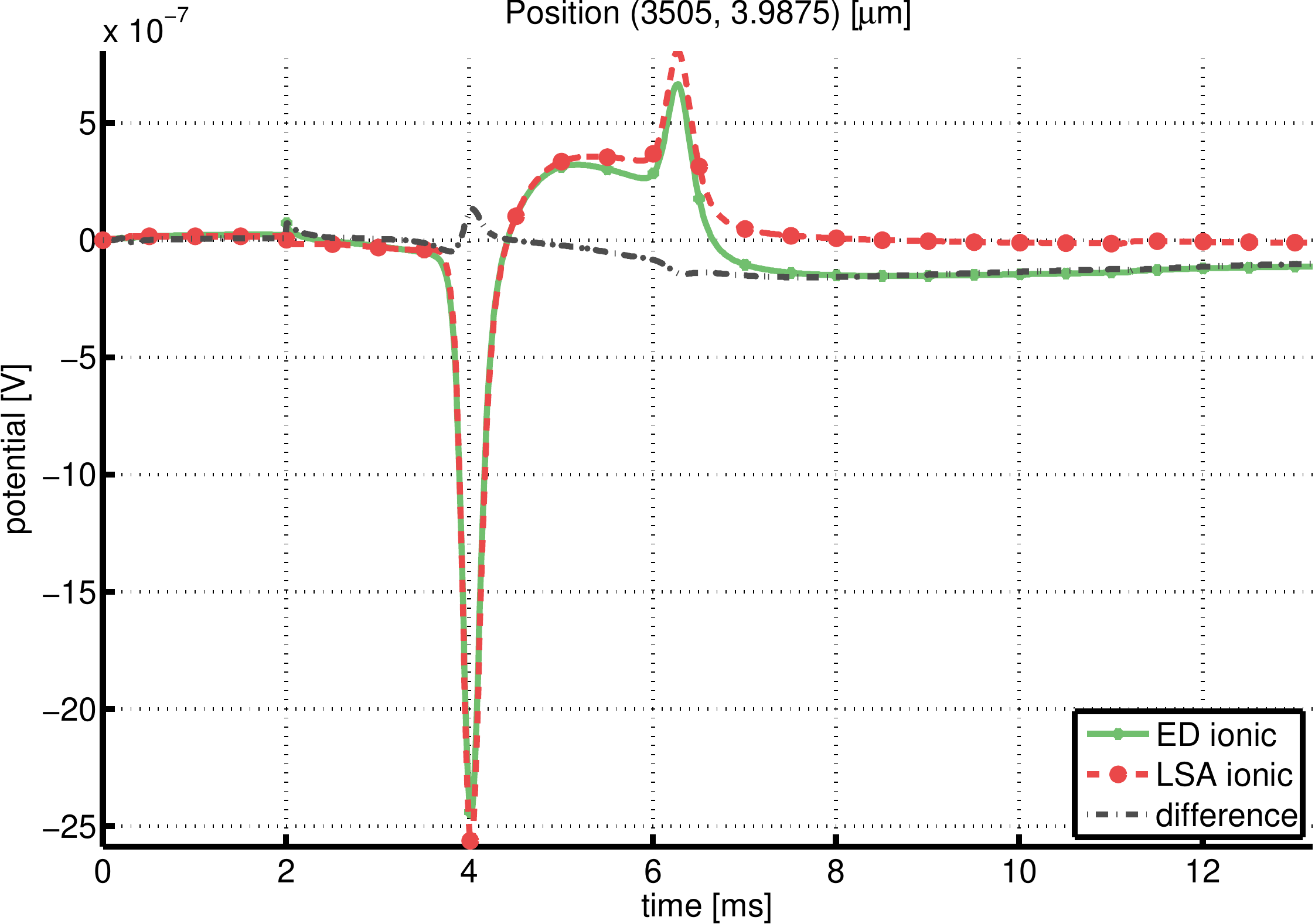}\label{fig:ap_echo-lsa_ed_ionic6}}%
\mycaption[Comparison of the ionic component of the EAP for ED and LSA]{This shows only the ionic component of the full \gls{EAP} in \cref{fig:lsa_ed} for \gls{LSA} (\textit{dashed lines}) and \gls{ED} (\textit{solid lines}) model, as calculated by subtracting results of the echo simulation from the source simulation.}%
\label{fig:ap_echo-lsa_ed_ionic}%
\end{figure}

Since now the potential from the source simulation as well as the reduced AP echo simulation are available at the same points in space and time, it is possible to simply subtract the potentials of both electrodiffusion simulations from each other, yielding a potential due to ionic currents only.
In analogy to the isolated capacitive component in \cref{fig:ap_echo-lsa_ed_capacitive}, this allows us to assess the isolated ionic component of the \gls{EAP} in \cref{fig:ap_echo-lsa_ed_ionic}.
In contrast to the capacitive component, the potential due to ionic currents shows bold deviations also for large distances.
It now becomes clear that the nearfield deviations between electrodiffusion and \gls{LSA} models can be tracked down to the ionic component of the \gls{EAP}.

We have now arrived at a point where we can try to explain the difference between the full electrodiffusion model and volume conductor models like \gls{LSA}.
The deviations have been found to originate in the ionic part of the extracellular potential, the main contributor.
The literature reveals several hints for an explanation: in \cite{frankenhaeuser1956after}, the authors describe a region of potassium accumulation around a nerve fiber with an approximate size of \SI{30}{\nano\metre}, which is referred to as a \emph{Hodgkin-Frankenhaeuser space} or simply \emph{Frankenhaeuser space} in later publications.
This might actually be a different name for the \emph{diffuse layer} from the \gls{EDL} field, see \cref{sec:theory.membrane}.
In his analysis of the \gls{PNP} equations, Mori \cite{mori2006three} describes the existence of an intermediate \emph{diffusion layer} of size $\sqrt{\dDebye}$ between Debye layer and bulk solution and also mentions Frankenhaeuser spaces in this context.
However, the dimension of his diffusive layer is orders of magnitude larger (about \SI{30}{\micro\metre} for a Debye length of $\dDebye = \SI{0.9}{\nano\metre}$) than the dimension mentioned in \cite{frankenhaeuser1956after}.
This difference in magnitude does not have to be contradictory: while the diffuse layer is the region where concentrations deviate from their bulk values, Mori's diffusion layer describes the region in which these concentration deviations have an influence on the potential, such that the electroneutrality assumption can not be used.
In any case, the existence of such a layer and its influence on the \gls{EAP} might be the very difference we see in \cref{fig:ap_echo-lsa_ed_ionic} and therefore the one that is not captured in volume conductor models.

\subsection{Numerical Solution of the Volume Conductor Equation}
In addition to the special case of the line source approximation, the underlying \gls{PDE} was also solved numerically, to which we will refer as \gls{VC} in the rest of this section, in dissociation to the analytical \gls{LSA}.
Our main goal was to have a second model to compare with, and to ensure our numerical methods work correctly, as in this case \gls{VC} should give the same results as \gls{LSA}.
The same computational grid and exterior boundary conditions as for the source simulation from the previous section were used for the \gls{VC} simulation, the only difference being that we only used the extracellular domain $\OmegaExtra$ of the source grid for faster computations, eliminating the cytosol unknowns from the system.
The internal membrane flux conditions were loaded from a previous \gls{PNP} simulation.
Large parts of the implementation infrastructure described in \cref{sec:implementation.components} could be reused for this, with the only major change being the usage of a different parameter class \lstinline!LaplaceParameters! for the operator \lstinline!Dune::PDELab::ConvectionDiffusionFEM!.

\begin{figure}%
\centering%
\subfloat[]{\includegraphics[width=0.33\textwidth]%
{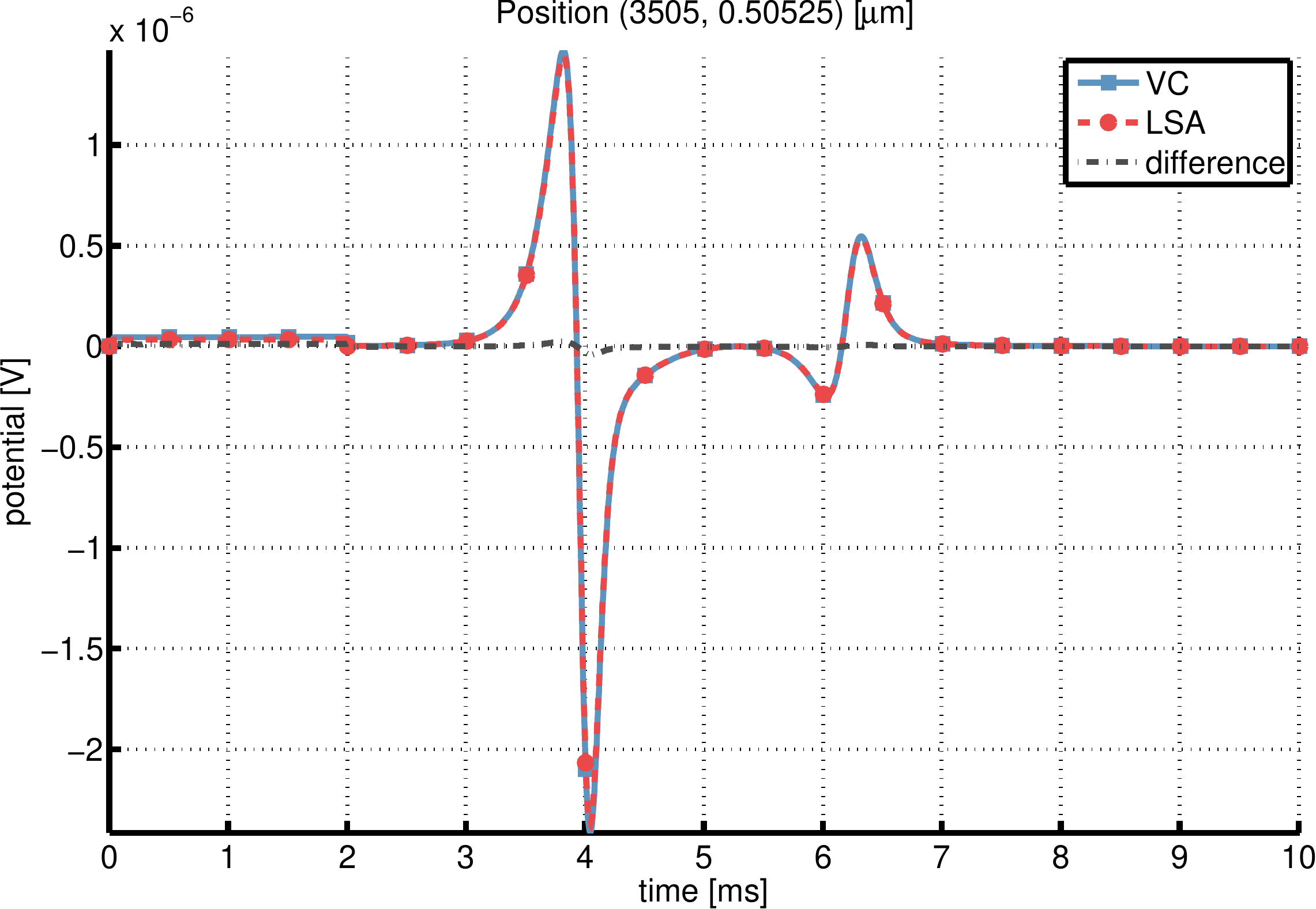}\label{fig:laplace_lsa1}}%
\subfloat[]{\includegraphics[width=0.33\textwidth]%
{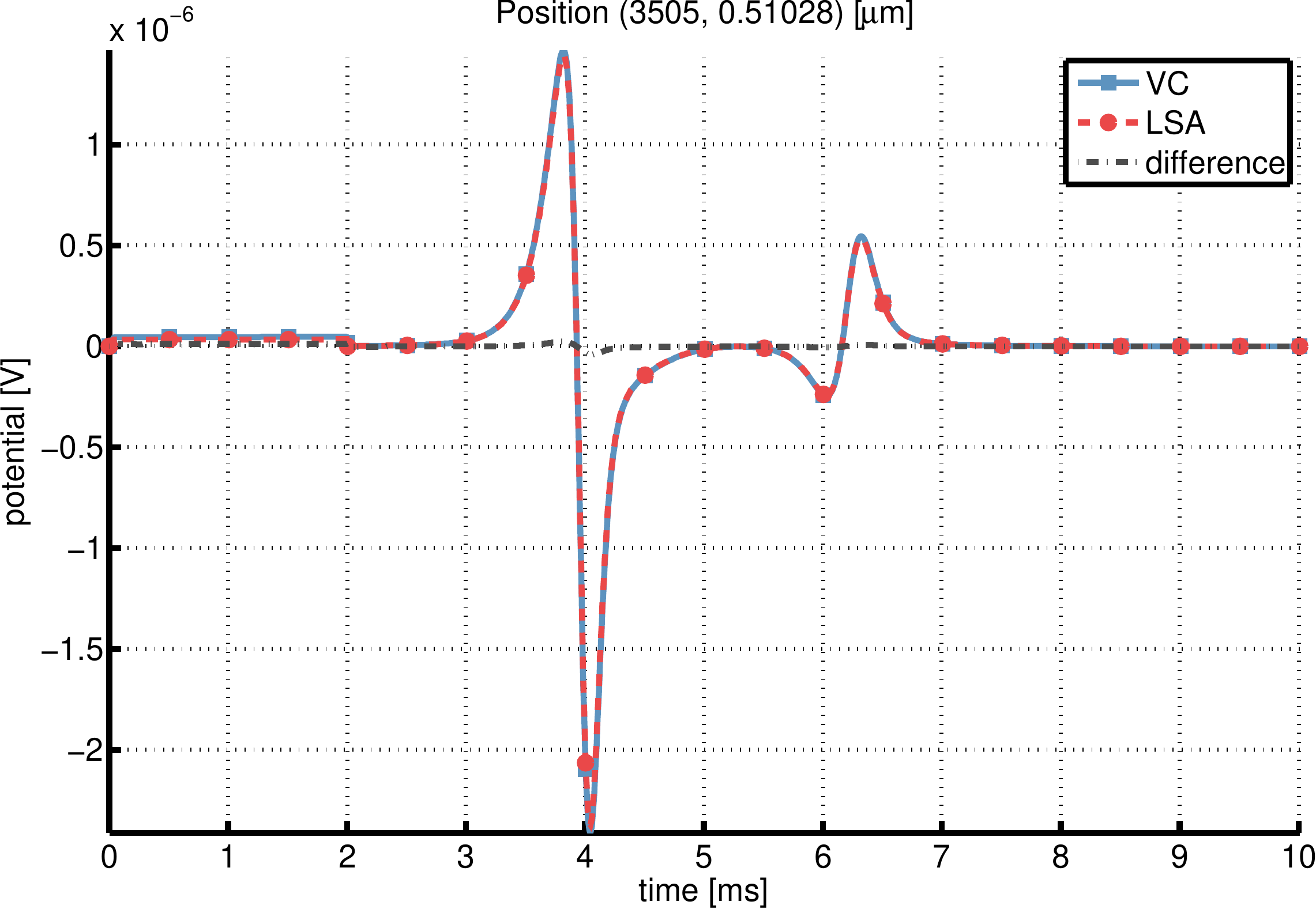}\label{fig:laplace_lsa2}}%
\subfloat[]{\includegraphics[width=0.33\textwidth]%
{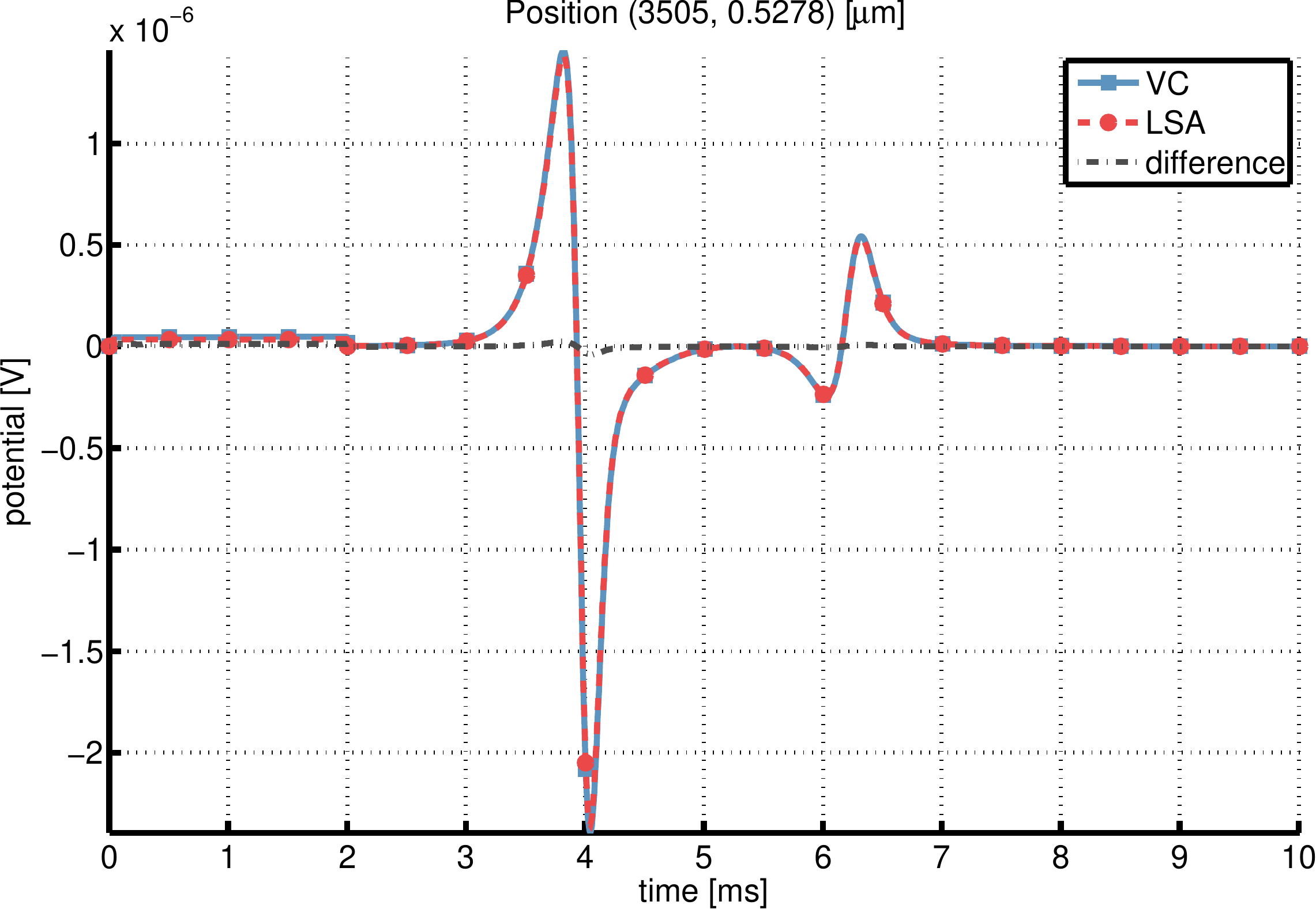}\label{fig:laplace_lsa3}}\\%
\subfloat[]{\includegraphics[width=0.33\textwidth]%
{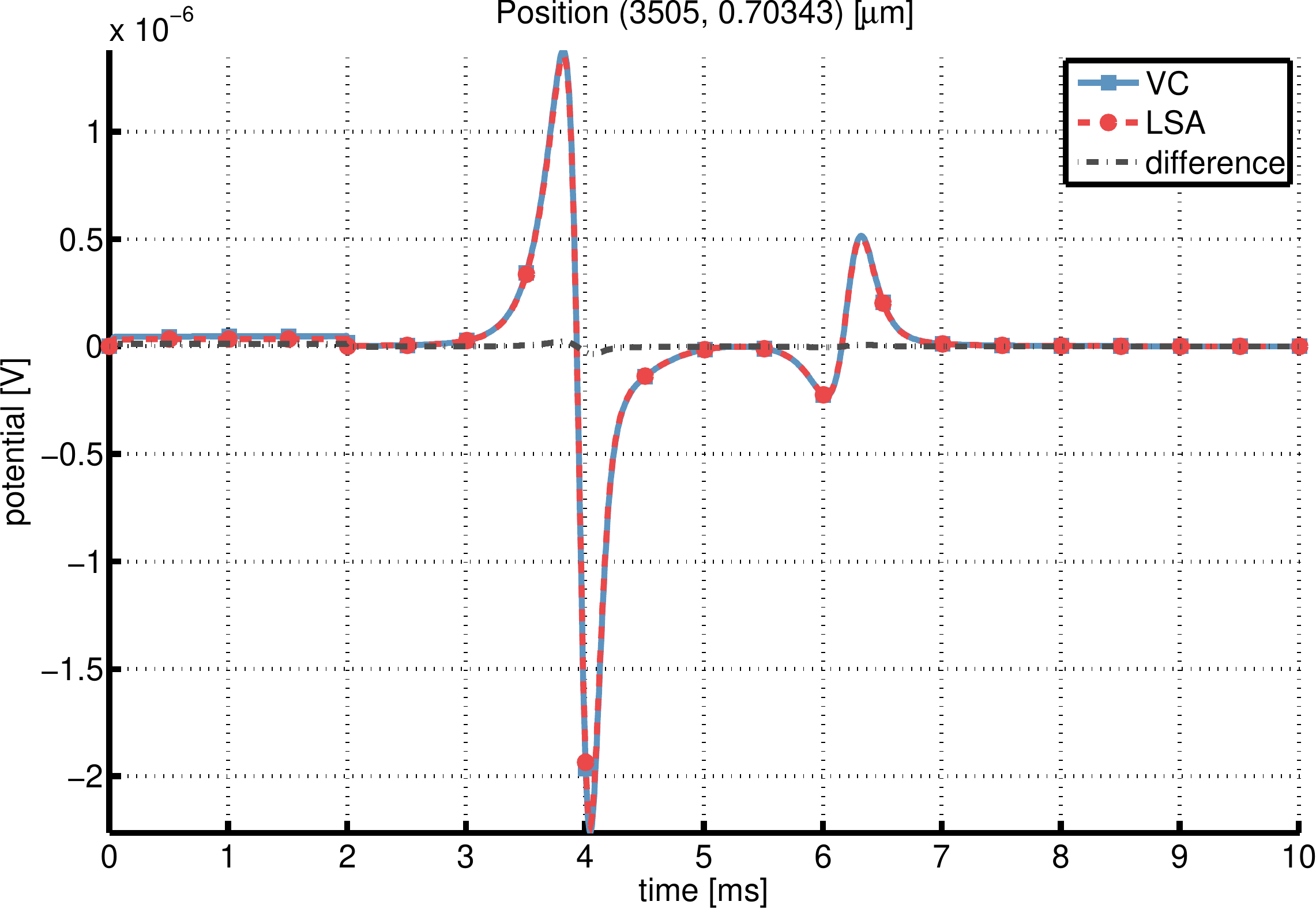}\label{fig:laplace_lsa4}}%
\subfloat[]{\includegraphics[width=0.33\textwidth]%
{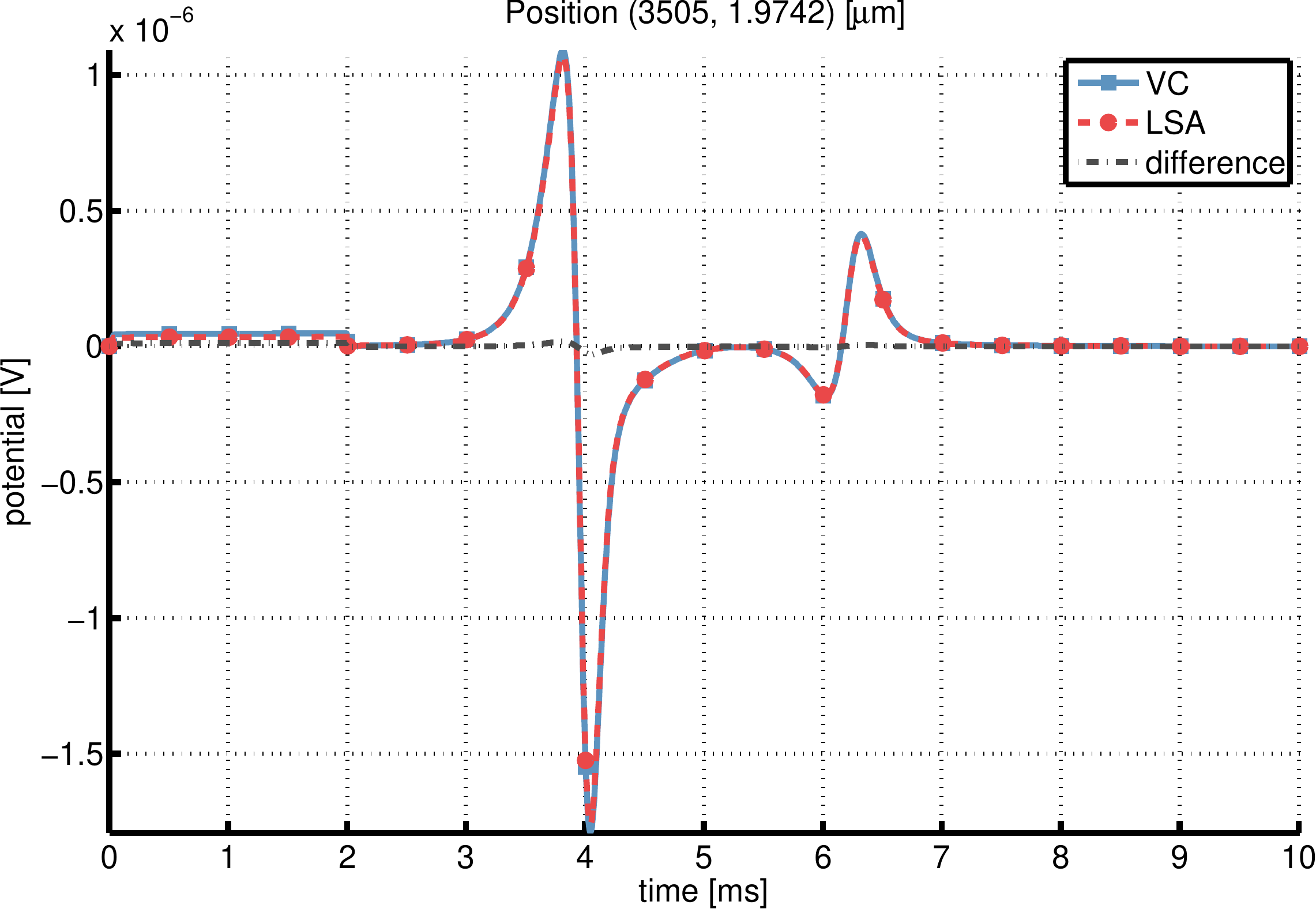}\label{fig:laplace_lsa5}}%
\subfloat[]{\includegraphics[width=0.33\textwidth]%
{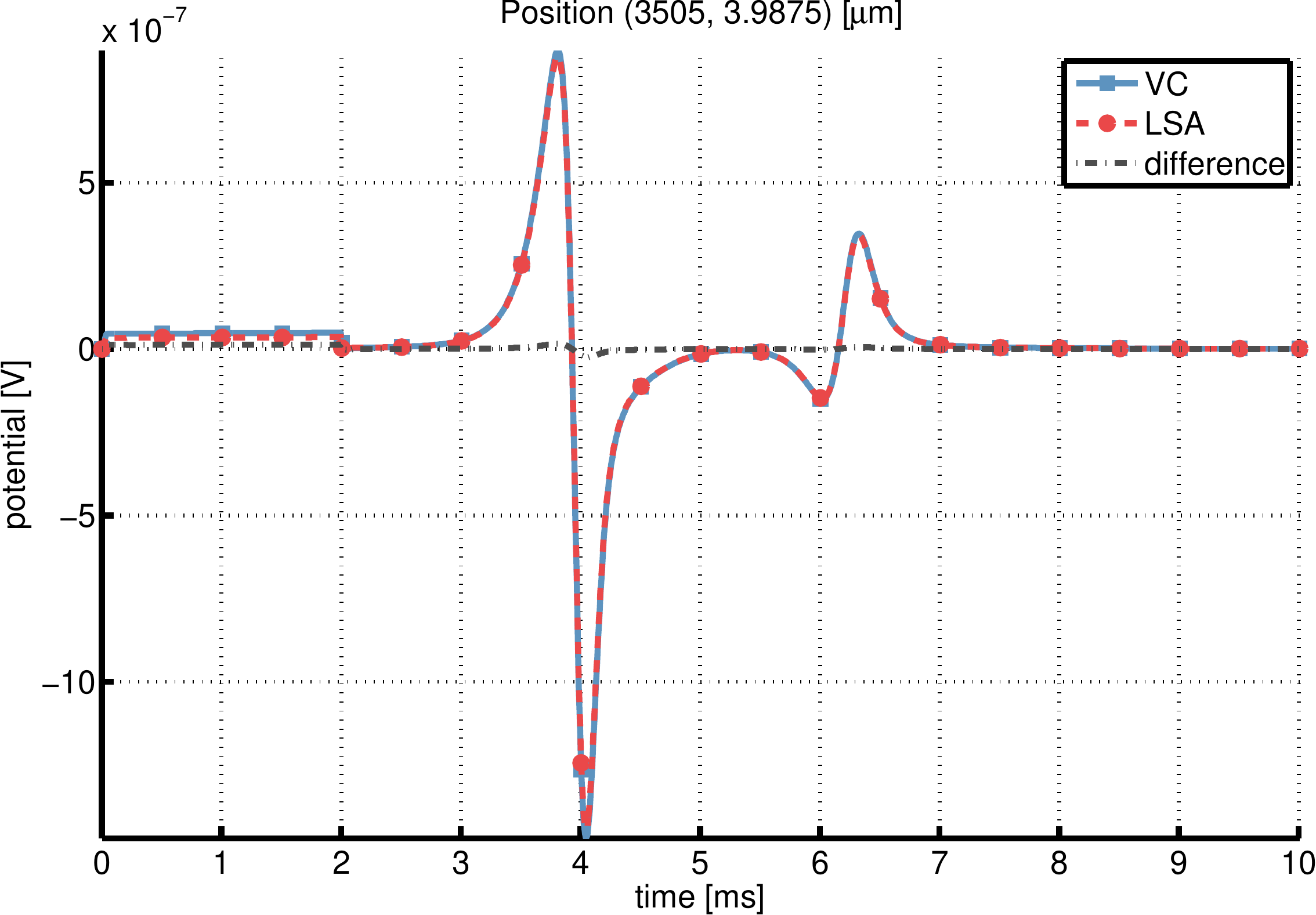}\label{fig:laplace_lsa6}}\\%
\subfloat[]{\includegraphics[width=0.33\textwidth]%
{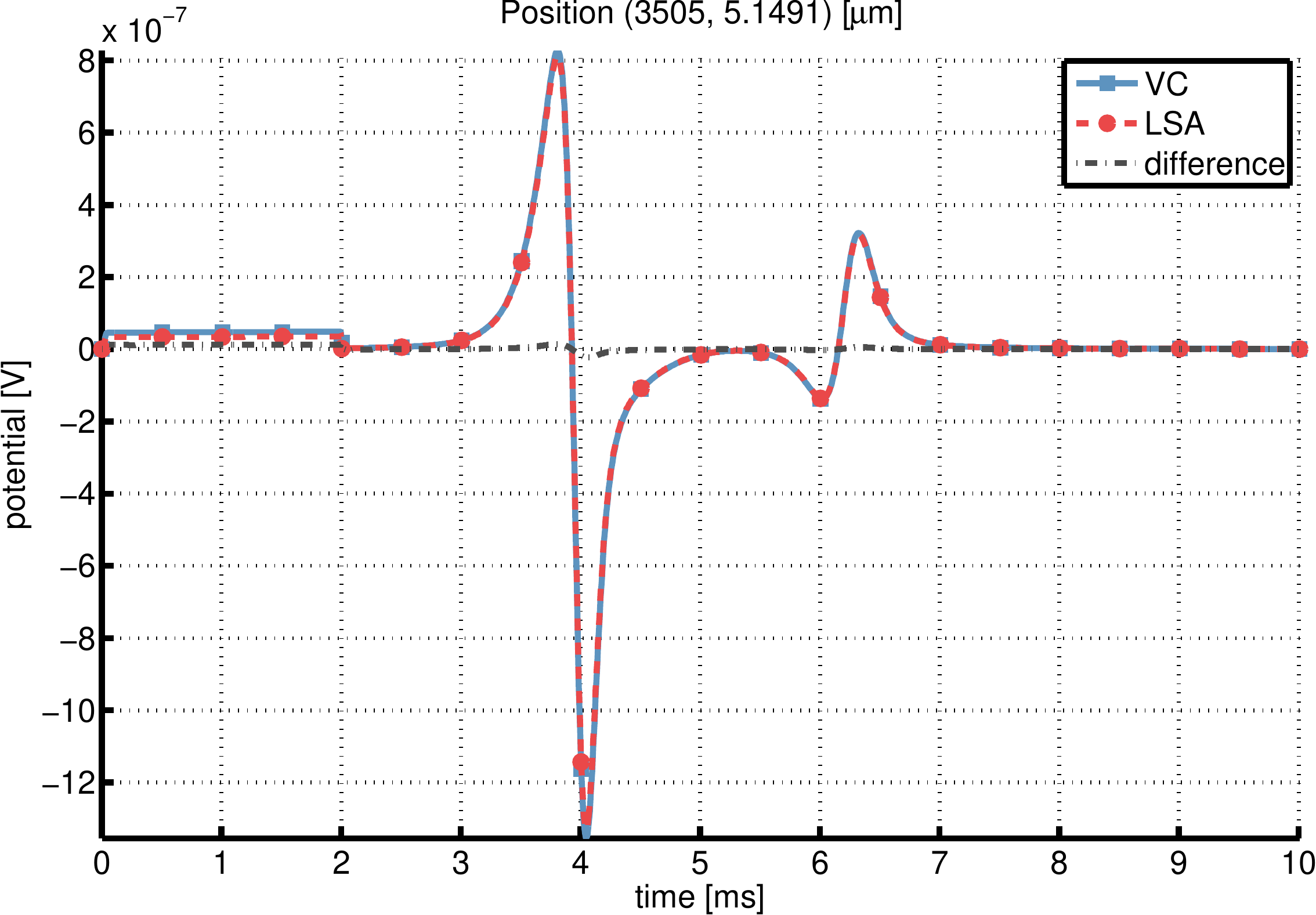}\label{fig:laplace_lsa7}}%
\subfloat[]{\includegraphics[width=0.33\textwidth]%
{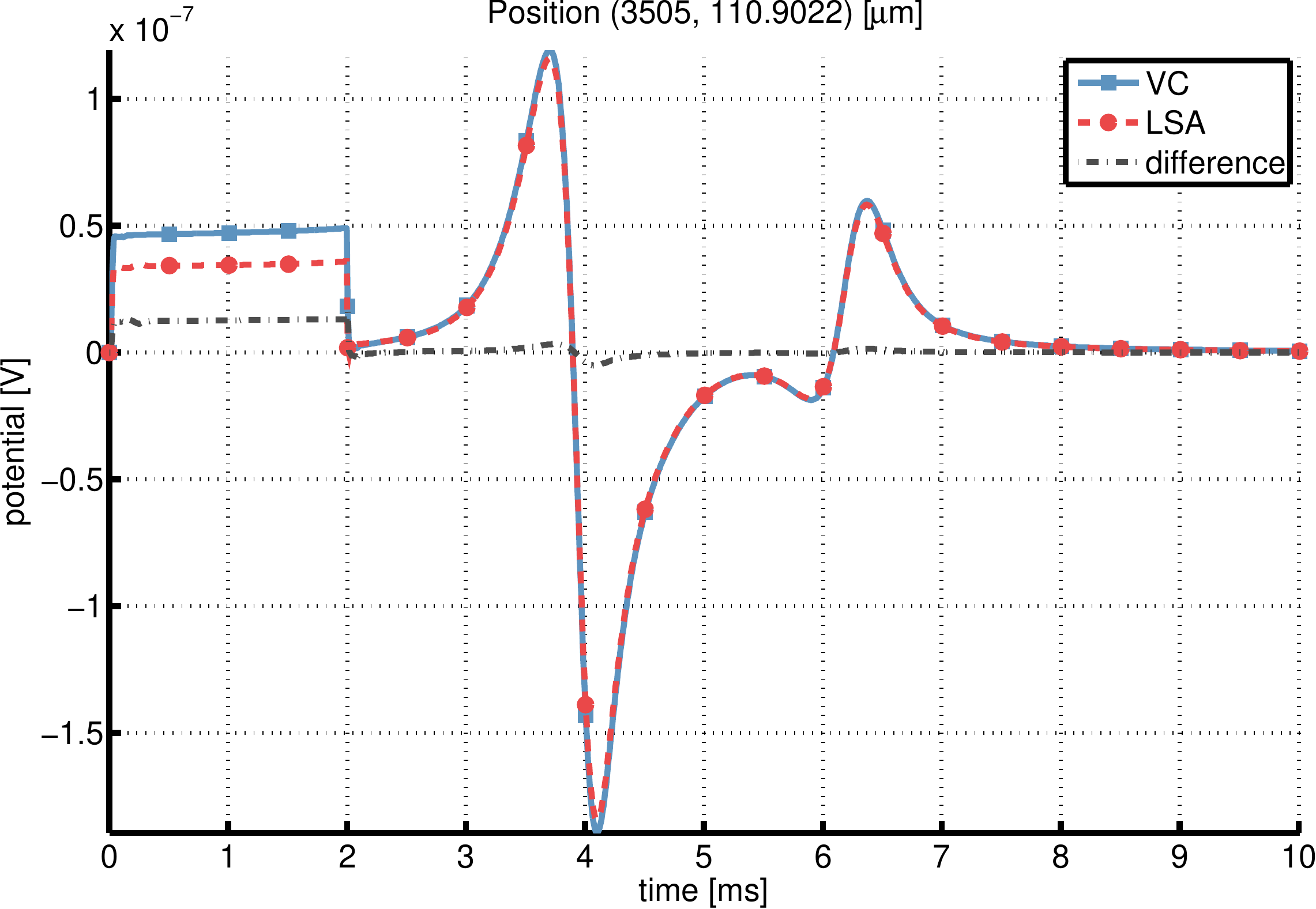}\label{fig:laplace_lsa8}}%
\subfloat[]{\includegraphics[width=0.33\textwidth]%
{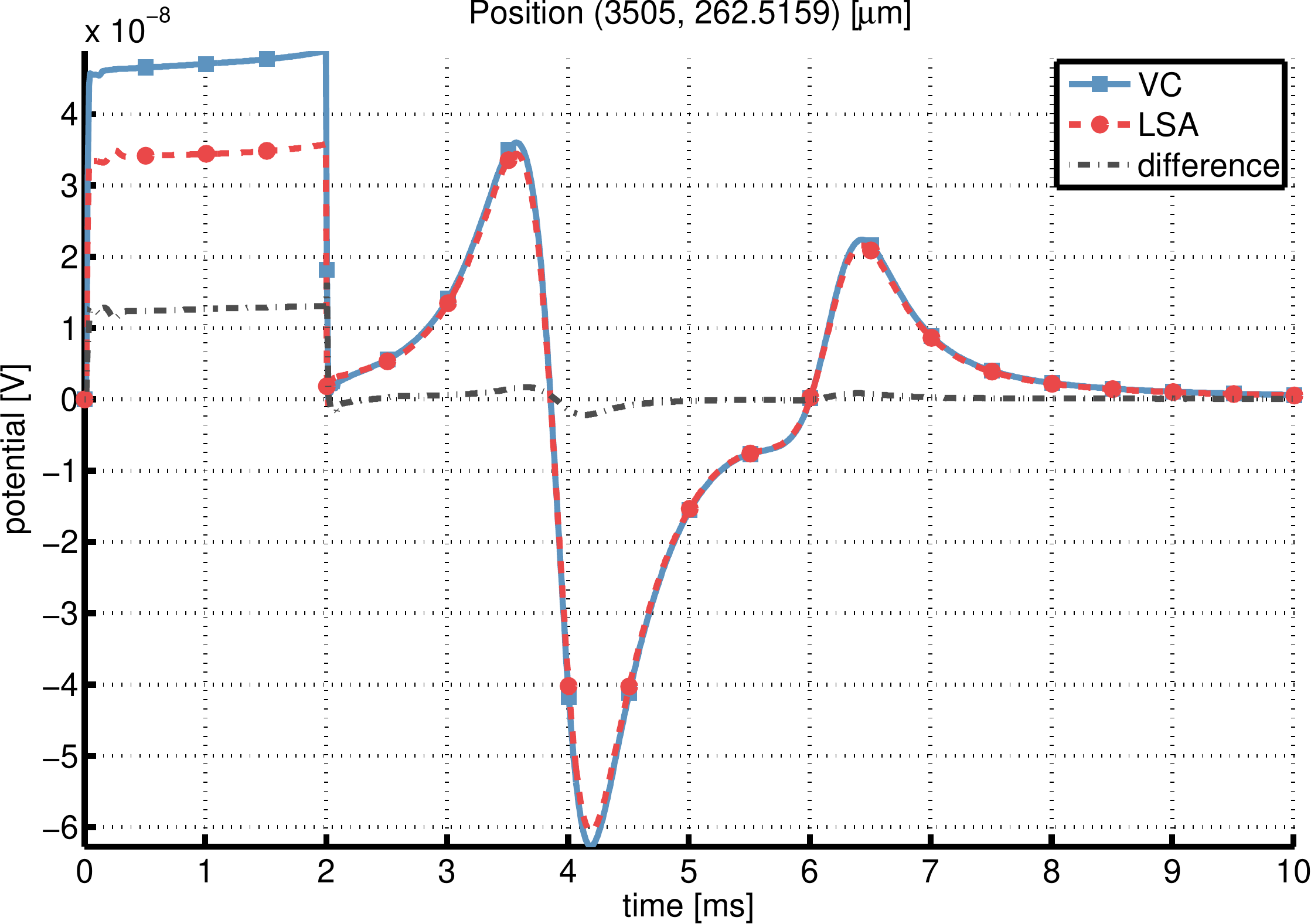}\label{fig:laplace_lsa9}}%
\mycaption[Comparison of numerical and LSA solution to the volume conductor equation]{%
A wide range of extracellular points was used to assess the accuracy of the numerical solution (\textit{solid lines}).
The match is very good even at small membrane distances, indicating that the line source approximation
(\textit{dashed lines}) provides a suitable reduction of the full cylinder geometry. At larger distances,
the influence of the upper Dirichlet boundary leads to notable deviations from \gls{LSA}.}%
\label{fig:laplace_lsa}%
\end{figure}

To validate the numerical results, they were first compared to the \gls{LSA} model, which should give an exact match within the tolerance of numerical and boundary errors present in the numerical algorithm.
\Cref{fig:laplace_lsa} shows this comparison.
As before, a wide range of the extracellular space was chosen, including very small distances from the membrane.
This serves to detect deviations close to the membrane current source, because the \gls{LSA} uses a collapsed line source, while the numerical simulation uses the full cylinder surface source.
No notable differences can be found even very close to the membrane, suggesting that the representation as a line source is a valid approximation.
At larger distances from the membrane, the artifact of the grounding Dirichlet boundary in \gls{VC} becomes visible.
Since this error is an absolute one (cf.~the elaboration in \cref{sec:model.boundary_conditions}), its relative influence grows with increasing distance from the membrane.
Within the considered range, i.e. at sufficient distances to the upper boundary, the error is still acceptable.

\begin{figure}%
\centering%
\subfloat[]{\includegraphics[width=0.33\textwidth]%
{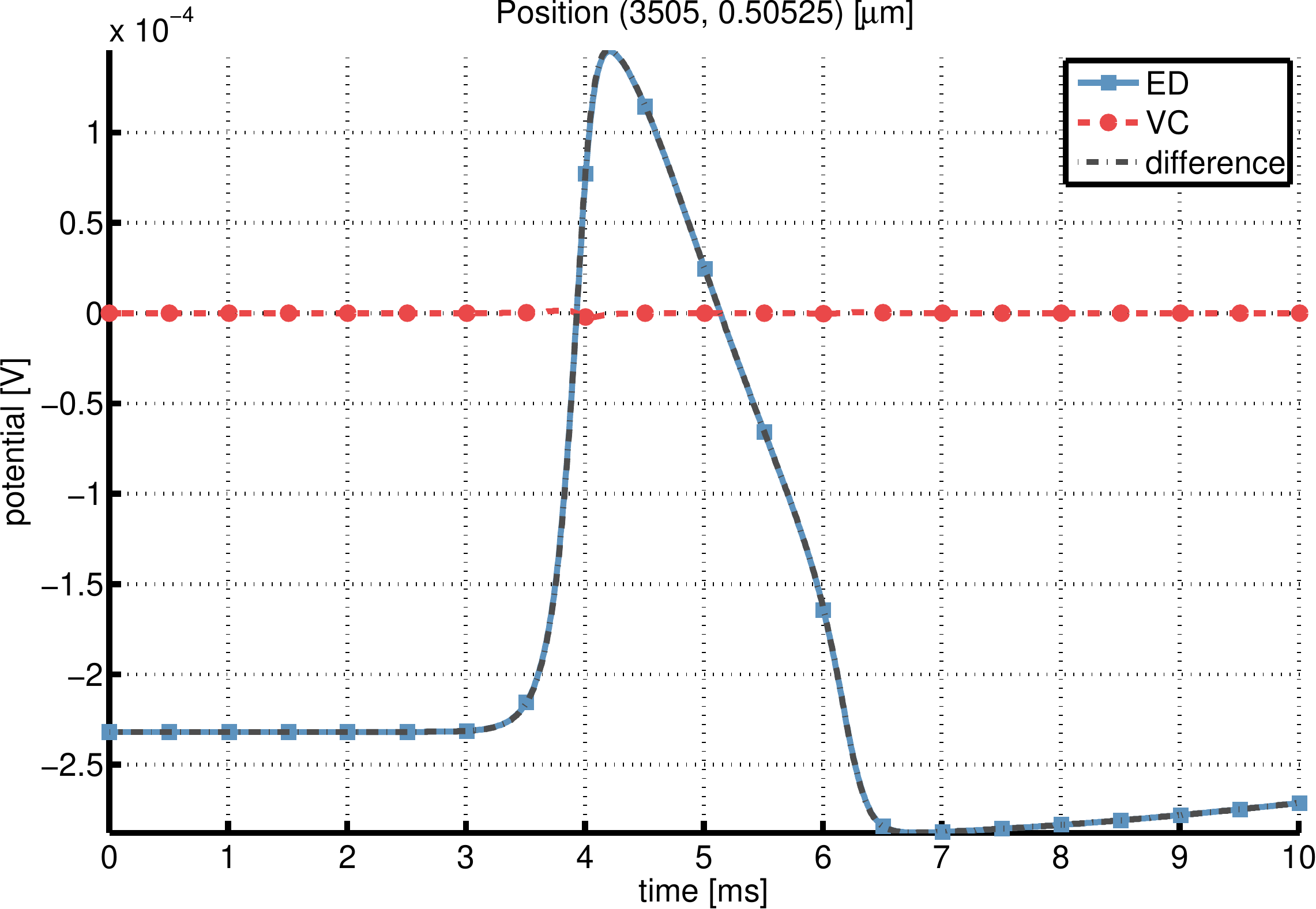}\label{fig:laplace_ed1}}%
\subfloat[]{\includegraphics[width=0.33\textwidth]%
{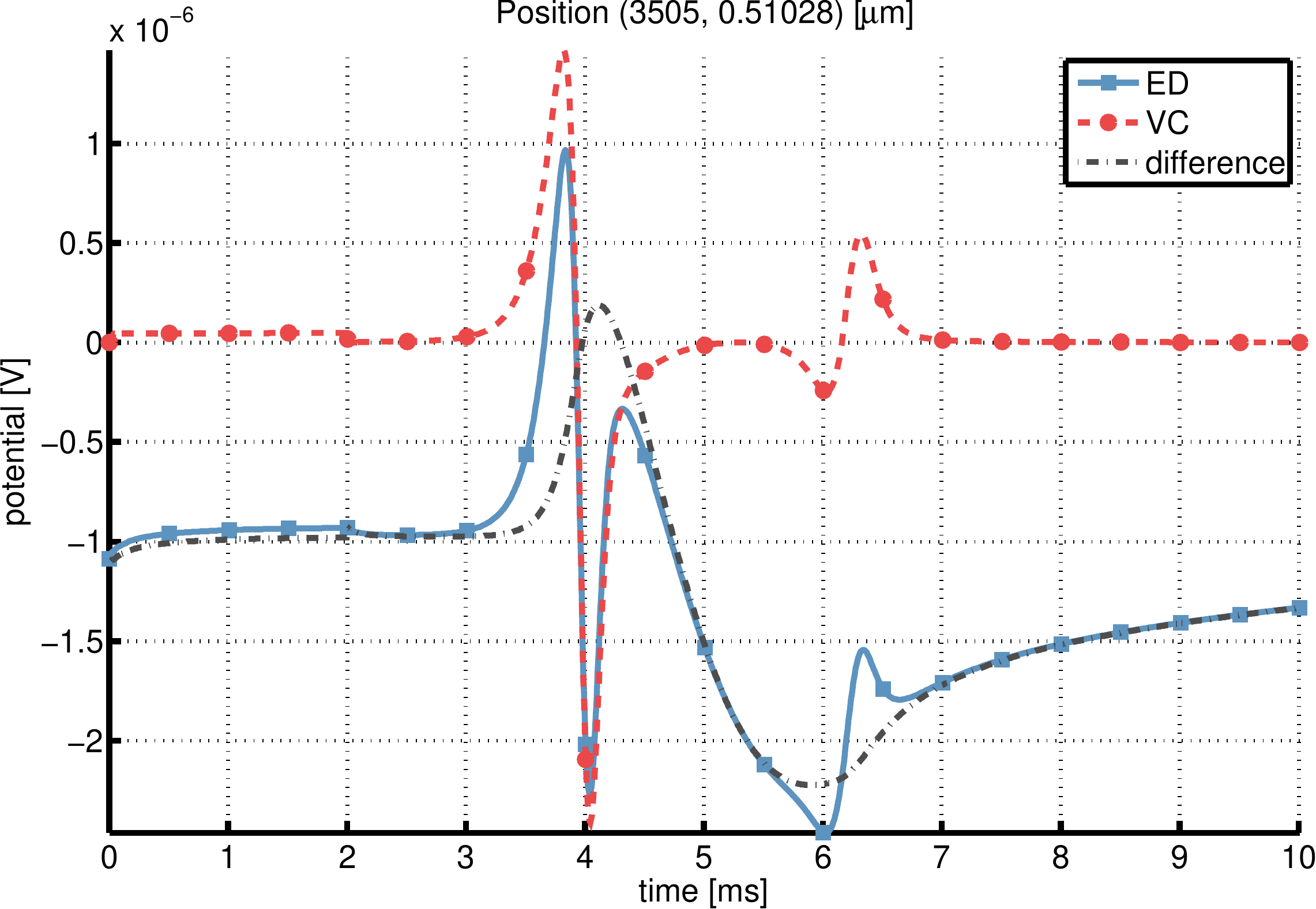}\label{fig:laplace_ed2}}%
\subfloat[]{\includegraphics[width=0.33\textwidth]%
{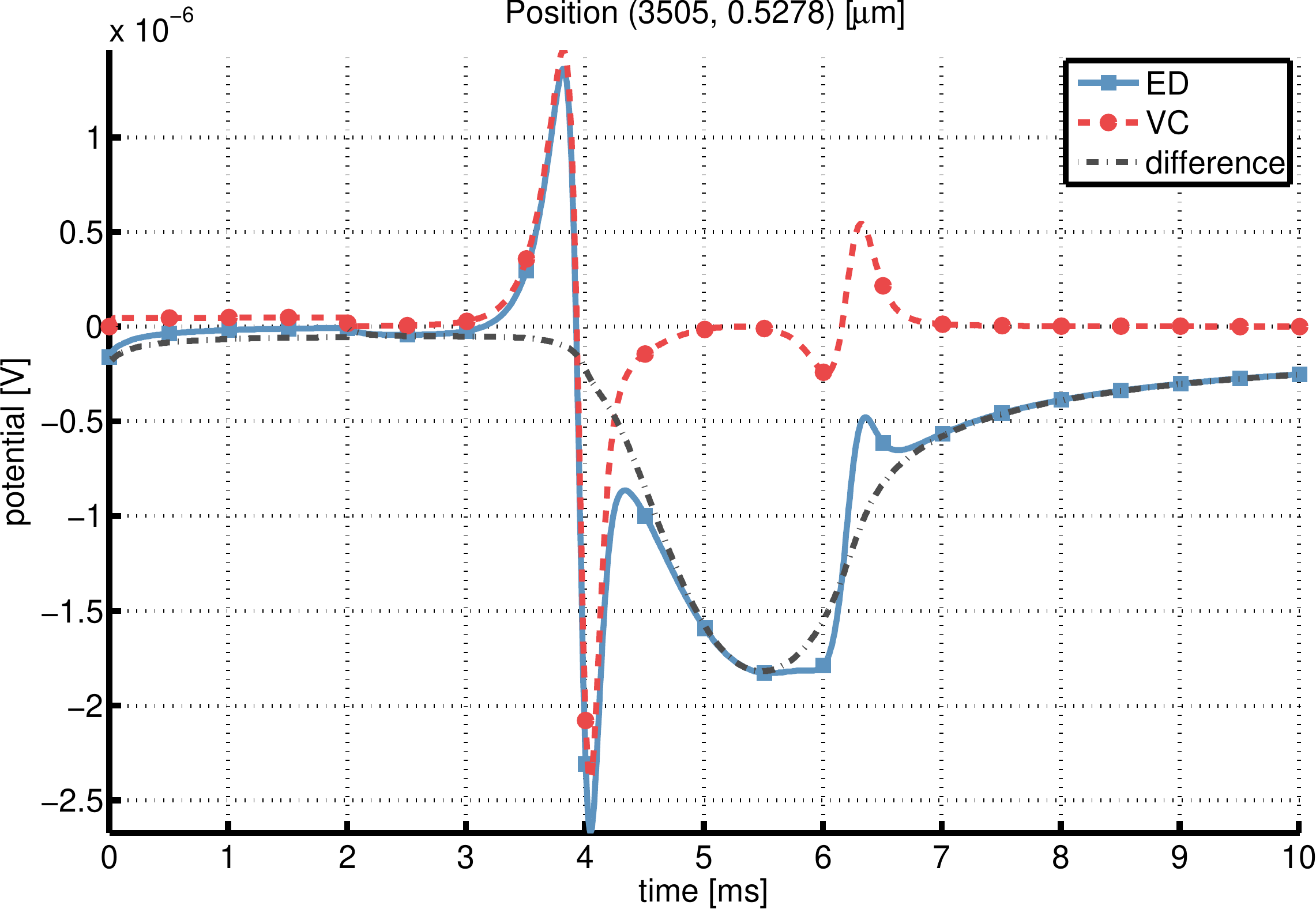}\label{fig:laplace_ed3}}\\%
\subfloat[]{\includegraphics[width=0.33\textwidth]%
{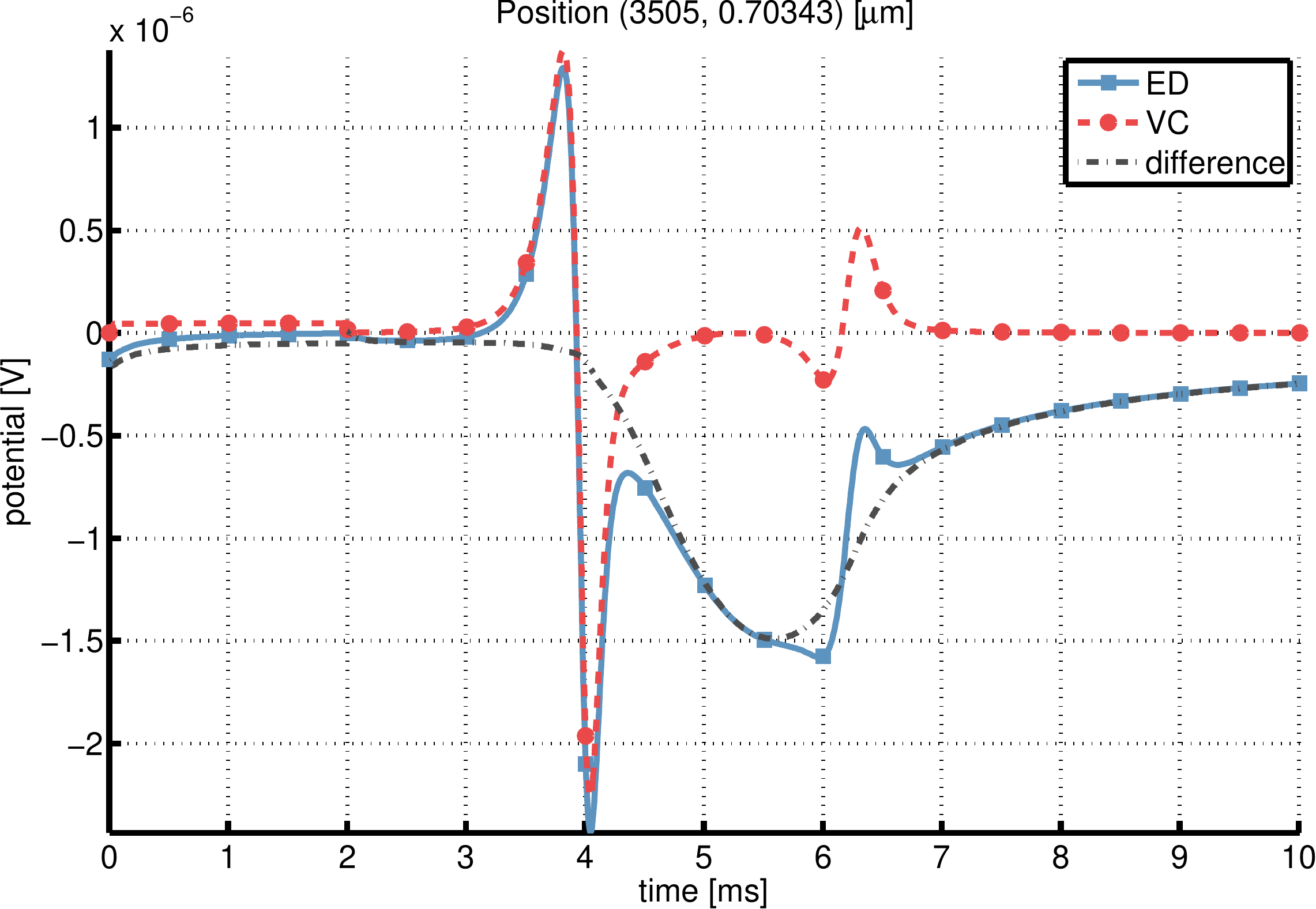}\label{fig:laplace_ed4}}%
\subfloat[]{\includegraphics[width=0.33\textwidth]%
{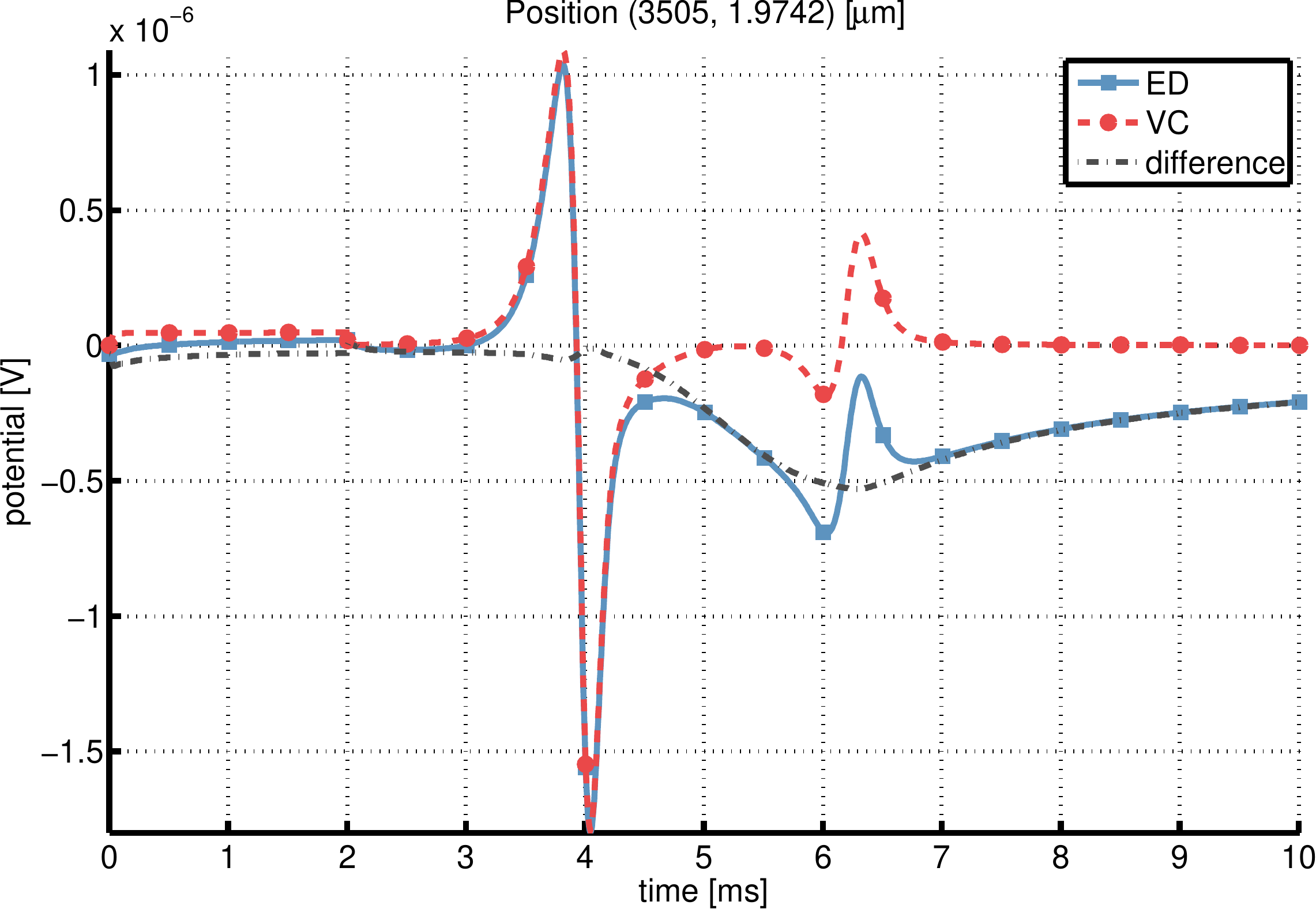}\label{fig:laplace_ed5}}%
\subfloat[]{\includegraphics[width=0.33\textwidth]%
{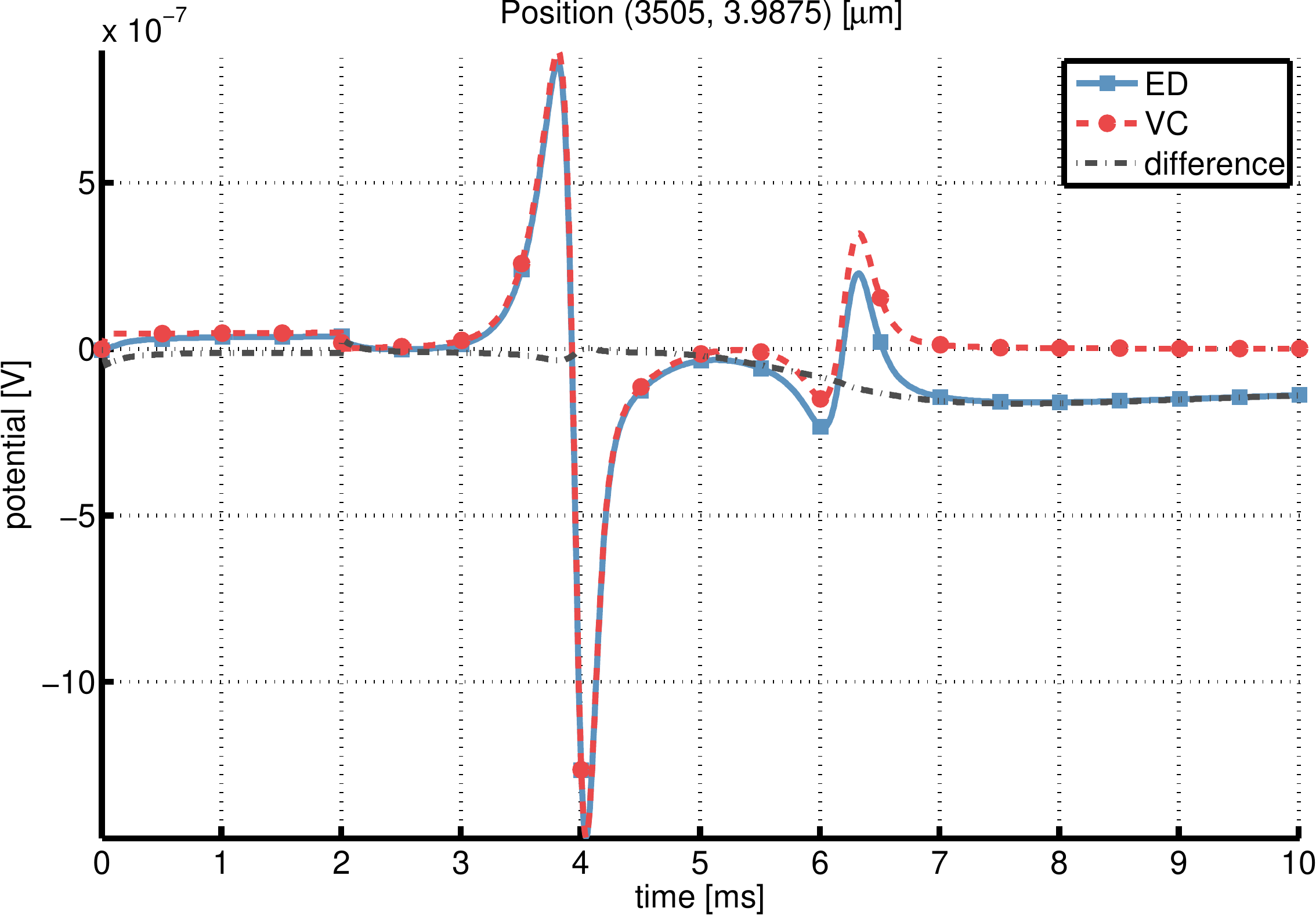}\label{fig:laplace_ed6}}\\%
\subfloat[]{\includegraphics[width=0.33\textwidth]%
{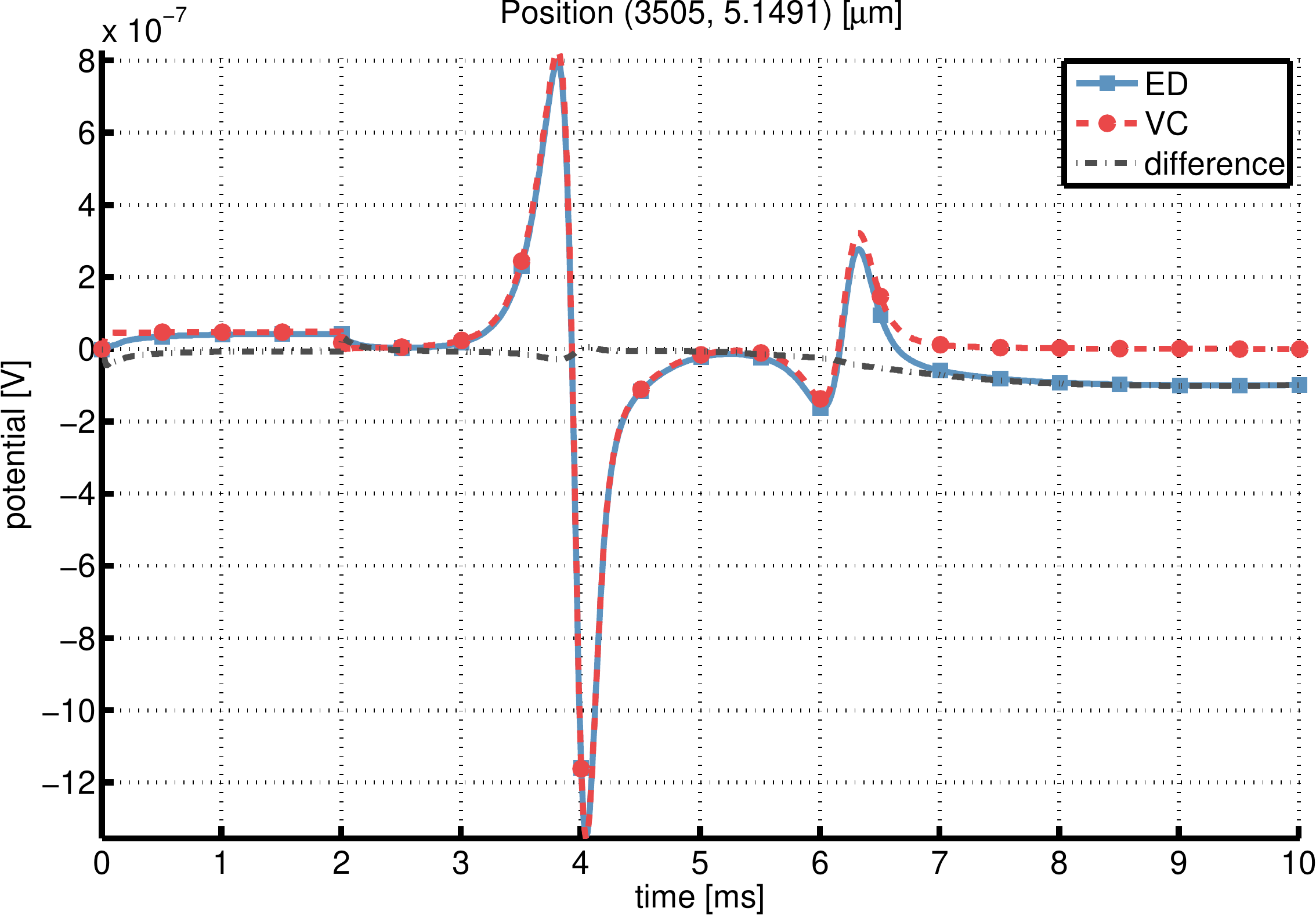}\label{fig:laplace_ed7}}%
\subfloat[]{\includegraphics[width=0.33\textwidth]%
{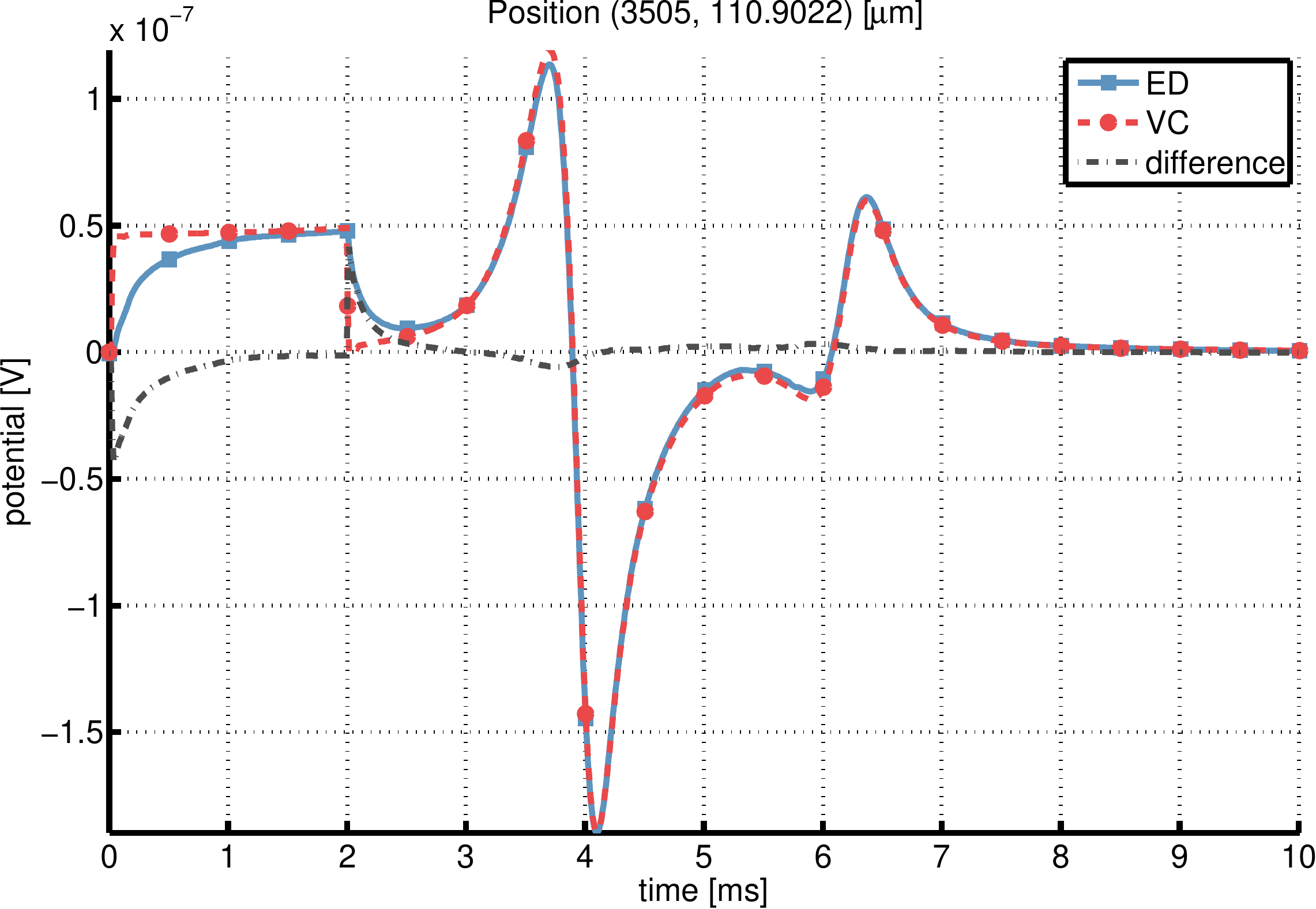}\label{fig:laplace_ed8}}%
\subfloat[]{\includegraphics[width=0.33\textwidth]%
{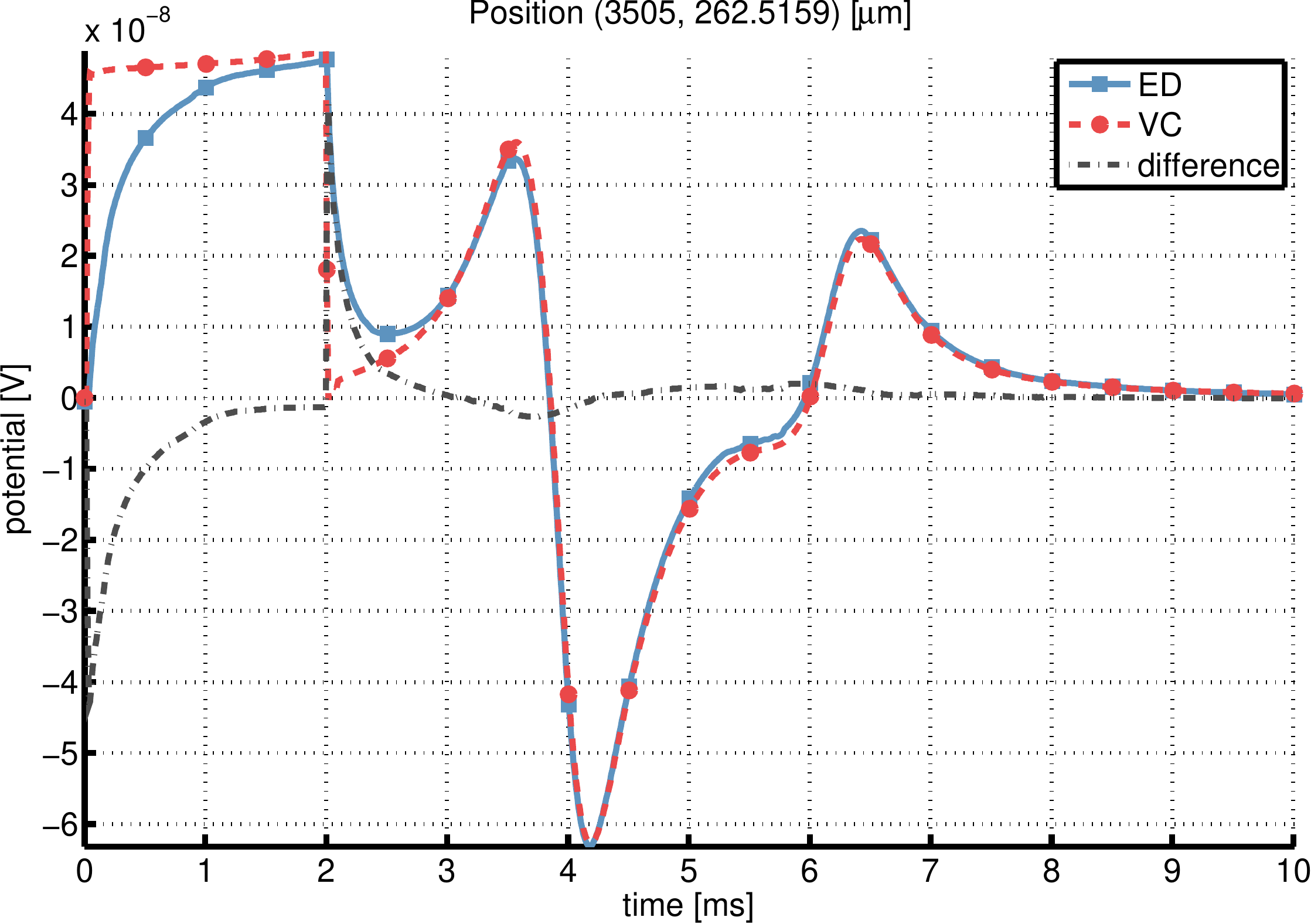}\label{fig:laplace_ed9}}%
\mycaption[Comparison of ED and VC results]{%
Both numerical simulations (\gls{ED}, \textit{solid lines}; \gls{VC}, \textit{dashed lines}) were generated on an identical computational grid and compared at the same positions as before in \cref{fig:lsa_ed}, showing essentially same differences (\textit{dash-dotted lines}).}%
\label{fig:laplace_ed}%
\end{figure}

After ascertaining the numerical solution is correct, it can be compared to the electrodiffusion results generated on the same computational grid, shown in \cref{fig:laplace_ed}.
Unsurprisingly, the plot shows the same situation as in \cref{fig:lsa_ed}.

\section{Summary}
The modeling approach of \cref{chap:model_general} was applied to an axon with homogeneous channel densities.
The resulting extracellular potentials can largely be attributed to the membrane currents following the \gls{HH} system, consistent with existing models based on volume conductor theory. 

However, close to the membrane, where the Debye layer was explicitly resolved, we found significant differences, as large concentration and potential gradients towards the membrane dominate the membrane currents.
In accordance with the Poisson-Boltzmann \cref{eq:pb}, the concentrations are screened exponentially, relaxing to their bulk values over the extent of the Debye layer.
The changes in concentration triggered by membrane fluxes here are small compared to the deviations of concentrations from their bulk values in the Debye layer, therefore this does not have a significant effect on the potential.
While this region with prominent characteristics is only about \SI{10}{\nano\metre} wide, its effect on the \gls{EAP} shape is clearly visible up to distances of at least \SI{5}{\micro\metre}, underlining the importance of this modeling approach.

We note that this is consistent with the \gls{PNP} system analysis in \cite{mori2006three}, where a diffusion layer with an extent on the order of $\sqrt{\dDebye}$ is found.
We attribute the deviations between \gls{LSA} and electrodiffusion model to the presence of such a layer with associated concentration redistributions and its effects on the potential.
On the contrary, the good agreement with the \gls{LSA} model in large parts of the domain can be regarded as a validation of our model.
The only free parameter in the \gls{LSA} model, the resistivity $\rho$, was fitted to yield the optimal agreement with our data, and turned out to match very well with reported values from the literature.

From the computational point of view, even this seemingly simple example of a homogeneous axon imposes a severe challenge on the numerical solution.
At the membrane, sharp discontinuities (in concentrations) as well as steep gradients (concentrations and potential) occur, which were accounted for by a sub-nanometer grid resolution in radial direction.
In longitudinal direction and off the membrane, large mesh sizes on the order of $10^2$--$10^3$ \si{\micro\metre} were used.
The usage of appropriate linear solvers was crucial to cope with the introduced grid anisotropy.
The multi-scale character of the system also shows in the generated solution.
For example, the amplitudes of the potential differ by several orders of magnitude between the intracellular values (about $\SI{100}{\milli\volt}$), the extracellular values close to the membrane (up to a few $\SI{100}{\micro\volt}$ in the Debye layer), and some microns away from the membrane (fractions of $\SI{1}{\micro\volt}$).

\setchapterpreamble[u]{%
\dictum[Ian Stewart]{If our brains were simple enough for us to understand them, we'd be so simple that we couldn't.}\bigskip}
\chapter{Model of a Myelinated Axon}\label{chap:myel}
To handle myelinated axons, some major changes are required in the model, causing further adjustments in the discretization and the numerical algorithms.
These changes will be presented in this chapter as well as their influence on the simulation results.

\section{Representation of Myelin in the Model}
The relevant changes to include a myelin sheath around the axon -- with respect to the general model from \cref{chap:model_general} -- concerns the membrane.
Following the model assumption of a cylindrical axon, the myelin wrapper can be seen as a hollow cylinder that encloses the axon, albeit with a larger radius than the axonal membrane.

In reality, myelin is not a homogeneous material, it rather consists of multiple repeating layers of lipid -- consisting of polar heads and hydrophobic tails -- and water (cf.~\cite[chapter 6.17]{hobbie2007intermediate}).
To simplify the myelin representation, the presence of these multiple layers with varying dielectric constants is neglected.
Instead, it is assumed to be a homogeneous material with an effective electric permittivity $\EpsMyelin$, which does not impose a problem if one is not interested in the fine-granular potential profile across the myelin sheath, but rather the effective membrane potential.

Estimated values for $\EpsMyelin$ are within the range of \numrange{3}{8.5} \cite{hobbie2007intermediate,northrop2000introduction}.
An increased myelin sheath thickness or tighter binding with the membrane might, however, result in a decreased value \cite{min2009interaction}.
Furthermore, the effective value should also respect the lower axonal membrane permittivity of $\EpsMembrane = 2$, which is located below the myelin sheath.
For the rest of this chapter, we made the more or less arbitrary choice of $\EpsMyelin = 6$, which can easily be changed in the application's config file.

\begin{figure}
\begin{center}%
\centering%
\includestandalone[width=\textwidth]{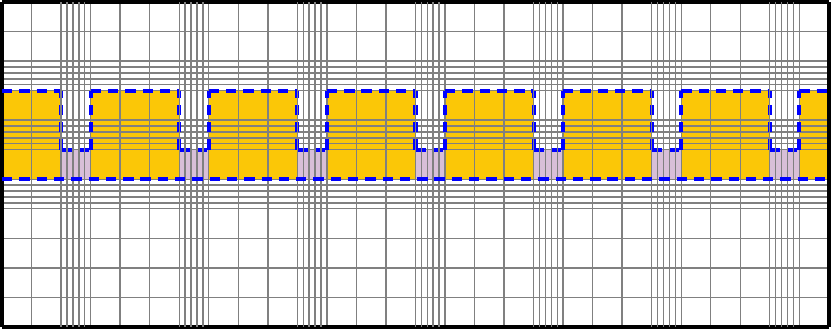}%
\end{center}%
\mycaption[Computational domain with explicit myelin representation]{Due to the varying membrane thicknesses, the resolution of the Debye layer would require an unnecessarily high number of unknowns when using a tensorial grid.}%
\label{fig:myelin_domain_explicit}
\end{figure}

The computational domain from \cref{sec:neuron_model.2d} needs to be adjusted to represent the myelin sheath as well, as shown in \cref{fig:myelin_domain_explicit}, where the myelin is included explicitly.
It can be seen that the dashed membrane interface is not a straight line anymore, it rather consists of multiple horizontal and vertical parts.
If the Debye layer was to be resolved for this membrane interface shape, as suggested in \cref{fig:myelin_domain_explicit}, the number of grid points would become very large, resulting in an intractable number of \glspl{DOF}.

Fortunately, a property of the potential can be utilized.
It can be observed that, in a first approximation, the potential shows a linear decay across the membrane.
This is due to the fact that the Debye length $\dDebye$ is much smaller than the membrane thickness $\dMemb$, meaning that charges on both sides of the membrane only have a limited effect on the electric field inside the membrane.
Arguing mathematically, the right-hand side of the Poisson \cref{eq:p} is close to zero, thus yielding a potential with zero curvature.

\begin{figure}
\begin{center}%
\centering%
\subfloat[][Equilibrium]{%
\includestandalone[width=0.46\textwidth]{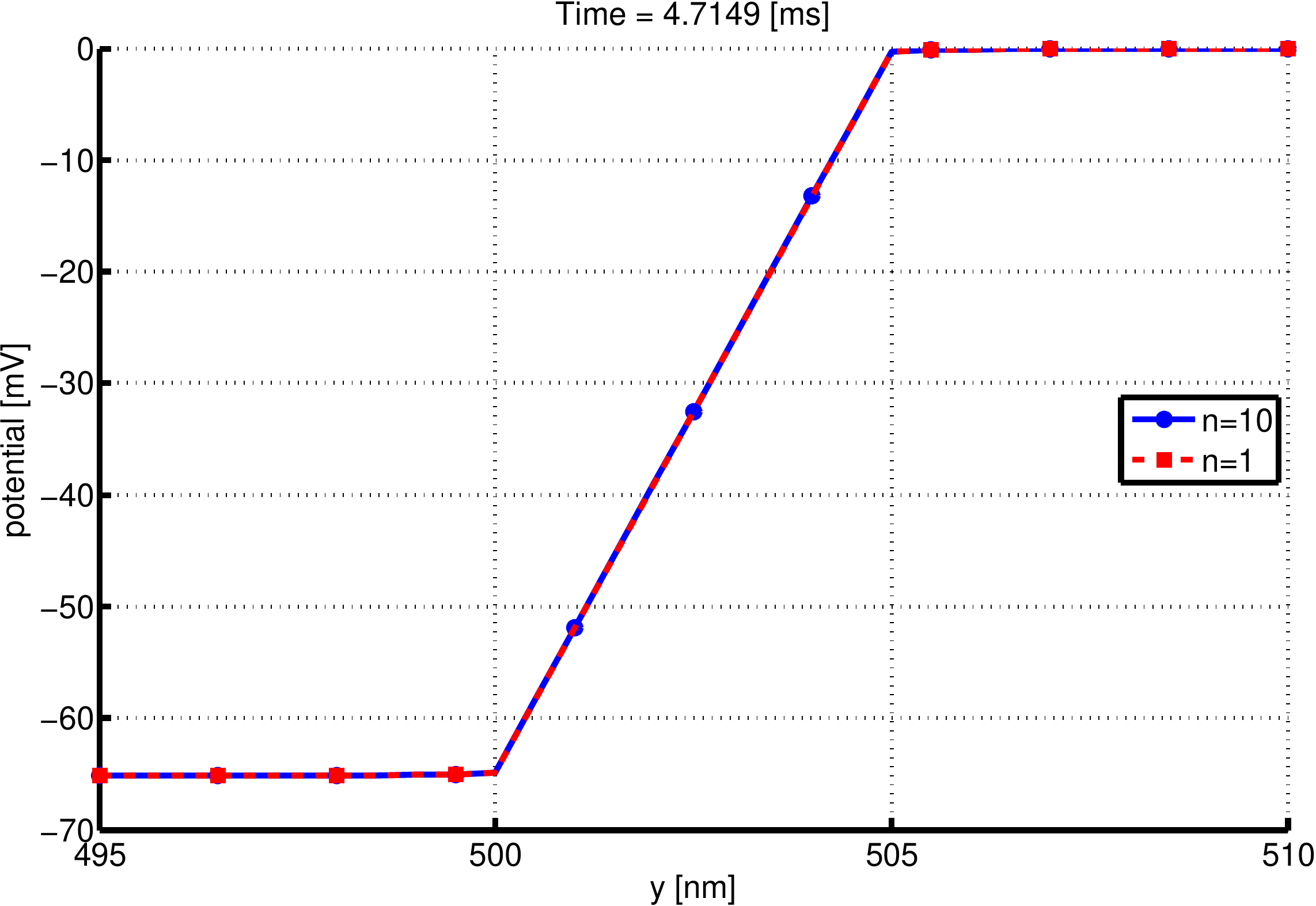}%
\label{fig:comparison_mme_n10_n1_i500}}%
\subfloat[][Action potential peak]{%
\includestandalone[width=0.46\textwidth]{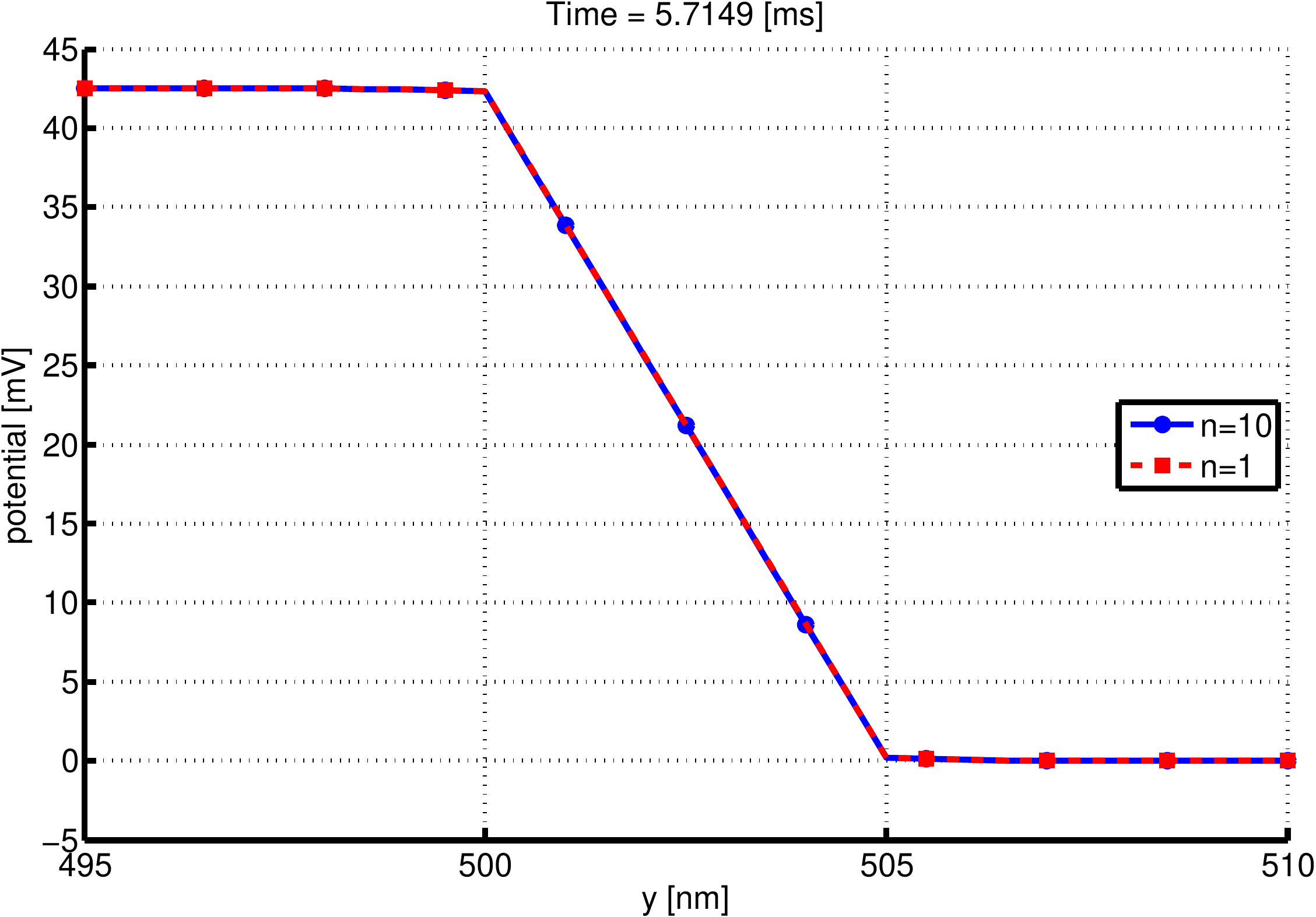}%
\label{fig:comparison_mme_n10_n1_i600}}%
\end{center}%
\mycaption[Comparison of potential profiles for different membrane resolutions]{ %
Potentials across the membrane at a fixed x-position of \SI{5}{\milli\metre} (center of the axon) at two exemplary time points: resting state
at $t=\SI{4.7}{\milli\second}$ (\emph{left}) and \gls{AP} peak at $t=\SI{5.7}{\milli\second}$ (\emph{right}). 
Solid lines show the profile for $n=10$ membrane elements 
and dashed lines for a single membrane element in $y$-direction. Differences are negligible at every point in time (not shown).}%
\label{fig:comparison_mme_n10_n1}%
\end{figure}

This was validated in simulations with a fine membrane resolution ($n=10$ membrane elements in $y$-direction) and compared to the case $n=1$, showing a negligible deviation in \cref{fig:comparison_mme_n10_n1}.
As described in \cref{sec:implementation.components}, this also enables blocking the system matrix, which is of great advantage for the performance of preconditioner and linear solver.

\begin{figure}
\begin{center}%
\centering%
\subfloat[][Thick membrane, high permittivity]{
\includestandalone[width=0.56\textwidth]{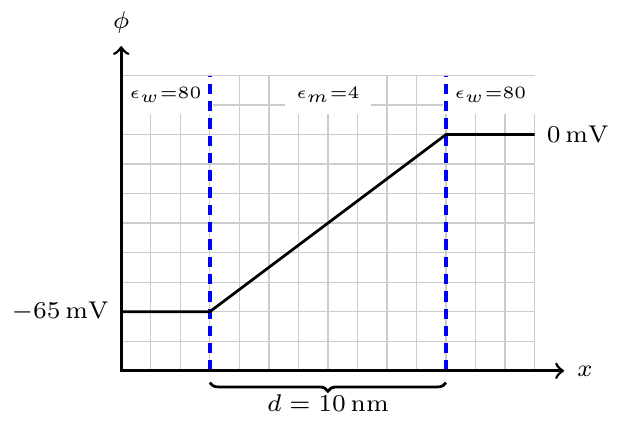}%
\label{fig:linear_potential_relationship_thick}}%
\subfloat[][Thin membrane, low permittivity]{
\includestandalone[width=0.44\textwidth]{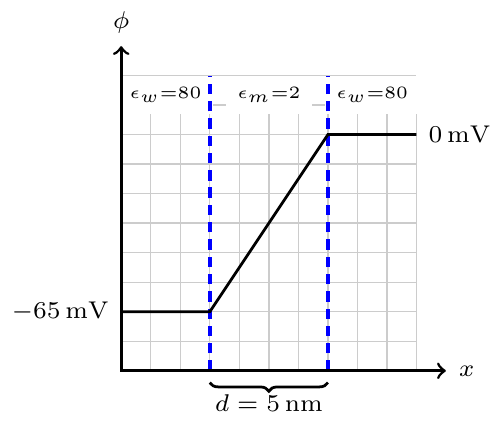}%
\label{fig:linear_potential_relationship_thin}}%
\end{center}%
\mycaption[Relation of linear potential decay and membrane thickness/permittivity]{Doubling membrane thickness and permittivity (\textit{left}) results in the same membrane potential difference when the potential is linear across the membrane.}%
\label{fig:linear_potential_relationship}
\end{figure}

\begin{figure}
\begin{center}%
\centering%
\includestandalone[width=\textwidth]{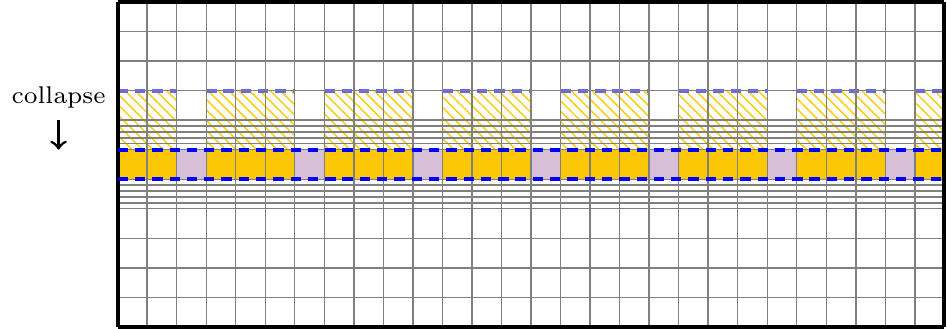}%
\end{center}%
\mycaption[Computational domain with collapsed myelin representation]{Myelination is represented implicitly, by collapsing the membrane to a homogeneous thickness and instead using effective permittivities that compensate for the thickness modification.}%
\label{fig:myelin_domain_collapsed}%
\end{figure}

When assuming a linear potential course over the membrane, a nice property can be utilized: a membrane with thickness $d_1$ and permittivity $\epsilon_1$ as in \cref{fig:linear_potential_relationship_thick} will have the same membrane potential decay as another membrane with thickness $d_2 = \frac{d_1}{2}$ and permittivity $\epsilon_2 = \frac{\epsilon_1}{2}$ in \cref{fig:linear_potential_relationship_thin}.
With this, the myelin parts of the membrane from \cref{fig:myelin_domain_explicit} can be ``collapsed'' onto the axonal membrane by changing the permittivity accordingly, see \cref{fig:myelin_domain_collapsed}.
This results conceptually in the same computational grid as for an unmyelinated axon, with only one finely resolved Debye layer on each side of the membrane. 

Following the previous explication, the collapsed myelin permittivity $\EpsMyelin'$ can be calculated as
\begin{align*}
 \EpsMyelin' = \EpsMyelin \frac{\dMemb}{\dMyelin} \ ,
\end{align*}
with the (nodal) membrane thickness $\dMemb$ as before and the myelin thickness $\dMyelin$, which was chosen to be \SI{500}{\nano\metre} in the following.
With this, we arrive at $\EpsMyelin' = 0.06$.

\section{Simulation Setup}
The changes in the axonal geometry described above require a number of changes in the numerical implementation, which will be presented in the following.

\subsection{Modified Grid Generation}
The grid spacing in $x$-direction can not be equidistant as before, since the different material properties of myelin and nodes of Ranvier have to be accurately represented without using a prohibitively small grid size.
A related problem arises through the jumping material coefficients, necessitating the introduction of fine transition intervals between myelin and nodes of Ranvier.
Inside the membrane groups -- where the permittivity is constant -- a coarser, equidistant mesh size may be used.
Analogous to the $y$-direction, abrupt changes in mesh width are avoided. 

This results in a mesh generation procedure as follows: the transition intervals between myelin and node of Ranvier -- as well as the node of Ranvier itself -- are finely resolved with a minimum mesh width of $\hxMin$.
This value should be a fraction of the node width $\lNode = \SI{1}{\micro\metre}$, here we used $\hxMin = \SI{100}{\nano\metre}$.
Outside the transition interval, the grid spacing is smoothly increased following a geometric series up to a maximum equidistant grid size of $\hxMax = \SI{10}{\micro\metre}$ for the myelin parts.
This procedure is mirrored towards the next node of Ranvier and repeated for each node until the end of the $x$-interval.
The length of an internode segment $\lMyelin = \SI{999}{\micro\metre}$ was chosen such that the starting coordinates of two successive nodes of Ranvier are always \SI{1}{\milli\metre} apart.

For an axon with 10 nodes of Ranvier and above parameters $\lNode$, $\hxMin$ and $\hxMax$, this will increase the overall number of grid points by a factor of 7-8 in comparison to the unmyelinated case.
This number is acceptable, especially when running the simulation in parallel, where even a much finer grid is still tractable.

In the following, we will consider a reference setup with a myelinated axon of length \SI{48}{\milli\metre} consisting of 48 nodes of Ranvier.
The number 48 derives from the fact that the simulation was run on a machine with 48 cores, allowing an easy partitioning that puts one node of Ranvier on each processor, see below.
All relevant parameters are listed in \cref{tab:sim_params_myel}.\newpage

\begin{longtabu} to \textwidth {X S s p{0.45\textwidth}}
\caption[Simulation parameters for the myelinated axon model]{\textbf{Simulation parameters for the myelinated axon model}.}\label{tab:sim_params_myel}\\
\toprule
  Parameter & {Value} & {Unit} & Description \\
\midrule
\endfirsthead
\toprule
  Parameter & {Value} & {Unit} & Description \\
\midrule
\multicolumn{4}{c}{\sffamily PHYSICS}\\
\midrule
\endhead
  \multicolumn{4}{c}{\sffamily GRID}\\
\midrule
  $\xMax$ & 48 & \si{\milli\metre} & Domain size ($x$-direction) \\
  $\yMax$ & 10 & \si{\milli\metre} & Domain size ($y$-direction) \\
  $\yMemb$ & 500 & \si{\nano\metre} & Radius of the axon \\
  $\dMemb$ & 5 & \si{\nano\metre} & Membrane thickness \\
  $\dMyelin$ & 500 & \si{\nano\metre} & (Implicit) myelin thickness \\
  $\lNode$ & 1 & \si{\micro\metre} & Width of one node of Ranvier \\
  $\lMyelin$ & 999 & \si{\micro\metre} & Width of myelin internode \\
  $\hxMin$ & 100 & \si{\nano\metre} & Minimum mesh size in $x$-direction (node of Ranvier)\\
  $\hxMax$ & 10 & \si{\micro\metre} & Maximum mesh size in $x$-direction (myelin)\\
  $\hyMin$ & 0.5 & \si{\nano\metre} & Minimum mesh size in $y$-direction (Debye layer)\\
  $\hyMax$ & 100 & \si{\micro\metre} & Maximum mesh size in $y$-direction\\
  \#\glspl{DOF} & 5250448 & 1 & Total number of unknowns \\
\midrule
  \multicolumn{4}{c}{\sffamily PHYSICS}\\
\midrule
  $\epsMemb$ & 2 & 1 & Membrane permittivity (node of Ranvier)\\
  $\EpsMyelin'$ & 0.06 & 1 & Collapsed membrane permittivity (myelin)\\
  $\epsElec$ & 80 & 1 & Electrolyte permittivity \\[0.5em]
  $\naCytosol$ & 12 & \si{\milli\molar} & Intracellular $\naIon$ bulk concentration \\
  $\kCytosol$ & 125 & \si{\milli\molar} & Intracellular $\kIon$ bulk concentration \\
  $\clCytosol$ & 137 & \si{\milli\molar} & Intracellular $\clIon$ bulk concentration \\
  $\naExtra$ & 100 & \si{\milli\molar} & Extracellular $\naIon$ bulk concentration \\
  $\kExtra$ & 4 & \si{\milli\molar} & Extracellular $\kIon$ bulk concentration \\
  $\clExtra$ & 104 & \si{\milli\molar} & Extracellular $\clIon$ bulk concentration \\
  $\DNa$ & 1.33e-9 & \si{\square\metre\per\second} & $\naIon$ diffusivity \\
  $\DK$ & 1.96e-9 & \si{\square\metre\per\second} & $\kIon$ diffusivity \\
  $\DCl$ & 2.03e-9 & \si{\square\metre\per\second} & $\clIon$ diffusivity \\
  $\gNav$ & 1200 & \si{\milli\siemens\per\centi\metre\squared} & Conductance of the voltage-gated Na channel \\
  $\gKv$ & 360 & \si{\milli\siemens\per\centi\metre\squared} & Conductance of the voltage-gated K channel \\
  $\gL$ & 0.5 & \si{\milli\siemens\per\centi\metre\squared} & Total leak conductance\\
\midrule
 \multicolumn{4}{c}{\sffamily NUMERICS}\\
\midrule
 \lstinline!reduction! & 5e-6 & & Newton reduction \\
 \lstinline!absLimit! & 1e2 & & Newton absolute limit \\
 $\tEnd$ & 20 & \si{\milli\second} & Simulated time \\
 $\dtMin$ & 0.05 & \si{\micro\second} & Minimum time step \\
 $\dtMax$ & 10 & \si{\micro\second} & Maximum time step \\
 $\dtMaxAP$ & 10 & \si{\micro\second} & Maximum time step during \gls{AP} \\
 $\itMin$ & 10 && Newton iteration threshold for time step increase \\
 $\itMax$ & 30 && Newton iteration threshold for time step decrease \\
\bottomrule
\end{longtabu}

\subsection{Permittivity Smoothing}\label{sec:myel.permittivity_smoothing}
\begin{figure}
\centering%
\subfloat[][Extracellular membrane permittivity profile]{
\includegraphics[width=0.5\textwidth]{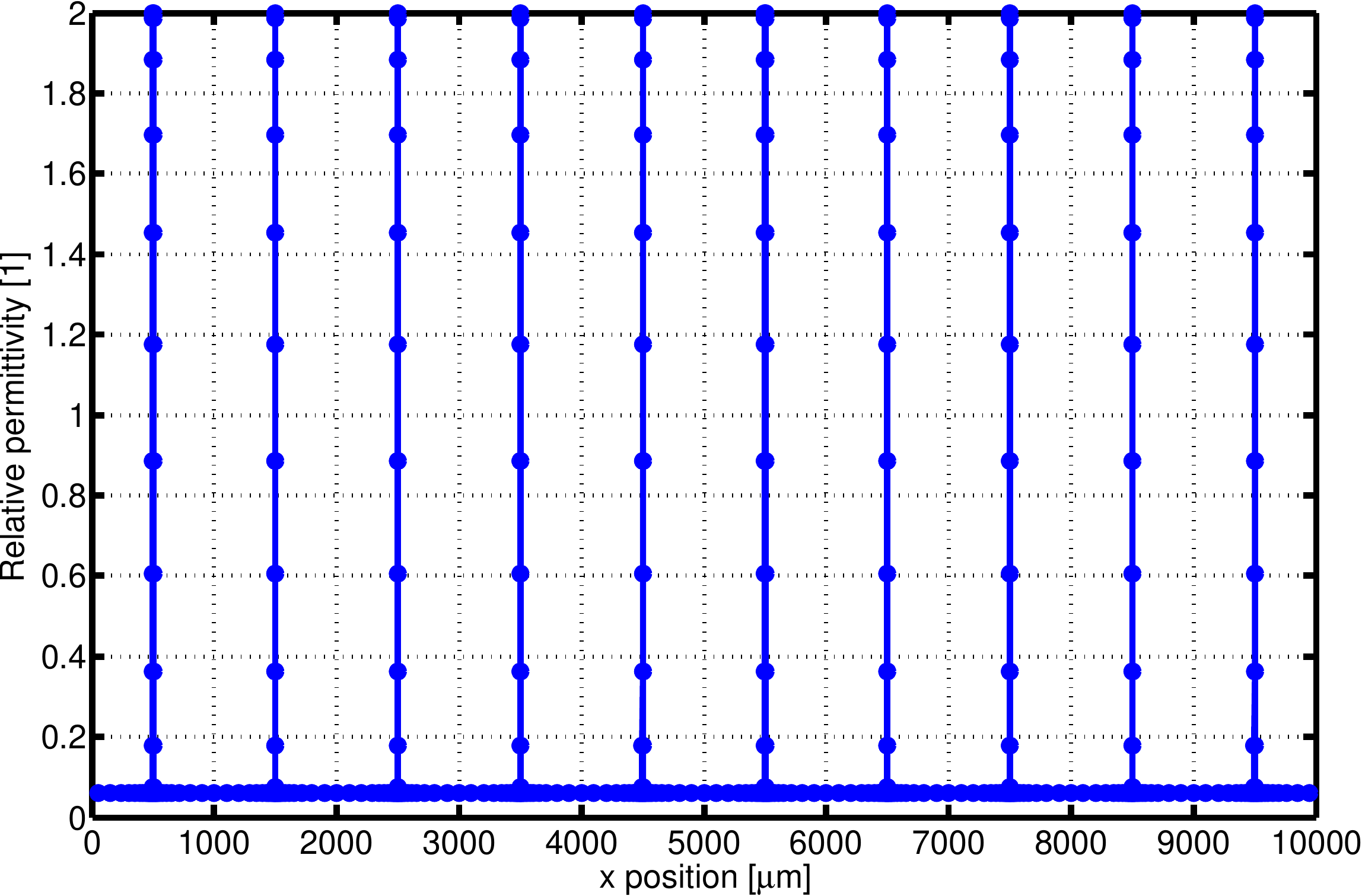}\label{fig:myelin_memb_permittivity_large}}%
\subfloat[][Zoom into one node of Ranvier]{
\includegraphics[width=0.5\textwidth]{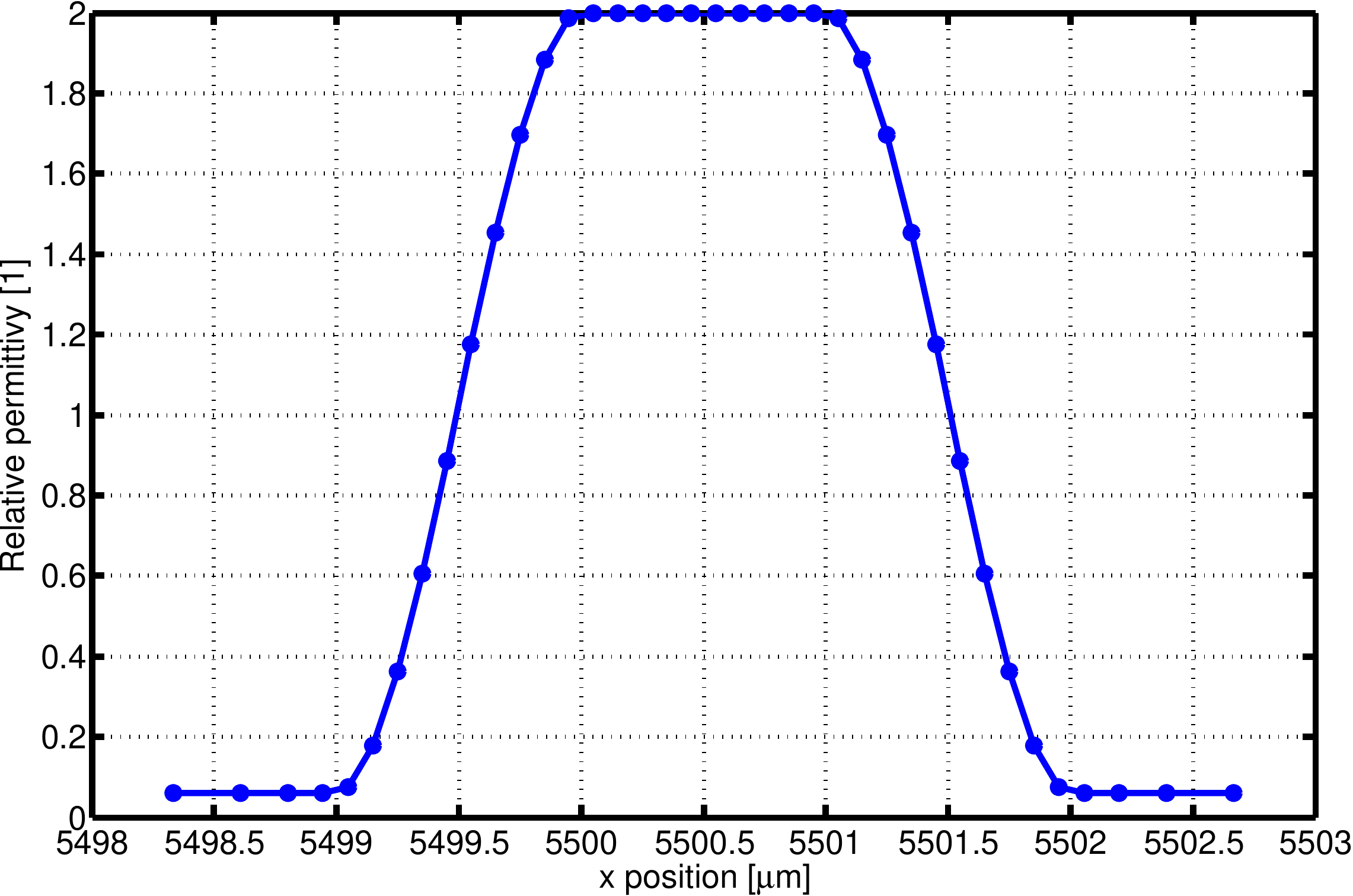}\label{fig:myelin_memb_permittivity_zoom}}%
\mycaption[Equilibrium permittivity profile]{ %
The relative permittivities exhibit jumps at membrane group boundaries, 
resulting in a comb-shaped profile (\textit{left}). 
The discontinuities were smoothed to prevent numerical oscillations, as can be seen in the zoomed-in transition region between myelin and node of Ranvier (\textit{right}).}%
\label{fig:myelin_memb_permittivity}
\end{figure}

Making sure the grid resolves the membrane group transitions is crucial, but the membrane permittivities will still exhibit a sharp discontinuity at these transitions.
Therefore, membrane permittivities are smoothed at the start of a simulation by applying the \lstinline!smoothstep! function 
\begin{align}
\operatorname{smoothstep}(x) = 3x^2 - 2x^3
\end{align}
to transitions intervals, effectively applying cubic Hermite interpolation.
The extent $d_T$ of a transition interval depends on the width of the nodes of Ranvier.
Here, it was chosen to be \SI{100}{\nano\metre}, i.e.~$\frac{1}{10}$ of the node width.
This results in the membrane permittivity profile in \cref{fig:myelin_memb_permittivity}; the effects of the smoothing operator can clearly be seen when zooming into a node transition interval in \cref{fig:myelin_memb_permittivity_zoom}.
This prevents numerical oscillations at sharp coefficient discontinuities, to which the standard finite element method is very susceptible.

\subsection{Parallelization and Adjustments of Numerical Methods}
\begin{figure}
\begin{center}%
\centering%
\includestandalone[width=\textwidth]{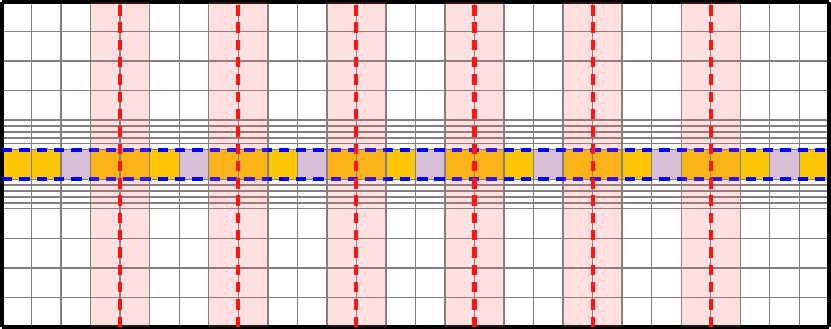}%
\end{center}%
\mycaption[Partition of the computational domain for the parallel case]{As in the unmyelinated case, the domain is divided by cuts in vertical direction. In the myelinated case, additional care has to be taken in order to not place the highly dynamic nodes of Ranvier close to a processor boundary, as suggested in the sketch.}%
\label{fig:myelin_domain_partitioning}
\end{figure}

Due to the increasing computational demands by the inclusion of myelin into the model, the numerical solution was only run in parallel.
As before, the grid was partitioned only in $x$-direction.
This not only ensures a joint treatment of membrane and adjacent electrolyte on one processor, it additionally avoids placing the highly dynamic nodes of Ranvier close to a processor boundary, which imposes a serious difficulty for the parallel solver.
We ensured that nodes of Ranvier were always close to the middle of each processor subdomain, as illustrated in \cref{fig:myelin_domain_partitioning}.

The fact that the computational grid now has strongly varying mesh sizes in both $x$- and $y$-direction introduces severe problems for the linear solver.
The \gls{BiCGStab} iterative solver that was used for the unmyelinated case exhibited \emph{breakdowns} when solving the problem in parallel, preferentially when using a higher number of processors.
This is a known problem of the default \gls{BiCGStab} algorithm \cite{simoncini2007recent}, which can apparently be circumvented by so-called \emph{look-ahead techniques} (see, e.g., \cite{brezinski1995look}).
These techniques have not yet been implemented in the \lstinline!dune-istl! module. 

But fortunately, there are a number of alternative linear solvers to choose from.
For the problem at hand, we used a \gls{GMRes} iterative solver, preconditioned by an \gls{AMG} with an \gls{ILU}0 smoother and using SuperLU on the coarsest grid. 
The \gls{GMRes} turned out to be very robust, albeit slightly less performant in comparison to \gls{BiCGStab}. Using the \gls{AMG} preconditioner in conjunction with an \gls{ILU} smoother takes care of the strong grid anisotropy.

\subsection{Solver Scaling Tests}
We again assessed the parallel performance of the numerical solution using the new linear solver backend. 
This time, however, we avoided strong scaling measurements, since this would inevitably place the processor boundaries close to nodes of Ranvier when choosing the number of nodes smaller than $p$.
On the other hand, taking a setup with the number of nodes of Ranvier equal to the highest processor count $p=48$ and using successively less processors to solve it was not an option either, since the sequential solution of such a large problem would have taken months.

Instead, a weak scaling test variant was used.
The domain size was increased together with the number of processors, and the scalability $S$ was calculated as
 \begin{align}
   S(p_1,p_2) = \frac{t_1}{t_2} \ ,
 \end{align}
where we implicitly assume that the number of \glspl{DOF} grows with a factor of $\frac{p_2}{p_1}$ and the number of unknowns is the same on all processors, although this is not exactly the case. 

Results obtained in this manner for a fixed simulation time length $\tEnd$ are, however, not comparable.
Because the \gls{AP} velocity is significantly higher compared to the unmyelinated case, the action potential wave has already left the computational domain on the right boundary before it has exhibited a full depolarization--repolarization--hyperpolarization cycle at the left boundary, for small domain sizes.
Only the largest simulation provides a sufficiently long axon of \SI{48}{\milli\metre} to represent one full spatial wave length of the \gls{AP}.

For smaller domain sizes, the \gls{AP} rushes through the small extent of the axon in a fraction of the total simulated time; the longer the axon gets, the more time is spent by the traveling action potential within the computational domain. 
Consequently, the fraction of the simulation time that the axonal membrane is at rest is getting smaller, and the membrane is active for a larger fraction of time.

\begin{figure}
\centering%
\includegraphics[width=0.6\textwidth]{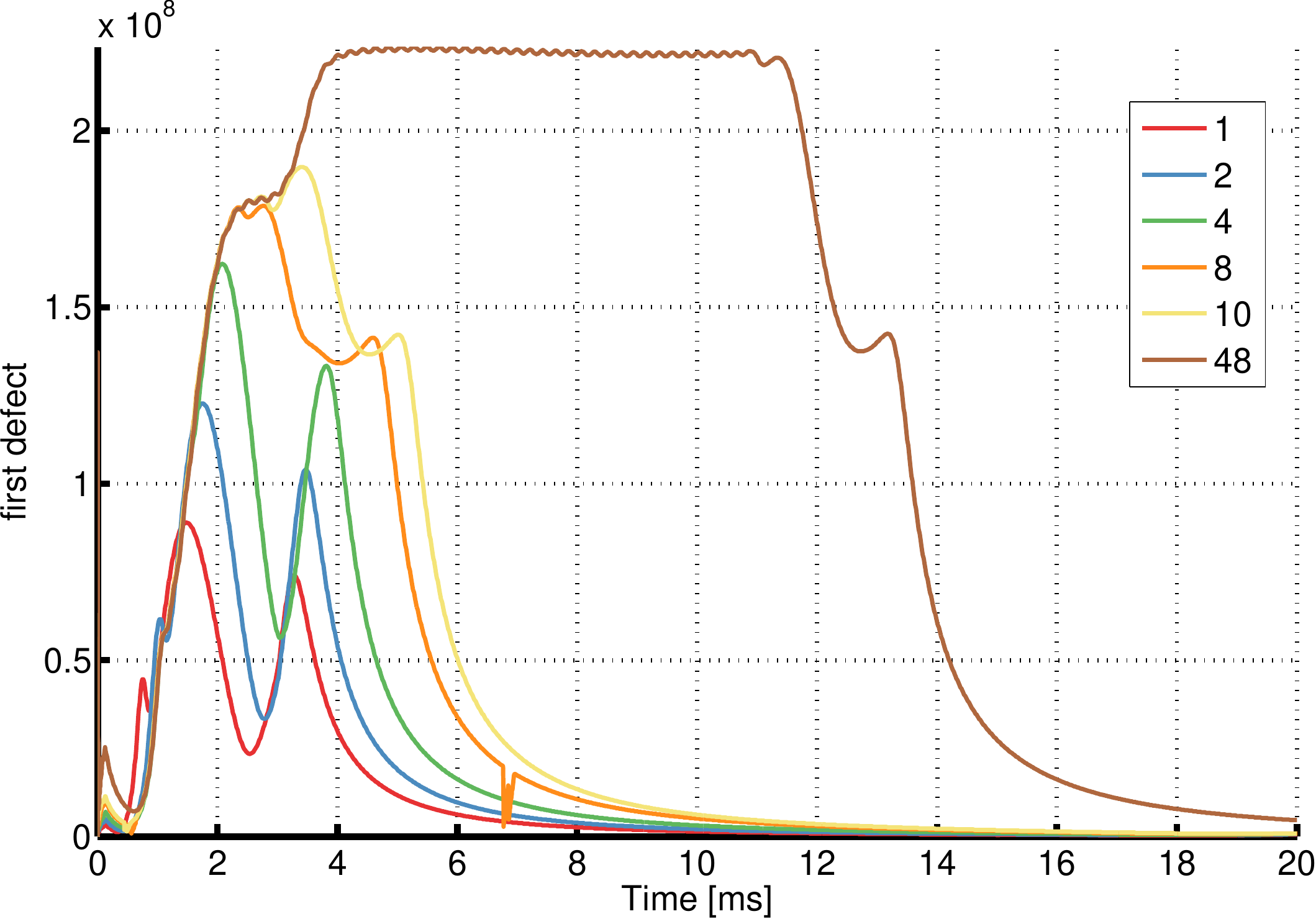}\label{fig:myelin_scaling_first_defect}%
\mycaption[Initial defect for different domain sizes]{The initial defect at the beginning of each time step is growing with processor count (and thus with domain size), since the \gls{AP} wave spends a larger fraction of time within a longer axon.}%
\label{fig:myelin_scaling_detailed}
\end{figure}

Of course, the active parts provide a much more difficult task for the numerical solution, as the system is more dynamic due to the large membrane currents, such that the solver needs more iterations and the total computation time increases.
This is reflected in the initial defect (i.e., the norm of the residual $\| r^0 \|$ at the beginning of each time step) in \cref{fig:myelin_scaling_detailed}, which shows an increase with the number of processors.

Instead we consider, for each different $p$, only the time interval where the rightmost membrane element has a membrane potential of at least \SI{60}{\milli\volt}.
With this, we can calculate the average solver statistics only for the \gls{AP} depolarization phase, shown in \cref{tab:myel.solver_statistics_part}.
This evaluation presents a normalization excluding the effects of different proportions of high or low system dynamics, as it only considers the time intervals of highest activity.

\begin{table}
\mycaption[Solver statistics per time step for the parallel solution using different processor counts $p$]{All values are averages over all time steps. The total time includes matrix and residual assembly as well as the actual solution time by Newton's method. The solver time includes both preconditioner setup and \gls{LS} time, the \gls{LS} time only the actual \gls{GMRes} solve. The number of Newton iterations as well as the number of linear solver iterations per Newton iteration increase with problem size.}\label{tab:myel.solver_statistics_part}
\centering
\begin{tabu} to \textwidth {@{} p{0.3cm} X[0.8,l] X[0.4,l] *5{X[1,l]} X[0.8,l] @{}}
\toprule
$p$ & \centering total time & \centering $S$ & \centering assembler time [s] & solver time [s] & \centering \gls{LS} time [s] & \centering \gls{LS} it. & \centering \gls{LS} time / it. [s] & \centering Newton it.\\
\midrule
1 & 139.1566 && 119.2886 & 3.6972 & 1.134798 & 6.467105 & 0.186419 & 2.6746\\
2 & 147.3648 & 0.94 & 115.6824 & 15.4106 & 4.185112 & 21.094705 & 0.206684 & 2.7416\\
4 & 160.3079 & 0.87 & 119.7298 & 23.6005 & 5.913261 & 23.816343 & 0.252483 & 3.0371\\
8 & 265.5756 & 0.52 & 132.6605 & 113.0225 & 28.407855 & 64.485612 & 0.412704 & 3.502\\
10 & 246.884 & 0.56 & 128.8954 & 99.4551 & 27.090431 & 69.478478 & 0.368952 & 3.4438\\
48 & 199.0493 & 0.70 & 85.2555 & 99.7457 & 31.620872 & 90.697419 & 0.332404 & 3.111\\
\bottomrule
\end{tabu}
\end{table}

Notable anomalies are the cases $p=8$ and $p=10$, which show inferior performance to the case $p=48$.
An inspection of the diagnostic output reveals that the time step size oscillates for these cases during the initiation of the \gls{AP}, caused by a series of non-converging Newton iterations.

The overall solver timings normalized to the depolarization time interval show acceptable scaling properties, particularly for the case $p=48$, which will be used to generate the simulation results considered in the following section.

\section{Simulation Results}
As before, equilibrium states and transient dynamics are considered separately.
Most of the following results were obtained from the largest simulation with parameters according to \cref{tab:sim_params_myel}, but some plots were also obtained from a smaller, coarser setup with $\xMax = \SI{10}{\milli\metre}$ and $\hxMax = \SI{100}{\micro\metre}$ when looking at point-data over time.
In these cases, the domain extent did not make any difference, and deviations due to the $x$-resolution were negligible\footnote{The reason for using the smaller dataset when possible is a purely technical one: the size of the large dataset is about \SI{419}{\giga\byte} on disk.
In order to visualize this data, each variable has to be loaded separately for a domain subset and time subinterval of interest, as otherwise the data would not fit into the \SI{8}{\giga\byte} of RAM on the author's workstation.
This results in tedious cycles of loading and clearing data for generating one single graph, which was much more comfortable with the smaller dataset of only \SI{44}{\giga\byte}.}.

\subsection{Some Notes on the Separate Visualization of Myelin and Node Values}
Because of the comb-shaped permittivity profiles -- as shown in \cref{fig:myelin_memb_permittivity} -- it is difficult to visualize the extracellular concentration and potential dynamics close to the membrane, as -- to stay in the picture of a comb -- the variation of these variables within ``shaft'' or ``tooth'' segments is much smaller than the difference of values across those two groups. 
This makes it impossible to see the variations within each group when plotting the complete domain.

Therefore, the vector of $x$-coordinates was partitioned into three disjoint sets: the set $\Xnodes$ of all node of Ranvier coordinates, the set $\Xtransition$ containing the coordinates at transition intervals between myelin and nodes (specified by the maximum distance of $d_T$ towards the nearest node coordinate), and the set $\Xmyelin$ containing the remaining $x$-coordinates of myelinated membrane cells, such that $x = \Xmyelin \cup \Xnodes \cup \Xtransition$ holds.
Note that this partition is done regardless of the $y$-coordinate, so the domain is cut into stripes parallel to the $y$-axis by this process.

With this partition of the $x$-axis, it is possible to visualize the values close to myelin and node of Ranvier parts separately by only using the data at $x$-coordinates given by $\Xmyelin$ or $\Xnodes$, respectively.
We will use this later on to compare our results with the \gls{LSA} model, for myelin and node stripes separately, and we will also take a close look at the highly dynamic transition interval between any of those two regions.
The partition into three disjoint subsets is also important for the calculation of the equilibrium state from two different, small equilibrium simulations, as will be explained in the following.

\subsection{Equilibrium States}
As in the unmyelinated case, the equilibration simulations are performed on a grid with only one element in $x$-direction to speed up the procedure. 
Since we now have different membrane permittivities, the equilibrium concentration and potential profiles are quite different between myelin and nodes of Ranvier.
Therefore, we need to carry out two different equilibration simulations, one for the myelin parts and one for nodes of Ranvier.
At the beginning of the actual \gls{AP} simulations, the values of both simulations are transferred to the fine grid, to the respective $x$-intervals of the domain associated with either myelin ($\Xmyelin$) or a node of Ranvier ($\Xnodes$).
At transition intervals $x \in \Xtransition$, the values of both equilibrium states are linearly interpolated with respect to the (smoothed) membrane permittivity at the current $x$-coordinate, calculated as described in \cref{sec:myel.permittivity_smoothing}. 
This yields a smooth equilibrium state that is valid with respect to the membrane permittivity distribution on the fine grid.
The system is then allowed to settle for a small number ($<20$) of time steps before the stimulation electrode is switched on.

\begin{figure}
\centering%
\subfloat[][Extracellular membrane charge density profile]{
\includegraphics[width=0.5\textwidth]{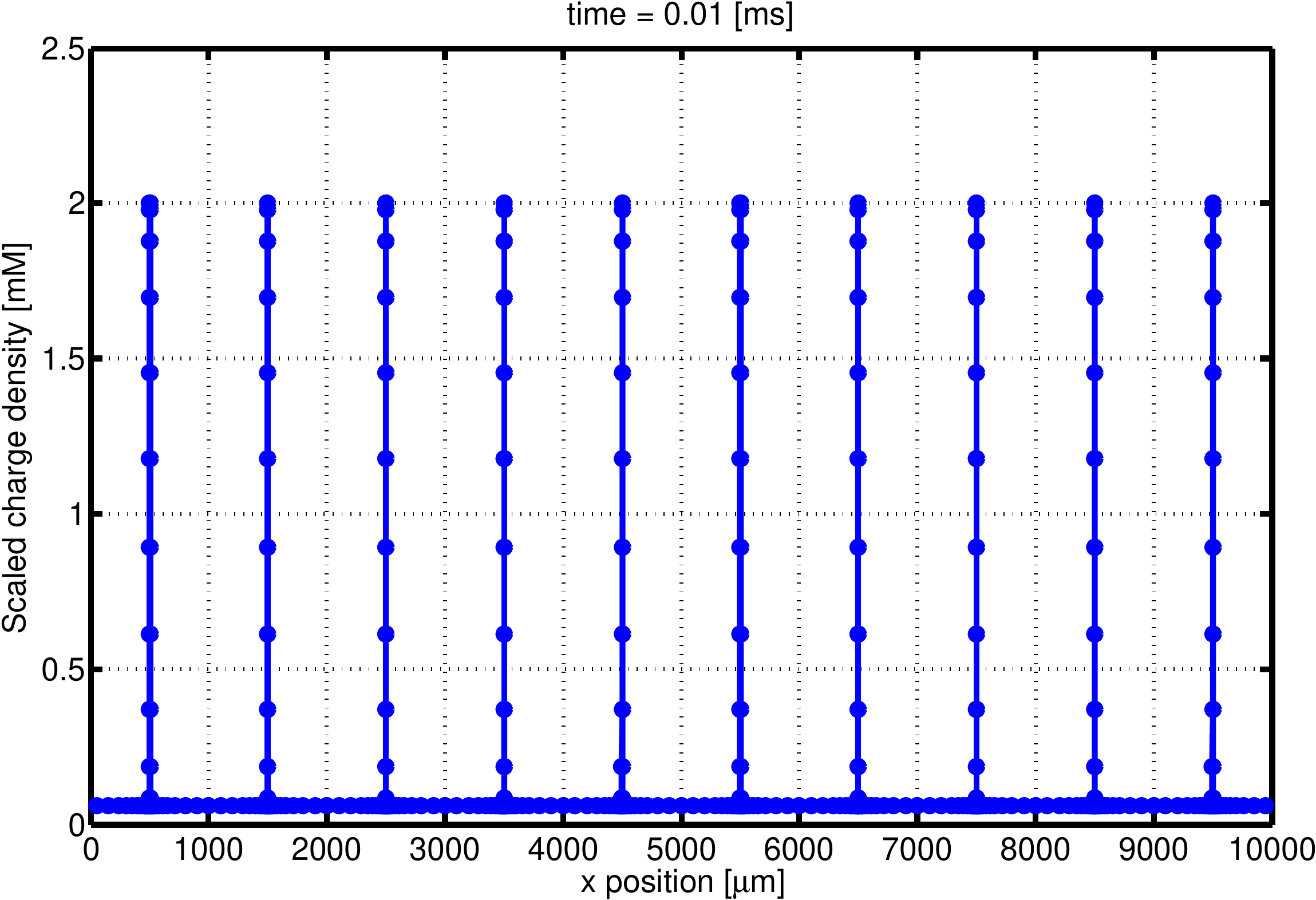}\label{fig:myelin_equi_conc_large}}%
\subfloat[][Zoom into one node of Ranvier]{
\includegraphics[width=0.5\textwidth]{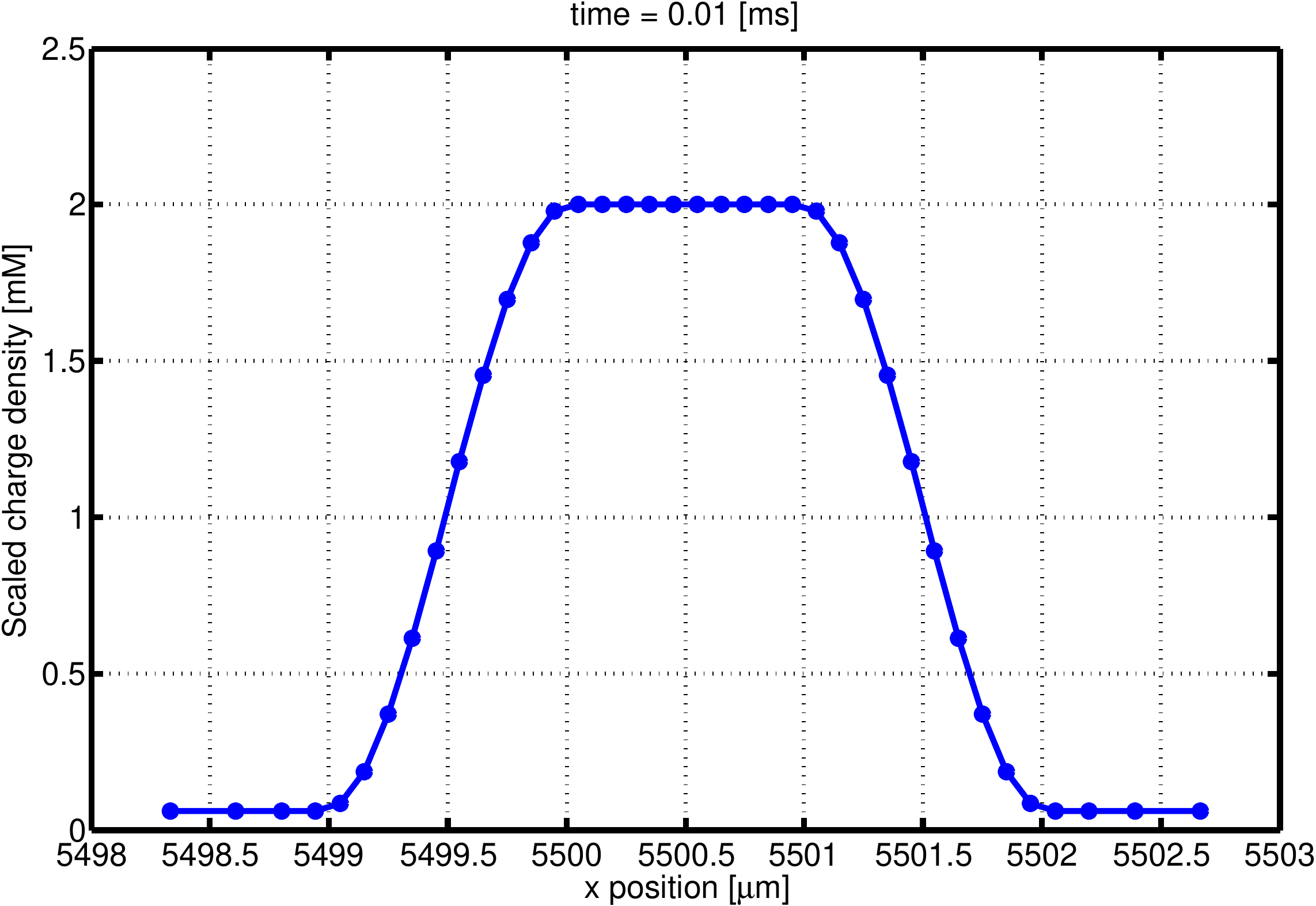}\label{fig:myelin_equi_conc_zoom}}%
\mycaption[Equilibrium charge density profile]{ %
The scaled charge densities $\sum_i z_i n_i$ at equilibrium follow a distinct, comb-shaped distribution along the membrane (\textit{left}).
On the right, a zoom into one node
of Ranvier reveals the jump in equilibrium concentrations due to different membrane permittivities at
myelin and nodes of Ranvier.}%
\label{fig:myelin_equi_conc}
\end{figure}

The equilibrium concentrations obtained hereby follow a distinct, comb-shaped distribution due to the variations of electrical permittivities along the axonal membrane, as depicted in \cref{fig:myelin_equi_conc}.
The charge density largely follows the permittivity profile from \cref{fig:myelin_memb_permittivity}.
The application of the smoothing operator to the permittivities turns out to be successful, as concentrations exhibit no oscillations, in contrast to the non-smoothed case (not shown).

\subsection{Action Potential}
As in the previous chapter, the stimulation electrode was mimicked by injecting a sodium rectangle pulse into the cell.
The stimulation site was placed at the starting coordinate of the first node of Ranvier at $\textbf{x}_\text{stim} = (\SI{500}{\micro\metre},\SI{0}{\micro\metre})$.
This time, a drastically reduced stimulation current of about \SI{10}{\pico\ampere} over a duration of \SI{3}{\milli\second} was enough to elicit an action potential, as the isolation by myelin reduced the current leak over the membrane.

\subsubsection{Intracellular Potential}
\begin{figure}
\centering%
\subfloat[][Unmyelinated axon]{
\includegraphics[width=0.5\textwidth]{img/matlab_temp/ap-crop}\label{fig:myelin_ap_unmyel}}%
\subfloat[][Myelinated axon]{
\includegraphics[width=0.5\textwidth]{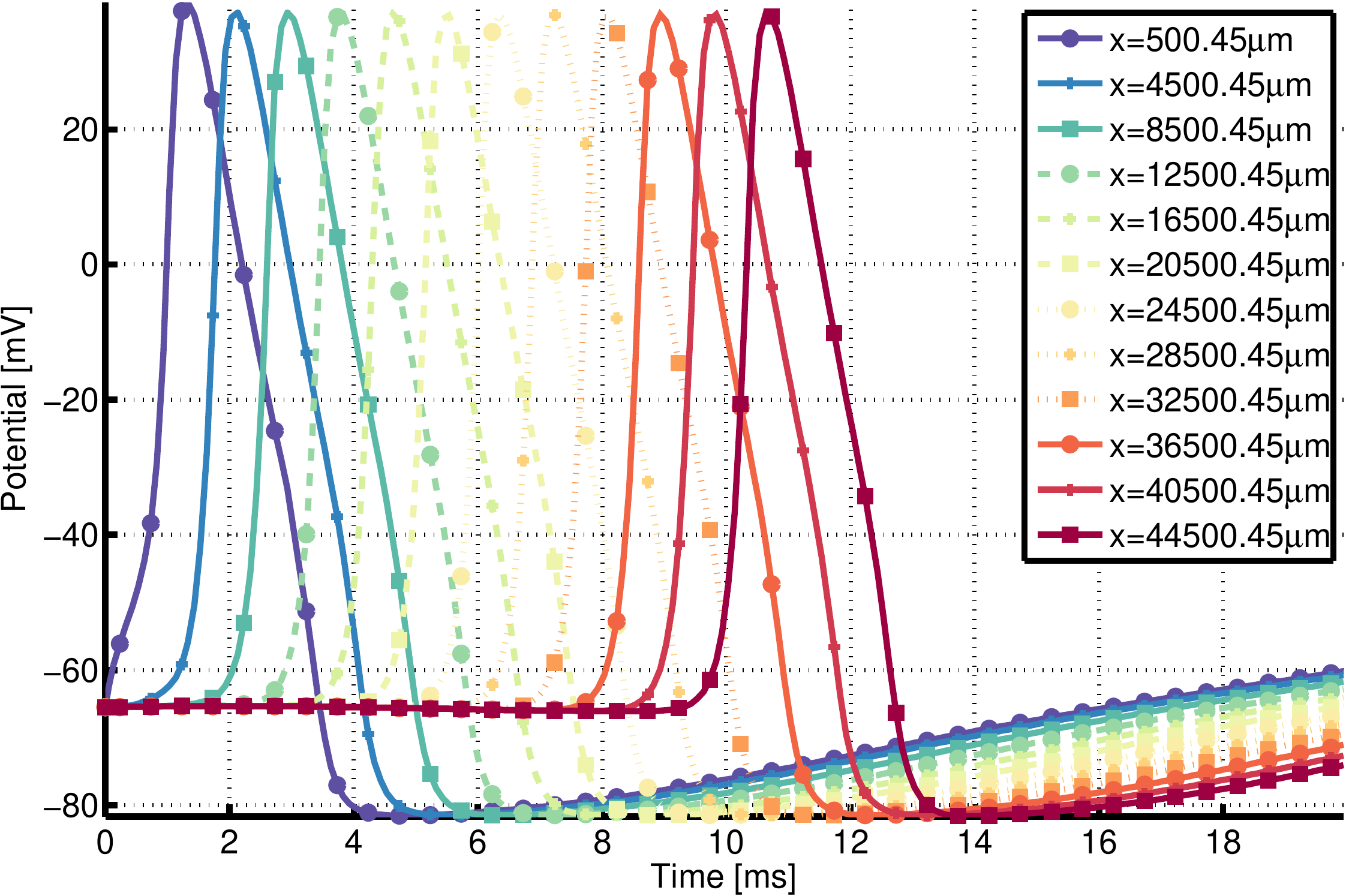}\label{fig:myelin_ap_myel}}%
\mycaption[Intracellular action potential propagation]{ %
The action potential of the current setup (\textit{right}, axon length \SI{48}{\milli\metre}) is
compared to the \gls{AP} of unmyelinated axon from the previous chapter (\textit{left}, length \SI{10}{\milli\metre}). Please note the difference in spacings 
between measurement points. The increase in propagation speed is especially obvious, which is about a factor of 5 here.}%
\label{fig:myelin_ap}
\end{figure}

In \cref{fig:myelin_ap}, the generated action potential at different positions along the axon is shown on the right, yielding an impressive speedup of about a factor of 5. 
Another subtle difference is that the last curve shows a slightly higher amplitude than the other \glspl{AP} because the potential can not freely exit at the rear end of the axon (Neumann boundary for concentrations). 
Interestingly, this does not have an effect in the unmyelinated example, probably because the \gls{AP} moves much slower, such that the excess in concentrations can drain off through membrane channels. 
In the myelinated example, the propagation is so fast that a swell of ions forms, showing a significant effect in the intracellular potential close to the right cell boundary. 
While this phenomenon clearly has to be regarded as a boundary artifact in this case, an accumulation of charge can be expected at axonal ``dead ends'' like synaptic boutons following an arriving action potential wave.

\subsubsection{Membrane Flux}
\begin{figure}
\centering%
\subfloat[][Node of Ranvier]{
\includegraphics[width=0.5\textwidth]{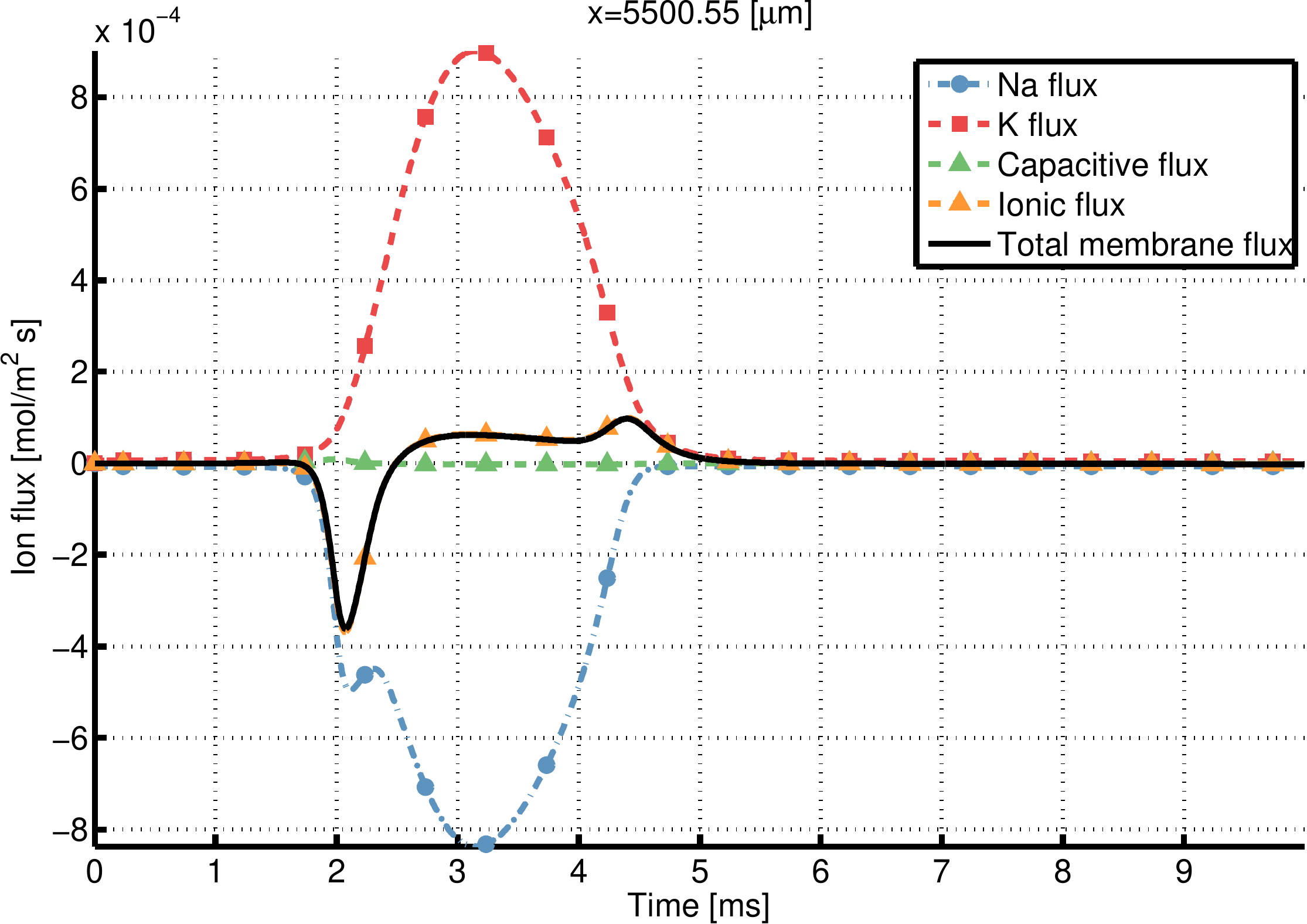}\label{fig:myelin_ap_memb_flux_node}}%
\subfloat[][Myelin]{
\includegraphics[width=0.5\textwidth]{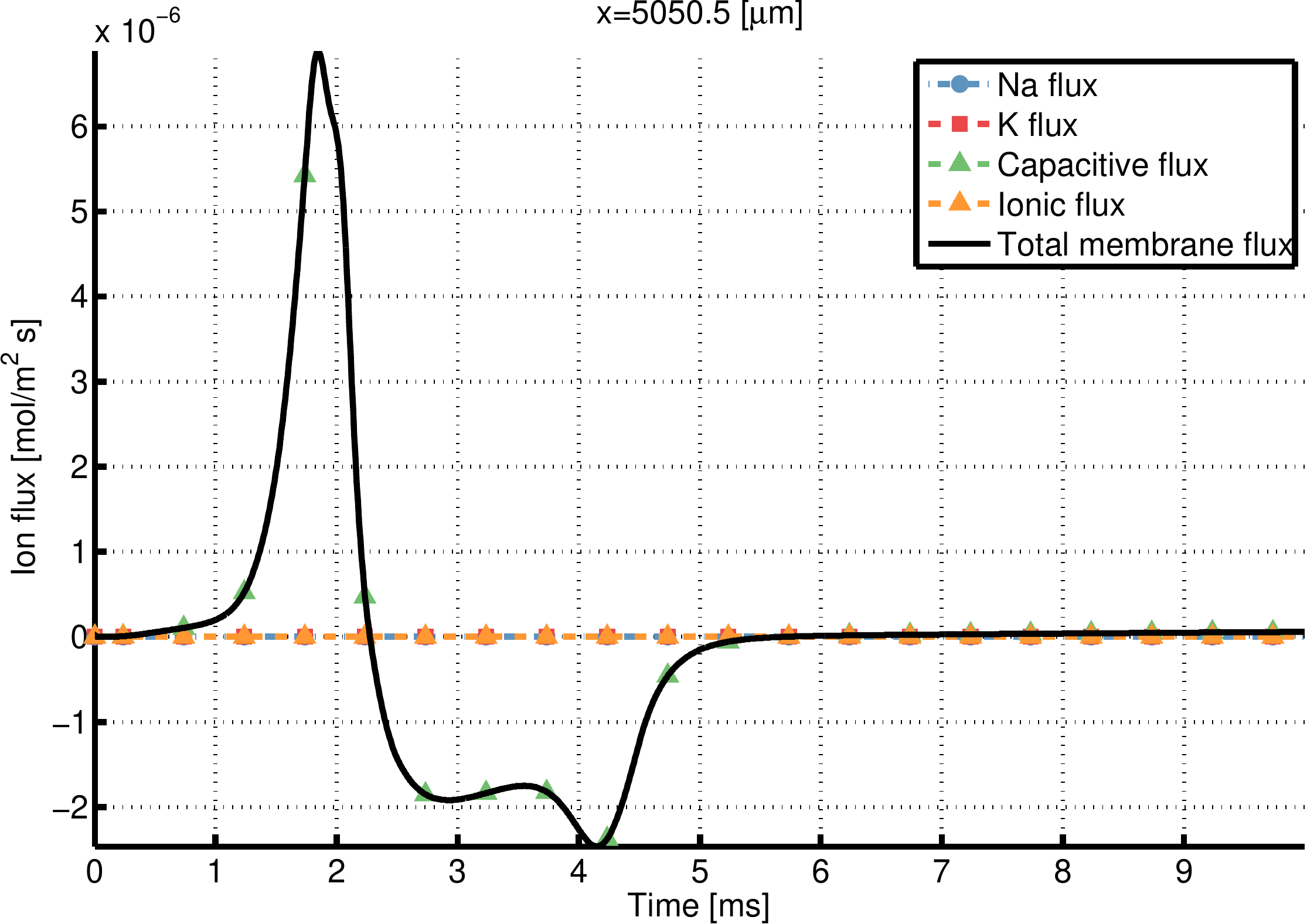}\label{fig:myelin_ap_memb_flux_myelin}}%
\mycaption[Membrane flux at node of Ranvier and myelin]{ %
The total membrane flux at a node of Ranvier (\textit{left}) shows differences to that in the unmyelinated axon in \cref{fig:ap_memb_flux}, as the capacitive component is very small. In contrast, at myelin parts (\textit{right}), there are no ionic fluxes, so the 
capacitive flux dominates, although it has a markedly reduced amplitude in comparison to the node of Ranvier flux.}%
\label{fig:myelin_ap_memb_flux}%
\end{figure}

The ionic fluxes at a node of Ranvier in \cref{fig:myelin_ap_memb_flux_node} show the same shape as in the unmyelinated example from \vref{fig:ap_memb_flux}.
This was expected, since the channel kinetics did not change. 
However, due to the factor of 10 in the channel densities, the amplitude is about a factor of 10 higher as well. 
This also explains why the capacitive flux is small against the others, resulting in a total membrane flux that is dominated by the ionic currents. 
A different picture shows at myelin parts, where no ion channels are present, resulting in zero ionic flux in \cref{fig:myelin_ap_memb_flux_myelin}. 
This causes the total membrane flux to be equal to the capacitive flux, which is smaller than at nodes of Ranvier due to the smaller permittivity (and therefore capacitance). 
The result is a markedly reduced amplitude of the flux in comparison with nodes of Ranvier.

\subsubsection{Debye Layer Concentrations}
\begin{figure}
\centering%
\subfloat[][Na]{
\includegraphics[width=0.5\textwidth]{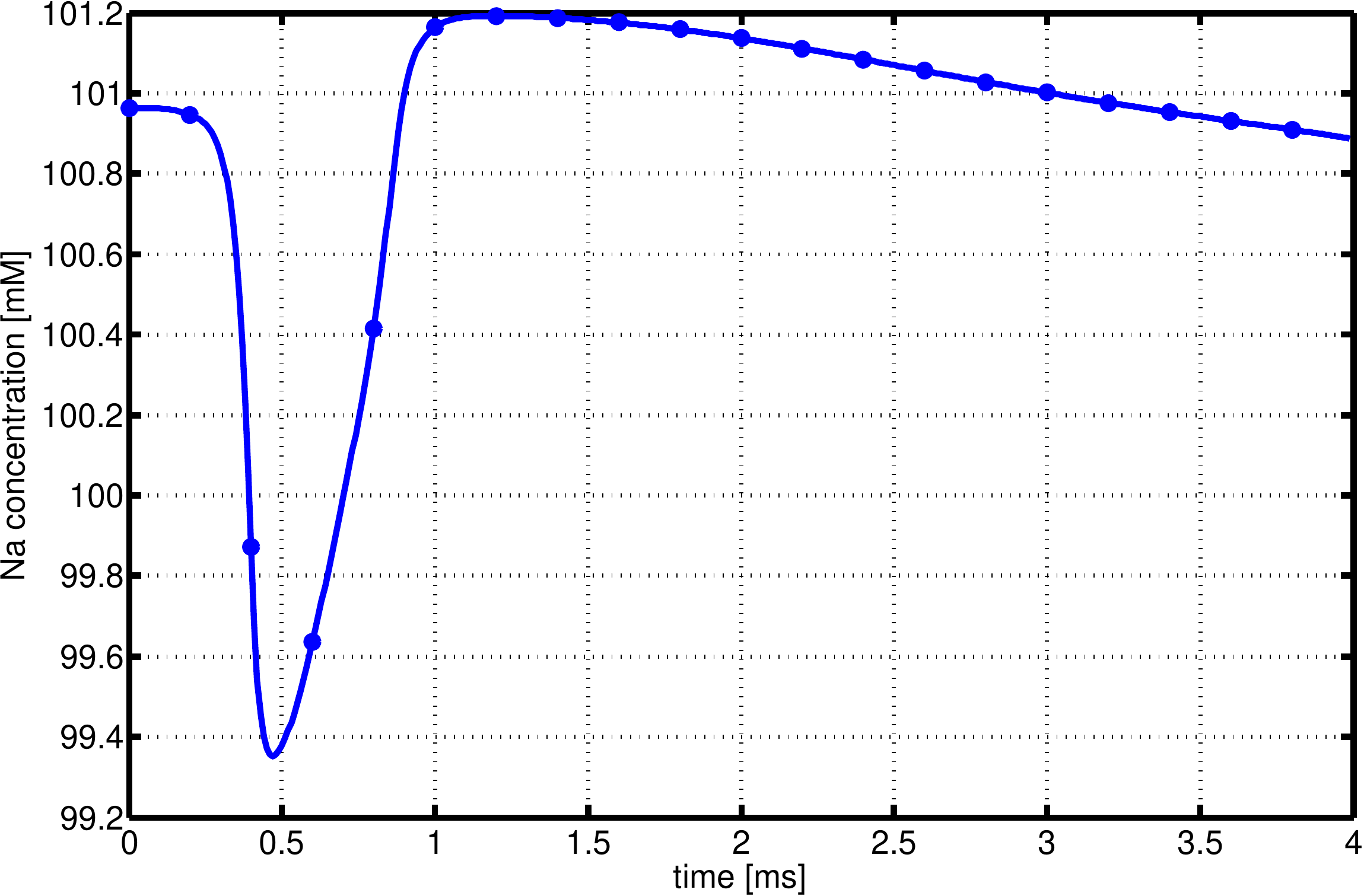}\label{fig:myelin_debye_conc_over_time_na}}%
\subfloat[][K]{
\includegraphics[width=0.5\textwidth]{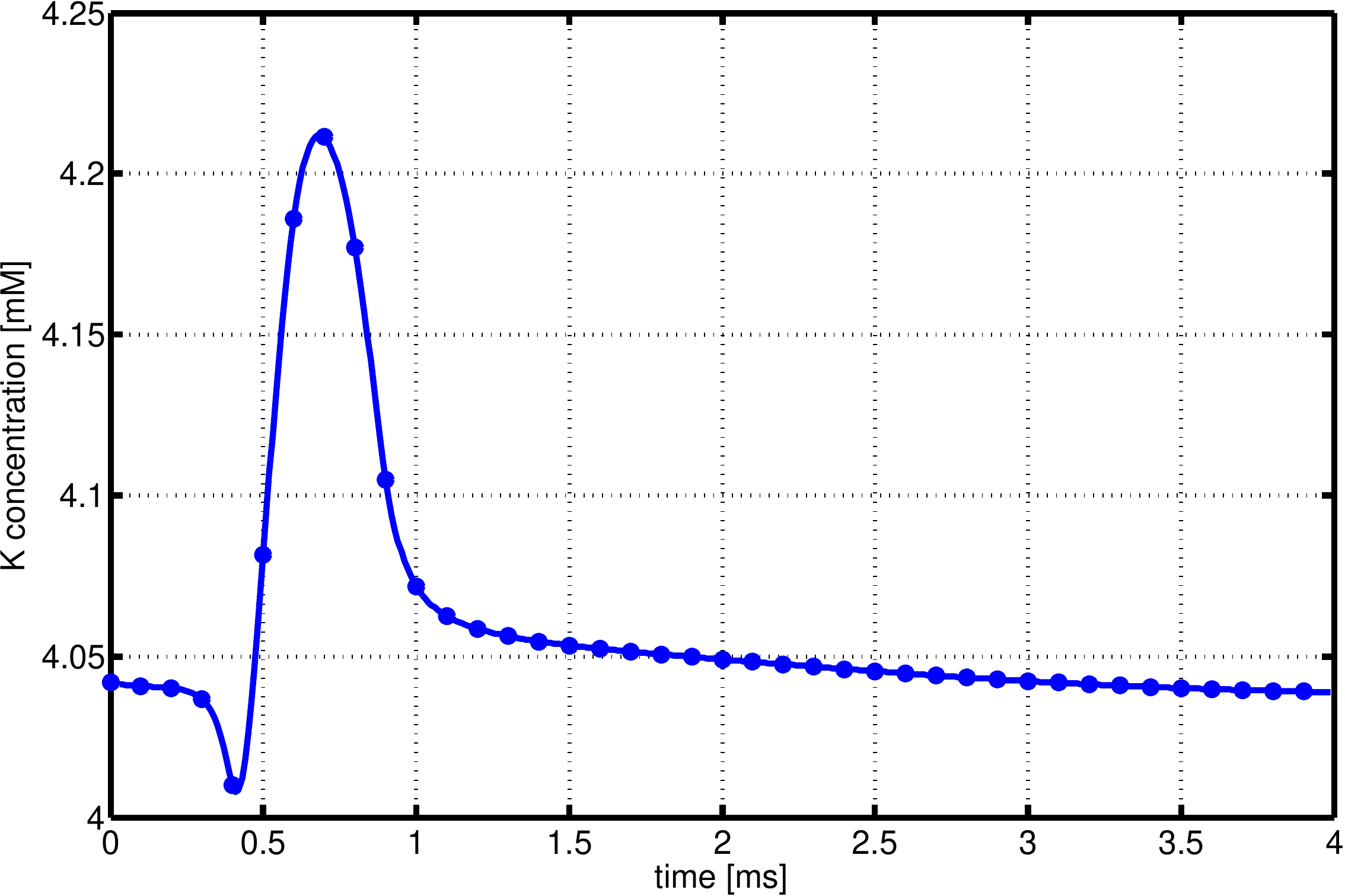}\label{fig:myelin_debye_conc_over_time_k}}\\%
\subfloat[][Cl]{
\includegraphics[width=0.5\textwidth]{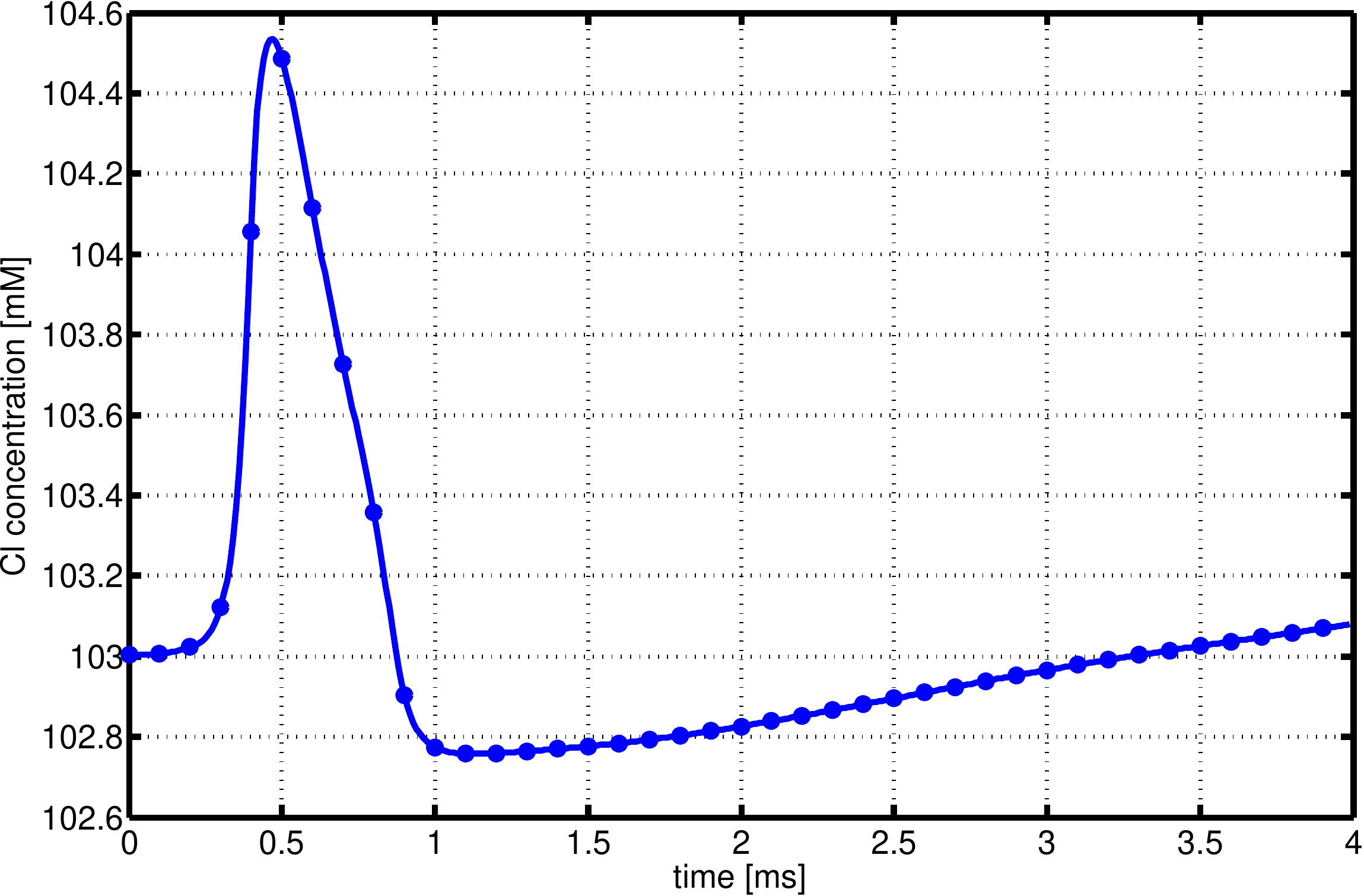}\label{fig:myelin_debye_conc_over_time_cl}}%
\subfloat[][Charge density]{
\includegraphics[width=0.5\textwidth]{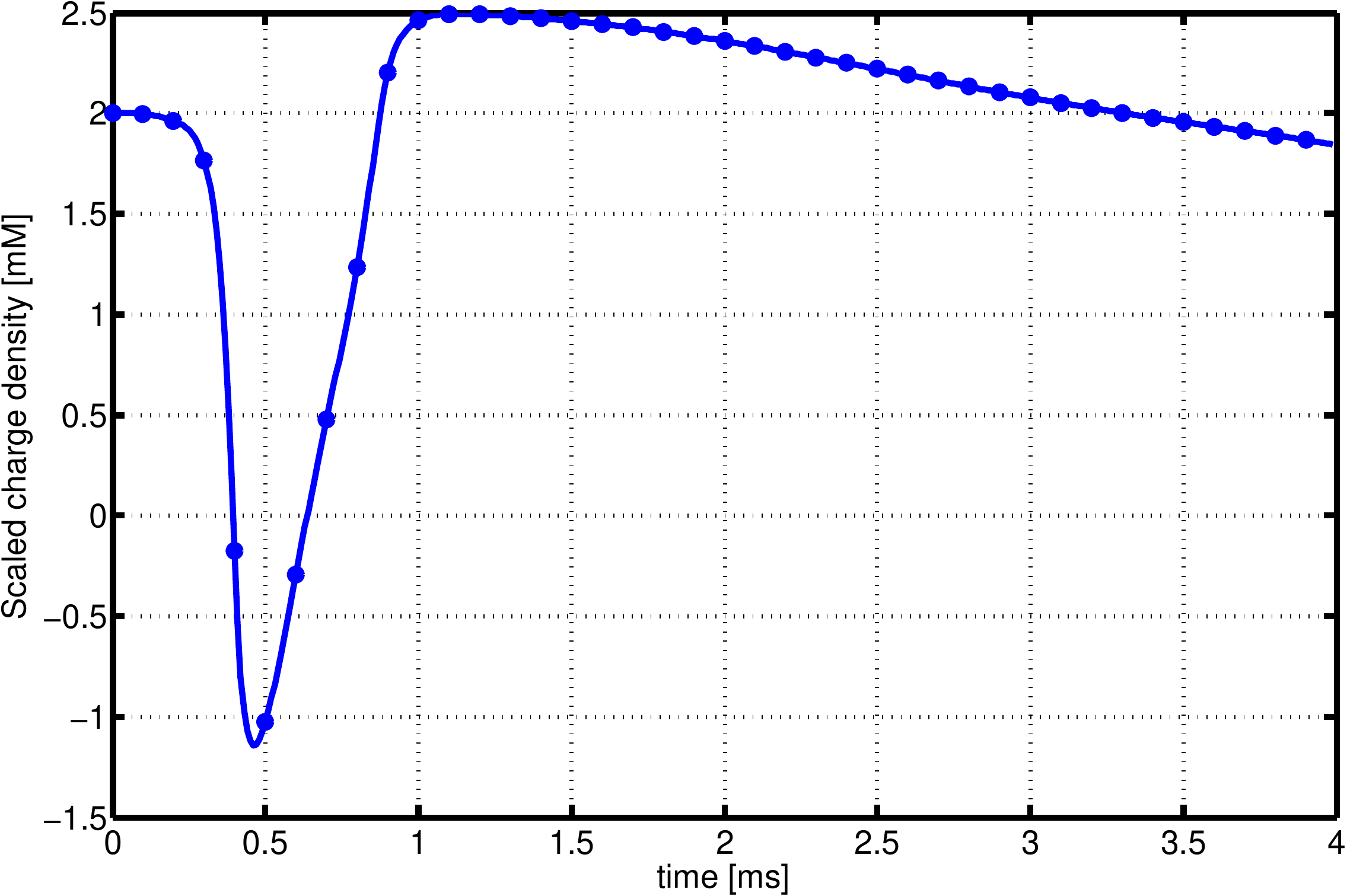}\label{fig:myelin_debye_conc_over_time_cd}}
\mycaption[Debye layer concentration time courses at node of Ranvier]{ %
When compared with the concentration snapshots
of the unmyelinated axon (cf.~\cref{fig:snapshot_conc}), the shapes are quite similar. Sodium concentrations
dominate the charge density and result in a chloride concentration with flipped sign. The potassium concentration
shows a negative peak due to the arriving \gls{AP} and a delayed positive response due to the activating 
voltage-gated $K_V$ channels.}%
\label{fig:myelin_debye_conc_over_time}
\end{figure}

The extracellular concentration time courses in the Debye layer are shown in \cref{fig:myelin_debye_conc_over_time}. 
The delay in the activation of voltage-gated potassium ($K_V) $ channels can be seen nicely in \cref{fig:myelin_debye_conc_over_time_k}, showing first a negative deflection due to the arriving \gls{AP} and then a larger positive peak as $K_V$ channels open.

\subsubsection{Debye Layer Extracellular Potential}
The most interesting part for the myelinated axon certainly is the extracellular potential.
One of the main questions is whether the \gls{AP} echo observed in the \gls{EAP} of the unmyelinated axon will be damped out by the introduction of a myelin insulation.

\begin{figure}
\centering%
\subfloat[][Debye layer extracellular potential]{
\includegraphics[width=0.5\textwidth]{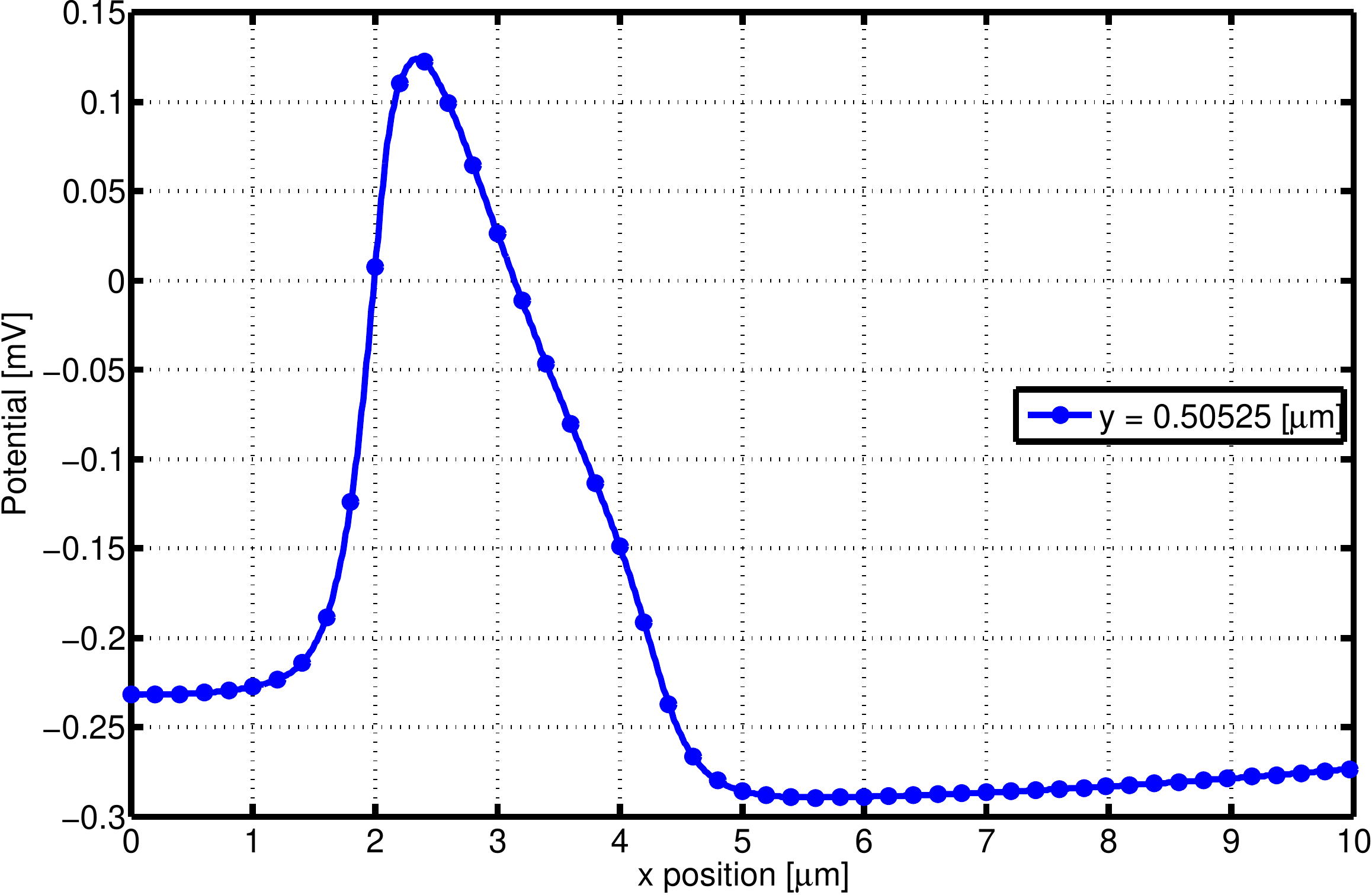}\label{fig:myelin_debye_pot_over_time_node_debye}}%
\subfloat[][Intracellular potential]{
\includegraphics[width=0.5\textwidth]{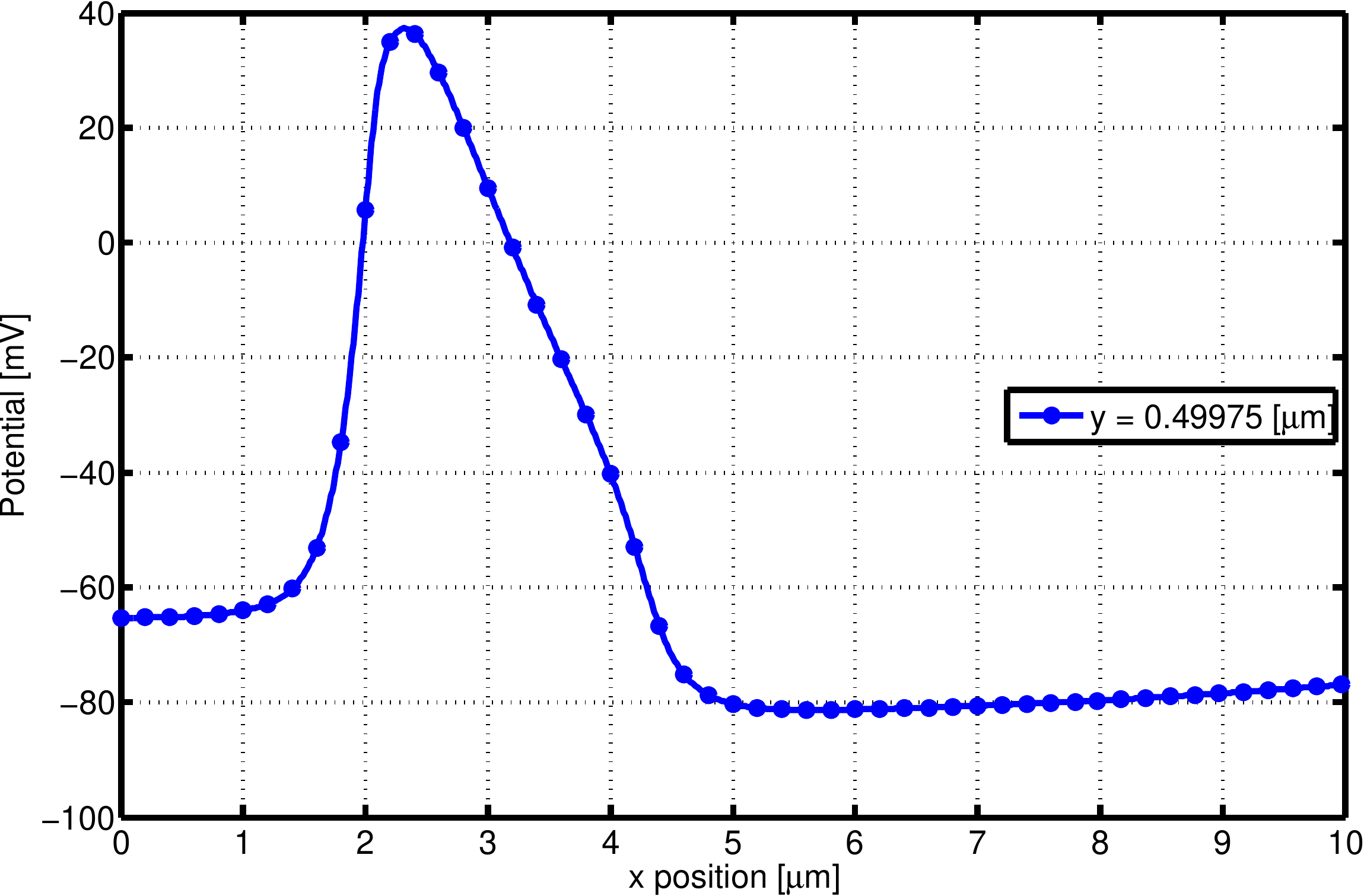}\label{fig:myelin_debye_pot_over_time_node_intra}}%
\mycaption[Debye layer potential time course at node of Ranvier]{ %
The Debye layer potential directly at the
extracellular membrane interface (\textit{left}) can be seen to directly correspond to the intracellular
potential on the opposite membrane interface (\textit{right}), the only difference being the amplitude, which
was significantly reduced by the membrane.}%
\label{fig:myelin_debye_pot_over_time_node}
\end{figure}

The Debye layer extracellular potential at nodes of Ranvier clearly shows the same shape as the intracellular potential in \cref{fig:myelin_debye_pot_over_time_node}, thereby not displaying a different behavior than in the unmyelinated case (cf~\vref{fig:lfp_debye_snapshot}). 
They even show the same amplitude, which is not too surprising -- the membrane has the same thickness and permittivity at nodes of Ranvier as in the unmyelinated case, and the intracellular potential amplitude has a comparable size as well, so a comparable extracellular potential according to \vref{eq:pot_in_out} could be expected.

\begin{figure}
\centering%
\includegraphics[width=0.7\textwidth]{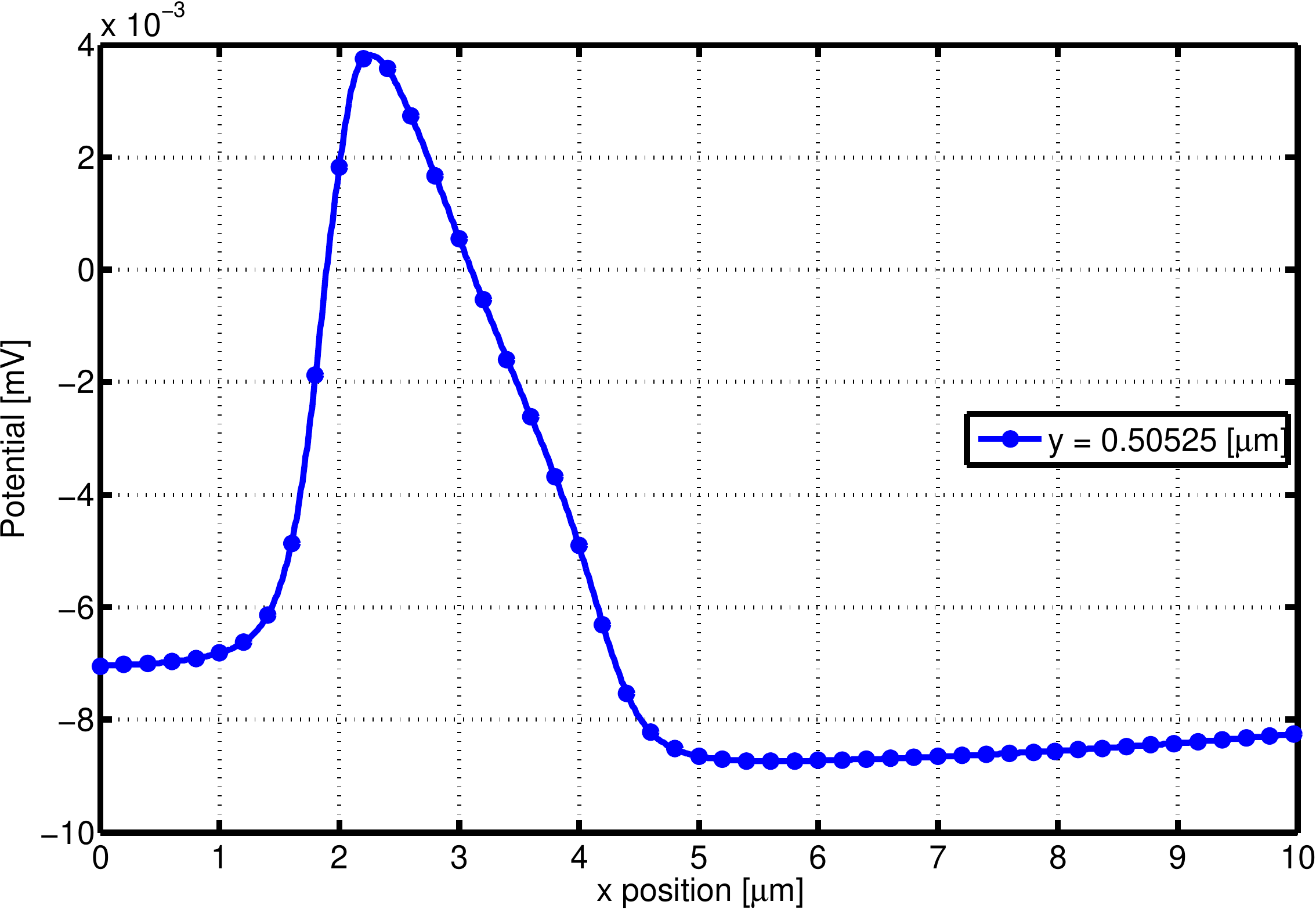}%
\mycaption[Debye layer potential time course at myelin]{ %
The effects of the lower membrane permittivity
are clearly visible, as the potential amplitude is greatly reduced compared to the one
in front of a node of Ranvier in \cref{fig:myelin_debye_pot_over_time_node_debye}}%
\label{fig:myelin_debye_pot_over_time_myelin}
\end{figure}

This does, of course, not hold for the myelin parts of the membrane, where the electric properties of the membrane have changed significantly. The result can be seen in \cref{fig:myelin_debye_pot_over_time_myelin}, where the effects of the lower myelin permittivity can be seen in the greatly reduced amplitude compared to the node of Ranvier in \cref{fig:myelin_debye_pot_over_time_node_debye}.

\subsubsection{Near- and Farfield Extracellular Potential}
\begin{figure}
\centering%
\includegraphics[width=\textwidth]{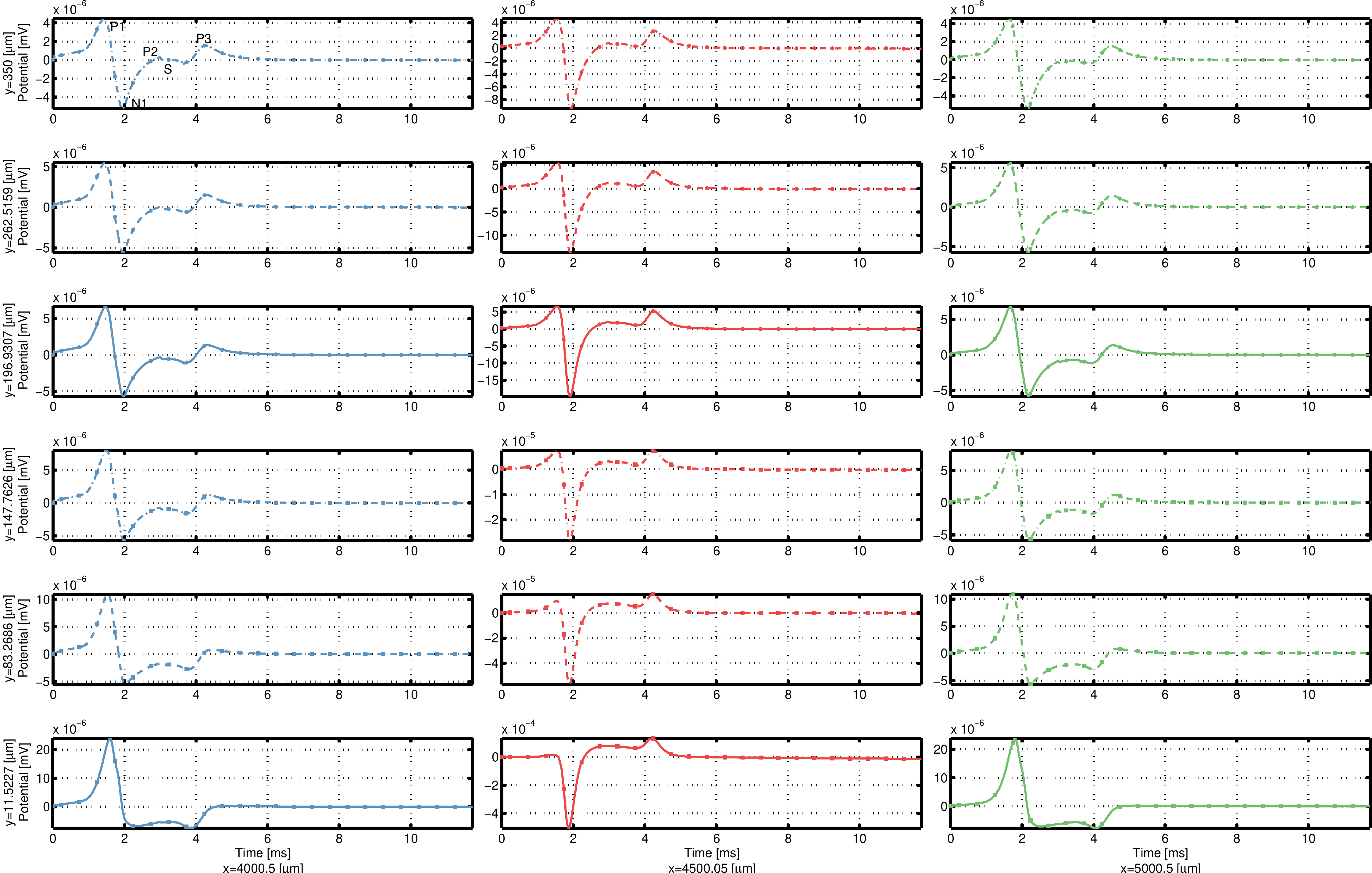}%
\mycaption[Extracellular potential time courses at various positions]{ %
Plots in the same row share a common $y$-coordinate and in the same column a common $x$-coordinate. 
They are arranged by ascending $x$-coordinates to the right and by ascending $y$-coordinates in the upward direction. 
Columns 1 and 3 correspond to myelin segments, column 2 to a node of Ranvier. 
While potential shapes are distinct close to the membrane in the lower rows, they become more and more similar in the upper rows, showing the distance dependence of contributions from different membrane compartments to the extracellular field. 
The different features of the characteristic triphasic shape are labeled as in the previous chapter, shown only for the upper left subplot.}%
\label{fig:myelin_lfp_grid}
\end{figure}

To get a better idea of how the extracellular potential looks like outside the Debye layer, the potential time courses at various positions were plotted in \cref{fig:myelin_lfp_grid}.
The arrangements of plots corresponds to the computational domain, i.e. plots in the same row have the same $y$-coordinate, and plots within the same column have identical $x$-coordinates.
The second column has its $x$-value at a node of Ranvier, while the first and third columns are located at the center of the neighboring internode (myelin) segments.
The difference in the \gls{EAP} curves close to the membrane is striking (bottom row), while at more distant locations (middle rows) the time courses at myelin and node positions become more and more alike.
At a distance of a few hundred micrometers from the membrane (top row), the signals look the same for each $x$-coordinate.

The interpretation is straightforward: close to the membrane, the potential mostly ``sees'' what is happening in the membrane compartments close-by and is therefore dominated by its activity.
This explains why nearfield potentials at myelin and nodes show such distinct shapes: at nodes, membrane currents are strongest and dominate the potential.
Conversely, at myelin compartments, there are no membrane currents at all, so the potential just shows the damped \gls{AP} echo -- albeit with a greatly reduced amplitude, which can be read from the scale of the $y$-axis between e.g. the bottom-left and bottom-middle plot.

The further away from the membrane, the more important become contributions from other membrane parts, resulting in a harmonization of potential curves at different $x$ positions.
Interestingly, the sheer distance dependence of the Poisson equation makes for a ``democratization'' of the membrane contributions.
The strong contributions of the small-extent nodes of Ranvier and the small contributions of the much longer myelin parts become equally important in the distant extracellular field. 

This also means that even in the myelinated case, the \gls{AP} echo and the resulting capacitive currents represent an important contribution to the \gls{EAP}, which was not intuitively clear from the beginning.

\section{Comparison with LSA}
Following up on the comparison of \glspl{EAP} at different distances from the membrane in \cref{sec:unmyel.lsa}, we now do the same for the myelinated axon.
The interesting question now is if we can find the same qualitative results, i.e.~a good agreement at larger distances and significant deviations close to the membrane.
Since now the axon is not homogeneous, we have to compare the potential time courses not only at different $y$ distances, but also at different $x$-coordinates.
For the total membrane current in \vref{eq:lsa_currents}, this also involves replacing the previously constant membrane capacitance by a position-dependent capacitance $C(x)$, because now $A(x)$, $l(x)$ and $\epsMemb = \epsMemb(x)$ vary with $x$ in \cref{eq:memb_capacity_cylinder,eq:memb_capacity_parallel}.
For the only free parameter in the \gls{LSA} \cref{eq:lsa}, the same value as before was chosen for the resistivity $\rho = \SI{72}{\ohm\centi\metre}$.

\begin{figure}
\centering%
\includegraphics[width=\textwidth]{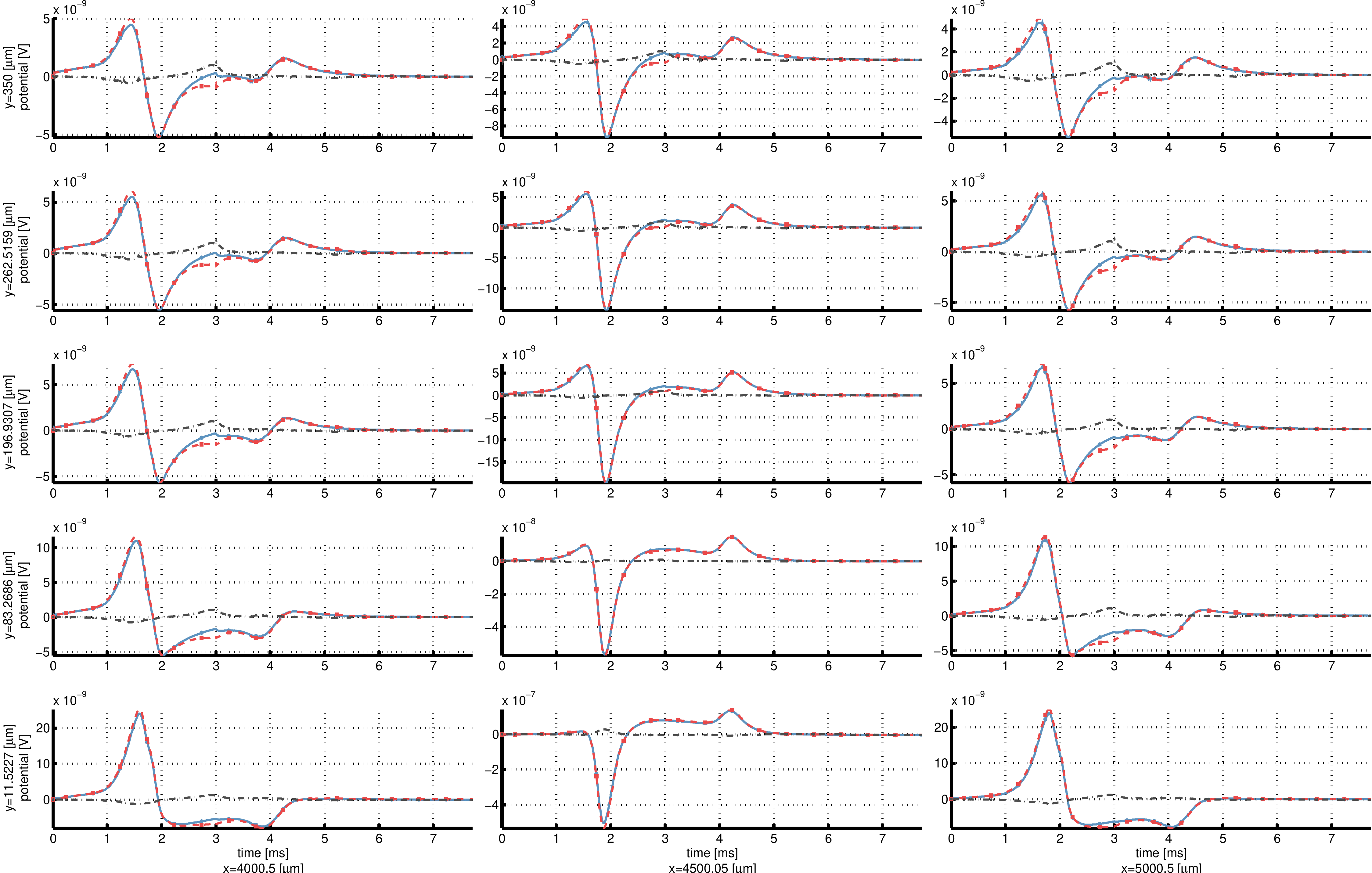}%
\mycaption[Comparison with LSA at various farfield positions]{ %
The same potential time courses as in \cref{fig:myelin_lfp_grid}
(\textit{solid lines}) are compared with the \gls{LSA} potential (\textit{dashed lines}). Additionally, the difference between both curves is shown (\textit{dash-dotted lines}). The agreement between both models generally is very good and shows notable deviations only between peaks N1 and P3.}%
\label{fig:myelin_lsa_ed_grid}
\end{figure}

In \cref{fig:myelin_lsa_ed_grid}, the time courses of the extracellular potentials are shown in a grid representation, where the position of each subplot roughly gives its position in the actual computational grid, as before in \cref{fig:myelin_lfp_grid}.
Again, points close to the membrane are found in the lower rows, points with larger distances in the upper rows.
The arrangement of columns and the positions of the individual subplots are the same as in \cref{fig:myelin_lfp_grid}.

\begin{figure}
\centering%
\includegraphics[width=\textwidth]{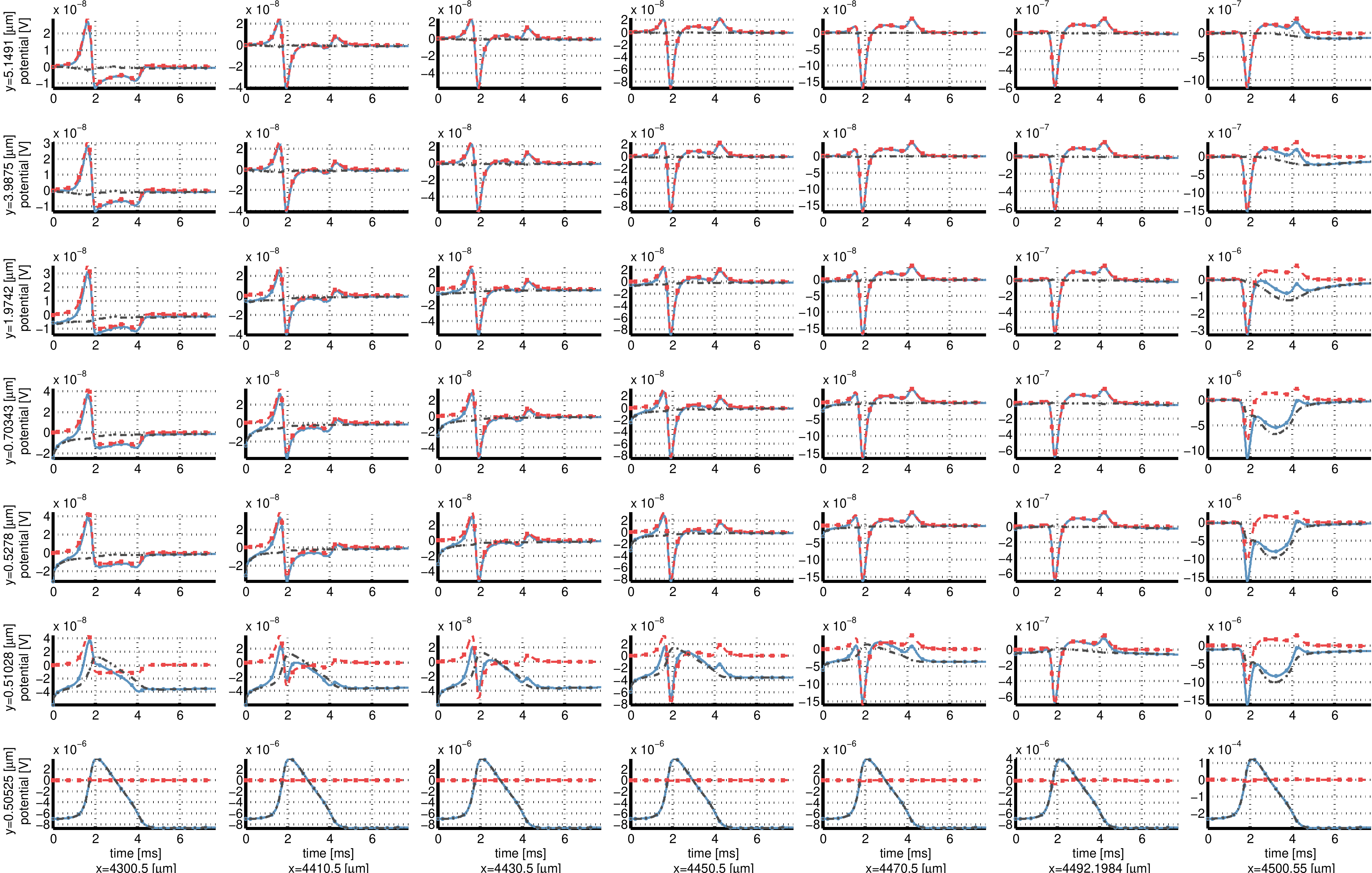}%
\mycaption[Comparison with LSA: zoom into myelin-node transition and nearfield]{ %
A closer look at the transition 
between myelin (\textit{left}) and node of Ranvier (\textit{right}) membrane parts in $y$-direction, plotted here for
nearfield distances up to about \SI{5}{\micro\metre}, reveals significant deviations (\textit{dash-dotted line}) between \gls{ED} (\textit{solid lines}) and \gls{LSA} (\textit{dashed lines}) especially at nodes of Ranvier. 
Columns in between mark coordinates in the transition interval between myelin and node of Ranvier. 
Myelin coordinates show strong deviations only up to a distance of about \SI{1}{\micro\metre}.}%
\label{fig:myelin_lsa_ed_grid_zoom}
\end{figure}

For larger distances as in \cref{fig:myelin_lsa_ed_grid}, the overall agreement between electrodiffusion and \gls{LSA} models is generally good, albeit in contrast to the unmyelinated case, notable differences can also be found in the farfield, e.g.~at peak P1 or between peaks N1 and P3.
Another comparison for the nearfield can be found in \cref{fig:myelin_lsa_ed_grid_zoom}, which shows a zoom into the transition interval ($x \in \Xtransition$, middle columns)
between myelin parts ($x \in \Xmyelin$, left columns) and one node of Ranvier ($x \in \Xnodes $, rightmost column).
Here, some interesting details can be made out.
Firstly, large parts of the plots at myelin coordinates show very good agreement, even for distances smaller than \SI{5}{\micro\metre}.
At distances $< \SI{1}{\micro\metre}$ and especially in the Debye layer (bottom row), however, the differences are large.
On the other hand, the node of Ranvier column (right) shows stronger deviations, even up to distances of \SI{5}{\micro\metre}.
This is consistent with our previous observation from \cref{sec:unmyel.lsa.separate}, where the ionic contributions accounted for the greatest deviations between electrodiffusion and \gls{LSA} models.
Following this reasoning, it is clear that the nodes of Ranvier with their large ionic currents and comparably small capacitive currents will show stronger deviations than myelin parts, which only contribute a single capacitive \gls{EAP} component.

\Cref{fig:myelin_lsa_ed_grid,fig:myelin_lsa_ed_grid_zoom} actually give insights into the ``big picture'' of the \gls{EAP} generation. It shows the distinct shapes due to the domination of either ionic membrane currents (nodes of Ranvier) or capacitive effects (myelin). 
Furthermore, it shows how those two shapes merge at greater distances from the membrane to constitute the characteristic triphasic ``up-down-up'' shape of the farfield \gls{LFP}.

\section{Summary}
The introduction of myelin required some fundamental changes in the model.
By recognizing that the transmembrane potential is essentially a linear function, we could avoid explicitly representing compartments with varying membrane thickness in the mesh.
Instead, myelin could be represented implicitly through adapted effective dielectric constants.
With this, we maintained the tensor grid structure in the spatial discretization, which is very beneficial for the computational efficiency.

The problem of jumping material coefficients between myelin and nodes of Ranvier was tackled in two stages: by resolving the transitions in $x$-direction geometrically in the grid, and by applying an additional smoothing operator to the discontinuity in permittivities.
We take the view that smoothing the permittivities is actually more realistic, since in real neurons (cf.~again the illustration in \vref{fig:theory.neuron_illustration}), the internodes do not show an abrupt increase in membrane thickness, but rather a gradual one, accompanied by a corresponding smoother change in the dielectric constant.

The presented changes in model and numerical methods allowed us to simulate a myelinated axon of length \SI{4.8}{\centi\metre} with 48 nodes of Ranvier.
The results show the expected \gls{AP} velocity increase and two different forms of extracellular potentials: one that is dominated by membrane currents at nodes of Ranvier, and one that is purely capacitive at internodes.
Interestingly, these two different \gls{EAP} shapes merge at larger distances and show the same triphasic shape as an unmyelinated axon.
This can be explained by the interplay of \gls{EAP} amplitude, which is considerably higher at nodes, and of the contributing membrane area, which is significantly higher for internodes.
At sufficiently large distances, the weighting of these two components balances and both components -- nodal ionic and internodal capacitive -- become equally important.

A comparison with the \gls{LSA} model as before showed deviations especially at small membrane distances, which could again to a large degree be attributed to the ionic components of the \gls{EAP} and their influence within the intermediate diffusion layer.
The extracellular potential dominated by capacitive currents at internodes generally shows a very good agreement. 
As before, an exception is the Debye layer, which exhibits fundamental differences in comparisons with \gls{LSA}.

\setchapterpreamble[u]{%
\dictum[Winston Churchill]{If you're going through hell, keep going.}\bigskip}
\chapter{Ephaptic Interaction Between Multiple Axon Fibers}\label{chap:multiple_fibers}
Now that the basic models for unmyelinated and myelinated nerve fibers have been derived, simulated, and validated, we would like to study interactions between nerve fibers. 
Of particular interest is the phenomenon of \emph{ephaptic coupling}, which describes the influence of one neuron on the other in the absence of chemical or electrical synapses (\emph{gap junctions}). 
Such a coupling is possible either directly through the electric field or indirectly by exchange of ions. 
In either case, the neurons have to be very close to each other to yield a notable effect, which is reflected in the adjective \emph{ephaptic} (from Greek $\epsilon\phi\alpha\pi\tau\omega$, ``to touch'') and the associated site \emph{ephapse}, the ``location of close contact or vicinity'' \cite{arvanitaki1942effects}, in distinction to a synapse.

\section{Previous Work}
While both chemical and electric synapses are traditionally believed to be the only possible source for eliciting an action potential, there is a surprisingly large body of studies on the ephaptic effects between two neighboring fibers, starting in the 1940s \cite{katz1940electric} and continuing until today, also involving the \gls{LFP} of a population of neurons as a possible source for \gls{AP} synchronization \cite{anastassiou2011ephaptic}.
An excellent overview on the existing literature with slightly different emphases is given in \cite{faber1989electrical} and \cite{jefferys1995nonsynaptic}.
A newer review is \cite{weiss2010field}, which focuses more on the electric fields of neuronal populations and their functional effects.

A special case in this context is the heart muscle, in particular \emph{cardiomyocytes} (cardiac muscle cells), which are considered to be excited via gap junctions \cite{aidley1998physiology,hall2010guyton}.
In \cite{sperelakis1977evaluation}, a modeling study was able to show that gap junctions are not necessary, since ephaptic coupling may serve to conduct the \gls{AP} from cell to cell over the cleft between cells.
In an experimental study using mice with suppressed expression of myocardial gap junctions \cite{gutstein2001conduction}, the animals were found to develop cardiac arrhythmia and eventually died by sudden cardiac arrest at about two months of age.
Until then, however, they showed normal heart development and function, indicating that the basic conduction functionality was maintained in the absence of gap junctions.

A number of modeling studies have addressed the topic of ephaptic coupling.
Holt \cite{holt1997critical} found the influence of the extracellular electric field of one axon onto the other to be a fraction of a millivolt, which he considered too small to elicit an \gls{AP} in a cell which is not already very close to threshold.
He used the \gls{LSA} model for this, neglecting concentration effects.
Mori \cite{mori2006three}, on the other hand, used a reduced version of the electrodiffusion equations (the ``electroneutral model'') and studied ephaptic conduction in the setting of myocardiac fibers mentioned above.
He was able to demonstrate that ephaptic coupling indeed can supplement or replace gap junction conduction \cite{mori2008ephaptic}.
However, the cleft width had to be at most \SI{5}{\nano\metre} in order for the conduction to be successful.

Holt \cite[chapter 2.3.1]{holt1997critical} also mentions a number of studies that address the interaction between neighboring axons with respect to phase-locking. Essentially, \glspl{AP} running in two parallel axons will tend to synchronize: if they are aligned, the extracellular current sources and sinks are also aligned, and each axon will try to hyperpolarize the other, resulting in a slowed-down \gls{AP} conduction. In staggered \glspl{AP} on the other hand, the outward currents of the leading axon will align with the inward currents of the following axon, which will enhance the following axon's \gls{AP} conduction. As a consequence, staggered, phase-locked \glspl{AP} are stable and faster, while aligned \gls{AP} will slow each other down. 

In the following chapter, we will use the electrodiffusion model developed over the course of the previous chapters to study ephaptic coupling between parallel axons in a relatively simple setting.

\section{Unmyelinated Axon Surrounded by an Axon Bundle}
\subsection{Geometry}
\begin{figure}
\subfloat[]{%
\centering%
\includestandalone[width=0.6\textwidth]{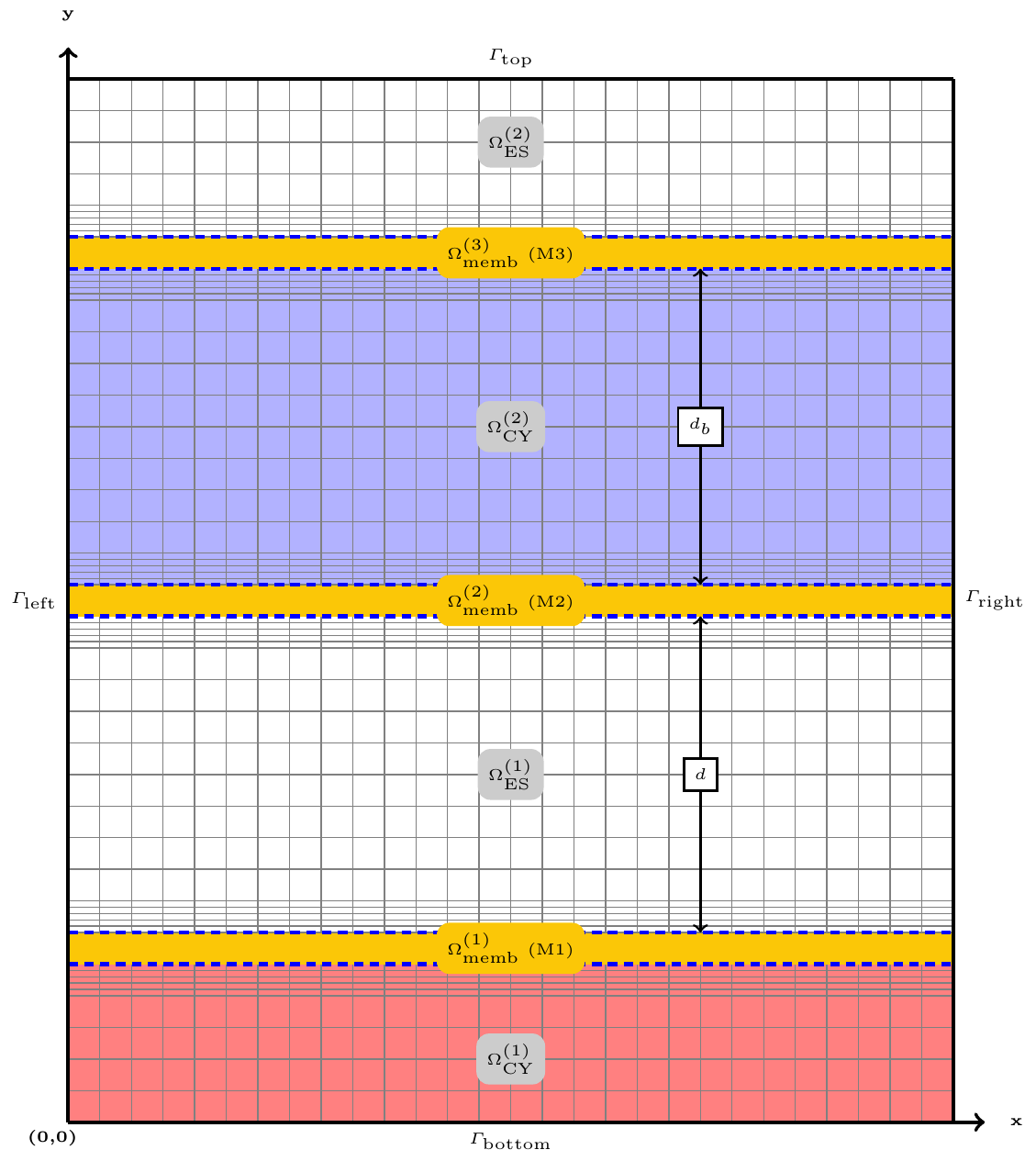}%
\label{fig:multiple_axons.geometry_2D}}%
\subfloat[]{%
\centering%
\includestandalone[width=0.4\textwidth]{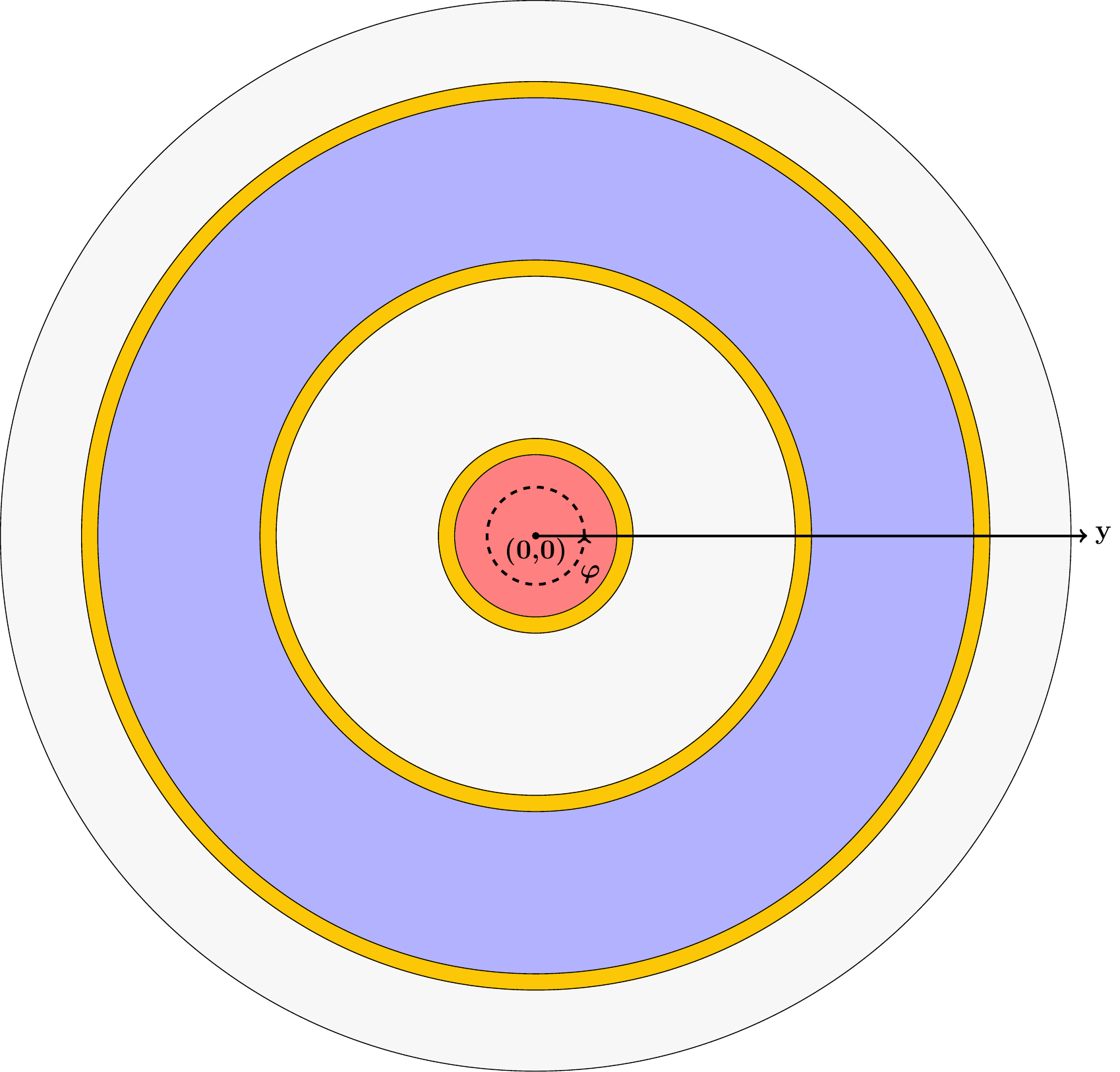}%
\label{fig:multiple_axons.geometry_cross}}%
\mycaption[Computational domain for multiple axons]{The two-dimensional computational domain (\textit{left}) was modified in order to include two additional membranes M2 and M3, enclosing a second cytosol domain. On the right, the equivalent cylinder cross section for this setup is shown, revealing a central axon that is surrounded by an interstitial space and an axon bundle, which is again surrounded by an extracellular space.}%
\label{fig:multiple_axons.geometry}%
\end{figure}

Due to the cylinder symmetry, we are limited in the choice of our geometry, since the angular direction was eliminated from the equations. This means we can not model two single axons next to each other, which is best exemplified in a picture, see \cref{fig:multiple_axons.geometry}.

It is clear that, when adding a second axon, this can not be modeled with one single membrane anymore, as we can only use the symmetry axis once.
Two additional membranes are needed to obtain an additional cytosol subdomain $\OmegaCytosolN{2}$ next to the obligatory $\OmegaCytosolN{1}$, as well as an additional subdomain $\OmegaExtraN{2}$, as shown in \cref{fig:multiple_axons.geometry_2D}.
$\OmegaExtraN{1}$ is now bounded by two membranes M1 and M2. 
In the cylinder geometry, however, this does not correspond to two adjacent axons, but rather to one cylindrical axon surrounded by a shallow cylinder axon (see the cross section in \cref{fig:multiple_axons.geometry_cross}).

We can think of this as an \emph{axon bundle} around a single axon. 
It is clear that this situation is not realistic, since in real tissue the axon bundle would not consist of a single intracellular domain, but of multiple cylindrical axons with a certain spacing in between them. 
We should keep this in mind when interpreting the results generated by this model and consider them as (at least partially) speculative.
For a basic phenomenological study, however, this setup shall prove useful.

A similar geometry is used in \emph{core conductor models} \cite{rall2011core}. 
When an axon is surrounded by a resistive sheath, as is often the case in peripheral nerves, e.g.~by an \emph{epineurium} or \emph{perineurium}, the extracellular space can be regarded to be one-dimensional, in analogy to the cable equation \cite{clark1966mathematical}.
This approximation, however, only holds for small distances between axon and sheath. 
Moreover, here we are not only interested in the extracellular field due to a restricted extracellular volume, but also on its impact on adjacent nerve fibers. 

\begin{figure}
\subfloat[rectangular array]{%
\centering%
\includestandalone[width=0.45\textwidth]{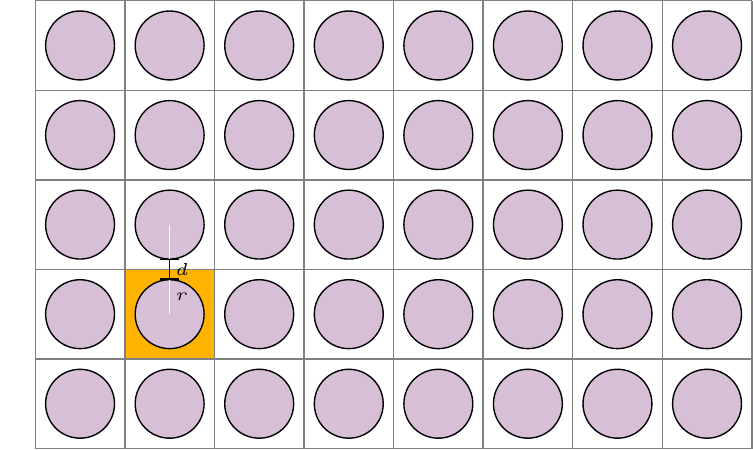}%
\label{fig:multiple_axons.packing_rectangular}}%
\subfloat[hexagonal array]{%
\centering%
\includestandalone[width=0.55\textwidth]{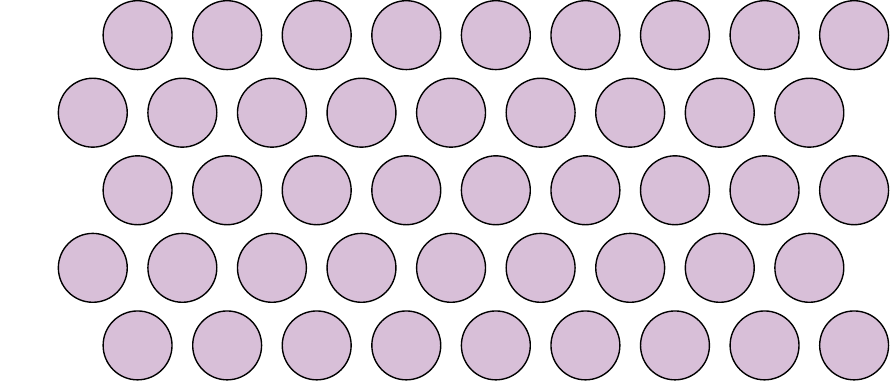}%
\label{fig:multiple_axons.packing_hexagonal}}%
\mycaption[Illustration of axonal packing in different geometric arrangements]{A regular rectangular packing (\textit{left}) leads to a non-optimal packing ratio. The optimal packing can be achieved in a hexagonal array (\textit{right}).}%
\label{fig:multiple_axons.packing}%
\end{figure}

A crucial parameter is the distance between axon and bundle, $d$. \citeauthor{holt1997critical} \cite{holt1997critical} considered the case of an axon bundle and estimated the spacing $d$ based on different arrangements of axons, shown as cross-sections in \cref{fig:multiple_axons.packing}.
In a rectangular assembly, as in \cref{fig:multiple_axons.packing_rectangular}, the packing is non-optimal.
The volume fraction 
\begin{align}
  \alpha = \frac{\VExtra}{\Vtotal}
\end{align}
between extracellular volume $\VExtra$ and the total volume $\Vtotal$ can be calculated as
\begin{align*}
  \alpha_r = \frac{(2r+2d)^2 - \pi r^2}{\pi r^2} = \frac{4r^2 + 8rd + 4d^2 - \pi r^2}{\pi r^2} \ .
\end{align*}
Even for touching axons with $d=0$, this gives a value of $\alpha_r = \frac{4-\pi}{\pi} \approx 0.2146$, which is larger than the observed average volume fraction of $\alpha=0.2$ \cite{sykova2008diffusion}. When using a hexagonal packing as in \cref{fig:multiple_axons.packing_hexagonal}, which Gauß proved to be optimal (see \cite{stephenson2005introduction}), the volume fraction reads
\begin{align*}
  \alpha_h = \frac{a^2 \frac{3}{2} \sqrt{3} - 3\pi r^2}{a^2 \frac{3}{2} \sqrt{3}} = \frac{(2r+d)^2 \frac{3}{2} \sqrt{3} - 3\pi r^2}{(2r+d)^2 \frac{3}{2} \sqrt{3}}
\end{align*}
with the hexagon side length $a = 2r+d$. For $d = 0$, this results in the densest possible packing with an extracellular volume ratio of $\alpha_h = \frac{2\sqrt{3}-\pi}{2\sqrt{3}} \approx 0.0931$. In the present case of an axon radius of $r= \SI{500}{\nano\metre}$, a volume ratio of 0.2 would be obtained for a spacing  of $d \approx \SI{65}{\nano\metre}$, which is already on the upper bound of the estimated average extracellular width of \SIrange{38}{64}{\nano\metre} \cite{sykova2008diffusion}.
This average membrane-to-membrane distance is remarkably small, and -- considering the results of the preceding chapters -- concentration effects may be assumed to play an important role within this regime, which is only one order of magnitude larger than the Debye length $\dDebye$, and definitely within the diffusion layer regime of order $\sqrt{\dDebye}$ (cf.~\cite{mori2006three}).

\subsection{Action Potential}
In an attempt to quantify the influence of one axon onto the surrounding bundle, an action potential was elicited in the central (unmyelinated) axon as before, while the axon bundle was not stimulated.
We largely used the same parameter set as in \cref{chap:unmyel}, as listed in \vref{tab:sim_params_unmyel}, with the exception of an adjusted mesh according to the changed geometry.
As in the previous setups, the Debye layers of all three membranes were fully resolved, and coarsening strategies were employed for the regions in between to minimize the total number of unknowns.

Simulations were carried out for various inter-fiber distances $d=\SI{1}{\micro\metre}$, \SI{500}{\nano\metre}, \SI{100}{\nano\metre} and \SI{50}{\nano\metre}.
Distances were deliberately chosen larger in tendency than the average \gls{ES} width of \SIrange{38}{64}{\nano\metre} in order to obtain an estimated upper bound on $d$ for which ephaptic interactions are possible.

\begin{figure}
\subfloat[fixed $d_b$]{%
\centering%
\includegraphics[width=0.5\textwidth]{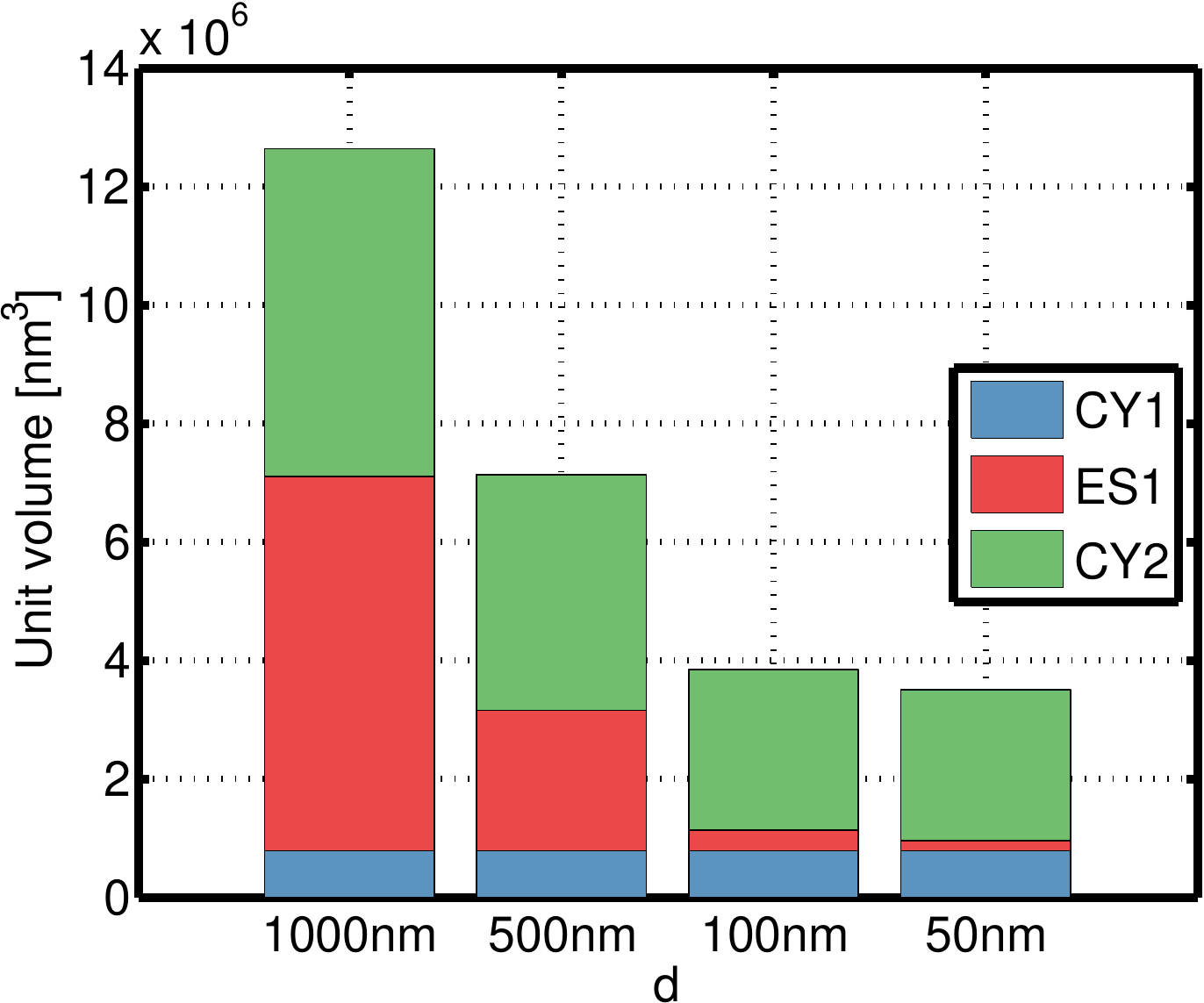}%
\label{fig:multiple_axons.volume_ratio_axonbundle_fixed}}%
\subfloat[volume-corrected $d_b$]{%
\centering%
\includegraphics[width=0.5\textwidth]{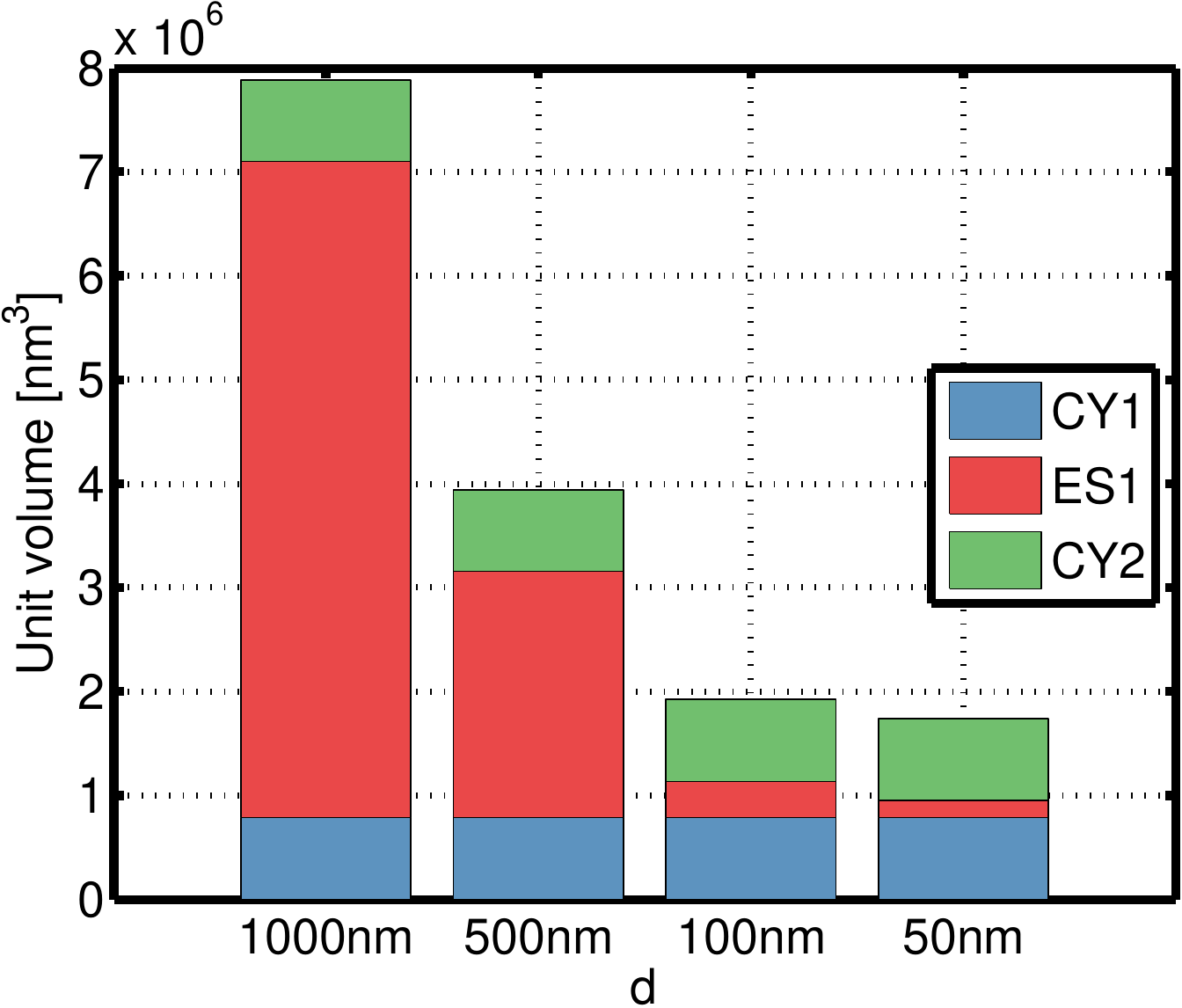}%
\label{fig:multiple_axons.volume_ratio_axonbundle_volumeCorrected}}%
\mycaption[Ratios of intra- and extracellular volumes for different axon-bundle distances]{In the fixed case (\textit{left}), the bundle volumes are growing with inter-fiber distance, thereby also changing relative volume fractions. In the volume-corrected case (\textit{right}), the bundle volume is kept fixed and equal to the central axon volume, such that only the interstitial volume in between is changing with $d$.}%
\label{fig:multiple_axons.volume_ratio_axonbundle}%
\end{figure}

The diameter of the axon bundle $d_b$ was chosen in two different ways: in the first set of simulations, it was fixed at $d_b = \SI{500}{\nano\metre}$, which means that for larger inter-fiber distances $d$, the volume of the axon bundle will be larger than for small distances due to the cylinder geometry.
Therefore, we also implemented a second version in which the bundle volume was kept constant (and equal to the source axon volume) by varying $d_b$ dependent on $d$, which is called the ``volume-corrected'' setup.
We will later on see the reason why the cell volumes and their relationship to the extracellular ones are of significance.
\Cref{fig:multiple_axons.volume_ratio_axonbundle} shows the fraction of the three volumes of interest, i.e.~those of the interstitial region $\OmegaExtraN{1}$, and of the two intracellular domains $\OmegaCytosolN{1}$ and $\OmegaCytosolN{2}$, for the two different variants.

\subsubsection{Fixed bundle diameter $d_b$}
\Cref{tab:multiple_fibers.times_fixed} shows the computation times\footnote{Note that the absolute computation times are rather high in comparison to \cref{chap:unmyel}. Accidentally, these simulations have been run using an executable that was compiled completely without optimizations, significantly impairing the performance.
In \vref{tab:multiple_fibers.times_vc}, the effect of using compiler optimizations can be clearly seen.} for each of the different choices of $d$, obtained from a parallel run with $p=10$ processors.
The overall performance is reasonable -- the smaller the distance, the more interaction between fibers, increasing the difficulty of the numerical problem.

\begin{table}
\mycaption[Simulation timings for the axon bundle setup using different inter-fiber distances $d$ and a fixed bundle diameter $d_b$]{
For each distance $d$, the number of unknowns, total computation time, the needed number of time steps, and the average
solution time per time step (full Newton iteration) are shown.}\label{tab:multiple_fibers.times_fixed}
\centering
\begin{tabu} to 0.9\textwidth {@{} *5{X[l]} @{}}
\toprule
$d$ [nm] & \# \glspl{DOF} & Total comp. time [s]& \# time steps & avg. time / time step [s]\\
\midrule
1000 & 116352 & 156071 & 2026 & 77.034\\
500 & 114332 & 164591 & 2026 & 81.24\\
100 & 114332 & 181175 & 2026 & 89.425\\
50 & 114332 & 184747 & 2026 & 91.188\\
\bottomrule
\end{tabu}
\end{table}

\begin{figure}
\subfloat[$d=\SI{1}{\micro\metre}$]{%
\centering%
\includegraphics[width=0.5\textwidth]{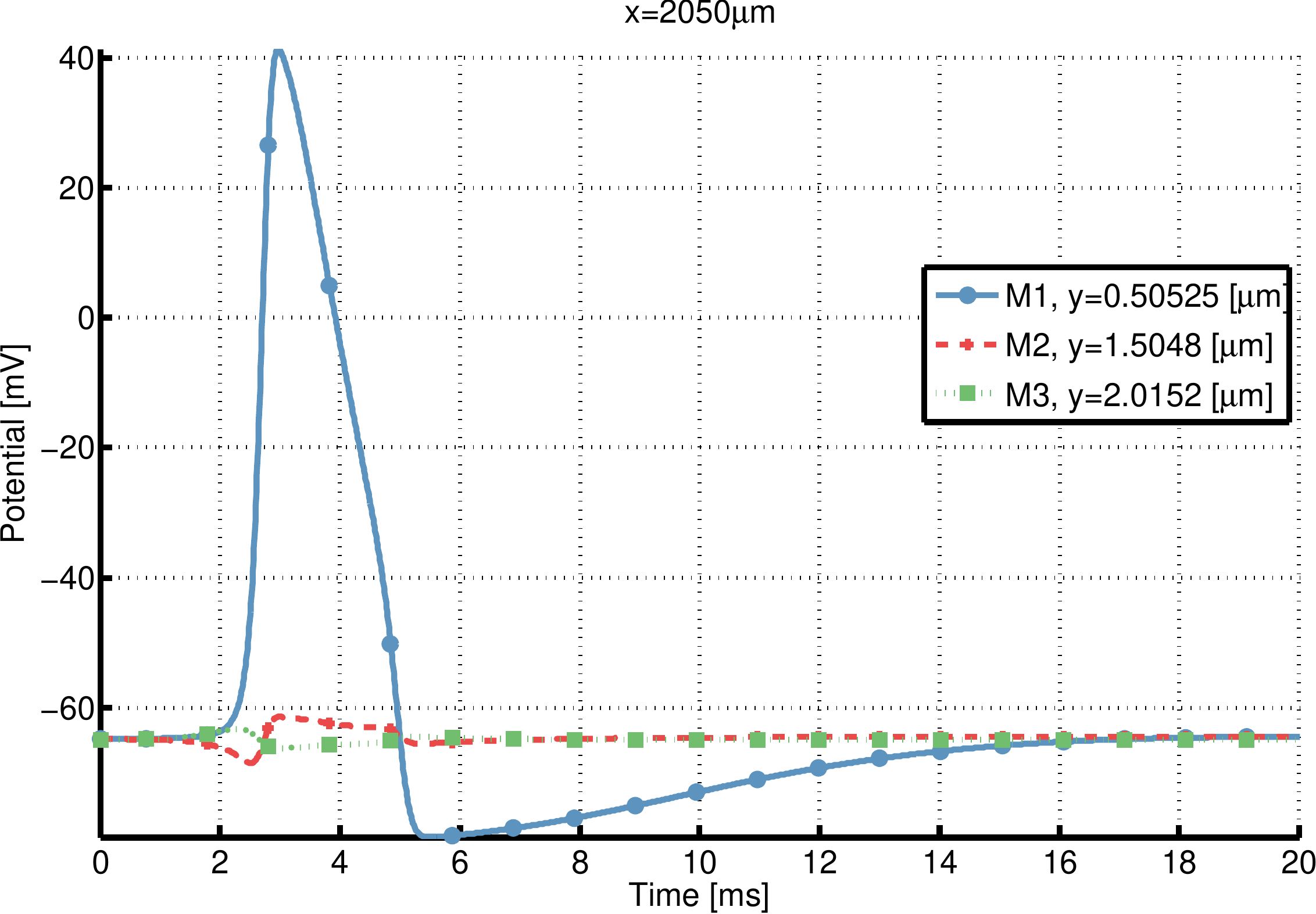}%
\label{fig:multiple_axons.memb_pot_1000nm_overview}}%
\subfloat[$d=\SI{500}{\nano\metre}$]{%
\centering%
\includegraphics[width=0.5\textwidth]{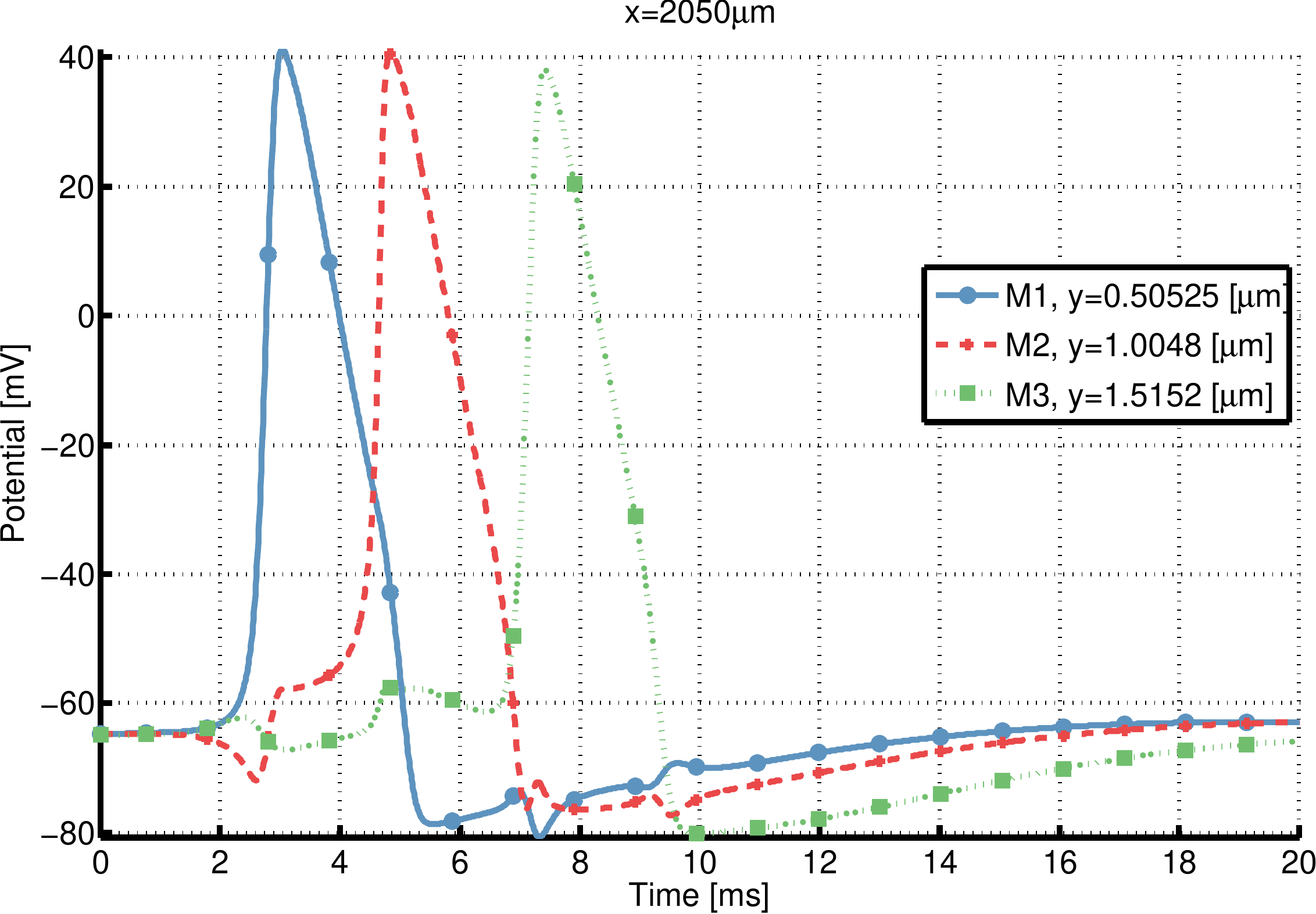}%
\label{fig:multiple_axons.memb_pot_500nm_overview}}\\%
\subfloat[$d=\SI{100}{\nano\metre}$]{%
\centering%
\includegraphics[width=0.5\textwidth]{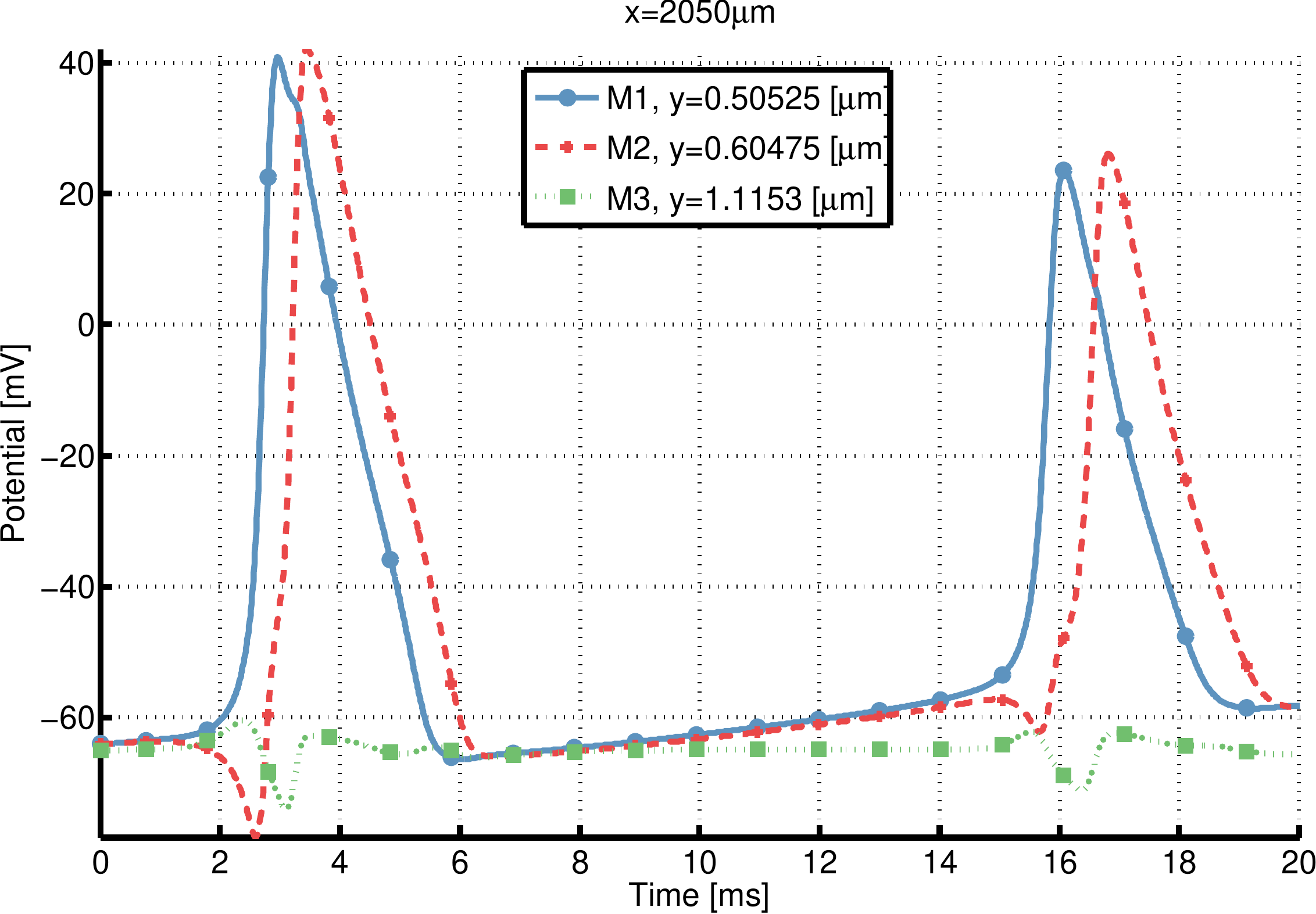}%
\label{fig:multiple_axons.memb_pot_100nm_overview}}%
\subfloat[$d=\SI{50}{\nano\metre}$]{%
\centering%
\includegraphics[width=0.5\textwidth]{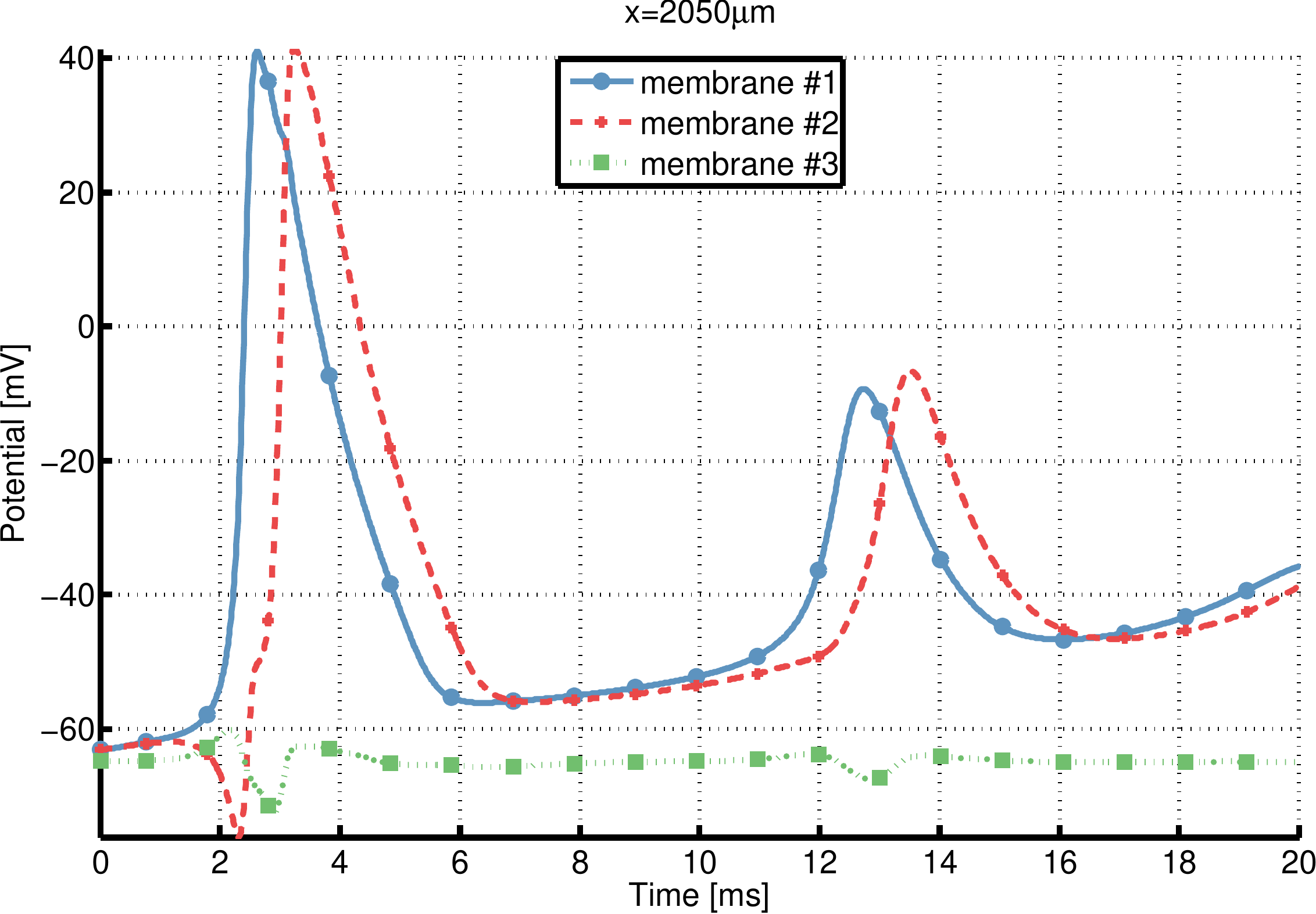}%
\label{fig:multiple_axons.memb_pot_50nm_overview}}%
\mycaption[Membrane potentials for different axon distances $d$ and fixed bundle diameter $d_b$]{For decreasing distances $d$, the ephaptic interactions of axon and bundle changes. For $d=\SI{1000}{\nano\metre}$, only sub-threshold potentials are induced. For smaller distances, an \gls{AP} is elicited in the two bundle membranes. For $d \leq \SI{100}{\nano\metre}$, this even leads to an oscillation between axon and bundle. For $d=\SI{50}{\nano\metre}$, the membrane potential drifts away from resting potential, as the extracellular concentration balance is disturbed.}%
\label{fig:multiple_axons.memb_pot}%
\end{figure}

\Cref{fig:multiple_axons.memb_pot} shows the membrane potential at each of the three membranes, for different distances $d$.
At a larger distance of \SI{1}{\micro\metre}, the effect on the axon bundle is small.
Already at a distance of \SI{500}{\nano\metre}, however, the axon bundle spikes in response to the source axon's \gls{AP}.
Interestingly, the outer membrane M3 fires with a notable delay towards the inner membrane M2.

In a real geometry, the outer axons would, of course, not have two distinct membranes, but outer and inner membrane would be a single connected structure.
By this direct coupling, the axonal membrane would probably fire as a whole instead of having a time delay between inner and outer parts of the membrane, consistent with the ``all-or-none'' principle of action potential generation. 

As axon and bundle come even closer in \cref{fig:multiple_axons.memb_pot_100nm_overview} and \cref{fig:multiple_axons.memb_pot_50nm_overview}, the picture changes. 
Membrane M2 fires earlier, but membrane M3 does not fire anymore.
Instead, membranes M1 and M2 generate an additional \gls{AP} of reduced amplitude, which is even smaller for the case $d=\SI{50}{\nano\metre}$.

These results are quite remarkable. How can they be explained?
For this, we look at the intra- and extracellular potential and charge density time courses directly adjacent to the three membranes.

\begin{figure}
\subfloat[Membrane potential]{%
\centering%
\includegraphics[width=0.5\textwidth]{memb_pot_axonbundle_1000nm-crop}%
\label{fig:multiple_axons.memb_pot_1000nm}}%
\subfloat[Extracellular charge density]{%
\centering%
\includegraphics[width=0.5\textwidth]{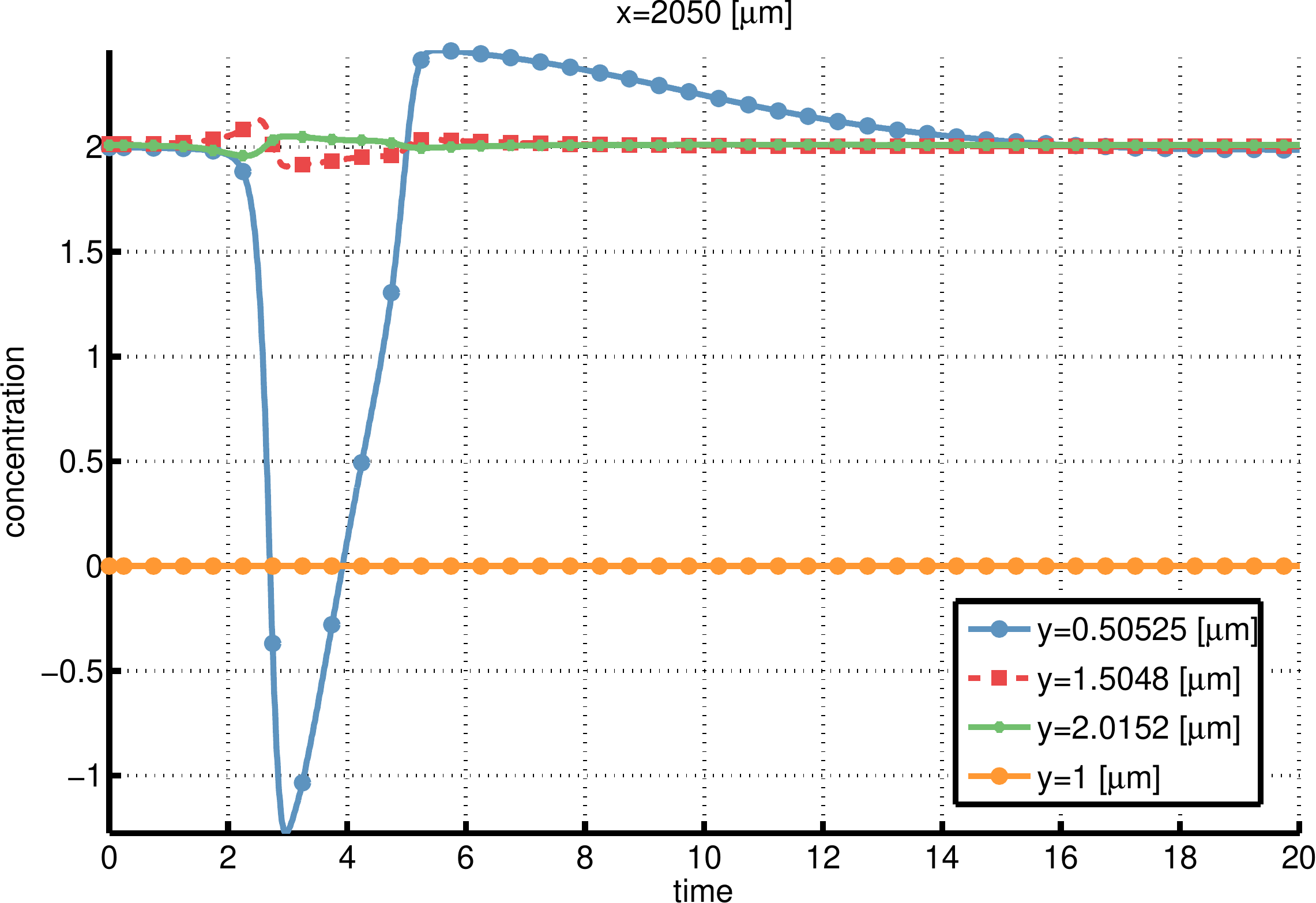}%
\label{fig:multiple_axons.cd_extra_1000nm}}\\%
\subfloat[Intracellular potential]{%
\centering%
\includegraphics[width=0.5\textwidth]{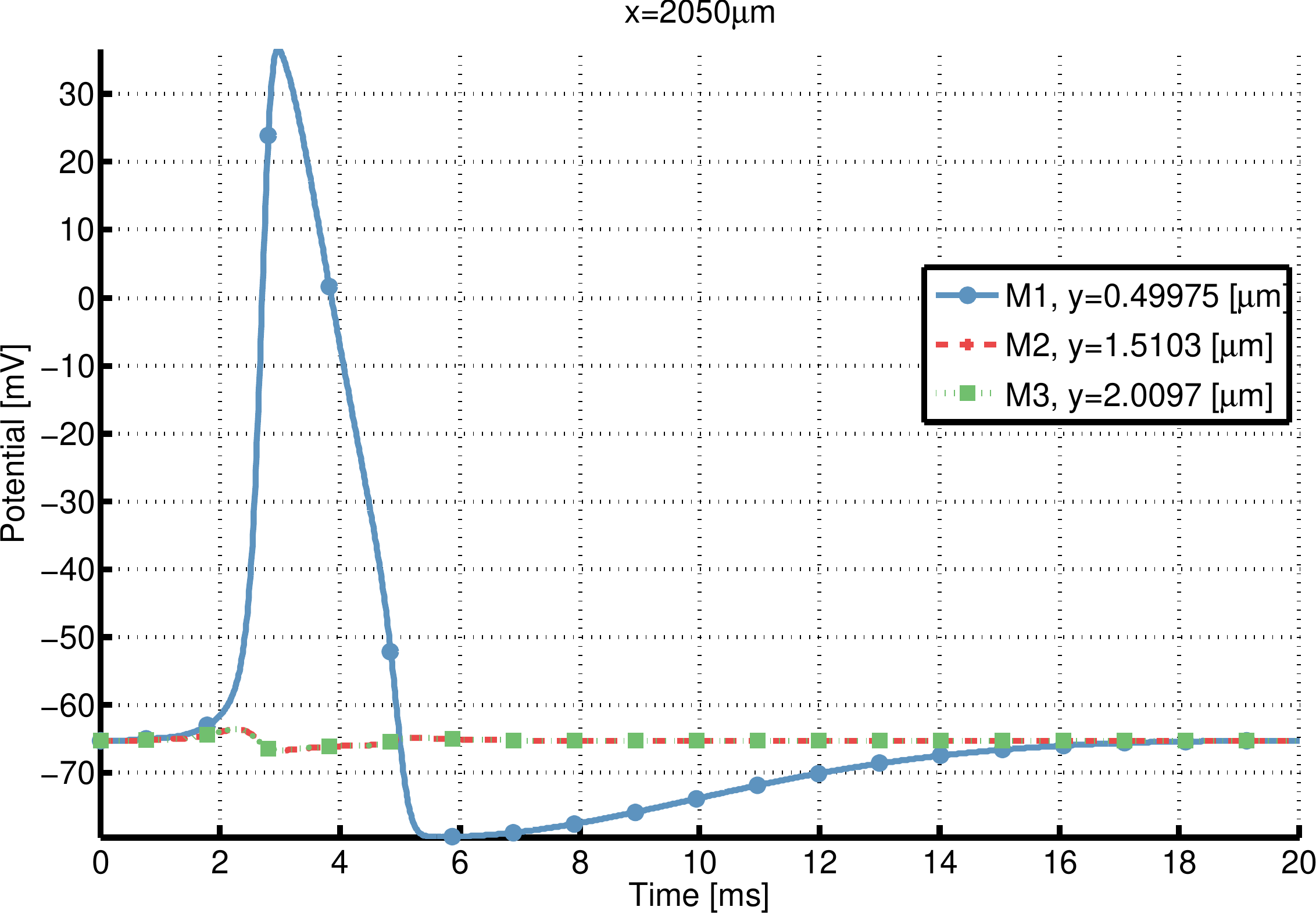}%
\label{fig:multiple_axons.pot_intra_1000nm}}%
\subfloat[Extracellular potential]{%
\centering%
\includegraphics[width=0.5\textwidth]{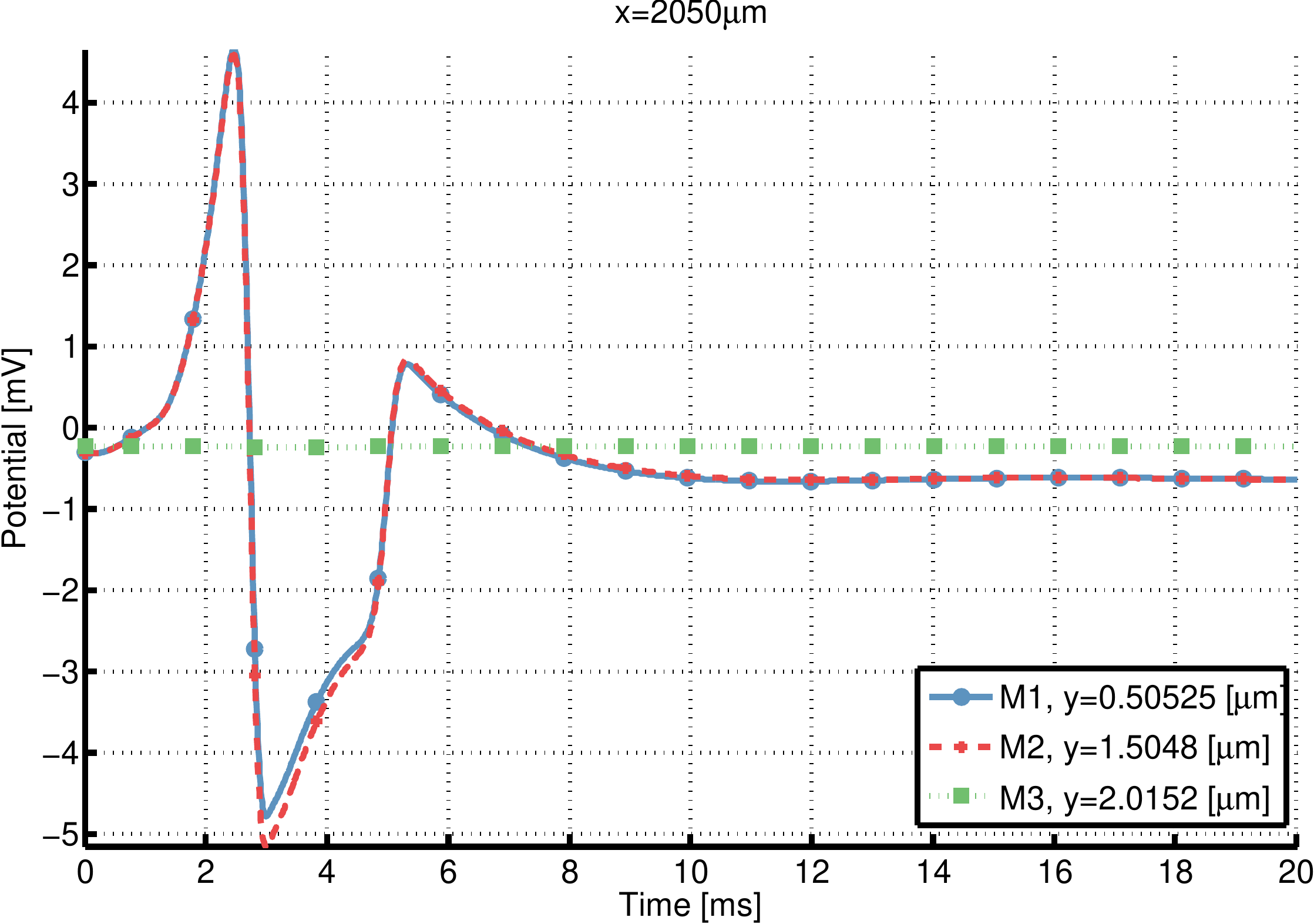}%
\label{fig:multiple_axons.pot_extra_1000nm}}%
\mycaption[Potential and charge density time courses for $d=\SI{1}{\micro\metre}$]{}%
\label{fig:pot_intra_extra_1000nm}%
\end{figure}

For $d = \SI{1}{\micro\metre}$, the intracellular potential in \cref{fig:multiple_axons.pot_intra_1000nm} takes on the expected form of an action potential, albeit with a reduced amplitude in comparison to the isolated unmyelinated axon from \vref{fig:ap}.
A different picture shows for the extracellular potential in \cref{fig:multiple_axons.pot_extra_1000nm}, which is not determined by the \gls{AP} echo, as in the case of an unrestricted extracellular space (cf.~\vref{fig:lfp_debye_snapshot}).
Instead, it reaches an amplitude of several millivolt.
It is clear that the relatively large extracellular potential contributes significantly to the total membrane potential in \cref{fig:multiple_axons.memb_pot_1000nm}.
Previously, the membrane potential was almost exclusively determined by the intracellular potential.
Now, the intracellular potential still contributes most of the membrane potential amplitude, but notable \SI{-5}{\milli\volt} are provided by the extracellular potential.

We note that it is the membrane potential \emph{difference} $\membPot = \phiIn - \phiOut$ which is ``seen'' by the ion channels, such that a depolarization of the cell can be reached by either an increase in the intracellular potential $\phiIn$ or by a decrease in the extracellular potential $\phiOut$.

How much of the membrane potential is contributed by the intracellular or extracellular regime here depends to a large degree on the ratio of intra- and extracellular volumes, see \cref{fig:multiple_axons.volume_ratio_axonbundle}.
We see that the volume of $\OmegaExtraN{1}$ is larger, but within the same order of magnitude as those of the two intracellular domains $\OmegaCytosolN{1}$ and $\OmegaCytosolN{2}$.
In smaller volumes, the same amount of charge results in a larger average charge density, i.e.~the right-hand side of the Poisson \cref{eq:p}, and will therefore cause a larger potential.

We once again acknowledge that the electric potential inherently is a \emph{relative} measure, such that a certain membrane potential difference, as generated by the \gls{HH} system, will be distributed differently onto the intra- and extracellular spaces depending on the respective volumes.
In \cref{chap:unmyel}, the intracellular volume was negligible in comparison with the extracellular one, so the membrane potential was approximately equal to the intracellular potential.
Now, the ratio between intra- and extracellular volumes is not negligible, and consequently part of the membrane potential is shifted to the extracellular potential.

We emphasize that this reasoning only applies to our very special geometry with a severely confined extracellular space, greatly reducing the extracellular conductivity.
In realistic geometries, the \gls{ES} is generally not isolated, but highly connected, such that the extracellular conductivity is not reduced proportionally with volume.

The influence of this large extracellular potential on the adjacent bundle can be seen in \cref{fig:multiple_axons.memb_pot_1000nm} already at this relatively large distance.
The bundle's membrane potential at membrane M2 responds with an initial hyperpolarization, followed by a depolarization of about the same size.
This pattern of a hyperpolarization followed by a depolarization in ephaptic interactions has been reported before using a different model \cite{barr1992electrophysiological}.
It is the consequence of the extracellular potential with its initial rise and the following trough, as visible in \cref{fig:multiple_axons.pot_extra_1000nm}.

\begin{figure}
\subfloat[Membrane potential]{%
\centering%
\includegraphics[width=0.5\textwidth]{memb_pot_axonbundle_500nm-crop}%
\label{fig:multiple_axons.memb_pot_500nm}}%
\subfloat[Extracellular charge density]{%
\centering%
\includegraphics[width=0.5\textwidth]{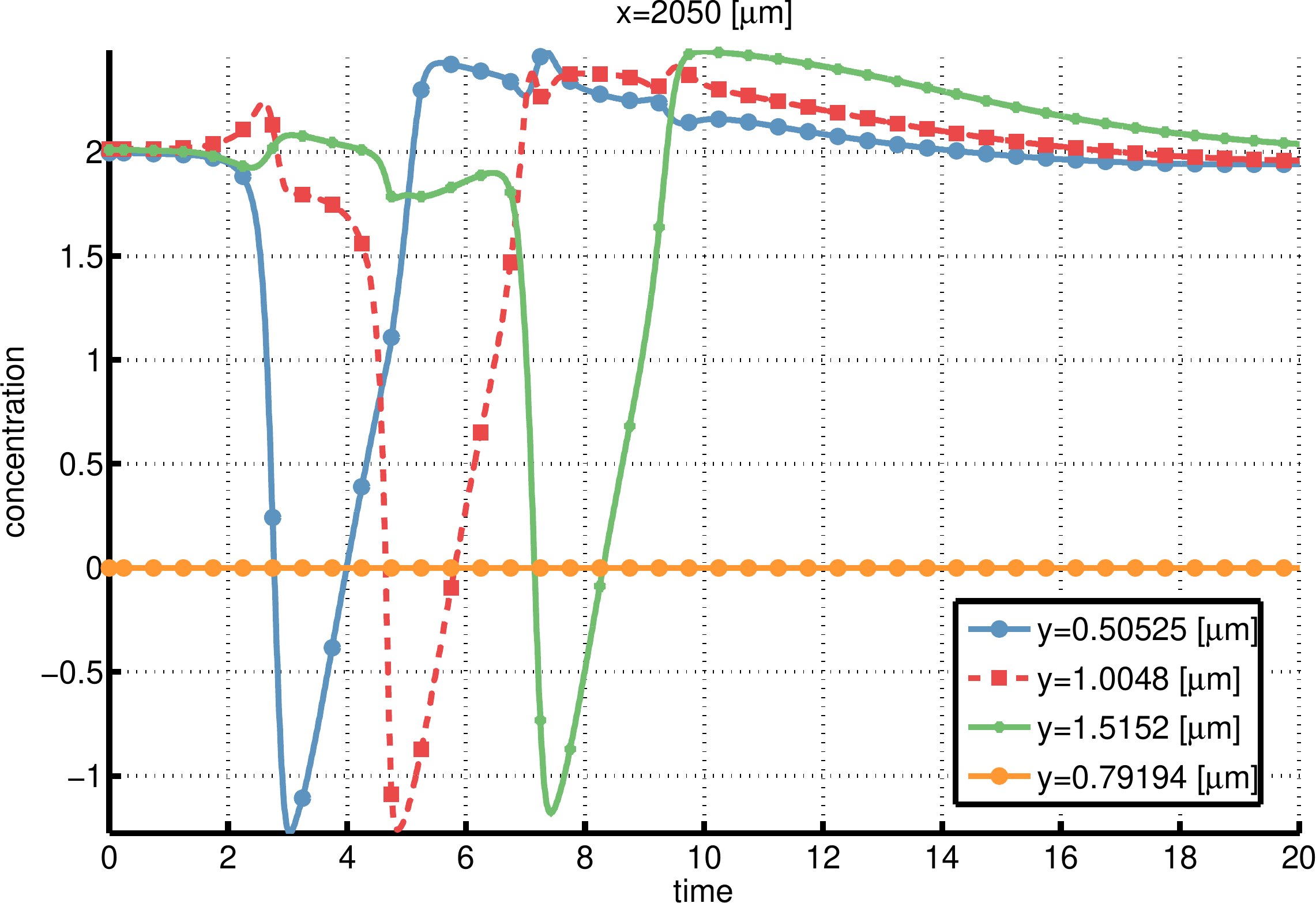}%
\label{fig:multiple_axons.cd_extra_500nm}}\\%
\subfloat[Intracellular potential]{%
\centering%
\includegraphics[width=0.5\textwidth]{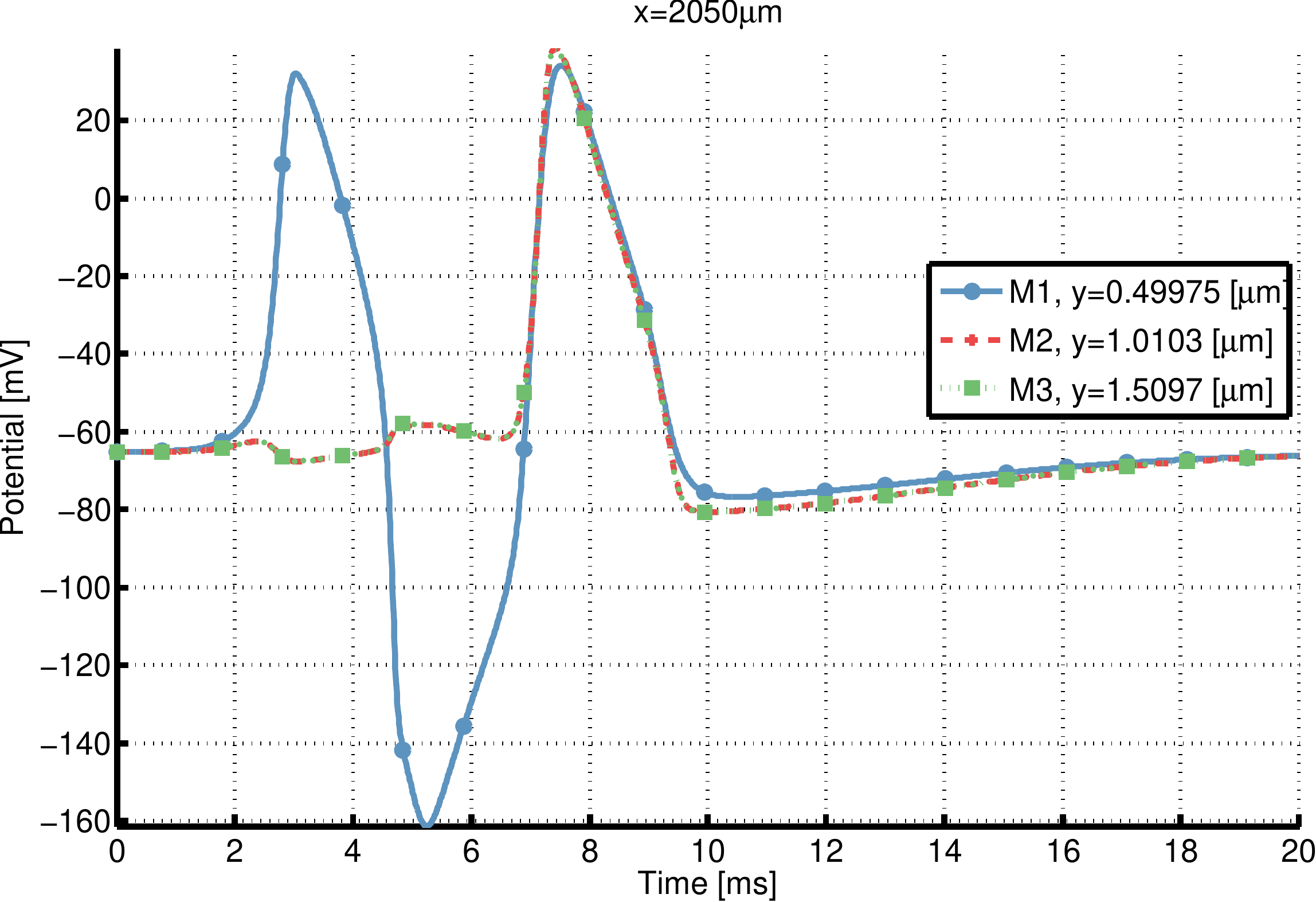}%
\label{fig:multiple_axons.pot_intra_500nm}}%
\subfloat[Extracellular potential]{%
\centering%
\includegraphics[width=0.5\textwidth]{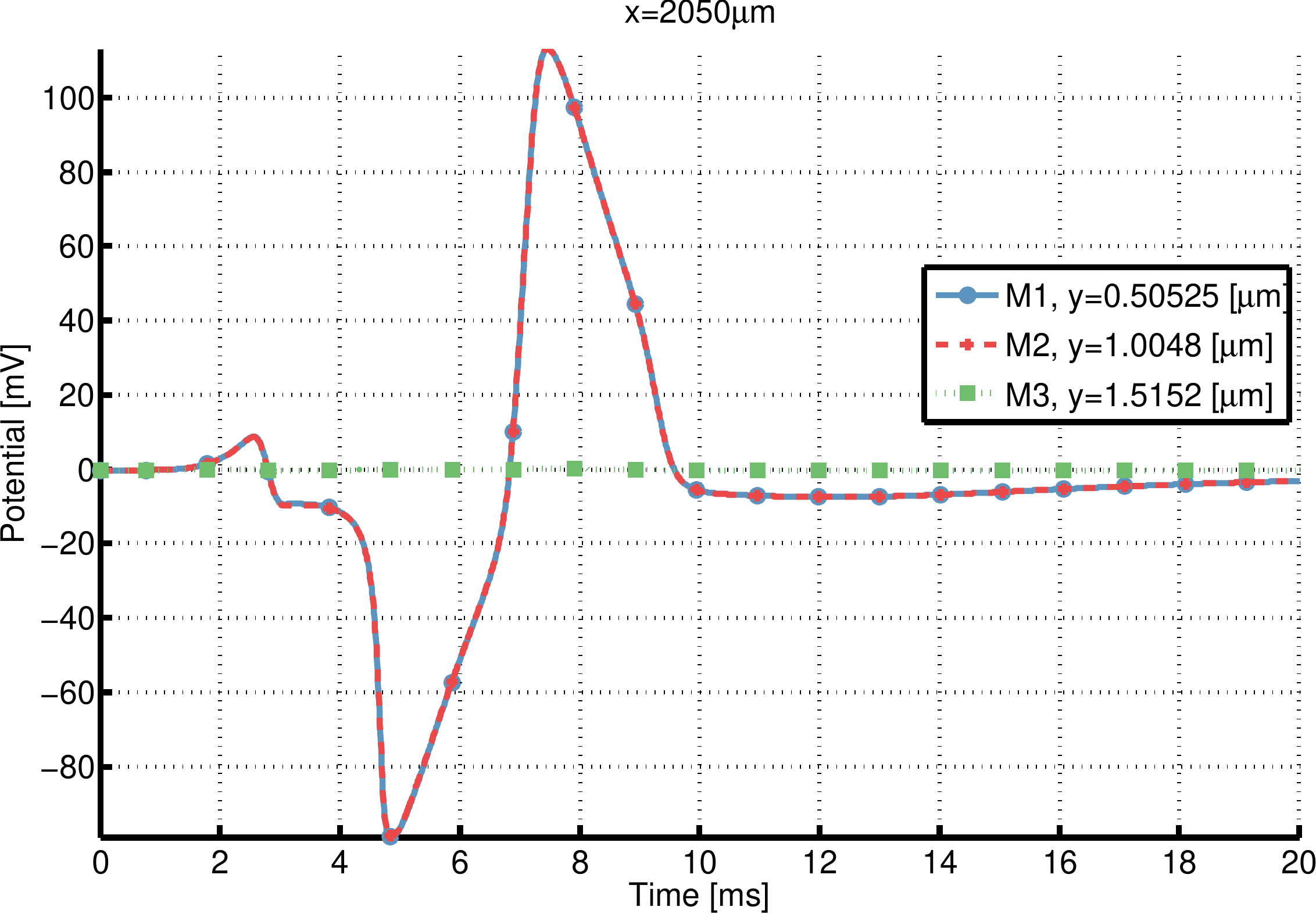}%
\label{fig:multiple_axons.pot_extra_500nm}}%
\mycaption[Potential and charge density time courses for $d=\SI{500}{\nano\metre}$]{}%
\label{fig:pot_intra_extra_500nm}%
\end{figure}

This becomes even more obvious if $d$ gets smaller, as the ratio between intra- and extracellular volumes further increases, as can be seen in \cref{fig:pot_intra_extra_500nm}.
The amplitude of the intracellular potential of the central axon is further reduced, and the \gls{EAP} amplitude next to membrane M1 is increased to about \SI{-10}{\milli\volt} for now.
The extracellular potential at membrane M2 is the same, since it is not significantly reduced over the small range of $d=\SI{500}{\nano\metre}$.
This large \gls{EAP} first causes a hyperpolarization of membrane M2, and then a depolarization, which is strong enough to elicit an action potential, as visible in \cref{fig:multiple_axons.memb_pot_500nm}.
At membrane M2, the ratio of intra- to extracellular potentials now is completely reversed, in accordance with the volume ratio: the extracellular volume is tiny compared to the intracellular volume of the axon bundle.

As a result, the depolarization of membrane M2 is mainly not achieved by an increase of the intracellular potential, but through a huge ``inverse spike'' in the extracellular potential with an amplitude of about \SI{-100}{\milli\volt}.
However, the small intracellular response is enough to trigger an action potential also on the opposite membrane M3, which fires in the normal mode, by an increase in the intracellular potential.
The giant \gls{EAP} generated by the second \gls{AP} is also visible in the intracellular potential of the central axon as a large hyperpolarization.

\begin{figure}
\subfloat[Membrane potential]{%
\centering%
\includegraphics[width=0.5\textwidth]{memb_pot_axonbundle_100nm-crop}%
\label{fig:multiple_axons.memb_pot_100nm}}%
\subfloat[Extracellular charge density]{%
\centering%
\includegraphics[width=0.5\textwidth]{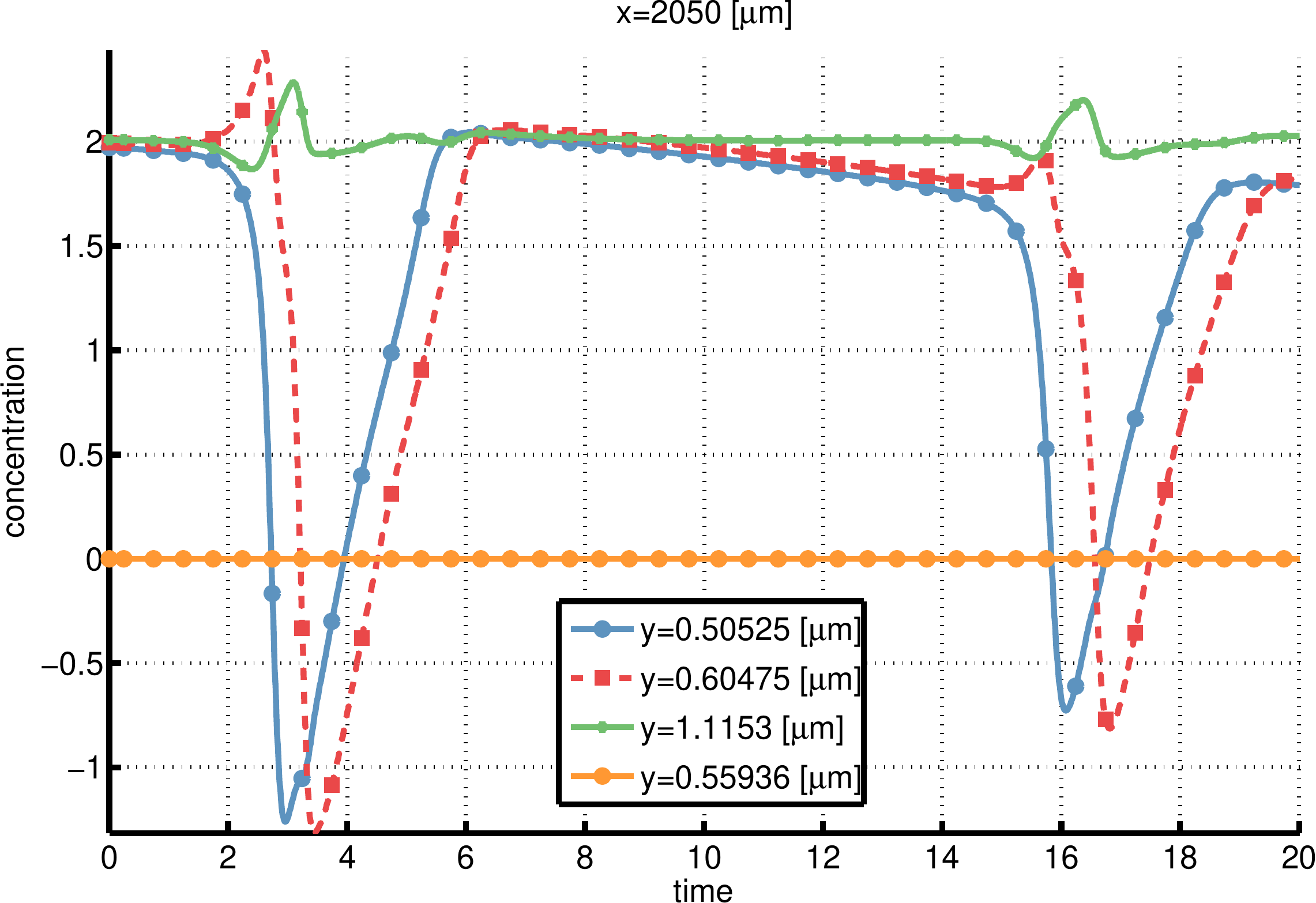}%
\label{fig:multiple_axons.cd_extra_100nm}}\\%
\subfloat[Intracellular potential]{%
\centering%
\includegraphics[width=0.5\textwidth]{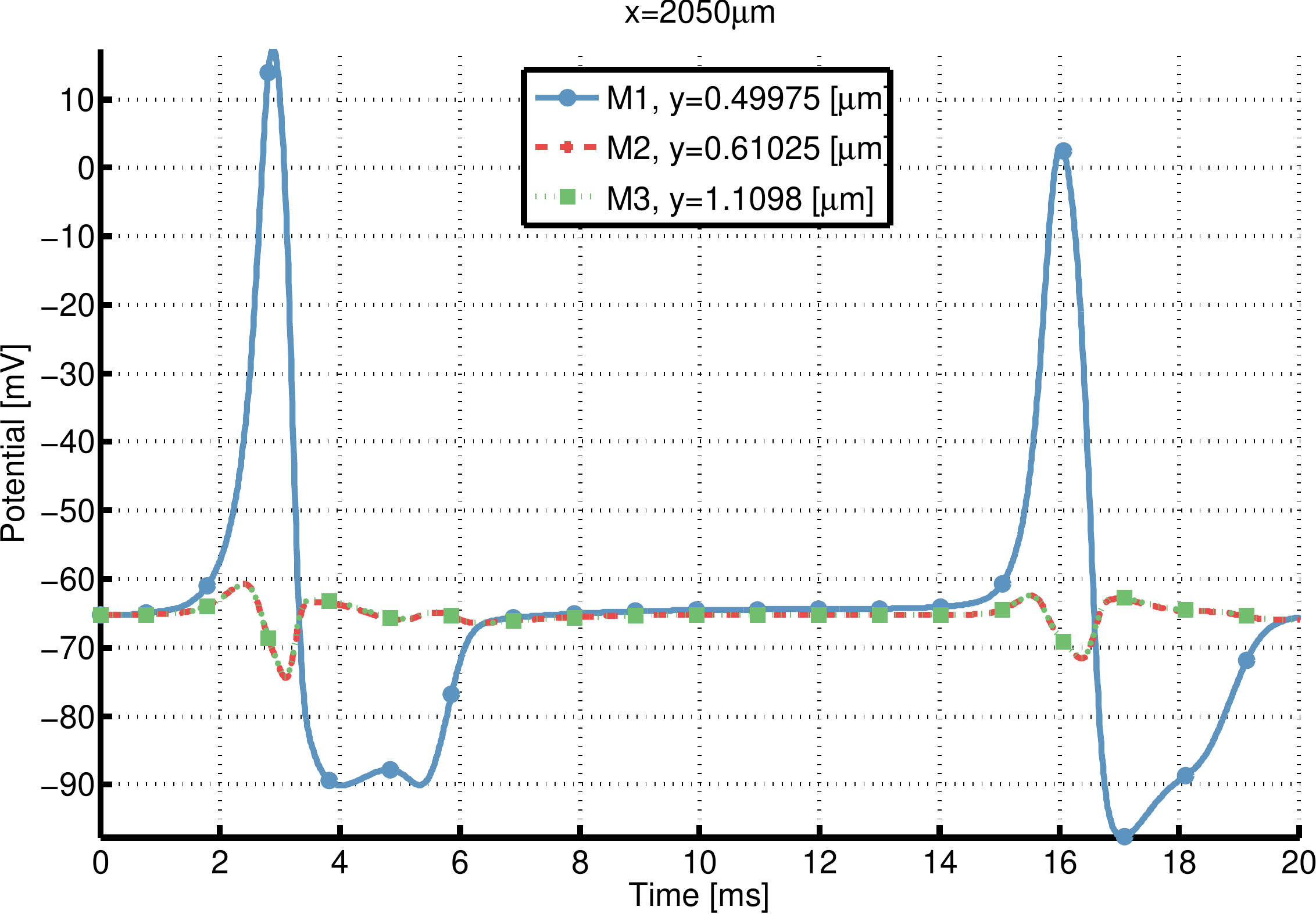}%
\label{fig:multiple_axons.pot_intra_100nm}}%
\subfloat[Extracellular potential]{%
\centering%
\includegraphics[width=0.5\textwidth]{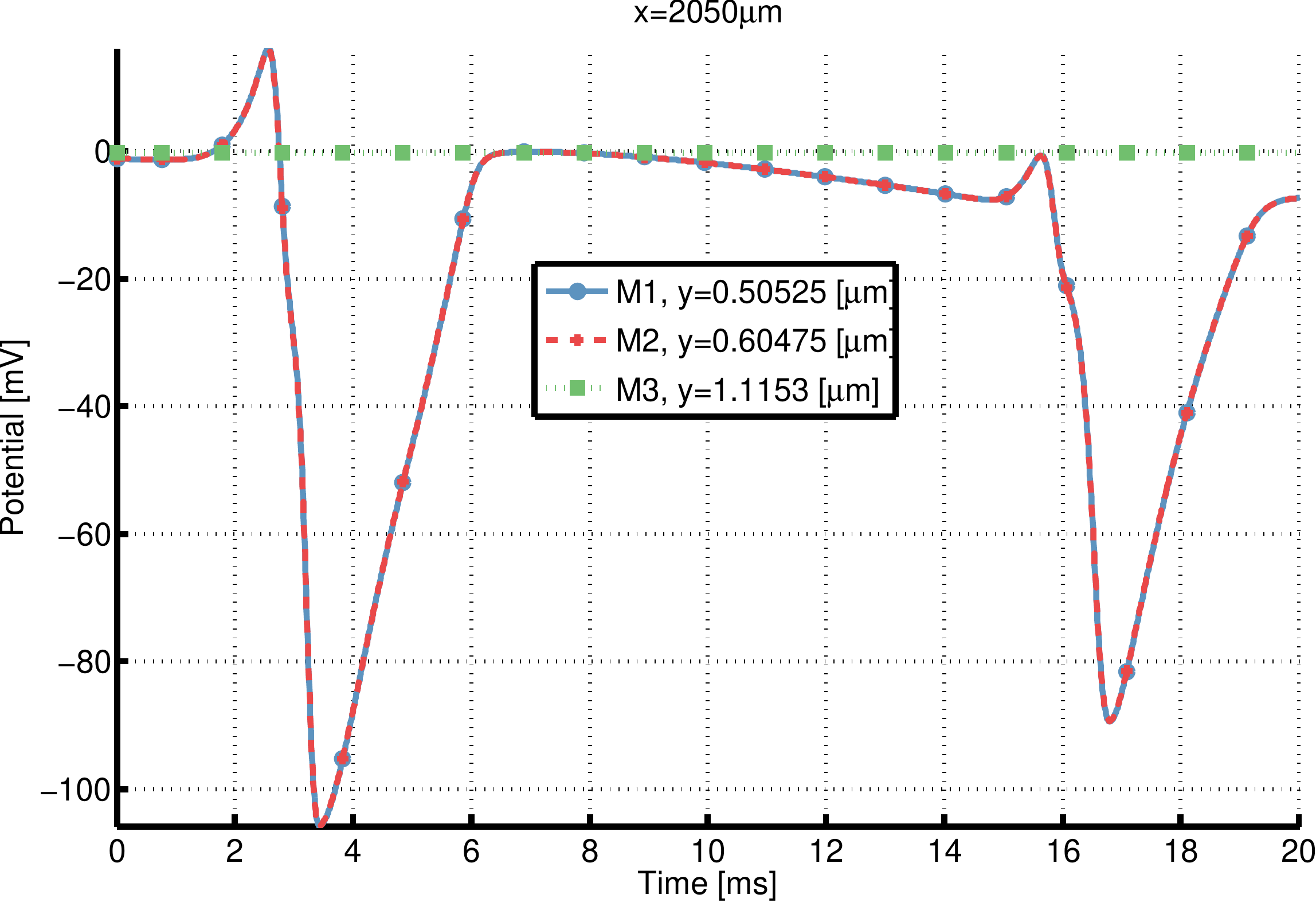}%
\label{fig:multiple_axons.pot_extra_100nm}}%
\mycaption[Potential and charge density time courses for $d=\SI{100}{\nano\metre}$]{}%
\label{fig:pot_intra_extra_100nm}%
\end{figure}

At an even smaller distance of $d=\SI{100}{\nano\metre}$, the extracellular volume is now significantly smaller than the central axon's volume, such that the \gls{AP} is completely generated by a large inverse spike in the extracellular space in \cref{fig:pot_intra_extra_100nm}, as before at membrane M2.
Consequently, an \gls{AP} is elicited also at membrane M2 through the large \gls{EAP}, after an initial hyperpolarization.

Since the volume ratio between $\OmegaExtraN{1}$ and $\OmegaCytosolN{2}$ is now highly in favor of the axon bundle, the intracellular response to the \gls{AP} is even smaller.
This response is not sufficient to depolarize membrane M3 enough to trigger an \gls{AP} there.
But another interesting effect can be observed: the effect of the ``hyperpolarization phase'' in the inverse extracellular spike is large enough to depolarize both membranes M1 and M2 and let them fire once again, albeit with a smaller amplitude.

\begin{figure}
\subfloat[Membrane potential]{%
\centering%
\includegraphics[width=0.5\textwidth]{memb_pot_axonbundle_50nm-crop}%
\label{fig:multiple_axons.memb_pot_50nm}}%
\subfloat[Extracellular charge density]{%
\centering%
\includegraphics[width=0.5\textwidth]{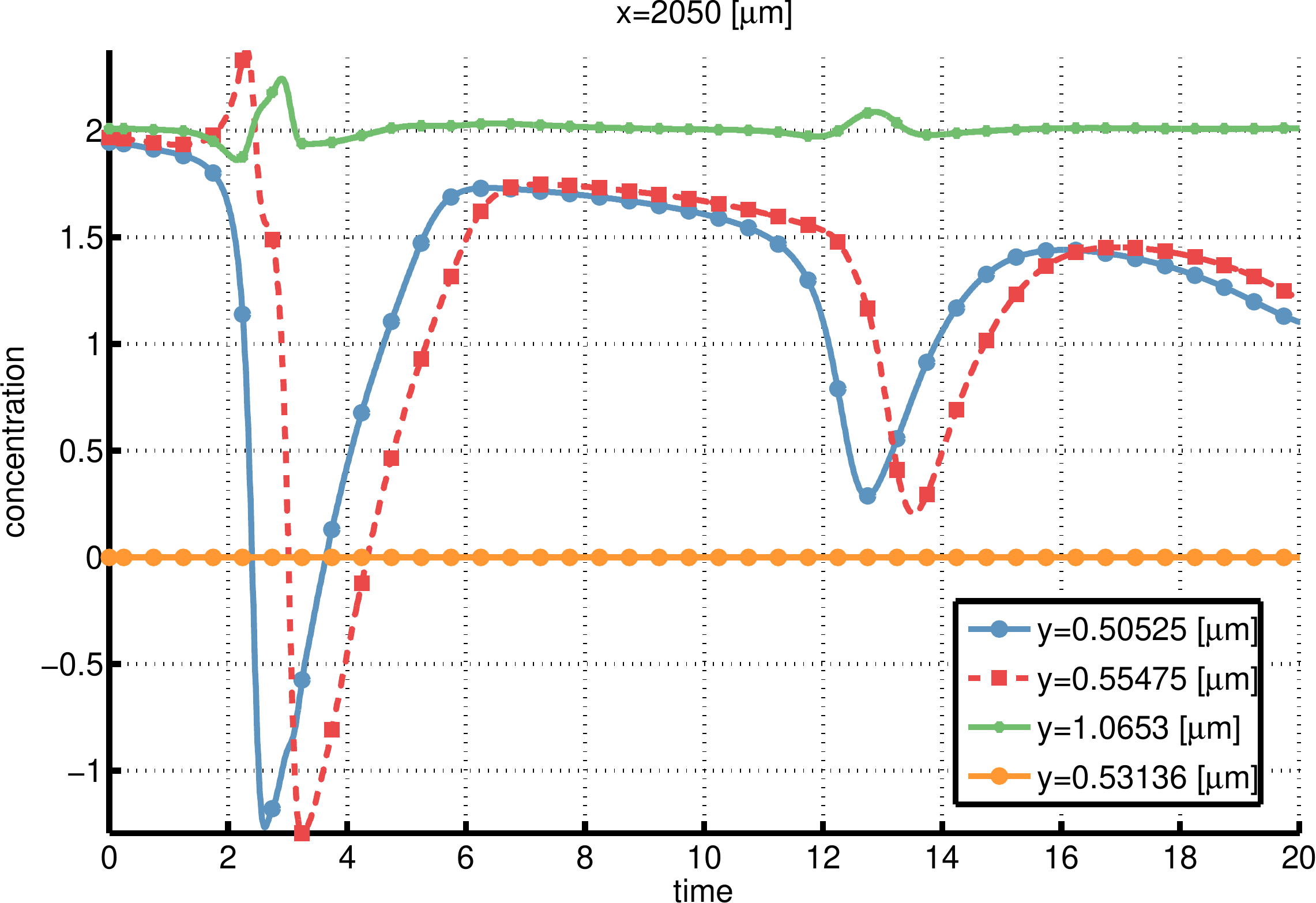}%
\label{fig:multiple_axons.cd_extra_50nm}}\\%
\subfloat[Intracellular potential]{%
\centering%
\includegraphics[width=0.5\textwidth]{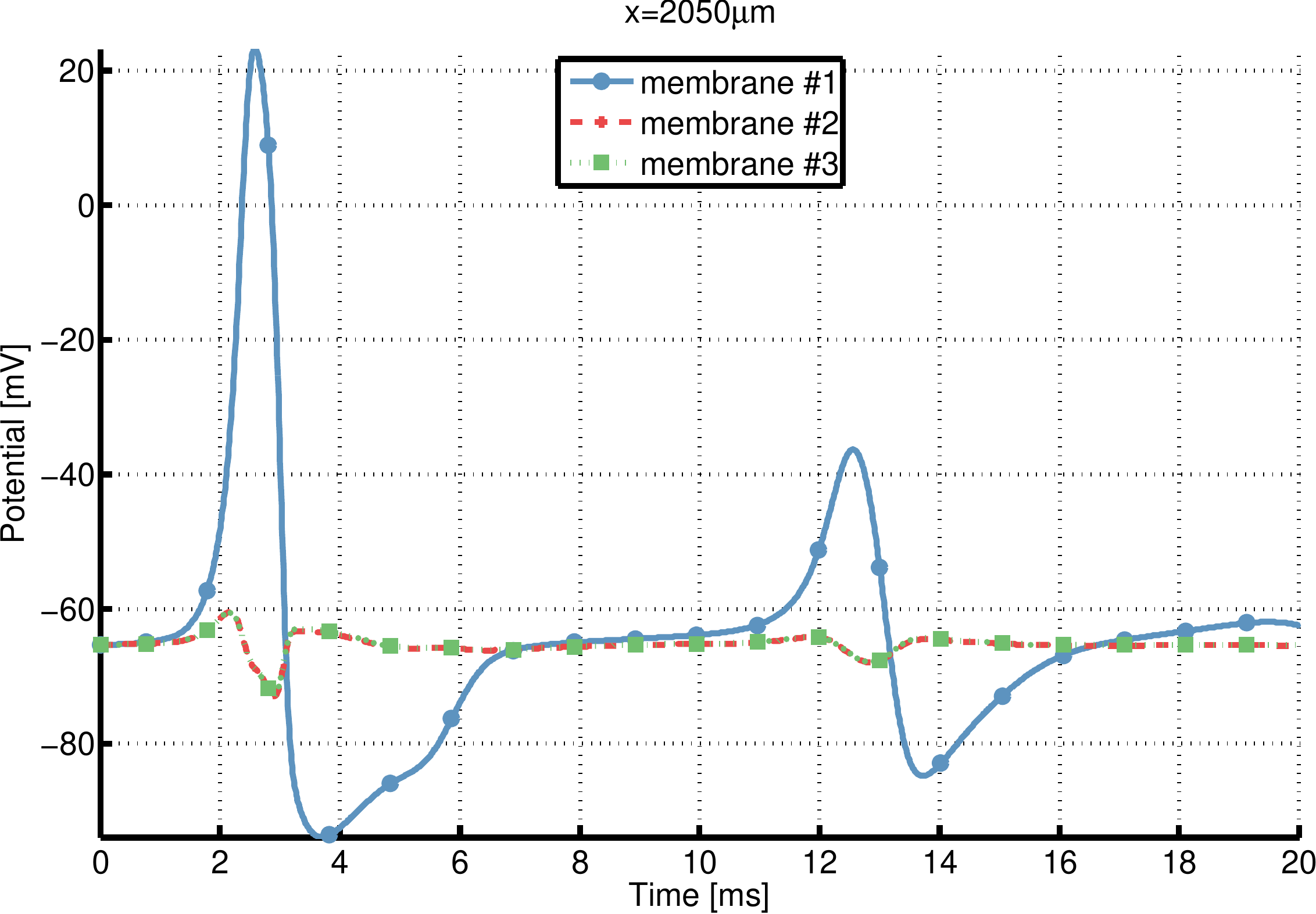}%
\label{fig:multiple_axons.pot_intra_50nm}}%
\subfloat[Extracellular potential]{%
\centering%
\includegraphics[width=0.5\textwidth]{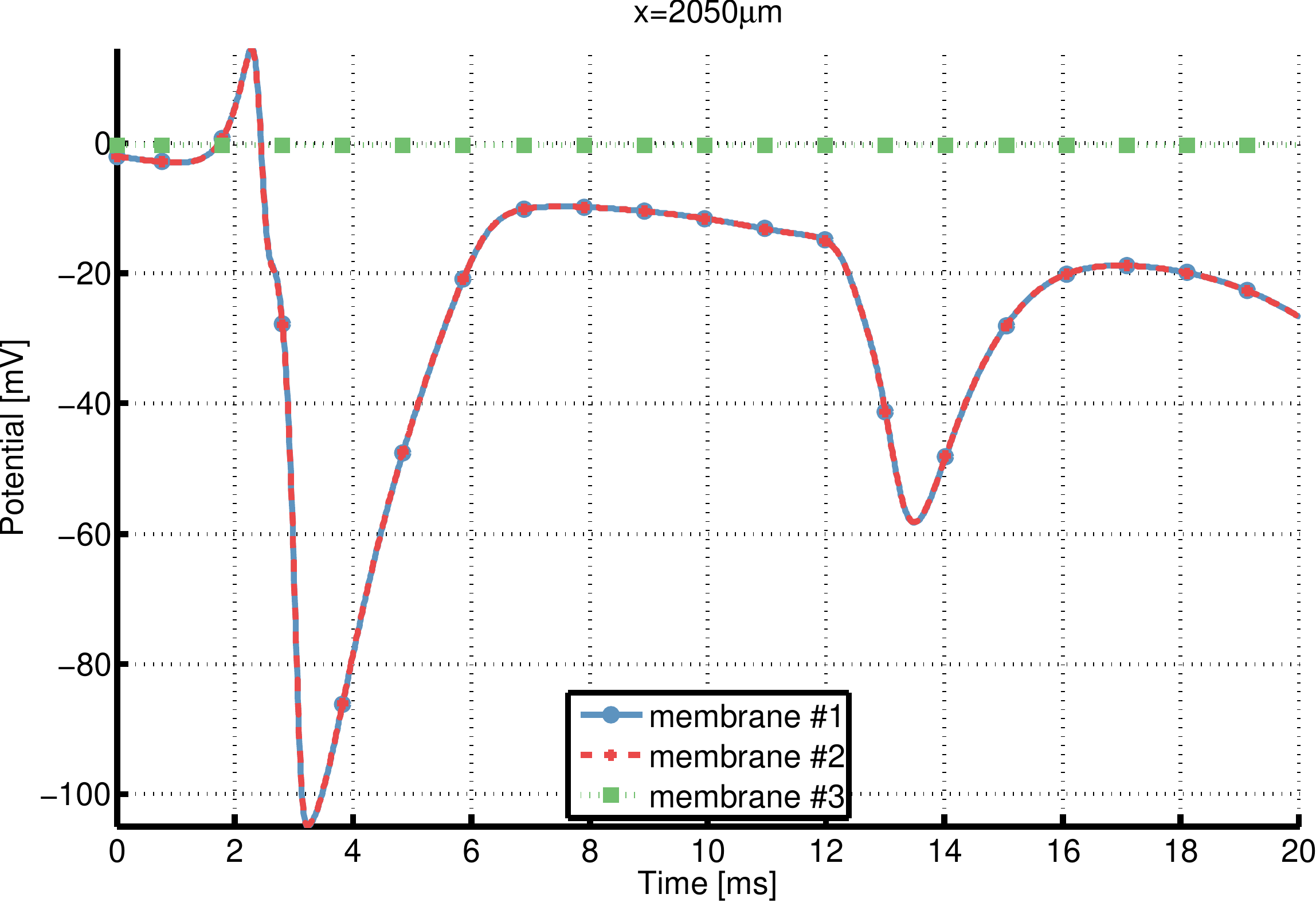}%
\label{fig:multiple_axons.pot_extra_50nm}}%
\mycaption[Potential and charge density time courses for $d=\SI{50}{\nano\metre}$]{}%
\label{fig:pot_intra_extra_50nm}%
\end{figure}

For the case $d=\SI{50}{\nano\metre}$, basically the same qualitative situation as for $d=\SI{100}{\nano\metre}$ can be observed.
However, the extracellular volume is now so small that an undersupply of concentrations is formed.
As a result, the membrane potentials at M1 and M2 do not fully return to their resting potential in \cref{fig:multiple_axons.memb_pot_50nm}.
Two spikelets are generated afterwards, and the membrane potential then further drifts away from its resting potential.
The concentration balance is now disrupted completely.

\subsubsection{Volume-corrected bundle diameter $d_b$}
\begin{figure}
\subfloat[$d=\SI{1}{\micro\metre}$]{%
\centering%
\includegraphics[width=0.5\textwidth]{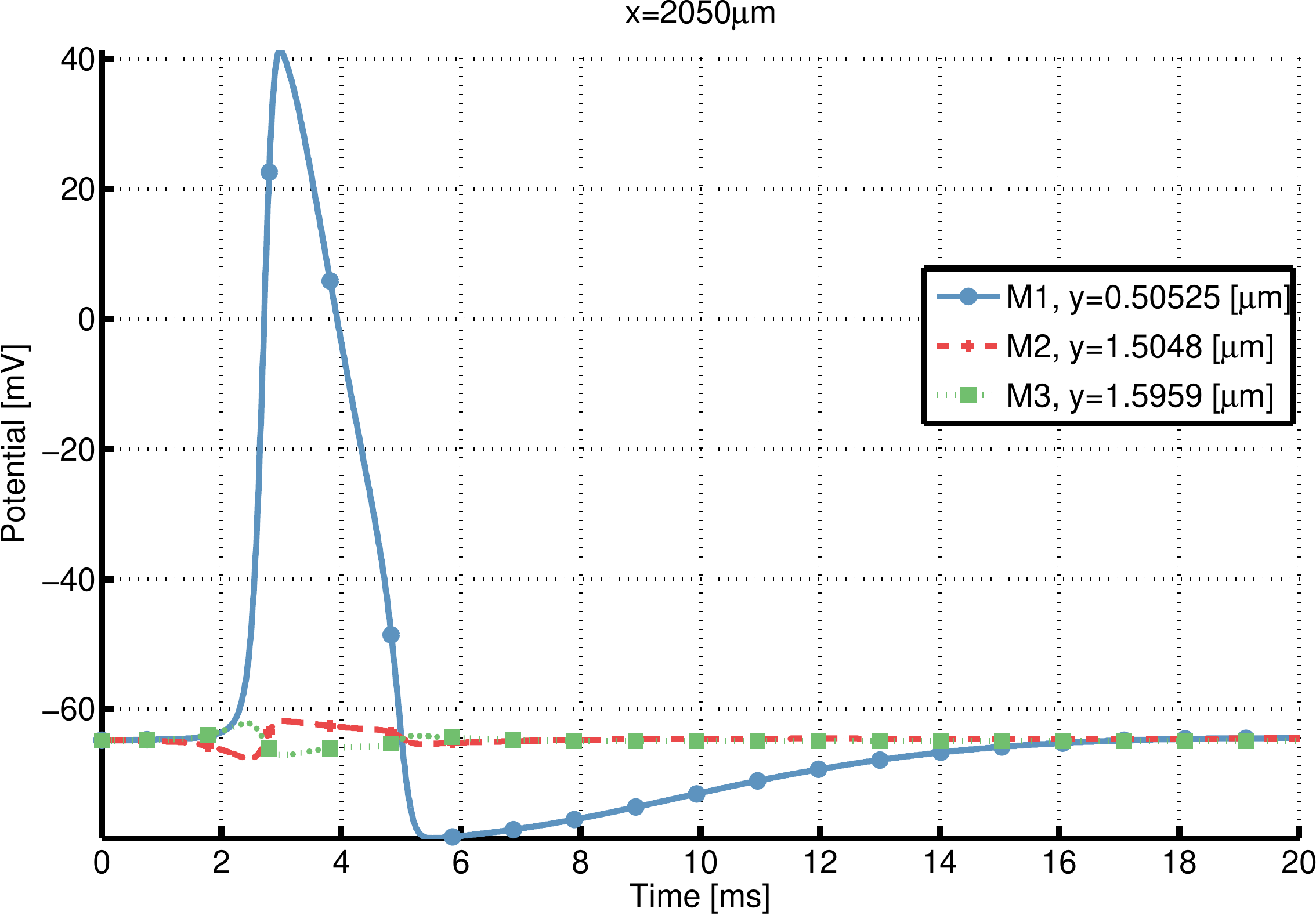}%
\label{fig:multiple_axons.memb_pot_1000nm_volumeCorrected_overview}}%
\subfloat[$d=\SI{500}{\nano\metre}$]{%
\centering%
\includegraphics[width=0.5\textwidth]{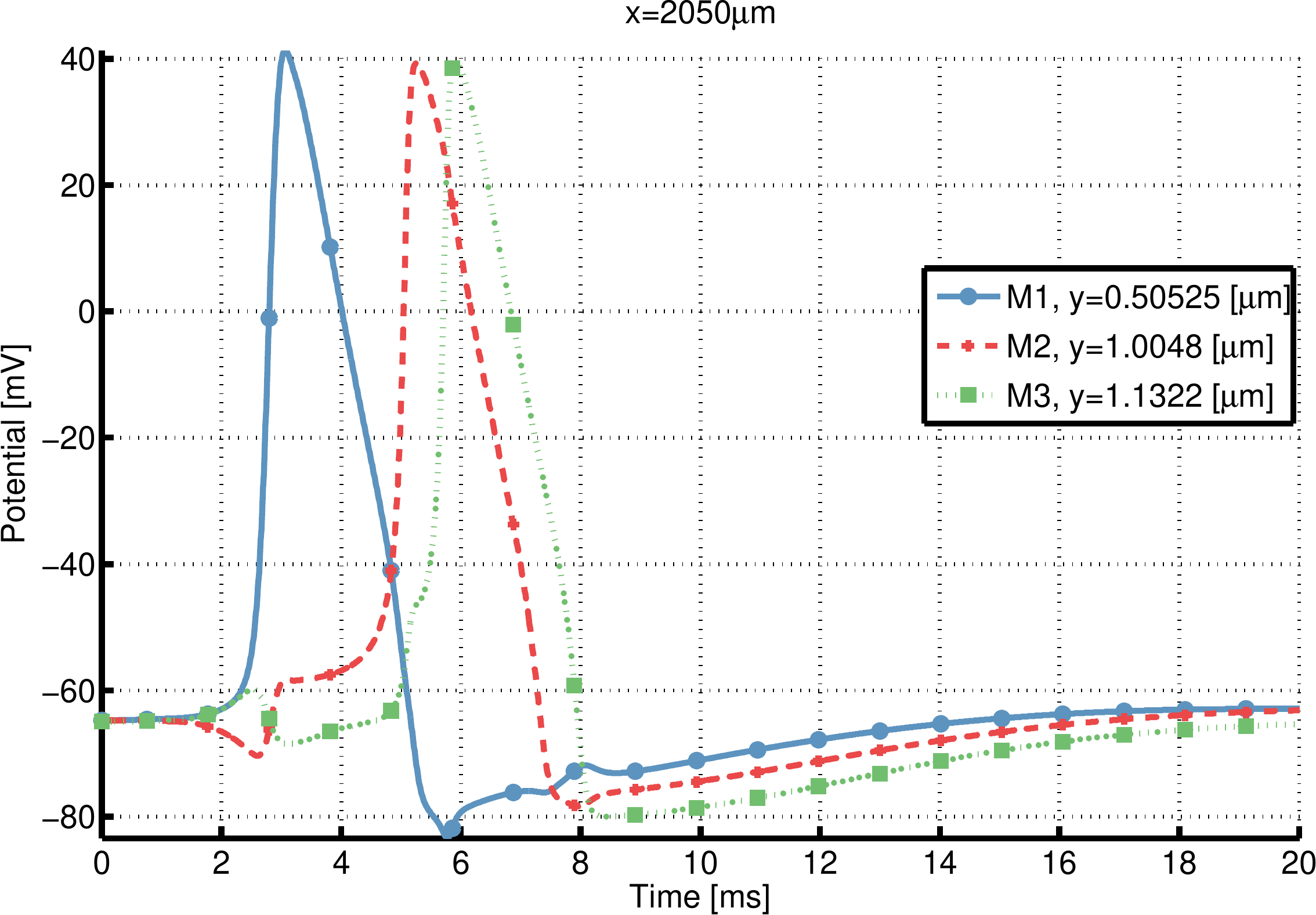}%
\label{fig:multiple_axons.memb_pot_500nm_volumeCorrected_overview}}\\%
\subfloat[$d=\SI{100}{\nano\metre}$]{%
\centering%
\includegraphics[width=0.5\textwidth]{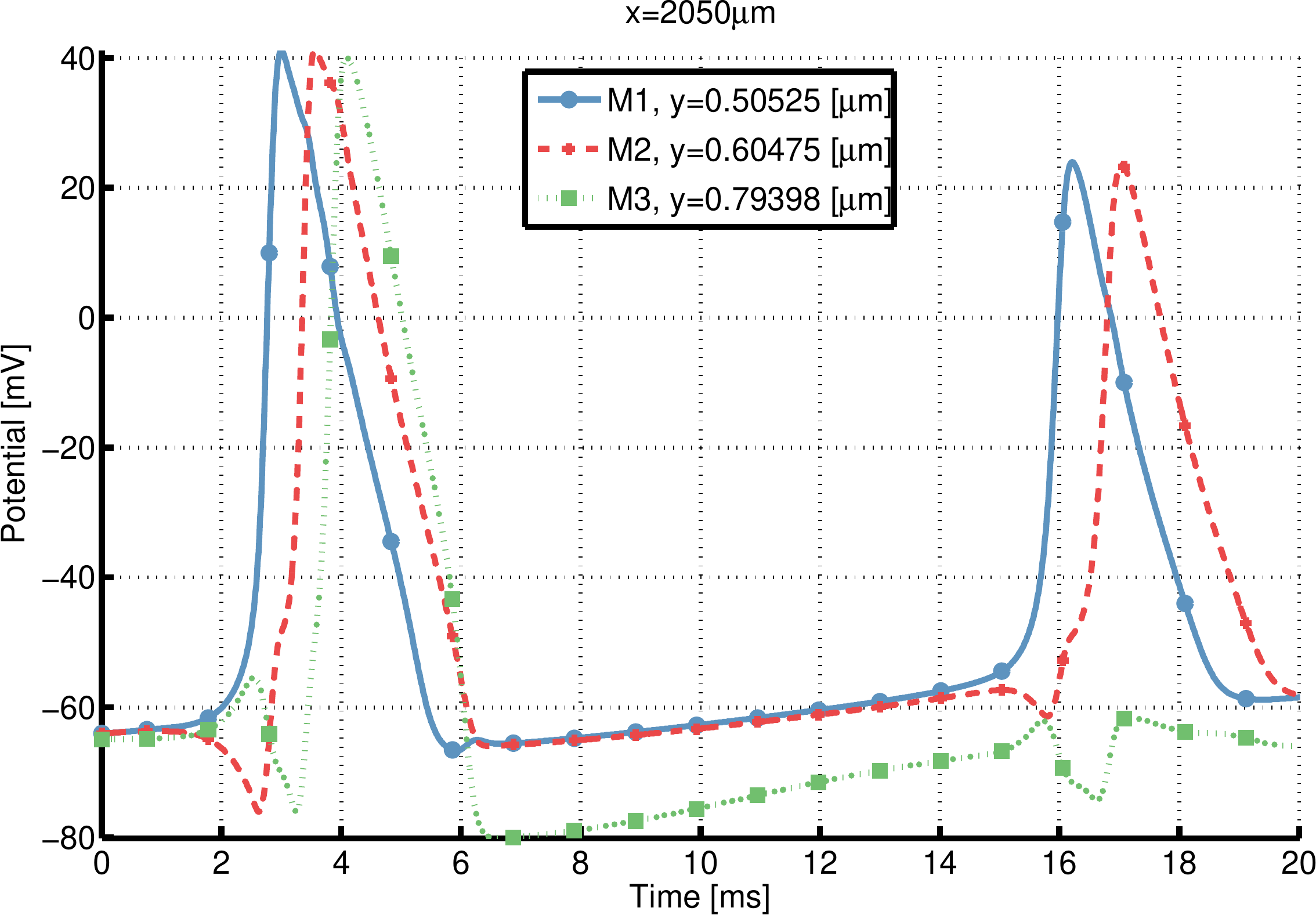}%
\label{fig:multiple_axons.memb_pot_100nm_volumeCorrected_overview}}%
\subfloat[$d=\SI{50}{\nano\metre}$]{%
\centering%
\includegraphics[width=0.5\textwidth]{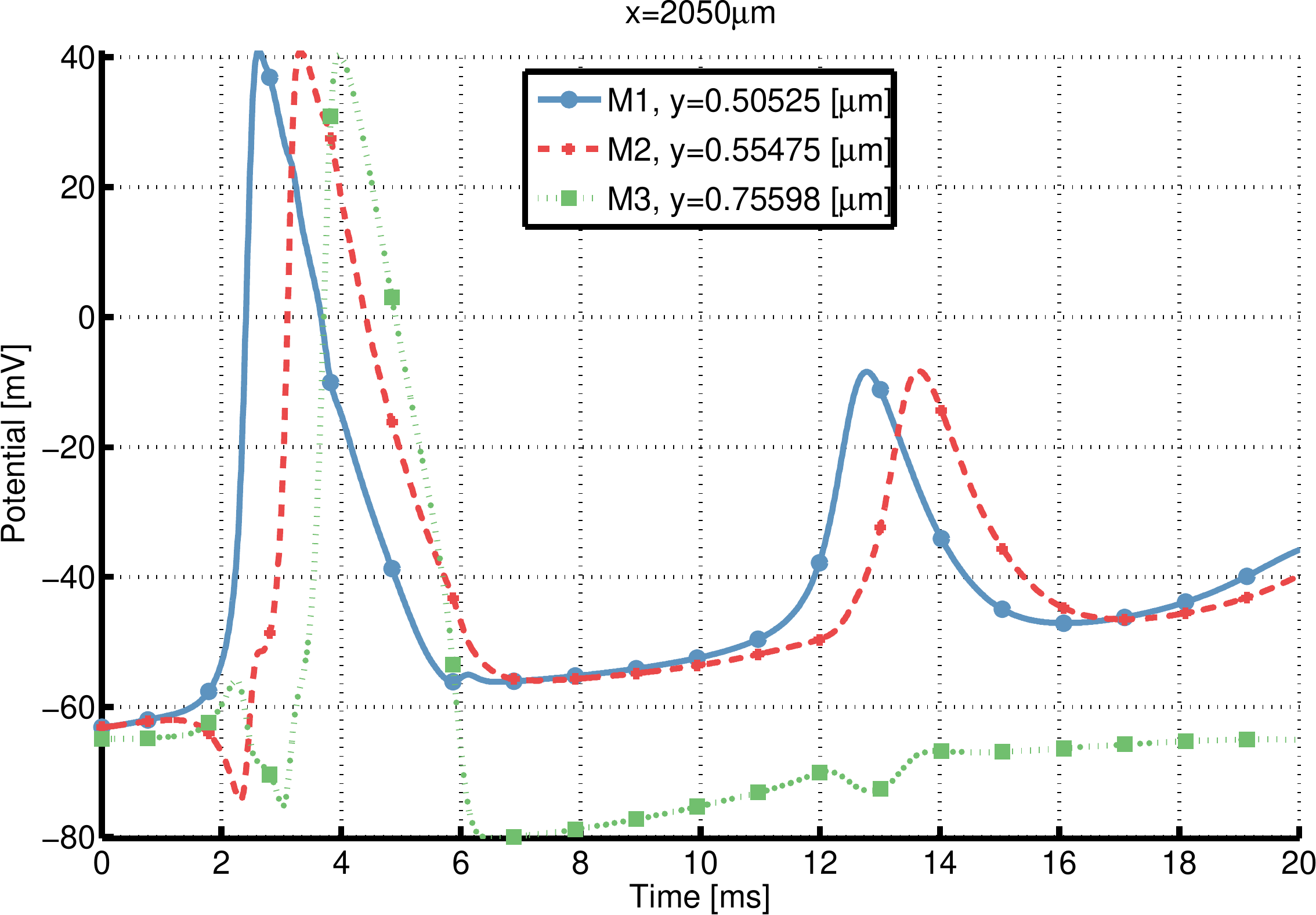}%
\label{fig:multiple_axons.memb_pot_50nm_volumeCorrected_overview}}%
\mycaption[Membrane potentials for different axon distances $d$ and adjusted bundle diameters $d_b$ (volume-corrected)]{}%
\label{fig:multiple_axons.memb_pot_volumeCorrected}%
\end{figure}

The results for the volume-corrected axon bundle in \cref{fig:multiple_axons.memb_pot_volumeCorrected} basically show the same picture as the previously shown findings for varying axon bundle volumes. 
Some minor differences can be made out, e.g.~the reduced delay between membrane M2 and M3 for the case $d=\SI{500}{\nano\metre}$ due to the smaller distance $d_b$.

According to the elaboration above, the ratio of volumes influences the distribution of membrane potential onto the intracellular and the extracellular potential contribution. 
Indeed, the intra- and extracellular potential time courses look completely different than those for the fixed $d_b$, although the membrane potential shows a similar behavior.
We refrain from showing the detailed results here and defer this to \vref{chap:appendix}, see \cref{fig:pot_intra_extra_1000nm_volumeCorrected,fig:pot_intra_extra_500nm_volumeCorrected,fig:pot_intra_extra_100nm_volumeCorrected}.

\section{Summary}
In this chapter, we used a somewhat artificial setting to study ephaptic effects of one axon on a surrounding axon bundle through an extremely isolated extracellular space.
Quite remarkably, the emerging extracellular potentials are large -- even larger than the intracellular ones -- due to the small extracellular volume fraction and the resulting high resistivity.
An action potential traveling along the central axon was able to elicit another \gls{AP} in the surrounding bundle already at a distance of \SI{500}{\nano\metre}, a multiple of the actual average cell-to-cell distance in the brain.
In a more realistic representation, the extracellular space would, of course, be connected and the resistivity would therefore be lower.
Accordingly, the maximum distance for inducing an \gls{AP} in another cell by ephaptic coupling would probably be much lower.

Nevertheless, some interesting effects can be seen in the above results. 
In some configurations, an oscillation between the two fibers can be seen, and the bulk concentrations in the extracellular volumes are thrown out of balance for small distances.
The following discussion will try to put these results into the experimental context and estimate its physiological relevance.

\setchapterpreamble[u]{%
\dictum[Helge Schneider]{Ich habe mich vertan.}\bigskip}
\chapter{Summary and Discussion}\label{chap:summary}
In this work we have set up the mathematical model of axon fibers embedded in extracellular space based on the \gls{PNP} equations of electrodiffusion. 
A suitable method for the numerical simulation of this nonlinearly coupled system of \glspl{PDE} has been presented.
It is designed to properly represent the crucial physical properties of the underlying equations in the numerical scheme.
The key steps to yield a stable and efficient algorithm were:
\begin{itemize}
   \item The grid had to resolve the Debye layer close to the membrane in the normal direction. For the $x$-direction, a much coarser mesh size was sufficient.
   \item The \gls{PNP} system had to be solved in a fully-coupled fashion, as an operator-splitting
approach dramatically reduced the time step size needed for computing a non-oscillatory solution.
  \item The choice of an implicit time-stepping scheme took advantage of higher stable 
time step values and accounted for the diffusion-dominance in the Nernst-Planck equation, 
which we observed for parameters in the physiological range.
   \item The system had to be carefully equilibrated before setting a stimulus, as the concentration
profile towards the membrane had to reach steady-state in order to get meaningful reversal
potentials for each ion channel.
  \item The application of a \emph{threshold volume scaling} was crucial to compensate the large
differences of residual magnitudes introduced by the strongly varying cell volumes in this 
cylinder geometry. With this, the Newton iteration was able to converge even for large domain sizes with an only
slightly lower average time step size compared to the 2D problem.
  \item The blocking of unknowns into dense 4x4 matrices at each grid vertex turned out to be mandatory for large numbers of \glspl{DOF}, since it enabled the exact inversion of diagonal blocks by the \gls{ILU} preconditioner.
  \item A suitable choice of linear solvers and matching preconditioners was critical for the convergence and
efficiency of the algorithm, especially in the parallel case. The application of state-of-the-art 
iterative solvers was a necessary requirement to cope with the grid anisotropy and to enable parallelization on
larger processor counts.
\end{itemize}
We emphasize that, when taking into account the above measures, the \gls{PNP} could be solved numerically with standard methods, i.e.~with conforming linear finite elements for the spatial and an implicit Euler method for the time discretization, despite the multi-scale character of the underlying model.

For the representation of myelin, effective permittivities were introduced.
The permittivities were scaled to correct for the increased myelin thickness of myelin sheaths.
The benefit is that the same tensorial grid structure as for the unmyelinated axon can be used.
The grid in $y$-direction did not have to be changed at all.
The $x$-direction, of course, had to be adapted to represent nodes of Ranvier in correct physical dimensions.
Furthermore, permittivities were smoothed at transitions between myelin and nodes of Ranvier.
This not only suppressed numerical oscillations at these transition zones due to a permittivity jump, it also represents the actual geometry of myelin sheaths, which smoothly increase in width near nodes of Ranvier instead of showing an abrupt change of thickness.

By means of this numerical solution strategy, we were able to simulate an action potential traveling along an active axonal membrane and its effects on the electric field and ion redistribution in the extracellular space.
In an effort to represent all relevant scales of the \gls{PNP} model, the Debye layer was explicitly resolved by the computational grid.

To our knowledge, this is the first application of a 3D electrodiffusion model in the context of neuronal excitation with an explicitly resolved Debye layer, which could be shown to be of capital importance for accurately modeling the \gls{LFP}.

We do not claim that our model is complete or fully realistic from the biological point of view.
Instead, it should be regarded as a first attempt to model the dynamics of neural systems on this detailed scale.
Model refinement concerning the detailed structure and properties of the active membrane (channel types and densities, membrane surface charges) will have to be done when comparing the results with experimental data, as here we largely used classical data from the squid giant axon with only two types of voltage-gated ion channels.

Moreover, a complete model of a neuron with a dendritic tree seems highly desirable, as this would also allow to quantify the influence of synaptic currents, which are thought to be the main contributors of \gls{EEG} and \gls{LFP} signals \cite{buzsaki2012,nunez2005electric}. 
In future models, also the representation of a realistic extracellular geometry -- either explicitly or as a homogenized model -- should be considered.
Particularly, the idealized setup of a single axon fiber in an extracellular space used in \cref{chap:unmyel,chap:myel} consisting exclusively of a homogeneous fluid will rarely occur in reality.
In our approach, the main limitation is the cylinder symmetry, which imposes a serious constraint on the geometry.
The setup for neighboring axons in \cref{chap:multiple_fibers} is somewhat artificial, but it is nevertheless useful to estimate the impact of ephaptic coupling between nerve fibers in a setting of a severely confined extracellular space.
For complex extracellular geometries, a full 3D model would be needed.
This would also allow to use reconstructed geometries by e.g.~electron microscopy imaging techniques.
However, the computational demand of such setups would be much higher than in the present case, and will probably require the use of massively-parallel simulations on supercomputers.
This will require some considerations about tailor-made linear solvers and preconditioners to obtain an optimal scalability of the problem.

Alternatively, using the approximation in \cite{mori2006three,mori2008ephaptic} on a 3D grid promises to be better tractable. The authors used a special boundary to include Debye layer effects without having to explicitly resolve it in the discretization, also eliminating the concentration variables from the equations.
Nevertheless, a full 3D electrodiffusion simulation seems desirable, if only to validate the accuracy of this approximation under physiological conditions. 

However, one of Mori's main results, the postulation of an intermediate diffusion layer between Debye layer and bulk solution on the order of $\sqrt{\dDebye}$, could already be confirmed by our simulations.
We found prominent deviations from volume conductor models for distances of at least \SI{5}{\micro\metre} from the membrane.
These are a consequence of violations of the electroneutrality assumption in volume conductor models in the Debye layer, which shaped the extracellular electric field to a large degree also in the adjacent diffusion layer.
This has been demonstrated for the unmyelinated case in \cref{chap:unmyel} as well as for the myelinated one in \cref{chap:myel}.
Both cases, unmyelinated and myelinated axon, share a common property: the Debye layer shows an extracellular potential that resembles the membrane potential, only reduced in amplitude, where the reduction is depending on different membrane and electrolyte material parameters. 
We called this the \gls{AP} echo.

Interestingly, the existence of this echo is also observed in experiments. 
Several experimentalists from Andreas Draguhn's lab in Heidelberg have confirmed that they see a sudden switch in potential shape when pushing the recording electrode against the neuronal membrane like it is done in juxtacellular recordings (Martin Both, Florian Bähner, Christian Thome, personal communication). 
This phenomenon can now be explained by the Debye layer dynamics present in our model.

At large distances, results were in good agreement with volume conductor models, consistent with an experimental validation \cite{gold2006origin}.
The fact that the LSA model is not a good representation at closer membrane distances has been reported before \cite{gold2006origin,holt1999electrical} with a critical distance of about \SI{1}{\micro\metre}.

Considering the typical dimensions of extracellular space (volume fraction $\alpha = 0.2$, width \SIrange{38}{64}{\nano\metre} \cite{sykova2008diffusion}), one has to conclude that each point of interest will necessarily always be very close to another cell membrane under physiological conditions.
Therefore, in reality Debye and diffusion layer effects will not be negligible.
Instead, the effects of ion concentrations have to be taken into account.
This indicates that representing the \gls{ES} as a homogeneous medium by a single scalar resistivity $\rho$ is not an accurate description.

Several previous models have dealt with this issue; in \cite{bedard2004modeling,bedard2009macroscopic}, the authors used an inhomogeneous conductivity in the \gls{ES} to account for the frequency filtering properties of extracellular space observed in experimental recordings.
This approach might also be used to find an effective conductivity distribution that implicitly incorporates the concentration effects.
However, it is unclear how to find the correct conductivities as a function of space. Even if the matching conductivities can be found, the question remains if such an approach will be successful in reproducing the full electrodiffusion dynamics and which spatial resolution will be required.
The closest approximation to the electrodiffusion equations without explicitly resolving the Debye layer and including concentration effects only implicitly is described in \cite{mori2006three}.
A comparison to the full electrodiffusion model in a 3D setting would be of great interest.

In \cref{chap:multiple_fibers}, we made an attempt to quantify ephaptic effects between an axon fiber and a surrounding bundle of fibers.
Due to the restrictions of the cylinder symmetry, this model does not represent the typical geometry of extracellular space, which is highly connected and tortuous, while it is rather isolated in our model.
It is, however, not completely unrealistic either, as there is some evidence for regions with a very isolated \gls{ES} or even ``dead ends'' in the extracellular matrix \cite[e.g.~Fig.~1]{sykova2008diffusion}.
In any case, the restricted volume and connectivity of the interstitial space in our model corresponds to a strongly reduced conductivity in terms of volume conductor theory.
Such an increase in electric resistivity favors the development of large extracellular potentials.
The simulation results show that already at relatively large inter-fiber distances of \SI{500}{\nano\metre}, one axon can elicit an action potential in the surrounding fiber bundle purely by electrical coupling.

We also showed the effect of different intra- and extracellular volume ratios: while the membrane potential did not change appreciably, the relative distribution of potential contributions by the intra- or extracellular side changed to a large degree.
This revealed another firing mode in which a large negative extracellular potential could depolarize the cell, without a large change in the intracellular potential, as in the classical firing mode.
This is plausible, since the ion channels only see the membrane potential difference $\membPot = \phiIn - \phiOut$, therefore a depolarization can be reached by either increasing $\phiIn$ or by decreasing $\phiOut$.
This mode of firing is interestingly not an artifact of our model geometry: it has, in fact, been reported before in \emph{in vitro} recordings of rat hippocampus CA1 cells (cf.~section \rom{5}.C in \cite{faber1989electrical}) and the squid giant axon (originally in \cite{ramon1978ephaptic}, and reviewed in \cite[Fig.~1]{jefferys1995nonsynaptic}). 

Our model shows large extracellular potentials ranging up to \SI{100}{\milli\volt}, e.g.~in \cref{fig:multiple_axons.pot_extra_50nm}. The emerging extracellular potentials are sometimes even larger than the intracellular ones, due to the severely restricted volume.
We also conducted some research on the experimental evidence for the large \gls{LFP} magnitudes we found in \cref{chap:multiple_fibers}.
To our surprise, comparable magnitudes have indeed been found in some cases. In \cite{barr1992electrophysiological}, a number of references are given for \emph{in vivo} measurements of large extracellular potentials with magnitudes larger than \SI{50}{\milli\volt}.

It is clear that such large \gls{LFP} magnitudes will suffice to elicit action potentials in neighboring cells, which is also the case in our model. When bringing the fibers closer together, we could even observe a ``ping-pong effect'': the induced \gls{AP} in the axon bundle in return produced a large extracellular response that was enough to elicit a second spike in the source axon.
Again, we found evidence for such a phenomenon, mentioned in \cite{barr1992electrophysiological}.

The authors also give a reason why such large potentials have only been measured in a few cases so far: they arise in areas where cells are most packed, i.e.~where the conductivity is very low.
These tightly packed areas are least accessible to electrodes, and any penetration by the electrode will inevitably increase the conductivity by destroying surrounding tissue, thereby diminishing the packing density and exposing the measurement site to regions with a higher extracellular volume.
In \cite{jefferys1995nonsynaptic}, again the rodent CA1 region is mentioned in this context -- together with a multitude of studies that assess the effect of electric fields within this area -- since it has an extremely high resistivity (more than twice the intracellular one), presumably due to the very dense packing with an unusually low extracellular volume ratio of $\alpha = 0.12$.

One more word on the experimental point of view: if the (invasive) measurement of large \glspl{LFP} is prevented by the measurement itself, this could mean that large extracellular potentials are a ubiquitous phenomenon that remains largely undetected with today's experimental methods. If we take this thought further, one might also argue that many action potentials would not be identified as such in patch-clamp measurements, as only the intracellular potential $\phiIn$ with reference to a distant grounding electrode is observed. 
In extreme cases, action potentials elicited by the second firing mode, i.e.~triggered by a large negative extracellular potential $\phiOut$ without any large changes in the intracellular potential, would not be recognized.
These ``extracellularly-induced spikes'' could only be seen by measuring both $\phiIn$ and $\phiOut$, in order to obtain the true membrane potential $\membPot$.

\bigskip

From a different perspective, the extracellular potentials generated by our model might also be of interest for another application. In \cite{anastassiou2013biophysics}, the authors mention the context of \emph{parameter estimation} for neuron models, particularly estimating spatially distributed channel conductances from intracellular measurements made only at a small number of points -- in the extreme case, from only one patch-clamp electrode.
This problem is known to be an underdetermined optimization problem, given the fact that many different parameter sets can produce the same voltage response, as is shown in a plethora of different studies \cite{achard2006complex,druckmann2007novel,druckmann2008evaluating,bhalla1993exploring,tabak2000parameter,keren2005constraining,vanier1999comparative,van2007neurofitter,van2008automated,banga2004global}.

The fact that there exist many local minima renders the application of classical, gradient-based methods extremely difficult, which is also the reason why many of the aforementioned studies utilized novel global optimization techniques like genetic algorithms or simulated annealing to overcome this problem -- although there is also recent progress when using classical methods, as long as the range of stimuli given to the model is sufficiently diverse to tickle out the influence of every single parameter in this model \cite{toth2011dynamical,kostuk2012dynamical}.

Given these difficulties in obtaining a meaningful and realistic parameter set for a given model, the authors in \cite{anastassiou2013biophysics} suggest to include extracellular measurements next to the commonly used intracellular ones and report a markedly decreased variability in the obtained parameter sets.
We note that using the presented electrodiffusion approach, extracellular signals may be used at any point in the \gls{ES}, while volume conductor models are restricted to certain minimum distances from the membrane, because they fail to reproduce the correct potentials close to the membrane.
Of course, using electrodiffusion models for such parameter estimation procedures will increase the computational complexity to a large degree, but it
might further increase the performance of the optimization strategy, as sharply varying (and therefore informative) juxtacellular signals may be used instead of the smoothed-out, less informative farfield signals.

\bigskip

In summary, we have shown that electrodiffusion models have proven useful in the context of neuroscience, especially when assessing the influence of variable concentration distributions on the extracellular electric potential.
We hope that this thesis will be useful for readers who want to employ electrodiffusion models in comparable or related contexts.

The reader should keep in mind that the work presented here is just a first modeling attempt on this scale of detail, and the results open a number of further unanswered questions. Ongoing interdisciplinary research on this topic is needed to answer those. 
If this thesis helps stimulate follow-up investigations at the interface between scientific computing and neuroscience, our main objective has been achieved.

\appendix
\newpage

\printbibliography[heading=bibintoc]
\newpage

\chapter{Results of the Volume-Corrected Axon Bundle Simulations}\label{chap:appendix}
\begin{figure}
\subfloat[Membrane potential]{%
\centering%
\includegraphics[width=0.5\textwidth]{memb_pot_axonbundle_volumeCorrected_1000nm-crop}%
\label{fig:multiple_axons.memb_pot_1000nm_volumeCorrected}}%
\subfloat[Extracellular charge density]{%
\centering%
\includegraphics[width=0.5\textwidth]{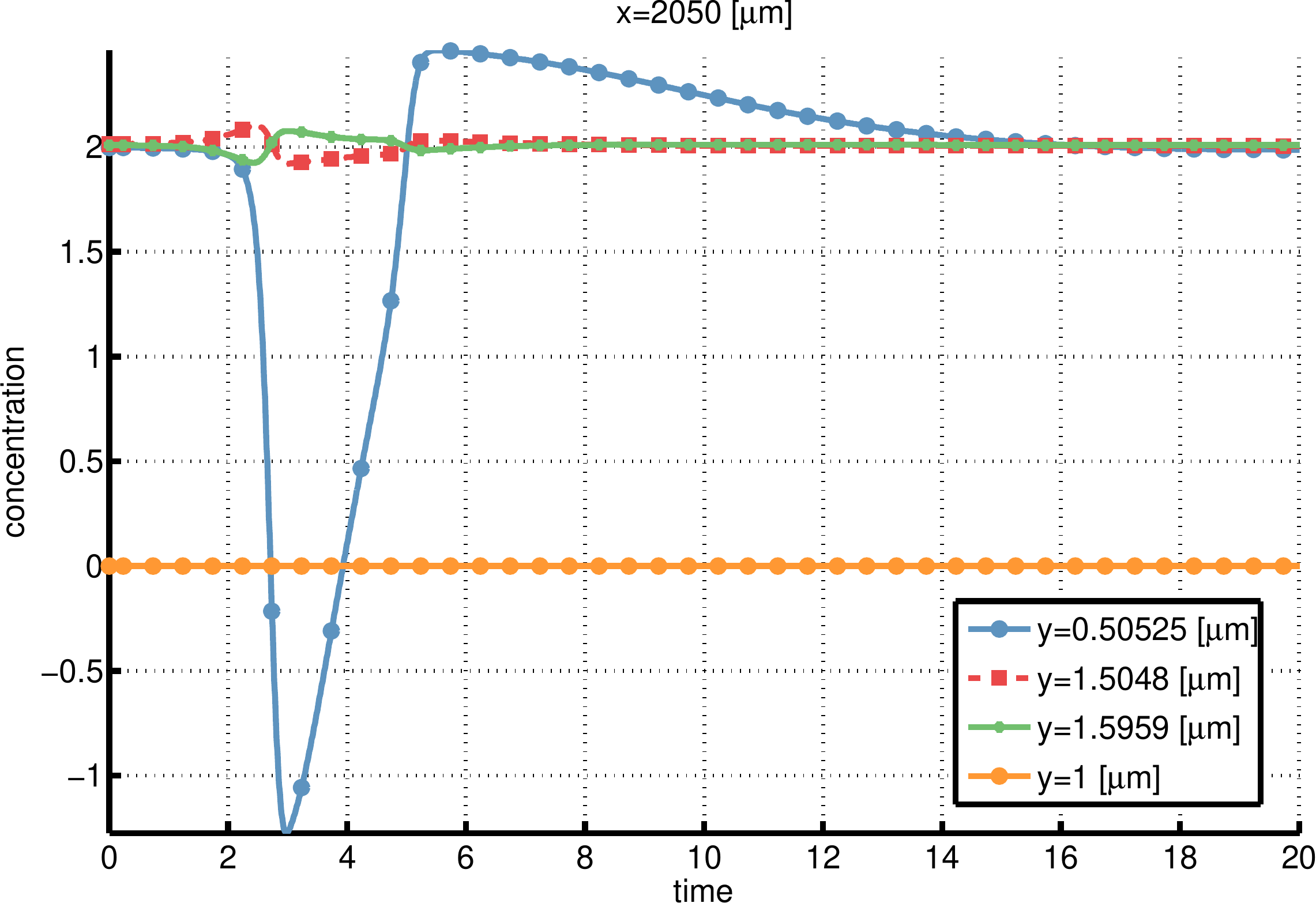}%
\label{fig:multiple_axons.cd_extra_1000nm_volumeCorrected}}\\%
\subfloat[Intracellular potential]{%
\centering%
\includegraphics[width=0.5\textwidth]{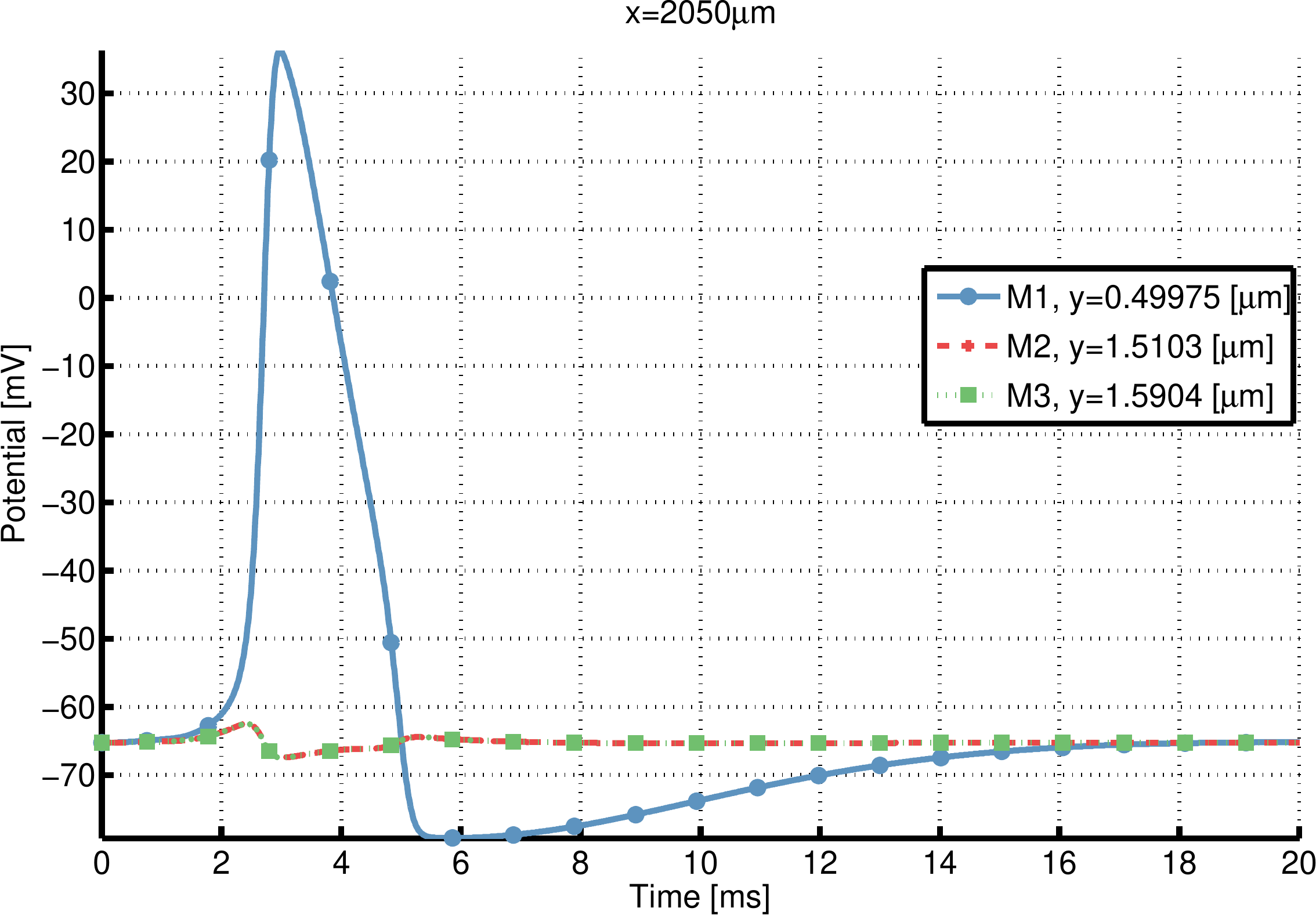}%
\label{fig:multiple_axons.pot_intra_1000nm_volumeCorrected}}%
\subfloat[Extracellular potential]{%
\centering%
\includegraphics[width=0.5\textwidth]{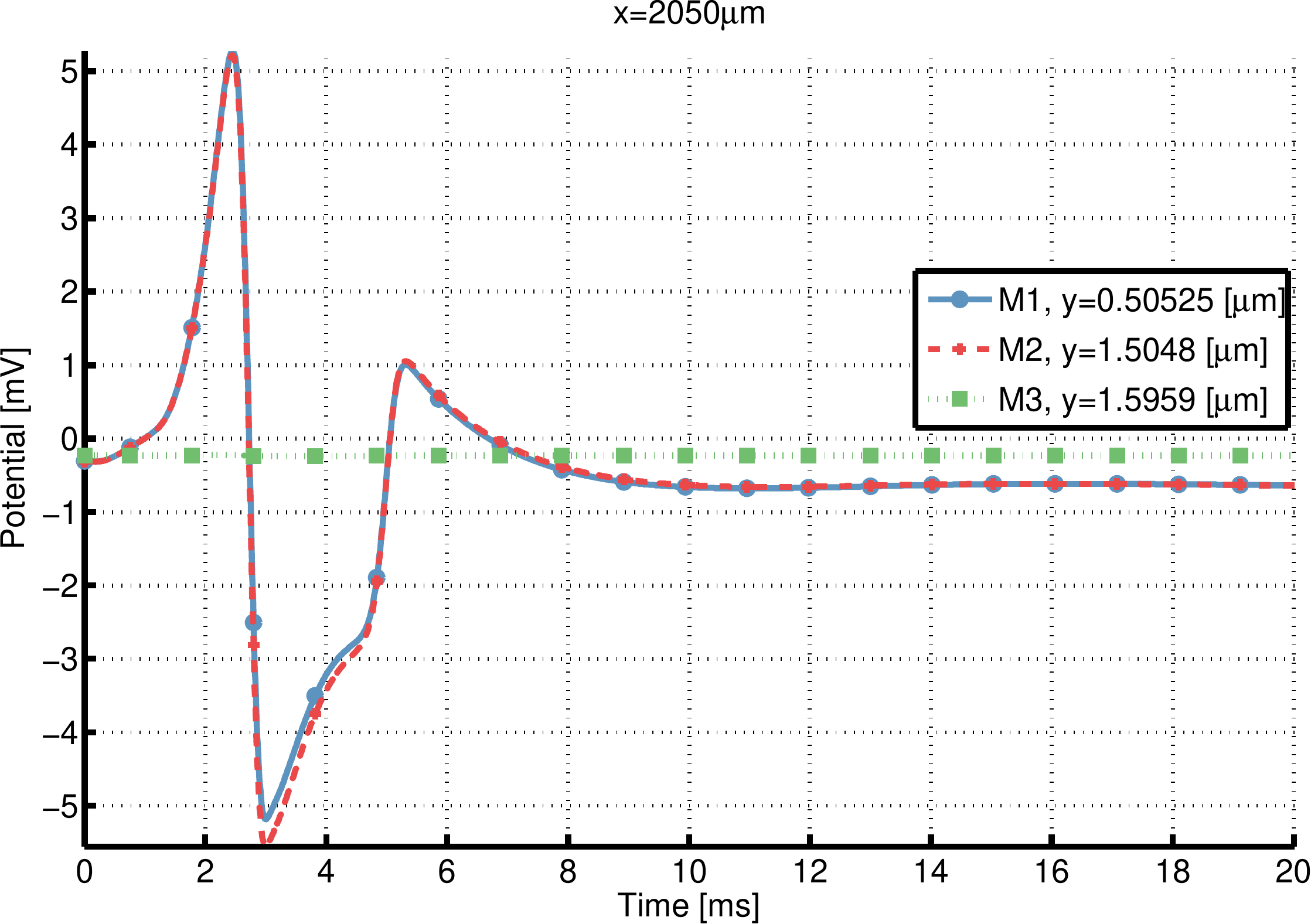}%
\label{fig:multiple_axons.pot_extra_1000nm_volumeCorrected}}%
\mycaption[Potential and charge density time courses for $d=\SI{1}{\micro\metre}$ (volume-corrected)]{}%
\label{fig:pot_intra_extra_1000nm_volumeCorrected}%
\end{figure}

\begin{figure}
\subfloat[Membrane potential]{%
\centering%
\includegraphics[width=0.5\textwidth]{memb_pot_axonbundle_volumeCorrected_500nm-crop}%
\label{fig:multiple_axons.memb_pot_500nm_volumeCorrected}}%
\subfloat[Extracellular charge density]{%
\centering%
\includegraphics[width=0.5\textwidth]{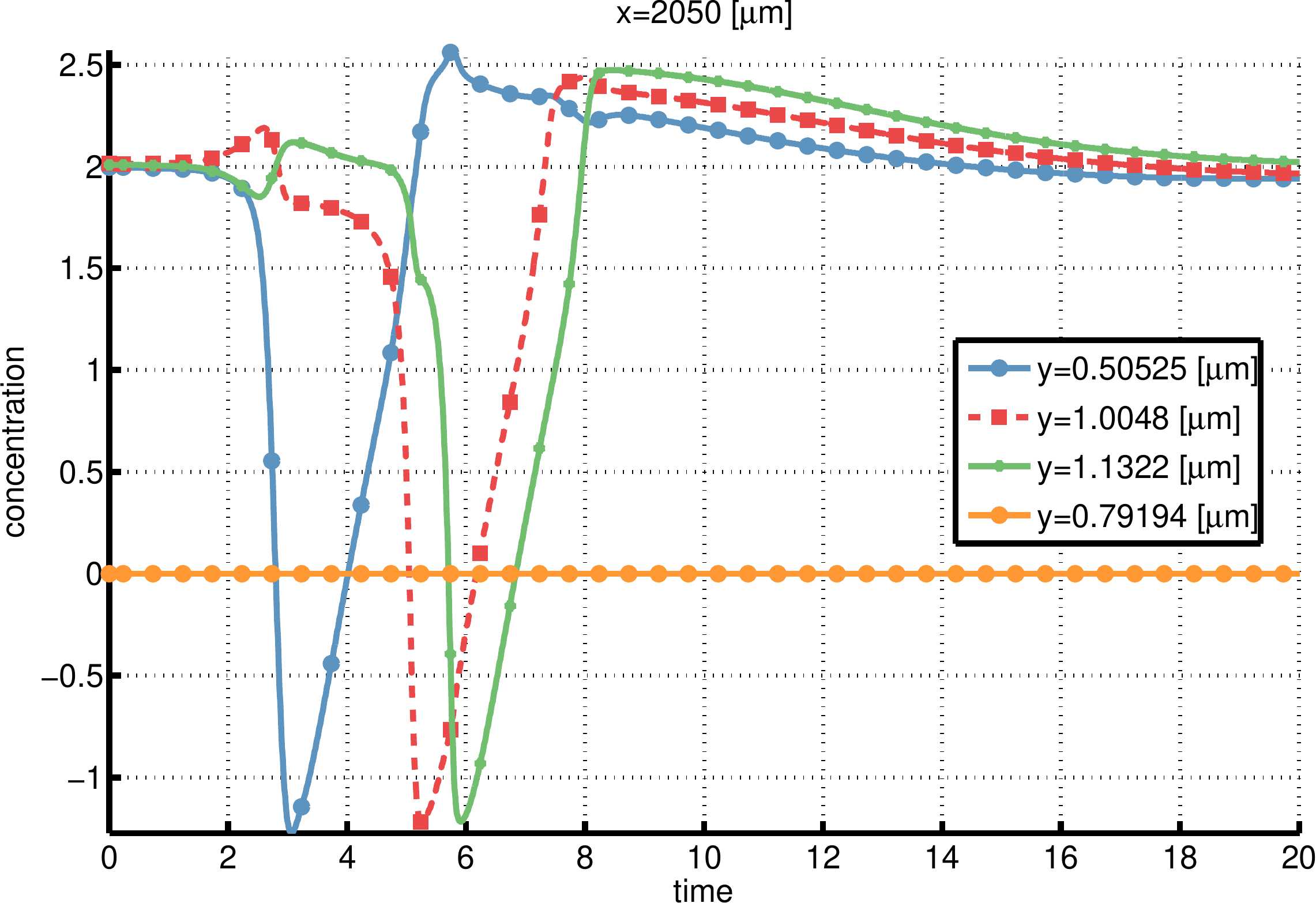}%
\label{fig:multiple_axons.cd_extra_500nm_volumeCorrected}}\\%
\subfloat[Intracellular potential]{%
\centering%
\includegraphics[width=0.5\textwidth]{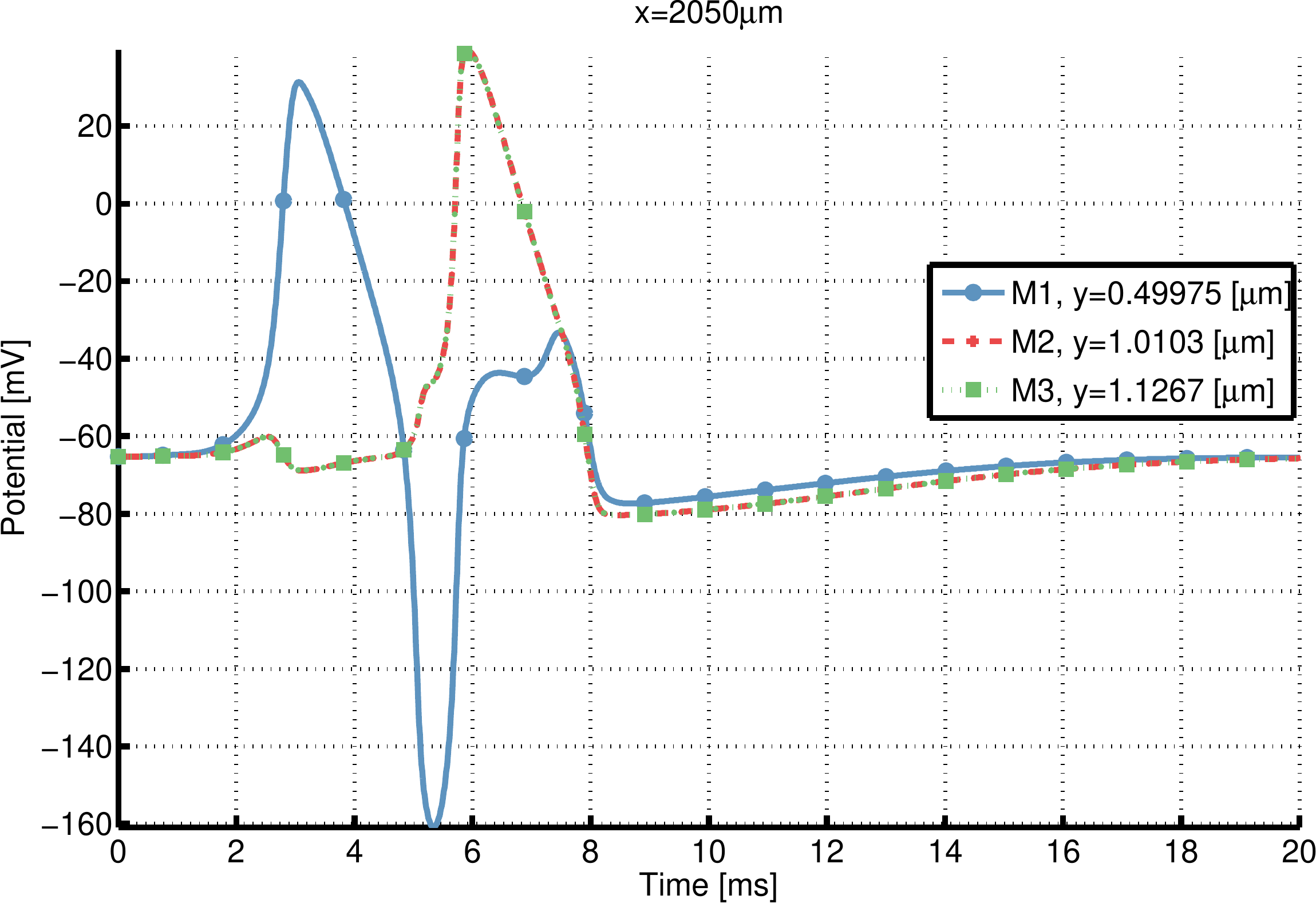}%
\label{fig:multiple_axons.pot_intra_500nm_volumeCorrected}}%
\subfloat[Extracellular potential]{%
\centering%
\includegraphics[width=0.5\textwidth]{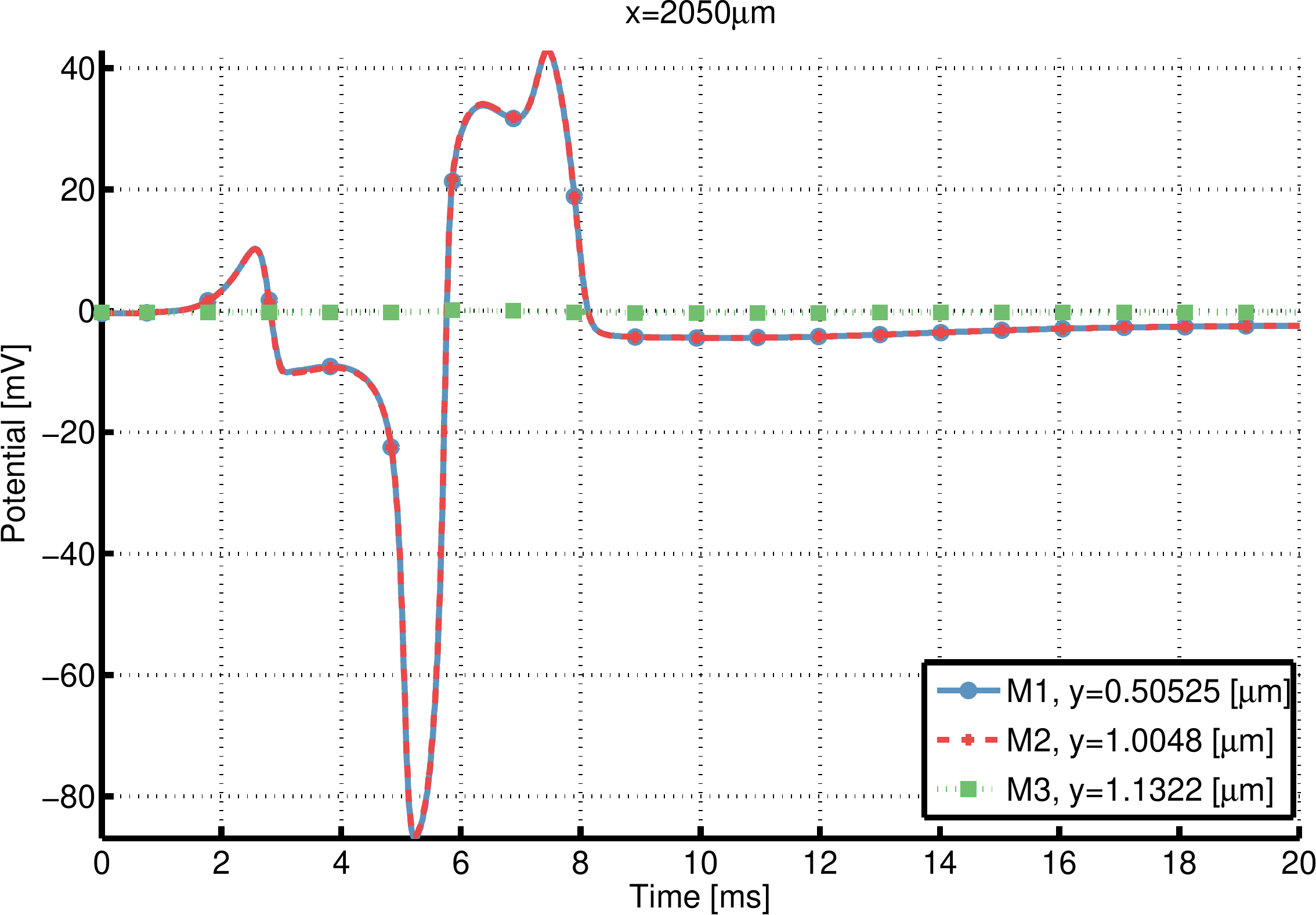}%
\label{fig:multiple_axons.pot_extra_500nm_volumeCorrected}}%
\mycaption[Potential and charge density time courses for $d=\SI{500}{\nano\metre}$ (volume-corrected)]{}%
\label{fig:pot_intra_extra_500nm_volumeCorrected}%
\end{figure}

\begin{figure}
\subfloat[Membrane potential]{%
\centering%
\includegraphics[width=0.5\textwidth]{memb_pot_axonbundle_volumeCorrected_100nm-crop}%
\label{fig:multiple_axons.memb_pot_100nm_volumeCorrected}}%
\subfloat[Extracellular charge density]{%
\centering%
\includegraphics[width=0.5\textwidth]{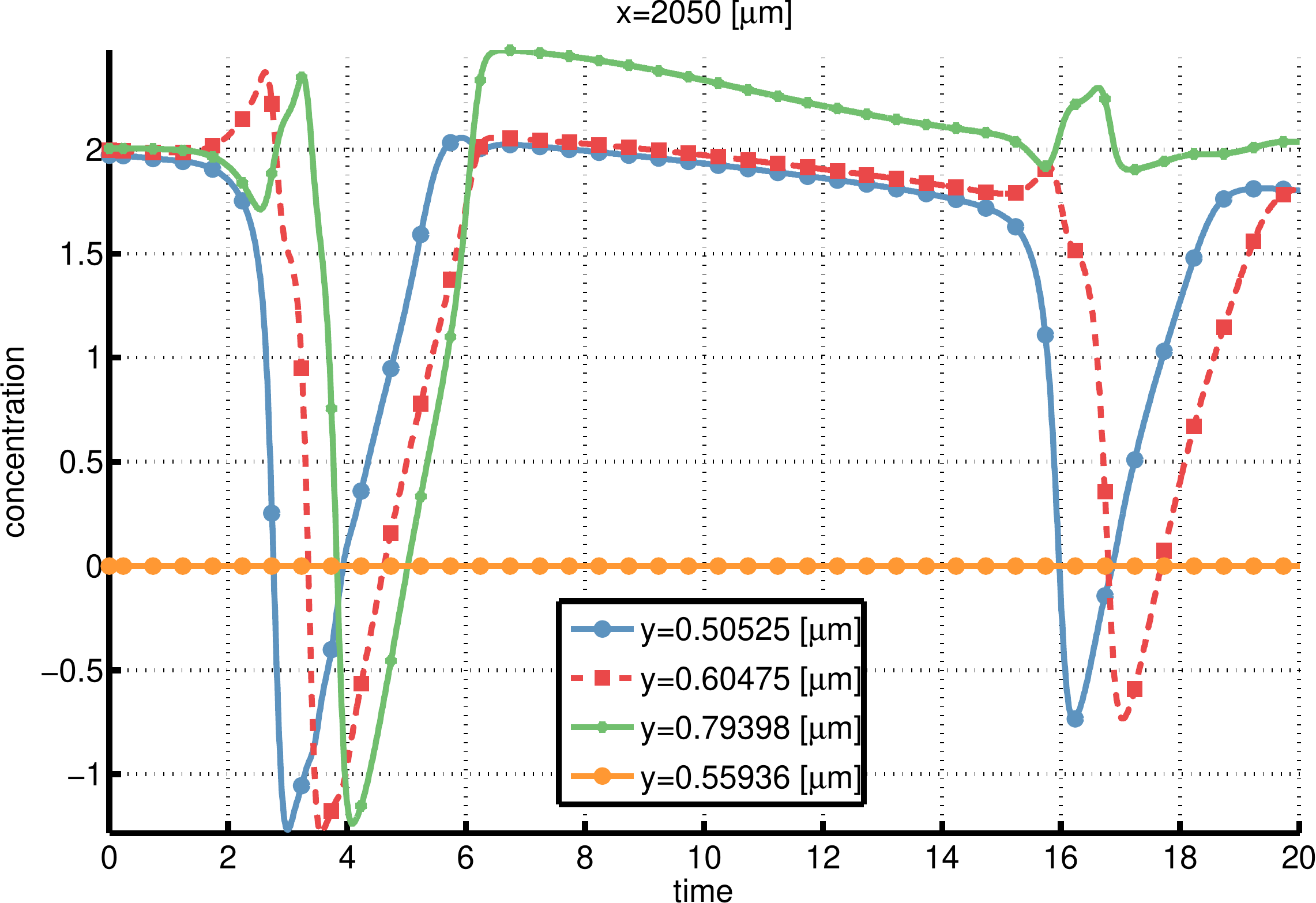}%
\label{fig:multiple_axons.cd_extra_100nm_volumeCorrected}}\\%
\subfloat[Intracellular potential]{%
\centering%
\includegraphics[width=0.5\textwidth]{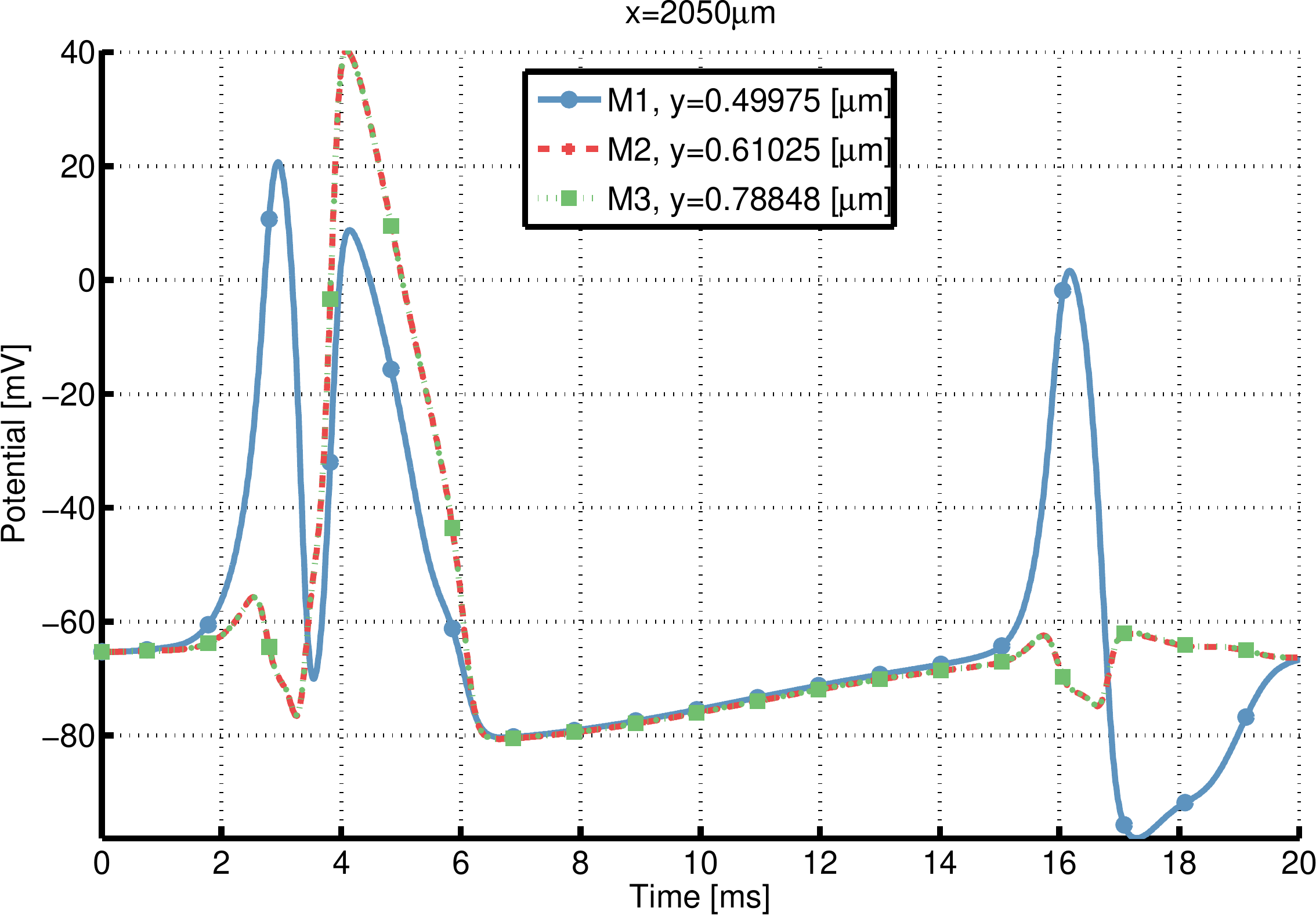}%
\label{fig:multiple_axons.pot_intra_100nm_volumeCorrected}}%
\subfloat[Extracellular potential]{%
\centering%
\includegraphics[width=0.5\textwidth]{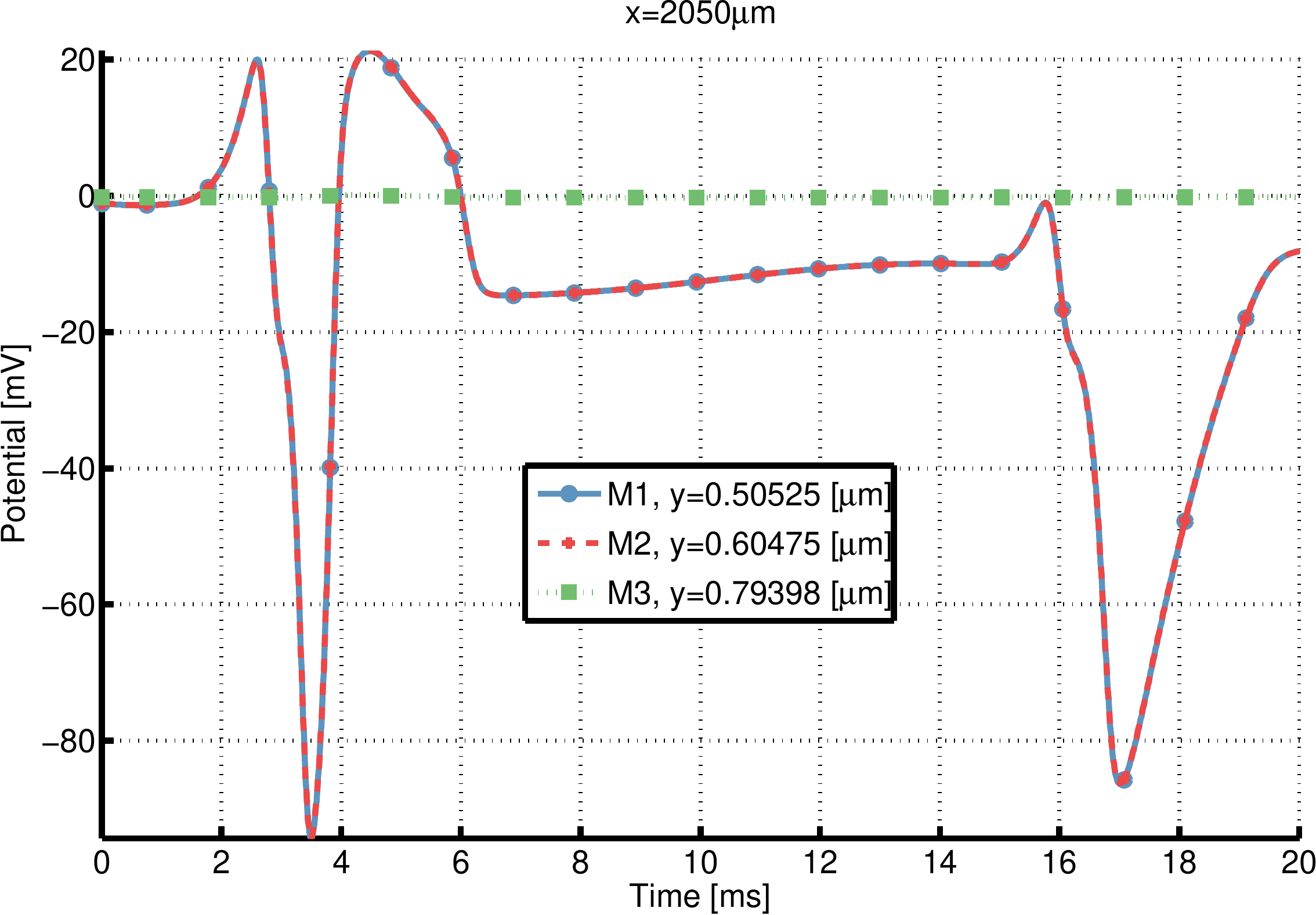}%
\label{fig:multiple_axons.pot_extra_100nm_volumeCorrected}}%
\mycaption[Potential and charge density time courses for $d=\SI{100}{\nano\metre}$ (volume-corrected)]{}%
\label{fig:pot_intra_extra_100nm_volumeCorrected}%
\end{figure}

\begin{figure}
\subfloat[Membrane potential]{%
\centering%
\includegraphics[width=0.5\textwidth]{memb_pot_axonbundle_volumeCorrected_50nm-crop}%
\label{fig:multiple_axons.memb_pot_50nm_volumeCorrected}}%
\subfloat[Extracellular charge density]{%
\centering%
\includegraphics[width=0.5\textwidth]{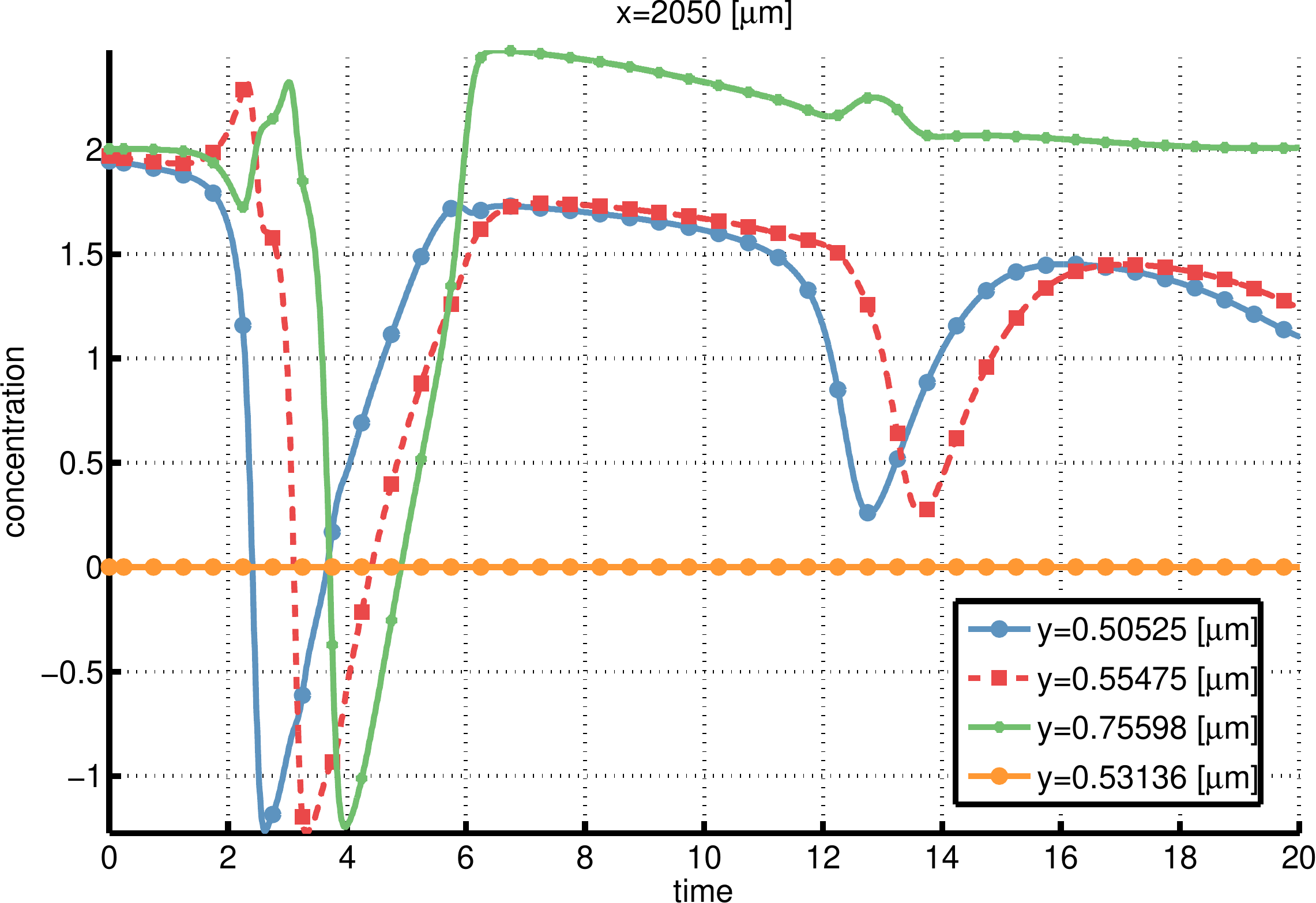}%
\label{fig:multiple_axons.cd_extra_50nm_volumeCorrected}}\\%
\subfloat[Intracellular potential]{%
\centering%
\includegraphics[width=0.5\textwidth]{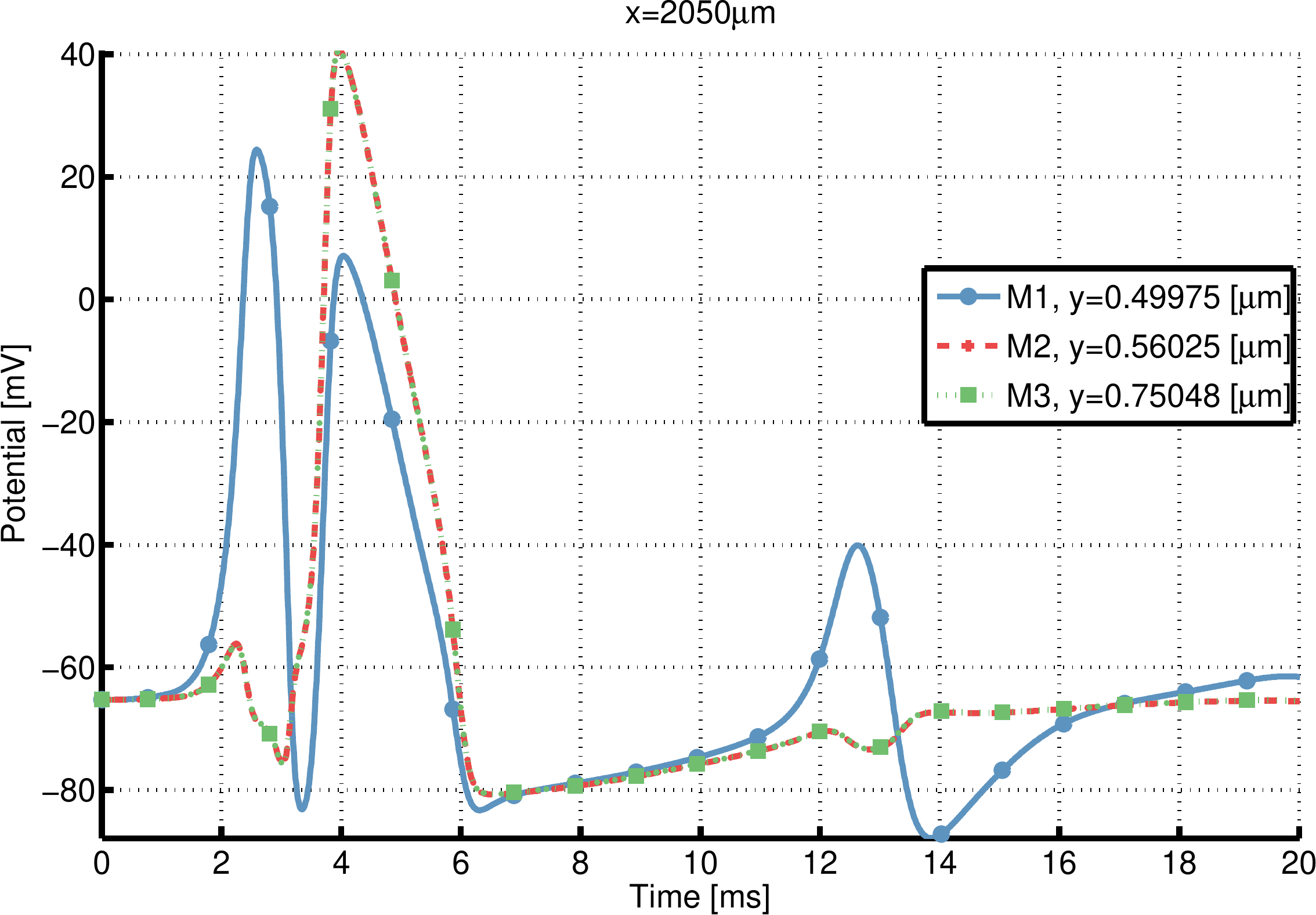}%
\label{fig:multiple_axons.pot_intra_50nm_volumeCorrected}}%
\subfloat[Extracellular potential]{%
\centering%
\includegraphics[width=0.5\textwidth]{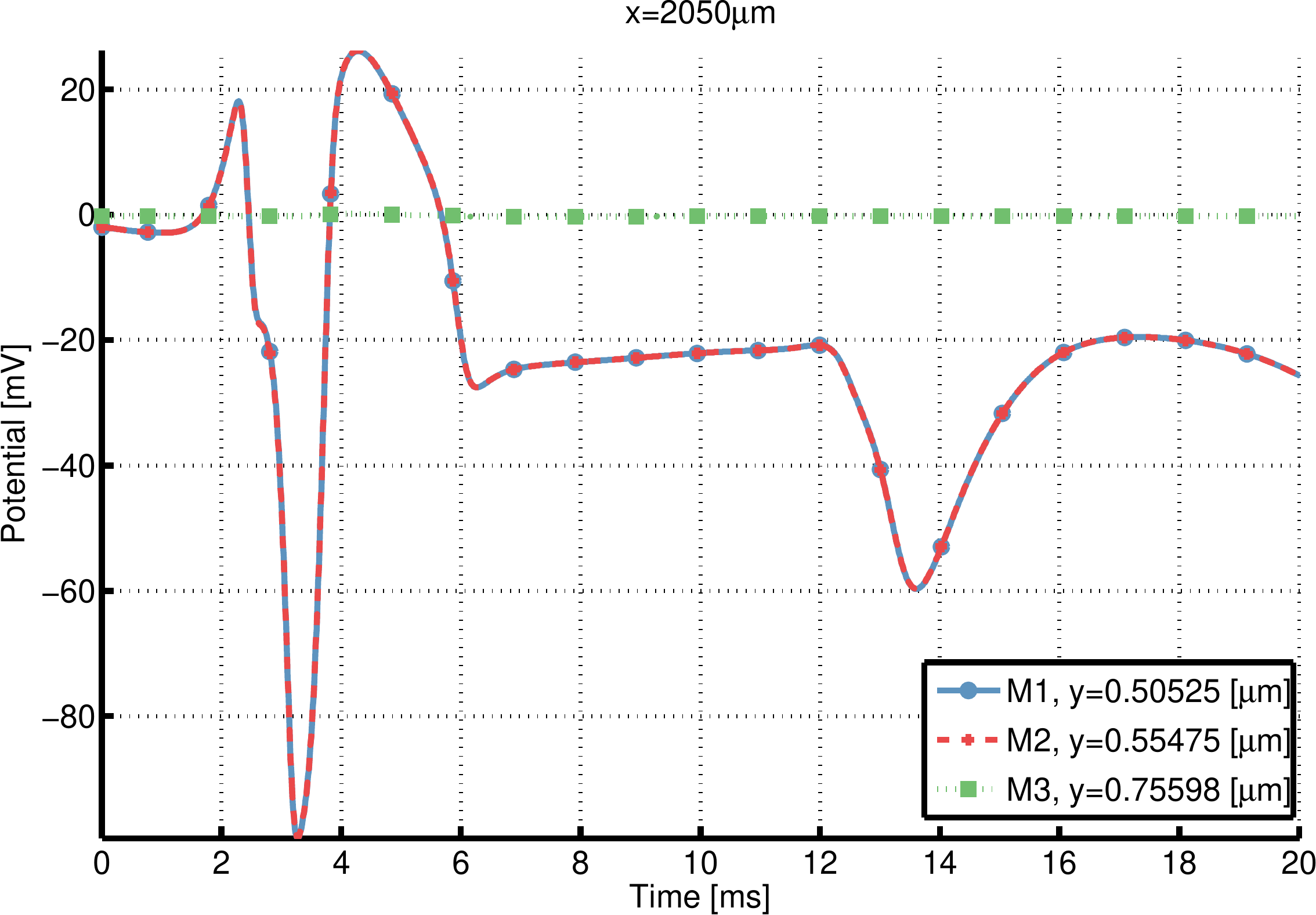}%
\label{fig:multiple_axons.pot_extra_50nm_volumeCorrected}}%
\mycaption[Potential and charge density time courses for $d=\SI{50}{\nano\metre}$ (volume-corrected)]{}%
\label{fig:pot_intra_extra_50nm_volumeCorrected}%
\end{figure}

The detailed results of the volume-corrected axon bundle simulations are shown in \cref{fig:pot_intra_extra_1000nm_volumeCorrected,fig:pot_intra_extra_500nm_volumeCorrected,fig:pot_intra_extra_100nm_volumeCorrected,fig:pot_intra_extra_50nm_volumeCorrected}. \Cref{tab:multiple_fibers.times_vc} shows the statistics for different inter-fiber distances $d$.
Note that only the setup $d = \SI{50}{\nano\metre}$ was compiled with full optimizations, the other cases were accidentally computed without compiler optimizations, resulting in a factor of 6-7 in the total computation time.

\begin{table}
\mycaption[Simulation timings for the axon bundle setup using different inter-fiber distances $d$ and a volume-corrected bundle diameter $d_b$]{
For each distance $d$, the number of unknowns, total computation time, the needed number of time steps and the average
solution time per time step (full Newton iteration) is shown.}\label{tab:multiple_fibers.times_vc}
\centering
\begin{tabu} to \textwidth {@{} *5{X[l]} @{}}
\toprule
$d$ [nm] & \# \glspl{DOF} & Total comp. time [s]& \# time steps & avg. time / time step\\
\midrule
1000 & 116352 & 157710 & 2027 & 77.8\\
500 & 114332 & 173738 & 2027 & 85.71\\
100 & 114332 & 183747 & 2027 & 90.65\\
50 & 114332 & 26027.7 & 2026 & 12.85\\
\bottomrule
\end{tabu}
\end{table}

\let\cleardoublepage\relax
\printglossary
\printindex

\pagestyle{empty}
\chapter*{Danksagung}
\thispagestyle{empty}
Geld ist (Arbeits-)Zeit, und daher danke ich zuallererst dem Bundesministerium für Bildung und Forschung für die Schaffung des BCCN Heidelberg/Mannheim und die damit verbundene Finanzierung meiner Arbeit (BMBF No. 01GQ1003A).

Weiterhin möchte ich Prof.~Peter Bastian dafür danken, dass er mir die Promotion zu diesem hochinteressanten Thema ermöglicht hat. 
Die parallele Einarbeitung in mehrere neue Themengebiete war herausfordernd und mit vielen Hindernissen verbunden. 
Für die Hilfestellung besonders im Bereich der numerischen Mathematik möchte ich mich ebenso bedanken wie für die angenehme Arbeitsatmosphäre, zu denen auch alle anderen Mitglieder der AG Paralleles Rechnen beigetragen haben. 
Allen Kollegen gilt daher mein herzlicher Dank für die vergangenen 3 1/2 Jahre.

Dem IWR und der Fakultät für Mathematik und Informatik der Universität Heidelberg möchte für die bereitgestellte Infrastruktur und die Möglichkeit einer interdisziplinären Promotion an der Schnittstelle zwischen Informatik, Mathematik und Biophysik danken.

Besonderer Dank geht an Dan Popović, der mich zu Beginn in die Untiefen von \Dune und der Heidelberger Kneipenkultur eingeführt hat.
Viele Gespräche mit oder ohne Fußballbezug haben geholfen, die schwierige Anfangszeit zu meistern.
Großer Dank gilt auch Johannes Schönke bzw.~Prof.~Henry, der mich thematisch in die richtige Spur gebracht hat, indem er mich zuerst auf den Beifahrersitz des Elektrodiffusionswagens gehievt hat und dann selbst bei laufender Fahrt abgesprungen ist. 
Ich denke, ich habe die Karre einigermaßen heil nach Hause gebracht, und neuen TÜV hat sie auch noch. 
Ich habe in unserer recht kurzen gemeinsamen Zeit viel gelernt über physikalische Intuition und einen anderen Blick auf die Welt der Wissenschaft.

Ich danke auch Prof.~Andreas Draguhn und den Mitarbeitern seines Labors für die Hilfsbereitschaft bei der Vermittlung von umfangreichem Hintergrundwissen im Bereich Neurophysiologie im Rahmen der Kooperation im Bernstein Zentrum. Ganz besonderes danke ich Martin Both für viele hilfreiche Gespräche, sowohl inner- als auch außerhalb des Whirlpools.

Im weiteren Sinne gilt mein Dank auch dem Fußballteam des IWR sowie der Capella und Camerata Carolina für den bitter nötigen Ausgleich.
Insbesondere meine ich damit Gerald und Martina, mit denen ich Freud und Leid der akademischen Welt in vielen Bergbahnsitzungen teilen konnte. Schöne polierte!

Für das fachkundige Korrekturlesen danke ich diesen beiden ebenso wie Tatjana. Ich bin zwar etwas enttäuscht, dass meine bahnbrechende Entdeckung des positiven Chlor-Ions sowie viele meiner geliebten Bandwurmsätze vor euch keinen Bestand hatten, aber die zahlreichen fachlichen und inhaltlichen Hinweise haben doch letztendlich sehr zum Gelingen beigetragen.

Für die regelmäßige Neujustierung meiner neuronalen Konfiguration bin ich Christoph, Philipp und Simon zu nachhaltigem Dank verpflichtet. Unsere diversen Trips nach Ferropolis, Natterthal oder zu Herrn Weber auf Balkonien haben mir geholfen, manch schwere Phase zu überwinden. Wer solche Freunde hat, braucht keine Feinde.

Ich danke meiner Familie, ganz besonders meiner Mutter\footnote{i.e.~diejenige welche, die mir das Leben schenkte}, für die bedingungslose Unterstützung. 
Die Urlaube im Heizcamp am Ende eines jeden Jahres waren Zäsuren, die es mir ermöglichten einmal abzuschalten und das vergangene Jahr mit einem gewissen Abstand zu betrachten, um dann das kommende mit neuer Kraft anzugehen.

Schließlich danke ich Pia, für einfach alles. Du hast alle Aufs und Abs, scharfen Kurven und die immer länger werdende Zielgerade mitgemacht und mir Halt gegeben. Ich kann mir gar nicht vorstellen, wie es die vergangenen Jahre ohne dich gewesen wäre, und das möchte ich auch nicht mehr.

\end{document}